\documentclass[a4paper,USenglish,cleveref, autoref, thm-restate]{lipics-v2021}

\pdfoutput=1 
\hideLIPIcs  

\newcommand{\citet}{\cite}
\newcommand{\refapp}{\Cref}
\newcommand*{\fullref}[1]{\hyperref[{#1}]{\Cref*{#1} (\nameref*{#1})}}

\title{Online Stochastic Matching:\texorpdfstring{\\}{} A Polytope Perspective}

\titlerunning{Online Stochastic Matching: A Polytope Perspective}

\author{C\'eline Comte\footnote{Corresponding author.}}{Eindhoven University of Technology, Eindhoven, The Netherlands \and CNRS and LAAS, Toulouse, France \and \url{https://homepages.laas.fr/ccomte/}}{celine.comte@cnrs.fr}{https://orcid.org/0009-0005-9413-7124}{}
\author{Fabien Mathieu}{Swapcard, Paris, France \and Sorbonne Université, CNRS, LIP6, Paris, France \and \url{https://balouf.github.io/}}{fabien@swapcard.com}{https://orcid.org/0000-0003-1362-0359}{}
\author{Sushil {Mahavir Varma}}{Industrial and Systems Engineering, Georgia Institute of Technology, Atlanta, US \and \url{https://sites.google.com/view/sushil-varma/home}}{sushil@gatech.edu}{https://orcid.org/0000-0001-7855-3447}{}
\author{Ana Bu\v{s}i\'c}{Inria, Paris, France \and DI ENS, PSL University, Paris, France \and \url{https://www.di.ens.fr/~busic/}}{ana.busic@inria.fr}{https://orcid.org/0000-0002-4133-3739}{}

\authorrunning{Comte, Mathieu, Varma, and Bu\v{s}i\'c}

\Copyright{Comte, Mathieu, Varma, and Bu\v{s}i\'c}



\supplement{Stochastic Matching Package}
\supplementdetails[linktext={Documentation}, cite=SM22, subcategory={Python, PyPi}]{Software}{https://balouf.github.io/stochastic_matching/companion/index.html} 

\nolinenumbers 

\allowdisplaybreaks  

\usepackage{amsmath}
\usepackage{verbatim}
\newcommand\N{\mathbb{N}}
\newcommand\Np{\N_{>0}}
\newcommand\R{\mathbb{R}}
\newcommand\Rp{\R_{>0}}
\newcommand\Rnn{\R_{\geqslant 0}}

\newcommand\A{\mathcal{A}}
\newcommand\B{\mathcal{B}}
\newcommand\C{\mathcal{C}}
\newcommand\I{\mathcal{I}}
\newcommand\Q{\mathcal{Q}}
\newcommand\cS{\mathcal{S}}
\newcommand\T{\mathcal{T}}

\DeclareMathOperator*{\argmax}{arg\,max}

\newcommand\PP{\mathbb{P}}

\renewcommand{\ge}{\geqslant}
\renewcommand{\geq}{\geqslant}
\renewcommand{\le}{\leqslant}
\renewcommand{\leq}{\leqslant}

\newcommand\Pol{\mathcal{P}}
\newcommand\Gre{\mathcal{G}}

\newcommand\La{\Pi}
\newcommand\Lap{\Pi_{>0}}
\newcommand\Lann{\Pi_{\geqslant 0}}
\newcommand\Lapol{\Pi_{\Pol}}
\newcommand\Lagre{\Pi_{\Gre}}

\newcommand\E{E}
\newcommand\V{V}
\newcommand\ind{\mathbb{I}}
\newcommand\KP{\mathit{KP}}
\newcommand\K{\mathit{K}}

\usepackage[acronym]{glossaries}
\newacronym{crpg}{CRPG}{Complete Resource Pooling Gap}
\newacronym{spp}{SPP}{Static Planning Problem}
\newacronym{gpc}{GPC}{General Position Condition}
\newacronym{gpg}{GPG}{General Position Gap}
\newacronym{ml}{ML}{Match-the-Longest}
\newacronym{vqml}{VQML}{Virtual-Queue Match-the-Longest}
\newacronym{fcfm}{FCFM}{First-Come-First-Match}
\newacronym{egpd}{EGPD}{Extended Greedy Primal-Dual}
\newacronym{hrf}{HRF}{Highest-Reward-First}
\newacronym{crpd}{CRPD}{Constant-Regret Primal-Dual}
\newcommand{\glsf}[1]{\glsreset{#1}\gls{#1}}
\glsdisablehyper

\usepackage{dsfont}
\newcommand{\one}{\mathds{1}}

\usepackage{amssymb}
\usepackage{pifont}
\newcommand{\cmark}{\ding{51}}
\newcommand{\xmark}{\ding{55}}

\usepackage{tikz}
\usetikzlibrary{calc,positioning}

\usepackage{ifthen}

\tikzset{
	class/.style = {draw, circle, fill=blue!20},
	hedge/.style = {draw, rounded corners},
	server/.style = {draw, circle, fill=blue!20},
	fcfs/.style={
		draw,
		rectangle split,
		rectangle split parts=#1,
		rectangle split horizontal,
		rectangle split empty part width=-.17cm,
		rectangle split empty part height=.65cm,
		inner ysep=.27cm,
	},
	spanner/.style={line width=.7mm},
	extra/.style={dotted, line width=.7mm},
	k1/.style={red, line width=.7mm},
	k2/.style={white!40!blue, line width=.7mm},
	dot/.style = {circle, fill=orange, minimum size=#1,
		inner sep=0pt, outer sep=0pt},
	dot/.default = 4pt, 
}

\newcommand{\placenodes}[1]{
	\node[class] (1) {$1$};
	\foreach \i/\s/\a in {#1}		
	\path (\s) ++(\a:\d) node[class] (\i) {$\i$};
}
\newcommand{\halfwidth}[2]{
	\node at ($(#1)-(#2,0)$) {};
	\node at ($(#1)+(#2,0)$) {};
}

\def\bifan{\placenodes{2/1/30,3/1/-30,4/3/30}}
\def\yshape{\placenodes{2/1/-90,3/1/-30,4/3/0}}
\def\threefan{\placenodes{2/1/0, 3/2/0, 5/1/-60, 4/5/0}}
\def\codomino{\placenodes{2/1/30, 6/1/-30, 3/2/0, 5/6/0, 4/5/30}}
\def\whirl{\placenodes{2/1/30, 3/2/30, 4/3/-30, 5/4/-30, 8/1/-30, 7/8/-30, 6/7/30, 9/1/90, 10/5/-90}}

\newtheoremstyle{informs}%
{}{}%
{\itshape}{}%
{\normalfont\sffamily\mdseries\scshape}{.}%
{.3em}{\thmname{#1}\thmnumber{ #2}\thmnote{ \normalfont\bfseries(#3)}}
\theoremstyle{informs}
\newtheorem{thm}{Theorem}
\crefname{thm}{Theorem}{Theorems}
\Crefname{thm}{Theorem}{Theorems}
\newtheorem{prop}{Proposition}
\crefname{prop}{Proposition}{Propositions}
\Crefname{prop}{Proposition}{Propositions}
\newtheorem{dfn}{Definition}
\crefname{dfn}{Definition}{Definitions}
\Crefname{dfn}{Definition}{Definitions}
\newtheorem{lem}{Lemma}
\crefname{lem}{Lemma}{Lemmas}
\Crefname{lem}{Lemma}{Lemmas}
\newtheorem{coro}{Corollary}
\crefname{coro}{Corollary}{Corollaries}
\Crefname{coro}{Corollary}{Corollaries}
\newtheoremstyle{conjecture}%
{}{}%
{\itshape}{}%
{\normalfont\sffamily\mdseries\itshape}{.}%
{.3em}{\thmname{#1}\thmnumber{ #2}\thmnote{ \normalfont\bfseries(#3)}%
\ \normalfont\sffamily(open)}
\theoremstyle{conjecture}
\newtheorem{conj}{Conjecture}
\Crefname{conj}{Conjecture}{Conjectures}
\crefname{conj}{Conjecture}{Conjectures}
\theoremstyle{informs}
\newtheorem{rem}{Remark}
\crefname{rem}{Remark}{Remarks}
\Crefname{rem}{Remark}{Remarks}
\newtheorem{exa}{Example}
\Crefname{exa}{Example}{Examples}
\crefname{exa}{Example}{Examples}
\newtheorem{propositionlowerbound}{Extended Proposition}
\crefname{propositionlowerbound}{Extended Proposition}{Extended Propositions}
\Crefname{propositionlowerbound}{Extended Proposition}{Extended Propositions}
\crefname{algocf}{algorithm}{algorithms}
\Crefname{algocf}{Algorithm}{Algorithms}
\crefname{appsec}{Appendix}{Appendices}
\Crefname{appsec}{Appendix}{Appendices}
\crefname{enumi}{Statement}{Statements}
\Crefname{enumi}{Statement}{Statements}

\newcommand\numberthis{\addtocounter{equation}{1}\tag{\theequation}}
\newcommand\inner[2]{\left \langle #1, #2 \right \rangle}
\newcommand\Ex[1]{\mathbb{E}\left[#1\right]}

\usepackage{pgfplots,pgfplotstable}
\pgfplotsset{compat=1.18,width=7cm,height=5.5cm}

\newcommand{\mathdefault}[1]{#1}

\newcommand\mles{\Phi_{\E^\star}}

\newcommand\mle{\Phi_\epsilon}
\newcommand\mlk{\Phi_k}
\newcommand\egpd{\Phi_\beta}
\newcommand\egpdp{\Phi'_\beta}
\newcommand\crpd{\Phi_\alpha}
\definecolor{fcolor}{RGB}{0,0,0}
\definecolor{ecolor}{RGB}{44,160,44}
\definecolor{kcolor}{RGB}{31,119,180}
\definecolor{egcolor}{RGB}{255,127,14}
\definecolor{gpcolor}{RGB}{214,39,40}
\definecolor{crcolor}{RGB}{148,103,189}

\pgfplotsset{
	filter/.style={
		semithick,
		dash pattern=on 5.55pt off 2.4pt,
		fcolor 
	},
	egpd/.style={
		semithick,
		mark=*, mark size=2,
		mark options={solid},
		egcolor
	},
	egpdp/.style={
		semithick,
		mark=+, mark size=3,
		mark options={solid},
		gpcolor
	},
	crpd/.style={
		semithick,
		mark=triangle*, mark size=2.5,
		mark options={solid},
		crcolor
	},
	efilter/.style={
		semithick,
		mark=square*, mark size=2,
		mark options={solid},
		ecolor 
	},
	kfilter/.style={
		semithick,
		mark=x, mark size=3,
		mark options={solid},
		kcolor 
	}
}

\usepackage[linesnumbered, noend]{algorithm2e}

\newenvironment{reusefigure}[2][htbp]
{\addtocounter{figure}{-1}
	
	\renewcommand{\thefigure}{\ref{#2} (repeated)}
	\renewcommand{\addcontentsline}[3]{}
	\begin{figure}[#1]}
	{\end{figure}}

\usepackage{float}
\NewDocumentCommand{\citep}{o o m}{\IfValueTF{#1}{\cite[#1]{#3}}{\cite{#3}}}
\newcommand{\Citet}{\cite}
\newcommand{\citealt}{\cite}
\newcommand{\citealp}{\cite}
\newcommand*{\refappnc}[1]{\Cref{#1}}
\providecommand{\Z}{\mathbb{Z}}
\providecommand{\Halmos}{\ensuremath{\square}}

\begin{document}

\maketitle

\begin{abstract}
Stochastic dynamic matching problems have recently gained attention in the stochastic-modeling community due to their diverse applications, such as supply-chain management and kidney exchange programs. In this paper, we study a matching problem where items of different classes arrive according to independent Poisson processes. Unmatched items are stored in a queue, and compatibility between items is represented by a simple graph, where items can be matched if their classes are connected.
We analyze matching policies in terms of stability, delay, and long-term matching rate optimization. Our approach relies on the conservation equation, which ensures a balance between arrivals and departures in any stable system. Our main contributions are as follows.
We establish a link between the existence of stable policies, the dimensionality of the solution set of the conservation equation, and the compatibility graph's structure.
We describe the convex polytope formed by non-negative solutions to the conservation equation,
and we design policies that can achieve or closely approximate the vertices of this polytope.
When a vertex can only be approximated, we quantify the resulting trade-off between regret and delay: our policies achieve arbitrarily small regret at the cost of increasing delay, and we prove that this trade-off is unavoidable.
Lastly, we discuss potential extensions of our results beyond the main assumptions of this paper.
\end{abstract}

 % shared/body
 % sections/01-introduction
\section{Introduction.} \label{sec:intro}

Stochastic dynamic matching problems,
in which items arrive at random instants
to be matched with other items,
have recently attracted much attention
in the stochastic-modeling community.
These challenging control problems
are highly relevant in many applications, including
supply-chain management,
pairwise kidney exchange programs,
and online marketplaces.
In pairwise kidney exchange programs for example,
each item represents a donor-receiver pair,
and two pairs can be matched
if the donor of each pair
is compatible with the receiver of the other pair.
In online marketplaces,
items are typically divided into two categories,
called demand and supply,
and the goal is
to maximize a certain long-term performance criterion
by appropriately matching
demand items with supply items.

In this paper, we consider the following
dynamic matching problem\footnote{Detailed definitions of the concepts discussed here will be given in \Cref{subsec:model}.}.
Items of different classes
arrive according to independent Poisson processes.
Compatibility constraints between items
are described by a simple graph on their classes,
such that two items can be matched
if their classes are neighbors in the graph.
Unmatched items are stored
in the queue of their class,
and a matching policy decides
which matches are performed and when.
All in all, a stochastic matching model
is described by a triplet $(G, \lambda, \Phi)$,
where $G = (\V, \E)$ is the compatibility graph,
$\lambda = (\lambda_1, \lambda_2, \ldots, \lambda_n)$
is the vector of per-class arrival rates,
and $\Phi$ is the matching policy.
Assuming stability, i.e. existence of a steady-state behavior, the matching-rate vector, denoted by~$\mu$, represents the average rates at which each type of match is performed.
In \Cref{fig:illustration} for instance,
there are four item classes
numbered from~1 to~4;
classes~2 and~3 are compatible with all classes,
while classes~1 and~4 are compatible only with classes~2 and~3.

\def\queueheight{1}
\def\queuewidth{.4}

\newcommand{\queue}[2]{
	\begin{tikzpicture}[rotate=#1, scale=.7]
		\draw (0, 0) -- (\queueheight, 0) -- (\queueheight, \queuewidth) -- (0, \queuewidth);
		
		\ifthenelse{#2>0}{
			\foreach \x in {1,..., #2}
			{\draw[fill=blue]  ({\queueheight*(1-(\x-1)/8)}, 0) -- ({\queueheight*(1-\x/8)}, 0) 
				--  ({\queueheight*(1-\x/8)}, \queuewidth) -- ({\queueheight*(1-(\x-1)/8)}, \queuewidth) -- cycle;}		
		}{}
	\end{tikzpicture}
}

\begin{figure}[ht]
	\centering
	\begin{tikzpicture}[scale=.8]
		\node[class] (1) {\queue{0}{2}};
		\path (1) ++(30:3.2cm) node[class] (2) {\queue{-90}{0}};
		\path (2) ++(-30:3.2cm) node[class] (3) {\queue{180}{6}};
		\path (2) ++(-90:3.2cm) node[class] (4) {\queue{90}{4}};
		
		\draw (2) ++(0,1.5) edge[->] node[right] {$\lambda_2$} ++(0,-1);
		\draw (4) ++(0,-1.5) edge[->] node[right] {$\lambda_3$} ++(0,+1);
		\draw (1) ++(-1.5, 0) edge[->] node[above] {$\lambda_1$} ++(1, 0);
		\draw (3) ++(1.5,0) edge[->] node[above] {$\lambda_4$} ++(-1,0);
		
		\draw (1) edge node[above, sloped] {$\mu_{1, 2}$} (2)
		(1) edge node[above, sloped] {$\mu_{1, 3}$} (4)
		(3) edge node[above, sloped] {$\mu_{2, 4}$} (2)
		(3) edge node[above, sloped] {$\mu_{3, 4}$} (4)
		(2) edge node[right] {$\mu_{2, 3}$} (4);
	\end{tikzpicture}
	\caption{Illustration of a matching model $(G, \lambda, \Phi)$ on the diamond graph. \label{fig:illustration}}
\end{figure}

\subsection{Motivation.} \label{sec:motivation}

A central challenge in matching models is to understand their long-run performance and control. A prerequisite for any such study is to determine whether the model is \emph{stabilizable}, that is, whether some policy keeps the queue lengths under control over time. This first question governs what can be done next: when the model is stabilizable, one can study and optimize its long-run behavior (the matching rates and delays achieved by stable policies), whereas otherwise one must turn to alternative formulations, such as a finite-time horizon, agents with finite patience, or paired arrivals in bipartite graphs. This paper provides a unified framework that settles this first question and, for stabilizable models, characterizes the whole set of achievable matching rates; the optimization and control results are then built on top of this characterization.
Stability, delay, and matching rates are the criteria we use to compare the long-run performance of matching policies, and they are of direct practical relevance in applications such as organ exchange and online marketplaces. Stability may also be imposed as a constraint in its own right, for instance for fairness; that role is related to, but distinct from, its role as the regime in which long-run averages are well defined.

Many studies adopt an optimization viewpoint, where each match of type $k\in E$ gets a reward $r_k$, and the objective is to find a policy~$\Phi$ that maximizes the long-run reward rate $\sum_{k \in \E} r_k \mu_k$.
This paradigm has clear practical appeal: for instance, in organ exchange programs, rewards capture metrics such as quality of life and survival rates after transplant~\cite{B21}.
Yet, natural reward-maximizing policies such as greedy priority-based schemes can jeopardize stability, as noted by \citet[Section~5]{MM16}. Moreover, there is no guarantee that   the ordering of the resulting matching rates is consistent with the edge priorities.

These issues gain further complexity when a network of matching models is considered. For example, the matching rates~$\mu$ from a stable first-level model $(G, \lambda, \Phi)$ can be the (non-Poisson) arrival rates in a second-level model with a compatibility graph $G^\prime = (\E, \E^\prime)$ (the nodes of the second-level model are the edges of $G$). Then, controlling the output matching rates~$\mu$ of the first-level model may be critical, if only to ensure stability of the second-level model.

These problems have been approached from two angles: one, inspired by queueing theory, centers on stability and delay; the other, rooted in linear optimization, seeks to maximize reward or minimize regret against an optimal solution.
We take the first, stability-oriented angle.
Consequently, our central object is the polytope of achievable matching rates, as defined by the conservation equation and the non-negativity constraints.
This \emph{polytope perspective} also organizes the optimization question, as maximizing a reward over stable policies amounts to selecting a vertex of this polytope. Optimization thus appears in our study both as a motivation and as an application of the structural results, rather than as its main goal.

\subsection{Contributions.}

The foundation of our approach is the conservation equation:
any stable policy must satisfy $A\mu=\lambda$, where $A$ is the incidence matrix of the compatibility graph $G$.  
The properties of the linear mapping $\mu\mapsto A\mu$, particularly injectivity and surjectivity, are shown to be fundamentally linked to the behavior of matching models. We also establish precise connections between these linear-algebraic properties and the structure of the underlying graph (\Cref{def:surjective,def:injective,def:bijective,def:only,prop:dimensions}). With some abuse of language, we say that a graph has a property, such as bijectivity, injectivity, or surjectivity, if the associated linear map does; for example, $G$ is said to be bijective if $\mu \mapsto A \mu$ is bijective.
Building on these concepts, our main contributions are as follows.

First, we connect stability and the conservation equation: There exists a policy $\Phi$ such that the matching model $(G, \lambda, \Phi)$ is stable if and only if the graph $G$ is surjective (\Cref{prop:stability-region-nonempty}) and the conservation equation has a solution with positive coordinates (\Cref{prop:stability-region-form}).
While equivalent results exist, our formulation provides a direct, intuitive method to verify stabilizability, in polynomial time with respect to the number of classes and edges.

Second, we describe the solutions to the conservation equation, that is, the affine set $\La$ of vectors $\mu \in \mathbb{R}^m$ such that $A \mu = \lambda$ (see~\eqref{eq:Pi-def} in \Cref{subsec:all-solutions}). When $G$ is bijective, we provide a closed-form expression of the unique solution (\Cref{prop:unicyclic}). When $G$ is surjective-only, we give a parametric expression for $\La$ (\Cref{prop:affine-space,def:edge-kernel-basis}) and characterize the polytope $\Lann$ of non-negative solutions. We show that the vertices of $\Lann$, i.e., its extreme points, correspond to injective solutions in the sense that the subgraph restricted to their support is injective (\Cref{prop:vertex}). Moreover, we show that the vertices are bijective with probability $1$ when the vector $\lambda$ is sampled randomly from a reasonable distribution (\Cref{prop:all-is-bijective}).

Third, we apply these results to the linear optimization of $\mu$, which for a stable policy amounts to achieving a vertex of $\Lann$ (\Cref{prop:vertices-optimal,prop:reward-is-vertex}).
If the target vertex is bijective, we propose an explicit optimal policy that attains it (\Cref{coro:achievable}).
In contrast, when the vertex is injective-only, a strong trade-off emerges between regret with respect to the optimal solution and delay (\Cref{prop: lower_bound}). Motivated by this case, we introduce a sequence of policies that provably achieve arbitrarily small regret, at the cost of increasing delay (\Cref{prop: converging_to_vertex}).
For a bijective vertex, this recovers in steady state the reward-optimal match-the-longest policy of \citet{KAG23}; our new contribution is the injective-only case, where we extend the explicit queue-length guarantees of \citet{KAG23} to vertices that their policy cannot stabilize, complementing the policy of \citet{NS19}, which reaches such vertices but offers no queue-length control.

At a high level, all the optimizing policies we consider share a common recipe: start from a simple base policy designed for stability, and then adapt its parameters to be ``vertex-aware''. While this insight may appear straightforward, it is highly effective in practice, especially for numerical evaluation. Building on this, we developed a modular simulation engine that enables easy integration of existing policies, their modification, and the creation of new heuristics. This flexible framework has been instrumental in validating our theoretical results through extensive simulations, where our policies are systematically compared against state-of-the-art alternatives.

We then apply our framework (\Cref{sec:applications}): we benchmark our policies against the state of the art, and we show that no \emph{greedy} policy can reach a vertex of $\Lann$ (\Cref{prop:greedypositive,coro:greedypositive}), quantifying on worked examples how far greedy policies fall short of the achievable region (\Cref{prop:greedy-complete,prop:greedy-diamond,prop:greedy-fish}).

Finally, we extend the framework in two directions (\Cref{sec:extensions}). A convexity argument shows that \emph{any} target rate in the interior of $\Lann$, not only reward-optimal vertices, can be achieved by a stable policy (\Cref{prop:convexity}), which allows non-linear reward functions to be optimized. We also outline the extension of our framework to \emph{hypergraphs}, that is, multi-way matching with the option to discard under-demanded items; in particular, the stabilizability characterization carries over in its support form, as established in a separate paper \citep{M26}.

Note that this paper builds upon the original results reported in~\cite{BCM21} as a preprint. The overlapping content reflects the continuity and development of our research, with the current manuscript providing a more comprehensive and peer-reviewed presentation.

\subsection{State of the art.}\label{sec:state-of-the-art}

We now review the relevant work
related to (static or dynamic) matching problems.

\paragraph*{Non-bipartite or general stochastic matching.}

Our work is part of a broader research effort
on the stochastic matching model
that was briefly discussed earlier
and will be described in detail
in \Cref{subsec:model}, see
\citet{BMM21,BMMR20,CDFB20,C22,JMRS20,MM16,MBM21}.
Among these works, the following are particularly relevant
because directly related to our results on stability.
The paper of~\citet{MM16} is the earliest work
on this matching model.
It derives necessary and sufficient
stability conditions that
are instrumental in several of our results,
in particular \Cref{prop:stability-region-nonempty,prop:stability-region-form}.
This work also proves that the \gls{ml} policy
is maximally stable (in the sense that it always leads to stability whenever the matching problem $(G, \lambda)$ is stabilizable), a result
that is also applied
in \Cref{prop:stability-region-form}.
\citet{C22} and \citet{MBM21} focus
on the \gls{fcfm} policy.
In particular, \citet{MBM21} proves
that the \gls{fcfm} policy
is maximally stable,
and~\citet{C22} provides
a new sufficient stability condition
we prove to be also necessary
in \Cref{prop:stability-region-form}.

The recent work of~\citet{BMM21}
is perhaps the closest to ours,
so we provide a detailed discussion
to highlight the relation with our paper.
The first statement
of \citet[Theorem~1]{BMM21}
coincides with the equivalence of
statements~\ref{cond:stability-region-form-2}
and~\ref{cond:stability-region-form-3}
in \Cref{prop:stability-region-form}.
Our proof is significantly shorter
because it relies more heavily
on existing results.
Our observation
in \Cref{sec:unicyclic}
that the conservation equation has a unique solution
if and only if the graph is bijective
(and not surjective-only)
summarizes \citet[Theorem~3]{BMM21}.
Some formulas derived in
\citet[Section~9]{BMM21}
are special cases of
the formulas derived in \Cref{prop:unicyclic}.
The non-bipartite matching model in~\citet{BMM21}
is slightly more general
because it considers graphs with self-loops,
that is, an edge can have identical endpoints,
and this paper also considers the bipartite model
that will be discussed below.
However, \citet{BMM21} does not adopt the polytope approach
that allows us to derive the necessary \emph{and sufficient} conditions of \Cref{sec:unicyclic,sec:general,sec:non-unicyclic}.
Furthermore, although we decided to focus
on non-bipartite matching models to alleviate the discussion,
the general approach we develop in \Cref{sec:stability}
equally applies to bipartite graphs
and can be used to derive results for bipartite matching models,
such as those studied in \citet{BMM21}.

Other variants of the model were studied recently.
In particular, \citet{JMRS20} consider item abandonment, \citet{BMMR20}
consider graphs with self-loops, and \citet{GW14}, \citet{NS19}, \citet{KAG24}, \citet{G24}, \citet{WXY23}, and \citet{RM21}
allow matches not limited to two items by replacing the graph with a hypergraph,
which is also called multi-way matching in the literature.
The generalization of our work to multi-way matching will be discussed in \Cref{sec:hypergraphs}.

\paragraph*{Reward maximization.}
A series of papers on multi-way,~\cite{NS19, WXY23, KAG24, G24}, and two-way matching,~\cite{KAG23}, focus on maximizing edge-weighted rewards, and are closely related to our optimization results in \Cref{sec:non-unicyclic}. These works differ in how ``greedy'' their policies are: the longest-queue policies of \citet{KAG23,KAG24} are \emph{truly greedy}, performing a match whenever a feasible one is available, whereas the policies of \citet{G24} and \citet{WXY23} are \emph{greedily scheduled}, planning matches greedily while possibly leaving some feasible matches unperformed at a given instant. We postpone a detailed, framework-by-framework comparison to \Cref{sec:comp_to_lit}, once our own results are in place.

\paragraph*{Bipartite stochastic matching.}

Our \emph{optimization} results concern stabilizable graphs, which are necessarily non-bipartite in the model we consider;
the stabilizability characterization of \Cref{sec:stability}, by contrast,
applies to all graphs and in fact identifies bipartite graphs as a prototypical non-stabilizable case.
There is also an extensive literature devoted specifically to the bipartite case.
A major complication there is that, with unrestricted arrivals,
the model is non-stabilizable, in the sense that the underlying Markov chain
is at best null-recurrent, or transient, but never positive recurrent.
Such non-stabilizable models are often studied instead over a finite time horizon.
To the best of our knowledge,
the first example in the literature
of a stochastic matching model with an infinite time horizon,
which predated the model we consider,
is the bipartite matching model
introduced in~\citet{CKW09},
and later studied
by~\citet{ABMW17,AW12,BGM13,BM16,CBD19,CD21}.
In this model, the compatibility graph is bipartite,
with two parts that correspond to supply and demand,
respectively.
Stabilizability is enforced by making a synchronization assumption:
time is slotted and, during each time slot,
one demand item \emph{and} one supply item arrive.
\citet{AW12} and \citet{BGM13} obtained results on stability
and \citet{AW12} focused on matching rates;
these results are similar to those described above
regarding the literature on non-bipartite matching.
Another work in the literature has obtained
a stabilizable matching model by
allowing reneging or abandonment~\cite{ADW21}.
The polytope and conservation-equation tools we develop
are not specific to non-bipartite graphs
and apply to these bipartite models as well.

\paragraph*{Static and fractional matching.}

The static matching problem,
in which the nodes of the graph
represent items (rather than classes),
has been extensively studied
in mathematics, computer science, and economy,
see~\citet{LP09}.
Although the questions raised
in static and dynamic matching are often different,
the conservation equation that we obtain
is reminiscent of several results on static matching.
For example,
finding a maximum-cardinality matching
in the graph~$G$
(that is, a maximum-cardinality set
of edges without common endpoints)
is equivalent to finding integers
$\mu_k \in \{0, 1\}$ for each edge~$k \in \E$
that maximize $\sum_{k \in \E} \mu_k$
while satisfying the conservation equation
with $\lambda_i = 1$ for each $i \in \V$.
The relaxation of this integer linear program
leads to the so-called fractional matching problem,
which has been studied
in the literature, see \citet[Section~7.2]{LP09}.
Therefore, the fractional matching polytope
defined by \citet[Section~7.5]{LP09}
is a special case of the convex polytope
that we consider in \Cref{sec:polypoly},
and our characterization of this convex polytope
is a natural generalization
of existing characterizations of the fractional polytope%
\footnote{The fractional matching polytope is actually defined using non-strict inequalities rather than equalities. However, one can verify that these two convex polytopes have the same non-zero vertices.}.

\subsection{Companion package.}

This paper is supported by the Python package \emph{Stochastic Matching} by \citet{SM22}, which provides an open-source implementation of all the concepts and results developed here, together with additional tools for experimentation. The package is freely available on \href{https://github.com/balouf/stochastic_matching}{GitHub} and \href{https://pypi.org/project/stochastic-matching/}{PyPI}, and can be readily installed on any recent Python distribution\footnote{For instance, \texttt{pip install stochastic-matching} or \texttt{uv add stochastic-matching}.}.

The package is designed to remain flexible for extensions. Its main features include:
\begin{itemize}
	\item Support for both simple graphs and hypergraphs, created either manually or through a built-in graph library, making it possible to test theoretical results on arbitrary structures;
	\item Automated categorization of graphs $G$ (e.g., surjective, injective) and verification of the stabilizability of any instance $(G, \lambda)$;
	\item Computation of the solution space $\Pi$ and of the vertices of $\Pi_{\geqslant 0}$, including the reward-optimal vertex;
	\item Visualization tools that produce graph-based representations of $\Pi$, vertices, and policy matching rates;
	\item A scalable simulation framework for evaluating the long-term performance of arbitrary policies (up to $10^{10}$ time steps), together with a curated library of policy implementations that includes the approaches proposed in this paper and state-of-the-art baselines.
\end{itemize}

The package builds upon \href{https://scipy.org/}{SciPy} for linear algebra and optimization routines, \href{https://visjs.github.io/vis-network/docs/network/}{VisJS Network} for graph visualization, and \href{https://numba.readthedocs.io/en/stable/index.html}{Numba} for efficient simulation.

While the package is designed as a standalone tool that can serve a broader community working on stochastic matching and related problems, a \href{https://balouf.github.io/stochastic_matching/companion/index.html}{dedicated section of its documentation} accompanies this article by reproducing all the results presented here and presenting additional examples\footnote{\url{https://balouf.github.io/stochastic_matching/companion/index.html}}. We believe that this package constitutes a valuable contribution on its own, as it provides a foundation for future experimental and theoretical work.

\subsection{Outline.}

The remainder of the paper is organized as follows.
\textbf{\Cref{subsec:model}}
gives a formal definition of the model.
\textbf{\Cref{sec:stability}}
introduces the conservation equation
and defines the notions of
surjectivity, injectivity, and bijectivity for a graph.
We use these definitions to formulate new necessary and sufficient stability conditions; in particular, we show that stability requires the compatibility graph to be surjective (that is, either bijective or surjective-only). It also treats the bijective case (\textbf{\Cref{sec:unicyclic}}), for which we write the unique solution of the conservation equation in closed form.
\textbf{\Cref{sec:general}} characterizes the solution set
of the conservation equation for surjective-only graphs.
\textbf{\Cref{sec:non-unicyclic}} focuses on the linear optimization of the matching rates (that is, optimizing a linear combination of the matching rates) under a stable policy.
\textbf{\Cref{sec:applications}} presents two applications: it evaluates the numerical performance of the policies through simulations, and it examines the limitations of greedy policies for rate optimization.
Lastly, in \textbf{\Cref{sec:extensions}}, we consider two extensions of our framework: non-linear optimization and hypergraph matching.

 % sections/02-model
\section{Stochastic dynamic matching.} \label{subsec:model}

We consider the following model: items arrive at random times to be matched with other items; each incoming item may be paired with any unmatched item of a compatible class. When a match occurs, the involved items disappear immediately, while unmatched items wait in a queue. In this paper, such a stochastic dynamic matching system is described by a triplet $(G, \lambda, \Phi)$, where $G$ is the \emph{compatibility graph}, $\lambda$ the vector of \emph{arrival rates}, and $\Phi$ the \emph{matching policy}.

The remainder of this section formalizes the model in detail. 
Our goal here is to state the model precisely so that the paper remains both rigorous and self-contained.
Readers already familiar with queueing theory and matching systems may safely skip ahead to \Cref{sec:stability}, as the definitions and notations introduced here do not extend beyond the standard framework.

For convenience, notation is summarized in \Cref{table:notation}.

\subsection{Compatibility graph.}

Compatibility constraints between items
are described by a graph $G=(V, E)$,
called the \emph{compatibility graph} of the model,
which is simple
(undirected and without self-loop).
The number of nodes is represented by $n$, while $m$ denotes the number of edges.

We use $\V = \{v_1, v_2, \ldots, v_n\}$ to denote the set of nodes, where each node corresponds to a class in the matching model. When no confusion can arise, we may refer to a class~$v_i$ simply by its index~$i$.
Following the intuition conveyed in \Cref{fig:illustration},
we will use the terms ``class~$i$'' and ``queue~$i$'' interchangeably,
and refer for instance
to the number of unmatched class-$i$ items
as the size of queue~$i$;
this is a convenience of terminology,
and this does not preclude the matching policy
from using information not captured by this state representation,
such as the arrival order of items of different classes
(see \Cref{subsubsec:policy} for more details).

The set of edges is denoted by $\E = \{e_1, e_2, \ldots, e_m\}$.
These edges represent compatibility constraints
between item classes, in the sense that
a class-$i$ item and a class-$j$ item
can be matched with one another
if and only if their classes are adjacent,
that is, if there is an edge with endpoints~$i$ and~$j$.
When there is no ambiguity,
we may refer to an edge $e_k \in \E$ with endpoints $i, j \in \V$
by its index~$k$ or its set $\{i, j\}$ of endpoints.
In \Cref{fig:illustration} for instance,
there are four item classes numbered from~$1$ to~$4$.
Classes~$2$ and~$3$ are compatible with all classes,
but classes~$1$ and~$4$ are only compatible
with classes~$2$ and~$3$.
The absence of self-loop means that an item of a given class
cannot be matched with other items of the same class.

Lastly, we let $\ind$ denote the family of independent sets of the compatibility graph~$G$,
where an independent set of~$G$ is a non-empty set of nodes that are pairwise non-adjacent.
The family of independent sets in the compatibility graph of \Cref{fig:illustration}
is $\ind = \{ \{1\}, \{2\}, \{3\}, \{4\}, \{1, 4\} \}$.

\subsection{Arrival process.} \label{sec:arrival}

Class-$i$ items arrive according to an independent Poisson process
with rate $\lambda_i > 0$, for each $i \in \V$.
The vector of arrival rates is denoted by
$\lambda = (\lambda_1, \lambda_2, \ldots, \lambda_n) \in \R_{> 0}^n$.
Scaling all coordinates of $\lambda$ by the same positive constant
is equivalent to changing the time unit,
so we can renormalize $\lambda$ without changing the dynamics.
For example, we will sometimes use the \emph{unit} normalization,
in which $\sum_{i \in \V} \lambda_i = 1$.
We also let $I = (I_t, t \in \N)$ denote the sequence
of independent and identically distributed (i.i.d.) item classes,
so that $I_t$ is the class of the $(t+1)$-th item,
equal to $i$ with probability $\lambda_i / (\sum_{j \in \V} \lambda_j)$,
for each $t \in \N$.
The couple $(G, \lambda)$ is called a \emph{matching problem} or simply a \emph{problem}.
Occasionally, when we need to specify the sequence of incoming items
and not merely its distribution,
we will also refer to the couple $(G, I)$ as a (matching) problem.

\subsection{Policy and matching dynamics.} \label{subsubsec:policy}

Most of the paper focuses on deterministic size-based policies,
that is, policies whereby matching decisions
are deterministic functions of the queue-size vector.
However, as we will briefly explain at the end of this section
(and discuss in more detail in the supplementary material),
our results also apply to a more general definition of a policy.
Throughout the paper, we assume that
the system is initially empty,
meaning that it starts with no unmatched item.

\subsubsection{Deterministic size-based policies.} \label{sec:deterministic-size-base-policies}

A (deterministic) size-based \emph{matching policy} is defined formally as a function $\Phi: \Q \times \V \to \V \cup \{\bot\}$, where $\Q$ is an infinite subset of $\N^n$
that contains the reachable states of the system.
In-depth discussion on $\Q$ will come
in \Cref{sec:greedy-policies,sec:nongreedy-policies}.
For each $q \in \Q$ and $i \in \V$,
an incoming class-$i$ item that finds the system in state~$q$
is matched with an item of class~$\Phi(q, i)$ if $\Phi(q, i) \in \V$
and is added to class-$i$ queue if $\Phi(q, i) = \bot$.
The matching policy is assumed to be \emph{adapted}
to the compatibility graph~$G$ in the sense that
\begin{align} \label{eq:Phi}
	\Phi(q, i) \in \{j \in \V_i: q_j \ge 1 \} \cup \{\bot\},
	\quad q \in \Q,
	\quad i \in \V,
\end{align}
where $V_i$ is the neighbor set of node~$i$ in~$G$.
The system dynamics are described by a Markov chain $Q = (Q_t, t \in \N)$,
called the \emph{queue-size process}.
For each $t \in \N$,
$Q_t = (Q_{t, 1}, Q_{t, 2}, \ldots, Q_{t, n})$
is an $n$-dimensional vector
giving the number of unmatched items of each class
right after the arrival of the $t$-th item,
with the assumption that the system is initially empty,
that is, $Q_0 = 0$.
The system dynamics satisfy the recursion
\begin{align} \label{eq:Q-rec}
	Q_{t+1} = \begin{cases}
		Q_t + \one_{I_t} &\text{if $J_t = \bot$,} \\
		Q_t - \one_{J_t} &\text{if $J_t \neq \bot$,}
	\end{cases}
\end{align}
where $J_t = \Phi(Q_t, I_t)$ for each $t \in \N$,
and $\one_i$ is the $n$-dimensional vector
with one in coordinate~$i$ and zero elsewhere,
for each $i \in \V$.
We assume that the policy~$\Phi$ is such that
the Markov chain $Q$ has state space~$\Q$ and is irreducible.
By unfolding the recursion~\eqref{eq:Q-rec},
we obtain that, for each $t \in \N$,
\begin{align} \label{eq:Q-unfolded}
	Q_{t, i} &= L_{t, i} - \sum_{k \in \E_i} M_{t, k},
	\quad t \in \N,
	\quad i \in \V,
\end{align}
where $E_i \subseteq \E$ is the set of edges that are incident to node~$i$
in the graph~$G$, for each $i \in \V$,
$L_{t, i}$ is the number of class-$i$ items among the first~$t$ arrivals,
for each $t \in \N$ and $i \in \V$,
and $M_{t, k}$ is the number of times that classes~$i$ and~$j$ are matched
over the first~$t$ arrivals,
for each $t \in \N$ and $\{i,j\} = e_k \in \E$:
\begin{align}
	\label{eq:L}
	L_{t, i} &= \sum_{s = 0}^{t-1} \one_{\{I_s = i\}},
	\quad t \in \N, \quad i \in \V, \\
	\label{eq:M}
	M_{t, k} &= \sum_{s = 0}^{t-1} \one_{\{ \{I_s, J_s\} = e_k \}},
	\quad t \in \N, \quad k \in \E,
\end{align}
with the convention that the sums are zero if $t = 0$.
The triplet $(G, \lambda, \Phi)$ is called a \emph{matching model}, or simply a \emph{model}.
Occasionally, when specifying the sequence of incoming item classes is useful, we will also refer to the triplet $(G, I, \Phi)$ as a (matching) model.

\subsubsection{Greedy policies.} \label{sec:greedy-policies}

A policy~$\Phi$ is called \emph{greedy}
if an incoming item is matched whenever possible,
that is, if there is an unmatched item that is compatible.
More formally, the policy~$\Phi$ is greedy if
\begin{align} \label{eq:Phi-greedy}
	\Phi(q, i) \neq \bot
	\text{ for each }
	(q, i) \in \Q \times \V
	\text{ such that }
	\{j \in \V_i : q_j \ge 1\} \neq \emptyset.
\end{align}
Equivalently, a policy~$\Phi$ is greedy
if the set of unmatched item classes under this policy
is an independent set of the compatibility graph, meaning that
the state space~$\Q$ of the queue-size process is equal to
\begin{align} \label{eq:Q-greedy}
	\Q_\Gre = \{q \in \N^n: q_i q_j = 0
	\text{ for each } i, j \in \V
	\text{ such that } \{i, j\} \in \E\}.
\end{align}
Here are two examples of greedy matching policies that will appear later in the paper:
\begin{itemize}
	\item \textbf{\acrfull{ml}}:
	For each $(q, i) \in \Q \times \V$
	such that $\sum_{j \in \V_i} q_j \ge 1$,
	we choose $\Phi(q, i) \in \argmax_{j \in \V_i} (q_j)$
	(ties are broken arbitrarily).
	This policy was considered by~\citet{MM16,BMMR20,JMRS20,BMM21}.
	\item \textbf{\gls{hrf}}:
	This policy selects matches according to rewards defined on edges.
	If $r$ denotes a vector of $\R^m$ (indexed by the edges) with distinct coordinates, we let
	$\Phi(q, i) = \argmax_{j \in \V_i: q_j \ge 1} r_{i, j}$
	for each $(q, i) \in \Q \times \V$ such that $\sum_{j \in \V_i} q_j \ge 1$.
	This definition remains valid if the coordinates of $r$ have ties that do not impact the decision, i.e., if all pairs of edges that are incident to the same node and are not part of a triangle have distinct rewards\footnote{If three classes form a triangle in the compatibility graph, at most one of them can be non-empty under a greedy policy, so at most one edge of the triangle can be in the input of $\argmax$.}. Note that the order induced by $r$ fully defines an \gls{hrf} policy.
\end{itemize}

\subsubsection{Non-greedy policies.} \label{sec:nongreedy-policies}

The state space~$\Q$ under non-greedy policies
is a strict superset of the set~$\Q_\Gre$ defined in~\eqref{eq:Q-greedy}.
In \Cref{sec:non-unicyclic}, non-greedy policies
are obtained by applying the following modifications to greedy policies:
\begin{itemize}
	\item \textbf{Filtering}:
	Given a subset $\E^\star \subsetneq \E$ of edges,
	replace $\V_i$ (the neighbors of node $i$) with
	$\V^\star_i = \{ j \in \V_i: \{i, j\} \in \E^\star \}$
	in the definition~\eqref{eq:Phi} of $\Phi$.
	Intuitively, we eliminate
	the edges of $\E \setminus \E^\star$
	and follow a (greedy or non-greedy) policy
	on the subgraph $G^\star = (\V, \E^\star)$.
	\item \textbf{Semi-filtering}:
	We consider a variant of filtering policies
	in which the restriction of selecting matches from a subset $\E^\star$
	is not always enforced. 
	Examples of semi-filtering policies
	will be given in \Cref{sec:non-unicyclic,sec:numerical-results}.
\end{itemize}

\subsubsection{Other policies.} \label{sec:extended-definition}

Although our results are stated for deterministic size-based policies because they are notationally convenient,
our results apply to a broader family of policies
that are either random or require a more complex state descriptor, or both.

The extension to random policies is standard.
A random (size-based) policy~$\Phi$ can be defined as a function
$\Phi: \Q \times \V \times (\V \cup \{\bot\}) \to [0, 1]$
such that, for each $t \in \N$,
$J_t$ is sampled according to the distribution $\Phi(q, i, \cdot)$
given that $Q_t = q$ and $I_t = i$.
Saying that the policy is adapted to the compatibility graph~$G$
is then equivalent to saying that,
for each $q \in \Q$ and $i \in \V$,
the support of $\Phi(q, i, \cdot)$ is
included into $\{j \in \V_i: q_j \ge 1\} \cup \{\bot\}$.
The policy is greedy if $\Phi(q, i, \bot) = 0$
for each $(q, i) \in \Q \times \V$
such that $\{j \in \V_i: q_j \ge 1\} \neq \emptyset$
or, equivalently, if $\Q = \Q_\Gre$.

An even broader set of policies that fit into our framework
is described in \refapp{app:extended-definition}.
In a nutshell, the extended definition of a policy
starts with a pair $(\cS, |\cdot|)$,
where $\cS$ is a countably infinite set
and $|\cdot|: \cS \to \N^n$
is a function that maps any state $s \in \cS$
to the corresponding queue-size vector.
The policy is then a function
$\Phi: \cS \times V \times (V \cup \{\bot\}) \times \cS \to [0, 1]$
such that $\Phi(s, i, j, s^\prime)$
is the conditional probability that,
given an incoming class-$i$ item finds the system in state~$s$,
the matching decision is~$j$ and the new state is~$s^\prime$.
The policy is assumed to be such that
the Markov chain $S = (S_t, t \in \N)$
defined on $\cS$ by the evolution of the system state
is irreducible.
The stochastic process $Q = (Q_t = |S_t|, t \in \N)$
is called the \emph{queue-size process}.
It no longer satisfies the Markov property in general,
but it does satisfy the evolution equations~\eqref{eq:Q-rec} and~\eqref{eq:Q-unfolded},
with $L_i = (L_{t, i}, t \in \N)$ and $M_k = (M_{t, k}, t \in \N)$
defined by~\eqref{eq:L} and~\eqref{eq:M} for each $i \in \V$ and $k \in \E$.
The state space of the queue-size process
is given by $\Q = \{ |s|, s \in \cS \}$.
The policy is greedy if $\Q = \Q_\Gre$
and non-greedy if $\Q \supsetneq \Q_\Gre$,
where $\Q_\Gre$ is still given by~\eqref{eq:Q-greedy}.
\acrfull{fcfm} (e.g., see \citet{MBM21,C22})
is a classical example of a deterministic policy
that requires an expanded state descriptor
(remembering the arrival order of items of different classes).

This extended policy definition can initially be ignored, with $S_t$ (resp.\ $\cS$) understood as $Q_t$ (resp.\ $\Q$). All results for extended policies remain valid if restated for deterministic size-based policies, except for the convexity result in \Cref{prop:convexity}, which specifically relies on extended policies.

\subsection{Performance.} \label{subsubsec:stability}

We consider two main performance criteria: on the one hand, stability with the related notion of delay, and on the other hand, matching rates.

\paragraph{Stability.}

We first define the notions of stability and stabilizability, which will be explored further in \Cref{sec:stability}.

\begin{dfn}[Stability and stabilizability] \label{def:stability} ~
	 \begin{enumerate}[(i)] 
	 	\item A model $(G, \lambda, \Phi)$ is \emph{stable} if the Markov chain $(S_t, t \in \N)$ is positive recurrent. In this case, we say that the policy $\Phi$ \emph{stabilizes} the problem $(G, \lambda)$. 
	 	\item A problem $(G, \lambda)$ is \emph{stabilizable} if there exists a policy that stabilizes it. 
	 	\item A compatibility graph~$G$ is \emph{stabilizable} if there exists a vector $\lambda \in \Rp^n$ such that the problem $(G, \lambda)$ is stabilizable.
		\end{enumerate}
\end{dfn}

\noindent The \gls{ml} greedy policy, discussed in \Cref{sec:greedy-policies}, has the property of stabilizing all stabilizable matching problems (\citet{MM16}).
The \acrfull{fcfm} greedy policy has the same property (see \citet{MBM21} for details). 
When the matching problem $(G, \lambda)$ is clear from the context, we will simply call a policy (adapted to $G$) \emph{stable} if the corresponding model $(G, \lambda, \Phi)$ is stable.

Stability is a central property in our framework. In addition to ensuring finite delays
(a desirable feature for dynamic matching applications such as organ donation, assemble-to-order systems, and online marketplaces), it supports a rigorous theoretical analysis: it is the regime in which the long-run matching rates are well defined.

\paragraph*{Matching rates.}

Consider a stable matching model~$(G, \lambda, \Phi)$. We define the \emph{matching rate} $\mu_k$ along an edge $e_k \in \E$, with endpoints~$i$ and~$j$, as the long-run average number of matches between a class-$i$ item and a class-$j$ item per unit of time, given by:
\begin{align} \label{eq:lambda}
	\left( \sum_{i \in \V} \lambda_i \right)
	\times \frac1t M_{t, k}
	\xrightarrow[t \to +\infty]{\text{almost surely}}
	\mu_k,
	\quad k \in \E.
\end{align}
This quantity is uniquely defined according to the ergodic theorem, see \citet[Theorem~1.10.2]{norris}. Note that $M_{t, k} / t$ represents the average number of matches \emph{per arrival} (out of the first $t$ arrivals), and the factor $\left( \sum_{i \in \V} \lambda_i \right)$ converts this into the average number of matches per \emph{unit of time}.
The vector of matching rates associated with the model $(G, \lambda, \Phi)$ is denoted by $\mu = (\mu_1, \ldots, \mu_m) \in \Rnn^m$. When necessary, we write $\mu(\Phi)$ or $\mu(G, \lambda, \Phi)$ to indicate the dependence explicitly.
\Cref{sec:stability,sec:general} will investigate the structural constraints that a vector of matching rates must satisfy.

\begin{table}[ht!]
	\begin{tabular}{|l|p{11cm}|}
		\hline
		\multicolumn{2}{|c|}{General notation} \\
		\hline \hline
		$\N, \Np, \R, \Rnn, \Rp$ & Sets of non-negative integers, positive integers, real numbers, non-negative real numbers, positive real numbers. \\
		\hline
		$\ge$, $\le$, $>$, $<$ & Coordinate-wise comparison in $\R^n$. \\
		\hline
		$|\A|$ & Cardinality of the set $\A$. \\
		\hline \hline
		\multicolumn{2}{|c|}{Graph notation} \\
		\hline \hline
		$G=(V, E)$ & Simple graph $G$ with $|V|=n$ vertices and $|E|=m$ edges. \\
		\hline
		$v_i$ & Vertex indexed by $i$ (denoted by~$i$ if there is no ambiguity). \\
		\hline
		$e_k$, $k$, or $\{i, j\}$ & Edge indexed by $k$, with endpoint vertices~$i$ and~$j$. \\
		\hline
		$V_i$, $V(\I)$ & Neighbor set of a node~$i$ and of node set~$\I$. \\
		\hline
		$E_i$ & Set of edges incident to node~$i$. \\
		\hline
		$\mathbb{I}$ & Family of independent sets of the graph~$G$. \\
		\hline \hline
		\multicolumn{2}{|c|}{Matching notation} \\
		\hline \hline
		$\lambda=(\lambda_i)_{1\leq i \leq n}$ & Vector of arrival rates of the item classes. \\
		\hline
		$\Phi$ & A matching policy. \\
		\hline
		$\mu =(\mu_k)_{1\leq k \leq m}
		=(\mu_{i, j})_{\{i, j\}\in E}$ & Vector of matching rates along the edges. \\
		\hline	\hline
		\multicolumn{2}{|c|}{Linear-algebra notation} \\
		\hline \hline
		$A = (a_{i, k})_{i \in \V, k \in \E}$ & Incidence matrix of the graph~$G$. \\
		\hline
		$A^\intercal = (a_{k, i})_{k \in \E, i \in \V}$
		& Transpose of the matrix~$A$. \\
		\hline
		$\ker(A) = \{y \in \R^m: Ay = 0\}$ & Right kernel of the matrix~$A$.
		Its dimension is called the nullity of~$A$. \\
		\hline
		$\ker(A^\intercal) = \{x \in \R^n: A^\intercal x = 0\}$
		& Left kernel of the matrix~$A$.
		Its dimension is the nullity of $A^\intercal$. \\
		\hline
		$d = m - n$ & Dimension of the right kernel of the matrix~$A$ if $G$ is surjective. \\
		\hline
	\end{tabular}
	\caption{Table of key notation. \label{table:notation}}
\end{table}

 % sections/03-stability
\section{Stabilizability and the conservation equation.} \label{sec:stability}

In this section, we establish the connection between stability and the structure of the compatibility graph.
\Cref{subsec:conservation}
introduces the conservation equation, a system of linear equations satisfied by all matching rate vectors.
\Cref{subsec:graph} introduces the related concepts of
surjectivity, injectivity, and bijectivity
that will play a key role throughout the paper.
In \Cref{subsec:stabilizability},
we combine these concepts
to formulate new necessary and sufficient conditions
under which a compatibility graph~$G$
or a matching problem~$(G, \lambda)$
is \emph{stabilizable} in the sense of \Cref{def:stability}.
\Cref{subsec:examples} illustrates these results
with early examples.

\subsection{Conservation equation.}
\label{subsec:conservation}

Computing the matching rate vector achieved by a given stable policy
or characterizing the set of matching rate vectors that can be achieved by stable policies
is a difficult problem \textit{a priori}.
To circumvent this difficulty, we first establish a necessary condition known as the \emph{conservation equation}.
This equation is satisfied by the matching rate vectors achieved by all stable policies.
It asserts that, in a stable system, the arrival of items and their departure due to matches balance each other in the long run.
More formally, given a stable matching model~$(G, \lambda, \Phi)$,
the conservation equation~\eqref{eq:system}
can be derived by dividing~\eqref{eq:Q-unfolded} by~$t$
and taking the limit as~$t$ tends to infinity,
and it can be written in two equivalent forms,
either as a system of linear equations~\eqref{eq:system-equations}
or in matrix form~\eqref{eq:system-matrix}:
\begin{subequations}
	\makeatletter
	\def\@currentlabel{\textsc{ce}}
	\makeatother
	\label{eq:system}
	\begin{align}
		\tag{\ref{eq:system}--1}
		\label{eq:system-equations}
		\sum_{k \in \E_i} \mu_k& = \lambda_i, \quad i \in \V, \\
		\tag{\ref{eq:system}--2}
		\label{eq:system-matrix}
		A \mu &= \lambda,
	\end{align}
\end{subequations}
where the $n \times m$ matrix
$A = (a_{i, k})_{i \in \{1, 2, \ldots, n\}, k \in \{1, 2, \ldots, m\}}$
is the \emph{incidence} matrix of the graph~$G$,
defined by $a_{i, k} = 1$ if edge~$e_k$ is incident to node~$v_i$
and $a_{i, k} = 0$ otherwise.
Using a conservation equation is common in queueing theory, but our contributions in this paper primarily stem from the novel mixed approach that combines graph theory and linear algebra, which we will elaborate on in the following sections.

The conservation equation \eqref{eq:system} holds significant importance throughout this paper. While our primary focus is to understand the matching rate vector, we often find it beneficial to temporarily depart from interpreting $\lambda$ and $\mu$ as vectors of arrival and matching rates in a matching model. Instead, we view the conservation equation as a linear equation with $\lambda$ as a free variable and $\mu$ as an unknown. This perspective allows us to explore various aspects of the equation and its implications.
In this context,
we will sometimes allow the coordinates of~$\lambda$ and~$\mu$ to be negative,
even if the vectors of arrival and matching rates
in a matching model have non-negative coordinates.

\subsection{Surjectivity, injectivity, and bijectivity.} \label{subsec:graph}

\Cref{def:surjective,def:injective,def:bijective,def:only} below
introduce the notions of \emph{surjectivity},
\emph{injectivity}, and \emph{bijectivity} of a graph.
In a nutshell, a compatibility graph~$G$ is said to be
surjective (resp.\ injective, bijective)
if the linear application $\mu \mapsto A \mu$
defined by its incidence matrix~$A$
is surjective (resp.\ injective, bijective).
Interestingly, simple equivalent conditions exist
in terms of the graph structure.
As we will see later,
these notions are fundamental to study
the stability of matching models
and the associated matching rate vector.
In particular, we will see that
(i) a compatibility graph~$G$ is stabilizable 
if and only if $G$ is surjective,
and (ii) the matching rates
in a stabilizable matching problem $(G, \lambda)$
are independent of the policy~$\Phi$
if and only if $G$ is bijective.
Toy examples are shown in \Cref{fig:surjective-injective},
and other illustrative examples are discussed in \Cref{subsec:examples}.
The equivalence of the conditions given in
\Cref{def:surjective,def:injective,def:bijective}
and the implications in \Cref{prop:dimensions}
are proved in~\refapp{app:proofs-of-fullrefsubsecgraph}.

\def\d{2.3cm}
\begin{figure}[ht]
	\centering
	\hfill
	\subfloat[Graph that is neither surjective nor injective.\\
	Both $A^\intercal$ and $A$ have nullity~1.]{%
		\begin{tikzpicture}[%
			every node/.style={, minimum size=.3cm}, scale=.7, transform shape]	
			\bifan
			\draw (1) edge (2) edge (3)
			(4) edge (2) edge (3);	
			\node at ($(1)-(2.3cm,0)$) {};
			\node at ($(4)+(2.3cm,0)$) {};
		\end{tikzpicture}
	}
	\hfill
	\subfloat[Surjective-only graph.\\
	The nullity of $A^\intercal$ is 0 and that of $A$ is 1.]{%
		\begin{tikzpicture}[%
			every node/.style={, minimum size=.3cm}, scale=.7, transform shape]	
			\bifan
			\draw (2) edge (1) edge (3) edge (4)
			(3) edge (1) edge (4);			
			\node at ($(1)-(2.3cm,0)$) {};
			\node at ($(4)+(2.3cm,0)$) {};
		\end{tikzpicture}
	}
\hfill
\\
	\hfill
	\subfloat[Injective-only graph.\\
	The nullity of $A^\intercal$ is 1 and that of $A$ is 0.]{%
		\begin{tikzpicture}[%
			every node/.style={, minimum size=.3cm}, scale=.7, transform shape]
			\yshape
			\draw (3) edge (1) edge (2) edge (4);	
			
			\node at ($(1)-(2.3cm,0)$) {};
			\node at ($(4)+(2.3cm,0)$) {};
		\end{tikzpicture}
	}
	\hfill
	\subfloat[Bijective graph.\\
	Both $A^\intercal$ and $A$ have nullity~0.]{%
		\begin{tikzpicture}[%
			every node/.style={, minimum size=.3cm}, scale=.7, transform shape]	
			\yshape
			\draw (3) edge (1) edge (2) edge (4)
			(1) edge (2);	
			
			\node at ($(1)-(2.3cm,0)$) {};
			\node at ($(4)+(2.3cm,0)$) {};
		\end{tikzpicture}
	}
	\hfill
	\caption{Examples of graphs. The nullity of a matrix is the dimension of its kernel.}
	\label{fig:surjective-injective}
\end{figure}

\begin{dfn}[Surjective graph] \label{def:surjective}
	Consider a simple graph $G = (\V, \E)$
	with $n$ nodes and $m$ edges.
	Let~$A$ denote the $n \times m$ incidence matrix of~$G$.
	The graph~$G$ is called \emph{surjective}
	if one of the following equivalent conditions is satisfied:
	\begin{enumerate}[(i)]
		\item \label{cond:surjective-1}
		The function $\mu \in \R^m \mapsto A\mu \in \R^n$ is surjective.
		\item \label{cond:surjective-2}
		For each $\lambda \in \R^n$,
		the equation $A \mu = \lambda$
		of unknown $\mu \in \R^m$
		has at least one solution.
		\item \label{cond:surjective-3}
		The left kernel of the matrix~$A$ is trivial.
		\item \label{cond:surjective-4}
		Each connected component of the graph~$G$ is non-bipartite; in other words, each connected component of~$G$ contains at least one odd cycle.
	\end{enumerate}
\end{dfn}

\begin{dfn}[Injective graph] \label{def:injective}
	Consider a simple graph $G = (\V, \E)$
	with $n$ nodes and $m$ edges.
	Let~$A$ denote the $n \times m$ incidence matrix of~$G$.
	The graph~$G$ is called \emph{injective}
	if one of the following equivalent conditions is satisfied:
	\begin{enumerate}[(i)]
		\item \label{cond:injective-1}
		The function $\mu \in \R^m \mapsto A\mu \in \R^n$ is injective.
		\item \label{cond:injective-2}
		For each $\lambda \in \R^n$,
		the equation $A \mu = \lambda$
		of unknown $\mu \in \R^m$
		has at most one solution.
		\item \label{cond:injective-3}
		The right kernel of the matrix~$A$ is trivial.
		\item \label{cond:injective-4}
		Each connected component of the graph~$G$
		contains at most one odd cycle and no even cycle;
		in other words, each connected component of~$G$
		is either a tree or a unicyclic graph with an odd cycle.
	\end{enumerate}
\end{dfn}

\begin{dfn}[Bijective graph] \label{def:bijective}
	Consider a simple graph~$G = (\V, \E)$
	with $n$~nodes and $m$~edges.
	Let $A$ denote the $n \times m$ incidence matrix of~$G$.
	The graph~$G$ is called \emph{bijective}
	if the following equivalent conditions are satisfied:
	\begin{enumerate}[(i)]
		\item \label{cond:bijective-1}
		The function $\mu \in \R^m \mapsto A\mu \in \R^n$ is bijective.
		\item \label{cond:bijective-2}
		For each $\lambda \in \R^n$,
		the equation $A \mu = \lambda$
		of unknown $\mu \in \R^m$
		has exactly one solution.
		\item \label{cond:bijective-3}
		The matrix~$A$ is invertible.
		\item \label{cond:bijective-4}
		Each connected component of the graph~$G$
		contains one cycle and this cycle is odd.
	\end{enumerate}
\end{dfn}

\begin{dfn}[Surjective-only graph, injective-only graph] \label{def:only}
	A simple graph~$G$ is called
	\emph{surjective-only} (resp.\ \emph{injective-only})
	if~$G$ is surjective but not injective
	(resp.\ injective but not surjective).
\end{dfn}
The following proposition
gives necessary conditions for surjectivity and injectivity
in terms of the number of nodes and edges in the graph.

\begin{prop} \label{prop:dimensions}
	Consider an undirected graph~$G = (\V, \E)$
	with $n$ nodes and $m$ edges.
	\begin{enumerate}[(i)]
		\item \label{cond:dimensions-1}
		If~$G$ is surjective, then $n \le m$.
		\item \label{cond:dimensions-2}
		If~$G$ is injective, then $n \ge m$.
		\item \label{cond:dimensions-3}
		If~$G$ is bijective, then $n = m$.
		\item \label{cond:dimensions-4}
		If $G$ is surjective,
		then $G$ is also injective
		if and only if $n = m$.
		\item \label{cond:dimensions-5}
		If $G$ is injective,
		then $G$ is also surjective
		if and only if $n = m$.
	\end{enumerate}
\end{prop}

\subsection{Stabilizability of a graph and of a matching problem.} \label{subsec:stabilizability}

Combining the conservation equation with the notions of \Cref{subsec:graph} yields
necessary and sufficient conditions for stabilizability, first for a compatibility
graph~$G$ and then for a matching problem~$(G, \lambda)$.

\subsubsection{Stabilizable compatibility graph.} \label{subsec:stability-region-nonempty}

\Cref{prop:stability-region-nonempty} characterizes stabilizable graphs. The
equivalence is due to \citet[Theorem~1]{MM16}; casting it through the alternative
characterizations of surjectivity of \Cref{def:surjective} (e.g., a trivial left
kernel) does not seem to have been exploited before in the stochastic-matching
literature, and will be convenient later.

\begin{prop} \label{prop:stability-region-nonempty}
	Let~$G$ be a compatibility graph. The following are equivalent:
	\begin{enumerate}[(i)]
		\item \label{cond:stability-region-nonempty}
		the graph~$G$ is stabilizable;
		\item \label{cond:stability-region-nonempty-1}
		the graph~$G$ is surjective (equivalently, by \Cref{def:surjective}, each connected component of~$G$ is non-bipartite).
	\end{enumerate}
\end{prop}

In the remainder, we use the words ``stabilizable'' and ``surjective''
interchangeably for a graph and, unless stated otherwise, we assume that the
graph~$G$ is surjective.

\subsubsection{Stabilizable matching problem: a unifying theorem.} \label{subsec:stability-region-form}

We now turn to the stabilizability of a matching problem~$(G, \lambda)$.
As recalled in \Cref{subsubsec:stability},
both \gls{ml} and \gls{fcfm} stabilize the model
whenever it is stabilizable~\citep{MM16,MBM21}.
\Cref{prop:stability-region-form} gathers the equivalent characterizations of stabilizability.
The equivalence
\ref{cond:stability-region-form-1}
$\Longleftrightarrow$ \ref{cond:stability-region-form-2} is due to \citet{MM16},
who already observe that these conditions can be met
only if the compatibility graph~$G$ is non-bipartite (i.e., surjective).
The implication
\ref{cond:stability-region-form-2}
$\implies$ \ref{cond:stability-region-form-3}
is due to \citet{C22}.
Our main contribution is the implication
\ref{cond:stability-region-form-3}
$\implies$ \ref{cond:stability-region-form-2},
concurrently with \citet{BMM21}
(also see discussion in \Cref{sec:state-of-the-art}),
plus the equivalent condition \ref{cond:stability-region-form-4},
which is a by-product of \ref{cond:stability-region-form-3}.

\begin{prop}[Unifying theorem: stabilizability] \label{prop:stability-region-form}
	Consider a matching problem $(G, \lambda)$ with $\lambda \in \Rp^n$; unlike
	elsewhere in this section, the graph~$G$ is \emph{not} assumed surjective here. The
	following conditions are equivalent:
	\begin{enumerate}[(i)]
		\item \label{cond:stability-region-form-1}
		The matching problem $(G, \lambda)$ is stabilizable.
		\item \label{cond:stability-region-form-2}
		For each independent set $\I \in \ind$,
		we have $\sum_{i \in \I} \lambda_i
		< \sum_{i \in \V(\I)} \lambda_i$, where $\V(\I) = \bigcup_{i \in \I} \V_i$ is the neighbor set of~$\I$.
		\item \label{cond:stability-region-form-3}
		The graph~$G$ is surjective and the conservation equation~\eqref{eq:system}
		admits a solution $\mu \in \Rp^m$ (with strictly positive components).
		\item \label{cond:stability-region-form-4}
		The conservation equation~\eqref{eq:system} admits a
		solution $\mu \in \Rnn^m$ whose support graph
		$G^\star = (\V, \{k \in \E : \mu_k > 0\})$ is surjective.
	\end{enumerate}
\end{prop}

\begin{proof}
Conditions~\ref{cond:stability-region-form-1}--\ref{cond:stability-region-form-4}
are proved equivalent in
\refapp{app:proofs-of-fullrefsubsecstability-region-form}, using results
from~\cite{MM16,C22}.
\end{proof}

Conditions~\ref{cond:stability-region-form-2} and~\ref{cond:stability-region-form-3}
can be verified in polynomial time
with respect to $n$ and~$m$.
One might expect that the time complexity to verify condition~\ref{cond:stability-region-form-2} is exponential in the general case, considering that the number of independent sets grows exponentially fast with the number~$n$ of classes. However, \citet[Proposition~1]{MM16} proved that there exists a polynomial algorithm for verifying this condition. Note that this verification process is indirect in that it involves constructing a bipartite double cover of~$G$.
In contrast, condition~\ref{cond:stability-region-form-3} offers a more direct, polynomial-time method to verify the stabilizability of a matching problem~$(G, \lambda)$.
To present this approach more explicitly, we differentiate between two cases based on whether the graph~$G$, assumed to be surjective, is surjective-only or bijective.

\begin{rem}
	\label{rem:degree-proportional}
	As observed by \citet[Lemma~12]{C22},
	if the graph~$G$ is surjective,
	condition~\ref{cond:stability-region-form-3} in \Cref{prop:stability-region-form}
	gives a simple way of generating vectors~$\lambda \in \R_{\ge 0}^n$ such that the problem $(G, \lambda)$ is stabilizable: it suffices to take $\lambda = A \mu$ for some $\mu\in \Rp^m$. For instance, if $\mu=(\beta, \ldots, \beta)$ for some $\beta>0$, then the coordinates of $\lambda$ are proportional to the degree of each node.
\end{rem}

\paragraph*{Verify stabilizability when $G$ is bijective.}

If the graph~$G$ is bijective, then the matrix~$A$ is invertible, and~\eqref{eq:system} has a unique solution, namely $A^{-1}\lambda$.
\Cref{prop:stability-region-form} implies that the matching problem $(G, \lambda)$ is stabilizable if and only if all coordinates of $A^{-1} \lambda$ are positive.
The special case of bijective graphs will be investigated in detail in \Cref{sec:unicyclic}, where we will provide a more direct expression for $A^{-1} \lambda$.

\paragraph*{Verify stabilizability when $G$ is surjective-only.}

If the compatibility graph~$G$ is surjective-only, \eqref{eq:system} has multiple solutions.
To determine if one of these solutions is positive and thus checks
\Cref{prop:stability-region-form}\ref{cond:stability-region-form-3},
it suffices to solve a linear optimization problem
that searches for a solution to~\eqref{eq:system}
whose smallest coordinate is as large as possible:
\begin{align} \label{eq:linear}
	\begin{aligned}
		\underset{z = (z_1, z_2, \ldots, z_{m+1}) \in \R^{m+1}}{\text{Maximize}}
		\quad &
		z_{m+1}, \\
		\text{Subject to}
		\quad &
		A (z_1, z_2, \ldots, z_m)^\intercal = \lambda, \\
		& z_i \ge z_{m+1}, \quad i \in \{1, 2, \ldots, m\}.
	\end{aligned}
\end{align}
Here, the first $m$ coordinates of the vector~$z$ are the coordinates of a vector~$\mu \in \R^m$
that satisfies~\eqref{eq:system},
and the last coordinate of~$z$
is the smallest coordinate of this vector~$\mu$.
Indeed, the equality constraint means that $\mu$ satisfies~\eqref{eq:system},
and the inequality constraint
means that the last coordinate of~$z$
is less than or equal to its other coordinates.
The value to maximize is
the last coordinate of the vector~$z$.
If $z$ is a solution to~\eqref{eq:linear}, we call the corresponding vector $\mu\in \R^m$ a \emph{maximin} solution to~\eqref{eq:system}.

The optimization problem~\eqref{eq:linear}
is a textbook linear optimization problem.
It can be solved
with a time complexity that is polynomial
in the number~$n$ of nodes
and the number~$m$ of edges using many methods,
for instance the interior-point method, see \citet{K84}.

The linear optimization problem~\eqref{eq:linear}
has a solution with positive coordinates
if and only if~\eqref{eq:system}
has a solution with positive coordinates.
According to \Cref{prop:stability-region-form},
this is equivalent to saying that
the matching problem $(G, \lambda)$ is stabilizable.
Therefore, to verify if
a matching problem $(G, \lambda)$ is stabilizable,
it suffices to find a solution to
the linear optimization problem~\eqref{eq:linear}
and to check if its last coordinate is positive.

\begin{rem}
	Observe that the optimization problem~\eqref{eq:linear}
	always has solutions with finite coordinates.
	Indeed, the set of vectors
	that satisfy the constraints
	of~\eqref{eq:linear}
	contains at least one valid solution with real-valued coordinates (this is again a consequence of the surjectivity of~$G$). We just need to consider an arbitrary solution $\mu$ of~\eqref{eq:system} (see \Cref{sec:particular-solution} for a concrete example using the Moore-Penrose inverse) and to let $z_\mu = (\mu_1, \mu_2, \ldots, \mu_m, \min(\mu))$, where $\min(\mu)$ is the smallest coordinate of the vector~$\mu$.
	Any solution better than $z_\mu$ has all its coordinates lower-bounded by $\min(\mu)$ and upper-bounded by
	$\max(\lambda)-\min(0,(n-2)\min(\mu))$. The latter bound is obtained by observing that, if edge~$k$ is incident to node~$i$ and if $(\mu^\prime_1, \mu^\prime_2, \ldots, \mu^\prime_m, x^\prime)$ is a solution to~\eqref{eq:linear} such that $x^\prime\geq \min(\mu)$, then $\mu^\prime_k=\lambda_{i}-\sum_{\ell \in E_i\setminus k}\mu^\prime_\ell$ by~\eqref{eq:system}. We then use the inequalities $\lambda_{i}\leq \max(\lambda)$ and $\sum_{\ell \in E_i\setminus k}\mu^\prime_\ell\geq \min(0,(n-2)x^\prime)\geq \min(0,(n-2)\min(\mu))$ (this latter inequality is obtained by distinguishing two cases, depending on whether $\min(\mu) \ge 0$ or $\min(\mu) < 0$). Therefore, the solutions better than $z_\mu$ belong to a compact set of $\R^{m+1}$, which ensures the existence of an optimal solution with finite coordinates.
\end{rem}

\subsection{Early examples.} \label{subsec:examples}

We now provide examples that illustrate the stabilizability results of \Cref{prop:stability-region-nonempty,prop:stability-region-form} as well as our definitions of surjectivity, injectivity, and bijectivity. These examples will also introduce useful notions that will be further explored in \Cref{sec:unicyclic,sec:general,sec:non-unicyclic}.

\subsubsection{Bijective graphs.} \label{subsubsec:bijective}

We first consider connected compatibility graphs~$G$
that are both surjective and injective:
the graph contains exactly one cycle, and this cycle is odd.
According to \Cref{def:bijective}, \eqref{eq:system}
has a unique solution for each vector $\lambda \in \R^n$ of arrival rates.
\Cref{prop:stability-region-form} implies that
the matching problem $(G, \lambda)$ is stabilizable
if and only if 
the coordinates of this solution are positive,
in which case this solution gives the matching rates
achieved by \emph{all} stable matching policies.
By \Cref{rem:degree-proportional},
one can always exhibit
a vector $\lambda \in \Rp^n$ of arrival rates
such that the matching problem $(G, \lambda)$ is stabilizable.

\def\d{3cm}
\begin{figure}[!htb]
	\centering
	\hfill
	\subfloat[Matching rates in the triangle graph $\C_3$.\label{fig:triangle}]{%
		\begin{tikzpicture}[every node/.style={, minimum size=.3cm}, scale=.7]
			\placenodes{2/1/0,3/1/-60}				
			\draw (1) edge node[above] {$\frac{\lambda_1+\lambda_2-\lambda_3}{2}$} (2) 
			(1) edge node[left] {$\frac{\lambda_1+\lambda_3-\lambda_2}{2}$} (3)
			(2) edge node[right] {$\frac{\lambda_2+\lambda_3-\lambda_1}{2}$} (3);
			\halfwidth{3}{4cm}
		\end{tikzpicture}
	}
	\hfill
	\subfloat[Matching rates in the paw graph.
	$\bar{\lambda}_3=\lambda_3-\lambda_4$ denotes the residual rate that class~$3$ can provide to classes~$1$ and~$2$.\label{fig:pan}]{%
		\begin{tikzpicture}[every node/.style={, minimum size=.3cm}, scale=.7]
			\yshape			
			\draw (1) edge node[left] {$\frac{\lambda_1+\lambda_2-\bar\lambda_3}{2}$} (2) 
			(1) edge node[above, sloped] {$\frac{\lambda_1+\bar\lambda_3-\lambda_2}{2}$} (3)
			(2) edge node[below, sloped] {$\frac{\lambda_2+\bar\lambda_3-\lambda_1}{2}$} (3)
			(3) edge node[above] {$\lambda_4$} (4);
			\halfwidth{3}{5cm}
		\end{tikzpicture}
	}
	\hfill
	\\
	\subfloat[Matching rates in the square graph $\C_4$ with the normalization $\lambda_1+\lambda_4=\lambda_2+\lambda_3=\frac 12$. This graph is not stabilizable.\label{fig:square}]{%
		\begin{tikzpicture}[every node/.style={, minimum size=.3cm}, scale=.7]
			\bifan			
			\halfwidth{3}{3cm}
			\draw (1) edge node[sloped, above] {$2\lambda_1\lambda_2+\alpha$} (2) 
			(1) edge node[sloped, below] {$2\lambda_1\lambda_3-\alpha$} (3)
			(2) edge node[sloped, above] {$2\lambda_2\lambda_4-\alpha$} (4)
			(4) edge node[sloped, below] {$2\lambda_3\lambda_4+\alpha$} (3);
			\end{tikzpicture}
	}
	\hfill
	\subfloat[Matching rates in the diamond graph with the normalization $\lambda_1+\lambda_4=\frac 12$. $2\beta = \lambda_2 + \lambda_3 - \lambda_1 - \lambda_4 = \lambda_2+\lambda_3 - \frac{1}{2}$ is the difference between the arrival rates of the inner part $\{2, 3\}$ and the outer part $\{1, 4\}$.
	$\bar\lambda_2= \lambda_2 - \beta$ and $\bar\lambda_3= \lambda_3 - \beta$ represent the residual rates that classes~$2$ and~$3$ can provide to classes~$1$ and~$4$, and they are such that $\lambda_1 + \lambda_4 = \bar\lambda_2 + \bar\lambda_3 = \frac12$, as in \Cref{ex:square}.\label{fig:diamond}]{%
		\begin{tikzpicture}[every node/.style={, minimum size=.3cm}, scale=.7]	
			\bifan			
			
			\draw (1) edge node[sloped, above] {$2\lambda_1\bar\lambda_2+\alpha$} (2) 
			(1) edge node[sloped, below] {$2\lambda_1\bar\lambda_3-\alpha$} (3)
			(2) edge node[sloped, above] {$2\bar\lambda_2\lambda_4-\alpha$} (4)
			(4) edge node[sloped, below] {$2\bar\lambda_3\lambda_4+\alpha$} (3)
			(2) edge node[right] {$\beta$} (3);
			\halfwidth{3}{6.2cm}
		\end{tikzpicture}
	}
	\\
	\subfloat[Matching rates in the kayak paddle $\KP_{3, 3, 1}$. \label{fig:kayak}]{%
		\begin{tikzpicture}[every node/.style={, minimum size=.3cm}, scale=.8]
			\def\d{3cm}
			\placenodes{2/1/90,3/1/30,4/3/0,5/4/30,6/4/-30}			
						
			\draw (1) edge node[left] {$\frac{\lambda_1+\lambda_2-\lambda_3+\alpha}{2}$} (2) 
			(1) edge node[sloped, below] {$\frac{\lambda_1+\lambda_3-\lambda_2-\alpha}{2}$} (3)
			(2) edge node[sloped, above] {$\frac{\lambda_2+\lambda_3-\lambda_1-\alpha}{2}$} (3)
			(3) edge node[above] {$\alpha$} (4)
			(5) edge node[right] {$\frac{\lambda_5+\lambda_6-\lambda_4+\alpha}{2}$} (6) 
			(4) edge node[sloped, above] {$\frac{\lambda_4+\lambda_5-\lambda_6-\alpha}{2}$} (5)
			(4) edge node[sloped, below] {$\frac{\lambda_4+\lambda_6-\lambda_5-\alpha}{2}$} (6);
		\end{tikzpicture}
	}
	\caption{Examples of \Cref{subsec:examples}.}
	\label{fig:examples}
\end{figure}

\begin{exa}[Triangle] \label{eq:triangle}
	\label{ex:triangle}
	
	If~$G$ is the triangle graph $\C_3$, and the vector $\lambda = (\lambda_1, \lambda_2, \lambda_3)$ is given, the solution to~\eqref{eq:system} is unique and shown in \Cref{fig:triangle}. Indeed, a triangle graph contains a unique cycle which is odd, hence it is bijective according to \Cref{def:surjective}. According to \Cref{prop:stability-region-form}\ref{cond:stability-region-form-3}, $(G, \lambda)$ is stabilizable if and only if all coordinates of the unique solution of~\eqref{eq:system} are positive. This condition is equivalent to \Cref{prop:stability-region-form}\ref{cond:stability-region-form-2}, which states that $\lambda_1 < \lambda_2 + \lambda_3$, $\lambda_2 < \lambda_1 + \lambda_3$, and $\lambda_3 < \lambda_1 + \lambda_2$. Alternatively, this condition can be expressed as $\lambda_1$, $\lambda_2$, and $\lambda_3$ being the lengths of the sides of a non-degenerate triangle (i.e., they satisfy the triangular inequality). Under these conditions, the model $(G, \lambda, \Phi)$ is stable if $\Phi$ is for instance the (unique) greedy policy adapted to~$G$ (\refapp{prop:Kn-greedy-policy} shows that a complete graph~$G$ admits a unique greedy policy~$\Phi$).
\end{exa}

\begin{exa}[Paw graph] \label{ex:paw}
	If $G$ is a paw graph, the solution to~\eqref{eq:system} is again unique and shown in \Cref{fig:pan}. Here, $\bar\lambda_3 = \lambda_3-\lambda_4$ is the residual rate of class~$3$ after accounting for the needs of class~$4$. After this subtraction, the matching rates along edges $\{1, 2\}$, $\{1, 3\}$, and $\{2, 3\}$
	are defined as in the triangle graph of \Cref{fig:triangle}.

It is important to note that, while the existence of a matching rate vector with positive coordinates guarantees that a greedy policy like \gls{ml} is stable, there may still exist greedy policies that are unstable. Consider, for instance, the paw graph of \Cref{fig:pan} and a \glsentrylong{hrf} policy where edges $\{1,3\}$ and $\{2,3\}$ offer higher rewards than $\{3,4\}$.
Even if the problem is stabilizable, this policy might select edge $\{3,4\}$ at a slower rate than class-4 items arrive, causing instability (see \cite[Section~5]{MM16}).
\end{exa}

\subsubsection{Bipartite graph (neither injective nor surjective).}

Besides explaining intuitively why bipartite graphs
cannot be stabilized,
the following example will prepare the ground
for \Cref{ex:diamond}.

\begin{exa}[Square graph] \label{ex:square}
	\Cref{fig:square} shows a square graph $\C_4$.
	This graph is not surjective because it is bipartite with parts $\{1, 4\}$ (called the outer part) and $\{2, 3\}$ (inner part). Therefore, according to \Cref{prop:stability-region-nonempty}, this graph is not stabilizable. Yet, given a vector $\lambda = (\lambda_1, \lambda_2, \lambda_3, \lambda_4)$ of arrival rates, \eqref{eq:system} may still have a solution with positive coordinates.
	This does not contradict \Cref{prop:stability-region-form},
	as \Cref{cond:stability-region-form-3}
	also requires the graph~$G$ to be surjective.
	Assuming unit normalization, the conservation equation \eqref{eq:system-matrix} has a solution if and only if 
	\begin{equation}
		\label{eq:square}
		\lambda_1 + \lambda_4 = \lambda_2 + \lambda_3 = \frac{1}{2}\text{.}
	\end{equation}
	If~\eqref{eq:square} is satisfied, the solutions to~\eqref{eq:system} can be described with a parameter $\alpha$ as shown in \Cref{fig:square}:
	starting from the particular solution
	$\mu =$ $(2 \lambda_1 \lambda_2,$ $2 \lambda_1 \lambda_3,$ $2 \lambda_2 \lambda_4,$ $2 \lambda_3 \lambda_4)$,
	all solutions can be generated by alternately adding and subtracting $\alpha$ along the (even) cycle 1--2--4--3.
	The positive solutions correspond to values of $\alpha$ such that  $-2\min(\lambda_1\lambda_2, \lambda_3\lambda_4)<\alpha<2\min(\lambda_1\lambda_3, \lambda_2\lambda_4)$.
	
	To understand why the matching problem $(\C_4, \lambda)$
	is not stabilizable even when \eqref{eq:system}
	has a solution with positive coordinates,
	let us focus on the system dynamics.
	We can use~\eqref{eq:Q-unfolded} to show that
	the difference in total queue size
	between the outer part $\{1, 4\}$
	and the inner part $\{2, 3\}$
	satisfies
	$Q_{t, 1} + Q_{t, 4} - (Q_{t, 2} + Q_{t, 3})
	= L_{t, 1} + L_{t, 4} - (L_{t, 2} + L_{t, 3})$
	for each $t \in \N$.
	Therefore,
	the stochastic process $(Q_{t, 1} + Q_{t, 4} - (Q_{t, 2} + Q_{t, 3}), t \in \N)$,
	which happens to be a Markov chain in this example,
	is a random walk on the integer number line
	$\{\ldots, -2, -1, 0, 1, 2, \ldots\}$,
	with transition probability proportional to
	$\lambda_1 + \lambda_4$ in the increasing direction
	and to $\lambda_2 + \lambda_3$ in the decreasing direction.
	If \eqref{eq:square} is not satisfied,
	this random walk is transient,
	and the difference between the queue sizes of the two parts grows linearly with time.
	On the other hand, if \eqref{eq:square} is satisfied, then the random walk does not have this bias, but the model is still unstable because the corresponding Markov chain is null recurrent\footnote{Some existing studies of matching in bipartite graphs solve this issue by coupling arrivals in the two components, see \citet{AW12,BGM13,CKW09}, or by assuming that items have a finite patience time, as in \citet{JMRS20}.}.
\end{exa}

\subsubsection{Surjective-only graphs.} \label{subsubsec:surjective-graphs}

We finally consider a compatibility graph~$G$ that is surjective but not injective.
In other words, the graph~$G$ is stabilizable
and~\eqref{eq:system}
has an infinite number of solutions.
The achievability of these solutions by a stable matching policy will be discussed in \Cref{sec:non-unicyclic}.

\begin{exa}[Diamond (double-fan) graph] \label{ex:diamond}
	\Cref{fig:diamond} shows the diamond graph~$D$, that is, a square graph with an additional edge between nodes~$2$ and~$3$. Compared to \Cref{ex:square}, this additional edge makes the graph non-bipartite, and therefore surjective, so that the graph is stabilizable according to \Cref{prop:stability-region-nonempty}. For ease of computation, we assume that the vector $\lambda = (\lambda_1, \lambda_2, \lambda_3, \lambda_4)$ of arrival rates is normalized so that $\lambda_1+\lambda_4=\frac12$.
	Under this assumption, and with $\beta = \frac12 (\lambda_2 + \lambda_3 - \lambda_1 - \lambda_4)
	= \frac12 (\lambda_2 + \lambda_3) - \frac{1}{4}$,
	$\bar\lambda_2 = \lambda_2 - \beta$, and $\bar\lambda_3 = \lambda_3 - \beta$,
	the general solution to~\eqref{eq:system} can be described with a parameter $\alpha$ as shown in \Cref{fig:diamond}.
	After subtracting~$\beta$ from $\lambda_2$ and $\lambda_3$, the solutions to~\eqref{eq:system} for all edges
	but~$\{2, 3\}$ are exactly the same as in the square graph of \Cref{ex:square}.
	
	According to \Cref{prop:stability-region-form}\ref{cond:stability-region-form-2}, the matching problem $(D, \lambda)$ is stabilizable if and only if
	\begin{align} \label{eq:diamond-stabilizability}
		\lambda_2 &< \lambda_1 + \lambda_3 + \lambda_4,
		&
		\lambda_3 &< \lambda_1 + \lambda_2 + \lambda_4,
		&
		\lambda_1 + \lambda_4 &< \lambda_2 + \lambda_3,
	\end{align}
	that is,
	$\bar\lambda_3>0$,
	$\bar\lambda_2>0$, 
	and $\beta>0$.
	If these inequalities are satisfied, the positive solutions correspond to values of $\alpha$ such that $-2\min(\lambda_1\bar\lambda_2, \bar\lambda_3\lambda_4)<\alpha<2\min(\lambda_1\bar\lambda_3, \bar\lambda_2\lambda_4)$.
	Intuitively, compared to the square graph, stabilizable matching problems $(D, \lambda)$ have a positive difference of $2\beta$ between the arrival rates of the inner part $\{2, 3\}$ and the outer part $\{1, 4\}$. This difference is absorbed by the central edge $\{2, 3\}$, which has matching rate $\beta$.
	Like \Cref{ex:triangle} and unlike \Cref{ex:paw}, the matching model $(D, \lambda, \Phi)$ is stable for every greedy policy~$\Phi$ provided that~\eqref{eq:diamond-stabilizability} is satisfied (as shown in \refappnc{cor:max-greedy} in~\refapp{app:minimal}).
\end{exa}

\begin{exa}[Kayak paddle graph] \label{ex:paddle}
	\Cref{fig:kayak} shows a kayak paddle $\KP_{3,3,1}$, consisting of two triangles linked by an edge.
	According to \Cref{prop:stability-region-form}, the matching problem $(G, \lambda)$ is stabilizable if and only if there exists $\alpha>0$ such that $(\lambda_1, \lambda_2, \lambda_3-\alpha)$ and $(\lambda_4-\alpha, \lambda_5, \lambda_6)$ are the arrival rate vectors of two stabilizable triangle graphs $\C_3$.
	The solutions to~\eqref{eq:system} can be described by varying $\alpha$ as shown in \Cref{fig:kayak}. Assuming the matching problem $(G, \lambda)$ is stabilizable, the solutions to~\eqref{eq:system} with positive coordinates correspond to the values of $\alpha$ such that 
	$$0 < \alpha < \min(\lambda_3-|\lambda_2-\lambda_1|, \lambda_4-|\lambda_5-\lambda_6|)\text{.}$$
	Intuitively, solutions with positive coordinates have a positive matching rate $\alpha$ along edge $\{3,4\}$. After subtracting this rate from $\lambda_3$ and $\lambda_4$, the subgraphs restricted to nodes $1$, $2$, and $3$ and to nodes $4$, $5$, and $6$, respectively, behave like the triangle of \Cref{fig:triangle}.
	Like \Cref{ex:paw} and unlike \Cref{ex:triangle,ex:diamond}, the fact that $(G,\lambda)$ is stabilizable does not guarantee the stability of all greedy policies.
\end{exa}

\subsection{Closed-form solution for bijective graphs.} \label{sec:unicyclic}

Among stabilizable (surjective) graphs, the bijective ones ($n = m$: each connected
component is unicyclic with an odd cycle) are special. By \Cref{def:bijective}, the
conservation equation~\eqref{eq:system} then has the unique solution
$\mu = A^{-1}\lambda$, which is the matching-rate vector of \emph{every} stable policy,
and $(G,\lambda)$ is stabilizable if and only if this solution is positive.
Surjective-only graphs ($n < m$), for which~\eqref{eq:system} has infinitely many
solutions, are studied in \Cref{sec:general,sec:non-unicyclic}.

\Cref{prop:unicyclic} gives a closed form for $\mu = A^{-1}\lambda$ that avoids the
matrix inversion and exposes the monotonicity of the matching rates in the arrival
rates; we take $G$ connected without loss of generality. A similar formula appears as \cite[Theorem~4.1]{KAG23} in the (stability-free)
reward-maximization setting; the two derivations are independent and roughly
contemporaneous (the first preprints of \citet{KAG23} and of our own \citet{BCM21}
both date to 2021). \Cref{prop:unicyclic} treats only the bijective case, whereas the formula of \citet{KAG23} also covers the tree components that the optimal basis of their static-planning problem may contain, the surplus of one node per tree being absorbed by a slack variable; the restriction to the bijective case simplifies both the statement and the proof.

\begin{prop} \label{prop:unicyclic}
	Consider a problem $(G, \lambda)$
	with a compatibility graph~$G = (\V, \E)$
	that is connected and bijective,
	and consider an edge~$k \in \E$.
	Let $\mu$ be the unique solution of~\eqref{eq:system}.
	\begin{enumerate}[(i)]
		\item \label{cond:unicyclic-1}
		If edge~$k$ does not belong
		to the (unique odd) cycle of the graph~$G$,
		then edge~$k$ separates the graph~$G$
		into two parts, namely a tree and a unicyclic graph.
		If $T_k \subset \V$ denotes
		the set of nodes that belong to the tree
		(including one endpoint of edge~$k$),
		then the matching rate along edge~$k$ is given by
		\begin{align} \label{eq:unicyclic-border}
			\mu_k
			= \sum_{i \in T_k}
			(-1)^{d_{i, k}} \lambda_i,
		\end{align}
		where $d_{i, k}$ is the distance between node~$i$ and edge~$k$,
		defined as the distance between node~$i$ and the closest endpoint of edge~$k$.
		\item \label{cond:unicyclic-2}
		If edge~$k$ belongs to the (unique odd) cycle of the graph~$G$,
		then the matching rate along edge~$k$ is
		\begin{align} \label{eq:unicyclic-cycle}
			\mu_k
			= \frac12 \left(
			\sum_{i \in \V}
			(-1)^{d_{i, k}} \lambda_i
			\right).
		\end{align}
	\end{enumerate}
\end{prop}

\begin{proof}
See \refapp{app:unicyclic}.
\end{proof}

One can check that \Cref{ex:triangle,ex:paw} in \Cref{subsubsec:bijective} follow the pattern predicted by \Cref{prop:unicyclic}. \refapp{app:unicyclic} provides two additional examples that illustrate the proposition. In general, we
remark that the influence of the arrival rate of a node on the matching rate along
an edge only depends on the parity of the distance between the edge and the node; the
actual distance does not. In particular, even in a very large bijective graph, a node
far away from an edge has the same (although possibly reversed) impact on this edge's
matching rate as an endpoint of the edge.
          
 % sections/04-general
\section{Polytope of solutions in surjective-only graphs.} \label{sec:general}

\Cref{sec:unicyclic} considered the case where the compatibility graph $G$ was bijective, so that the vector of matching rates for any stable policy was the unique solution to~\eqref{eq:system}. Let us now assume that $G$ is surjective-only, meaning that: each connected component of~$G$ contains at least one odd cycle; at least one component contains either an even cycle or a pair of odd cycles (cf.\ \Cref{def:surjective,def:injective,def:only}). In this case, the solution set of \eqref{eq:system} is infinite. The matching rate vector under any stable policy belongs to this set and has non-negative coordinates.

To better control the matching rates, we first need to understand the structure of the solution set of~\eqref{eq:system}. To this end, \Cref{subsec:all-solutions} generalizes the earlier examples from~\Cref{subsec:examples} by systematically characterizing the affine space of all real-valued
solutions to~\eqref{eq:system}.
Following this, \Cref{sec:polypoly} describes the convex polytope formed by the solutions with non-negative coordinates, in particular its vertices, that is, its extreme points. Whether these solutions can be achieved through a stable matching policy is explored in~\Cref{sec:non-unicyclic}. This section develops the structural machinery underlying our results; a reader mainly interested in the optimization results may
skim ahead to~\Cref{sec:non-unicyclic}, referring back to \Cref{sec:general} as needed.

\subsection{Affine space of real-valued solutions.} \label{subsec:all-solutions}

We first consider the set of solutions to~\eqref{eq:system} with real-valued (positive, zero, or negative) coordinates:
\begin{align} \label{eq:Pi-def}
	\La = \left\{ \mu \in \R^m: A \mu = \lambda \right\}.
\end{align}
We now study the properties of $\La$.
In \Cref{sec:coordinate-spaces}, we recall that
$\La$ is an affine space of dimension $d = m - n$ 
that can be described as a translation of the kernel of the incidence matrix~$A$ by a particular solution to~\eqref{eq:system}. \Cref{sec:particular-solution} gives examples of particular solutions that can be used. \Cref{sec:basis-of-the-kernel-of-the-incidence-matrix} gives an algorithm to construct a basis for the kernel of the incidence matrix using a spanning tree of the graph~$G$.

\subsubsection{Edge basis, kernel basis.} \label{sec:coordinate-spaces}

The following proposition characterizes the solution set~$\La$ of~\eqref{eq:system} using the incidence matrix of the compatibility graph. \Cref{eq:Pi} shows in particular that, up to translation, this solution set depends only on the structure of the compatibility graph~$G$, while the arrival rate vector~$\lambda$ impacts only the translation vector.

\begin{prop} \label{prop:affine-space}
	Consider a matching problem $(G, \lambda)$
	with a surjective-only compatibility graph~$G$,
	and let $A$ denote the incidence matrix of~$G$.
	The solution set $\La$
	of~\eqref{eq:system}
	is the affine space obtained by translating
	the kernel $\ker(A)$ of the matrix~$A$
	by a particular solution $\mu^\circ$
	of~\eqref{eq:system}, that is,
	\begin{align} \label{eq:Pi}
		\La = \left\{ \mu^\circ + \mu: \mu \in \ker(A) \right\}.
	\end{align}
	Furthermore, the vector space $\ker(A)$
	and the affine space $\La$
	have dimension $d = m - n$.
\end{prop}

\begin{proof}
That the set~$\La$ is of the form~\eqref{eq:Pi}
is a well-known result in linear algebra.
\Cref{def:surjective} about surjectivity implies
that the rank of~$A$ is~$n$, and we conclude from the rank-nullity theorem that the nullity of~$A$ is
$d = m - n$. 
The affine space $\La$ has the same dimension according to~\eqref{eq:Pi}.
\end{proof}
Thanks to \Cref{prop:affine-space},
given a particular solution~$\mu^\circ$
of~\eqref{eq:system}
and a basis $\B = (b_1, b_2, \ldots, b_d)$ of $\ker(A)$,
we can rewrite the affine space~$\La$ as
\begin{align*}
	\La = \left\{
	\mu^\circ + \alpha_1 b_1 + \alpha_2 b_2
	+ \ldots + \alpha_d b_d :
	\alpha_1, \alpha_2, \ldots, \alpha_d \in \R
	\right\}.
\end{align*}
Observe that the solutions in \Cref{ex:diamond,ex:paddle} are written in this form. In general,
we can define the following affine isomorphism between the coordinate space~$\R^d$ and the $d$-dimensional affine space~$\La$:
\begin{align} \label{eq:linear-bijection}
	\alpha = (\alpha_1, \alpha_2, \ldots, \alpha_d)
	\in \R^d
	\mapsto
	\mu = \mu^\circ
	+ \alpha_1 b_1
	+ \alpha_2 b_2
	+ \ldots
	+ \alpha_d b_d \in \La.
\end{align}
This allows us to define two coordinate systems,
\emph{edge coordinates} and \emph{kernel coordinates}.

\begin{dfn}[Edge basis, kernel basis] \label{def:edge-kernel-basis}
	Consider a matching problem~$(G, \lambda)$
	with a sur\-jec\-ti\-ve-only graph~$G$.
	Given a particular solution $\mu^\circ$ of~\eqref{eq:system}
	and a basis $\B = (b_1, b_2, \ldots, b_d)$ of $\ker(A)$,
	there are two natural bases to represent vectors in $\La$:
	\begin{itemize}
		\item \emph{Edge basis}:
		A vector of $\La$
		is described by its canonical coordinates
		$\mu = (\mu_1, \mu_2, \ldots, \mu_m) \in \R^m$,
		solution to~\eqref{eq:system},
		where $\mu_k$ represents
		a candidate matching rate 
		along edge~$k$,
		for each $k \in \E$.
		\item \emph{Kernel basis}:
		A vector of $\La$
		is described by its coordinates
		$\alpha = (\alpha_1, \alpha_2, \ldots, \alpha_d) \in \R^d$
		in the basis~$\B$,
		where $d = m - n$
		is the dimension of the affine space~$\La$.
	\end{itemize}
	If~$B$ is the $m\times d$ matrix giving the coordinates of the vectors of the basis~$\B$ in the edge basis, the change-of-basis formulas are as follows:
	\begin{itemize}
		\item A vector of $\La$ with coordinates $\alpha$ in kernel basis has coordinates $\mu = \mu^\circ+B\alpha$ in edge basis;
		\item A vector of $\La$ with coordinates $\mu$ in edge basis has coordinates $\alpha = B^+(\mu-\mu^\circ)$ in kernel basis, where $B^+$ is the pseudo-inverse%
		\footnote{The columns of the matrix~$B$ are linearly independent because~$\B$ is a basis, so that the pseudo-inverse~$B^+$ has the simple expression $B^+ = (B^\intercal B)^{-1} B^\intercal$, where the $d \times d$ matrix $B^\intercal B$ is invertible because $\ker(B^\intercal B) = \ker(B) = \{0\}$.}
		(or Moore-Penrose inverse) of $B$.
	\end{itemize}
\end{dfn}

Both bases have advantages. The edge basis, by definition, gives directly the candidate matching rates. The kernel basis allows us to work in lower dimension ($d$ instead of $m$) and to ignore the conservation equation \eqref{eq:system}, which is implicitly enforced.
In the remainder, we will often use interchangeably
the edge coordinates $\mu = (\mu_1, \mu_2, \ldots, \mu_m)$
and the kernel coordinates $\alpha = (\alpha_1, \alpha_2, \ldots, \alpha_d)$
to describe a given vector in $\La$.
Which basis we are actually using will be made clear
by our choice of letters
(either $\mu$ or $\alpha$).

For graphs that have a low kernel dimension $d$, it is convenient to mix both approaches by representing a generic vector of $\La$, i.e., a generic solution to~\eqref{eq:system}, in the form
$\mu^\circ + \alpha_1 b_1 + \alpha_2 b_2 + \ldots + \alpha_d b_d$.
The solutions in \Cref{ex:diamond,ex:paddle} displayed in \Cref{fig:diamond,fig:kayak} follow this convention. This representation, along with the possibility to switch between edge basis and kernel basis, will be used extensively in \Cref{sec:polypoly,sec:non-unicyclic}.

\subsubsection{Particular solution.} \label{sec:particular-solution}

We propose two ways of computing a particular solution to~\eqref{eq:system}.

\paragraph*{Maximin solution.}

A solution to the linear optimization problem~\eqref{eq:linear} from~\Cref{subsec:stability-region-form} is a particular solution.
Recall that such a solution allows us to determine whether the matching problem $(G, \lambda)$ is stabilizable by checking if all its coordinates are positive.

\paragraph*{Pseudoinverse.}

Alternatively, a standard approach to simultaneously finding a particular solution $\mu^\circ$ and characterizing $\ker(A)$ makes use of the pseudoinverse (or Moore-Penrose inverse) of the matrix~$A$. Since \Cref{def:surjective} on surjectivity implies that the rows of~$A$ are linearly independent, the pseudoinverse~$A^+$ of~$A$ has the following simple form:
\begin{align*}
	A^+ = A^\intercal (AA^\intercal)^{-1},
\end{align*}
where the $n \times n$ matrix $AA^\intercal$ is invertible because
$\ker(AA^\intercal) = \ker(A^\intercal) = \{0\}$.
We can then describe
a particular solution $\mu^\perp$
and the kernel $\ker(A)$ as follows:
\begin{align} \label{eq:pseudoinverse}
	\mu^\perp &= A^+ \lambda,
	&
	\ker(A) &= \left\{ (\mathrm{Id}_{m \times m} - A^+ A)\mu: \mu \in \R^m \right\},
\end{align}
where $\mathrm{Id}_{m \times m}$ denotes 
the $m$-dimensional identity matrix.
The vector $\mu^\perp$
is the \emph{least-squares solution} to~\eqref{eq:system},
and it is orthogonal to~$\ker(A)$.
In general, some coordinates of this solution can be negative even if non-negative solutions exist.
For example, if $G$ is the diamond graph~$D$ of \Cref{ex:diamond},
then the matching problem $(D, \lambda)$
with $\lambda = (4,5,2,1)$ is stabilizable ($\mu=(\frac72,\frac12,1,\frac12,\frac12)$ is a solution to~\eqref{eq:system} with positive coordinates),
but the solution given by the pseudoinverse is
$\mu^\perp = (\frac{11}4, \frac54, 1, \frac54, -\frac14)$.

Equation~\eqref{eq:pseudoinverse} shows that the pseudoinverse also provides an implicit characterization of $\ker(A)$, though this is not very convenient, as it relies on a projection from $\R^m$ to $\ker(A)$.
\Cref{sec:basis-of-the-kernel-of-the-incidence-matrix} offers a more direct characterization by constructing a basis for $\ker(A)$ based on the structure of the compatibility graph~$G$.

\subsubsection{Basis of the kernel of the incidence matrix.} \label{sec:basis-of-the-kernel-of-the-incidence-matrix}

Recall that a vector $\mu \in \R^m$
belongs to $\ker(A)$
if and only if $A \mu = 0$, which reads
$\sum_{k \in \E_i} \mu_k = 0$
for each $i \in \V$.
In other words, a vector $\mu \in \R^m$ belongs to $\ker(A)$ if and only if, for each $i \in \V$, the sum of the coordinates of~$\mu$ associated with the edges that are incident to node~$i$ is zero.
Using this observation,
we first give examples of vectors that belong to~$\ker(A)$,
and then we give an algorithm
that generates a basis
$\B = (b_1, b_2, \ldots, b_d)$ of $\ker(A)$.

\begin{figure}[htb]
	\subfloat[Vector of the kernel space of the diamond graph.\label{fig:kernel}]{%
		\hfill
		\begin{tikzpicture}[scale=.7]		
			\def\d{2cm}
			\bifan			
			\draw (1) edge node[above left] {$1$} (2) 
			(1) edge node[below left] {$-1$} (3)
			(2) edge node[above right] {$-1$} (4)
			(3) edge node[below right] {$1$} (4)
			(2) edge node[right] {$0$} (3);
			\halfwidth{2}{4cm}
		\end{tikzpicture}
	}
	\hfill
	\subfloat[Vector of the kernel space of the kayak paddle $\KP_{3, 5, 2}$.\label{fig:kernel2}]{%
		\begin{tikzpicture}[scale=.7]
			\def\d{2cm}
			\placenodes{2/1/90, 3/1/30, 4/3/0, 5/4/0, 6/5/54, 7/6/-18, 8/7/-90, 9/5/-54}						
			\draw (1) edge node[left] {$1$} (2) 
			(1) edge node[below right] {$-1$} (3)
			(2) edge node[above right] {$-1$} (3)
			(3) edge node[above] {$2$} (4)
			(4) edge node[above] {$-2$} (5)
			(5) edge node[left] {$1$} (6)
			(6) edge node[above] {$-1$} (7)
			(7) edge node[right] {$1$} (8)
			(8) edge node[below] {$-1$} (9)
			(9) edge node[left] {$1$} (5)
			;
		\end{tikzpicture}
	}
	\hfill
	\caption{Base vectors of kernel spaces of two toy graphs.}
	\label{fig:kernel-examples}
\end{figure}

First observe that an even cycle, if it exists,
is the support of a vector in $\ker(A)$:
it suffices to assign alternatively the values $+1$ and $-1$
to the edges of this cycle
and the value $0$ to all other edges.
In the diamond graph of \Cref{ex:diamond},
if edges are numbered in lexicographical order, then
$y = (1, -1, 0, -1, 1)$ is a vector of the unidimensional kernel ($d = m - n = 1$),
with support the even cycle 1--2--4--3 (see \Cref{fig:kernel}).
Intuitively, even cycles can be used to move weight between ``odd'' and ``even'' edges of the cycle without modifying the value of the product~$A\mu$.
Actually, in this example, \Cref{fig:diamond} shows that the only way to increase the matching rate along edges $\{1, 2\}$ and $\{3, 4\}$ is if we reduce the matching rate along edges $\{1, 3\}$ and $\{2, 4\}$, and conversely.

Apart from even cycles, other structures of interest are kayak paddles $\KP_{\ell, r, p}$, made of two odd cycles (of lengths~$\ell$ and~$r$) connected by a path (of length~$p$). Such graphs have a unidimensional kernel, and a base vector can be found by assigning properly the values $+1$ and $-1$ along the cycles and the values $+2$ and $-2$ along the path. \Cref{fig:kernel2} shows such an assignment for $\KP_{3, 5, 2}$.

Surprisingly, for any surjective graph~$G$, one can build a basis of $\ker(A)$ using only subgraphs of~$G$ that are even cycles and kayak paddles. \Cref{algo:spanning}, derived from~\cite{D73}, describes such a construction. \refapp{app:spanning} gives a more detailed description of this algorithm and proof that it terminates and returns the desired result.

\begin{algorithm}[htb]
	\SetInd{.5em}{1em}
	\KwData{A connected surjective-only compatibility graph~$G = (\V, \E)$}
	\KwResult{A basis $\B$ of the kernel of the incidence matrix~$A$ of $G$}
	$\T \leftarrow$ Edges of a spanning tree of~$G$\\
	$a \leftarrow$ An edge in $\E \setminus \T$
	such that $\T \cup \{a\}$
	contains an odd cycle\\
	$\B \leftarrow \emptyset$ \\
	\For{$s \in \E \setminus (\T \cup \{a\})$}{
		Select from $\T \cup \{a, s\}$ an even cycle or a kayak paddle made of two odd cycles\\
		Weight the selected edges like in \Cref{fig:kernel-examples} (unselected edges have weight $0$)\\
		Add the resulting kernel vector to $\B$\\	
		}
    \textbf{return} $\B$
	\vspace{.2cm}
	\caption{High-level description of the construction of a basis of the kernel of the incidence matrix~$A$ of a compatibility graph~$G$. See~\refapp{app:spanning} for detailed description.}
	\label{algo:spanning}
\end{algorithm}

\def\d{1.1cm}
\begin{figure}[thb]
	\centering
	\subfloat[Selection of $\mathcal{T}$ and $a$.\label{fig:codomino-a}]{
		\begin{tikzpicture}		
			\codomino			
			\draw (1) edge[spanner] (2)
			(2) edge[spanner] (3)
			(3) edge[spanner] (4)
			(4) edge[spanner] (5)
			(6) edge[spanner] node[below, white] {0} (5)
			(3) edge[extra] node[left] {$a$} (5)
			(1) edge[k1] node[below] {$s_1$} (6)
			(2) edge[k2] node[right] {$s_2$} (6);
			
	\end{tikzpicture}}
	\hfill
	\subfloat[Kernel vector~$b_1$ derived from $s_1$ ($\C_6$).\label{fig:codomino-b}]{
		\begin{tikzpicture}
			\codomino

			\draw (1) edge[k1] node[above, xshift=-.1cm] {$-1$} (2)
			(2) edge[k1] node[above] {$1$} (3)
			(3) edge[k1] node[above, xshift=.1cm] {$-1$} (4)
			(4) edge[k1] node[below] {$1$} (5)
			(5) edge[k1] node[below, xshift=-.1cm] {$-1$} (6)
			(6) edge[k1] node[below] {$1$} (1)
			(2) edge node[right] {$0$} (6)
			(5) edge node[left] {$0$} (3);

			\node at ($(1)-(1cm,0)$) {};
			\node at ($(4)+(1cm,0)$) {};
	\end{tikzpicture}}
	\hfill
	\subfloat[Kernel vector~$b_2$ derived from $s_2$ ($\C_4$).\label{fig:codomino-c}]{
		\begin{tikzpicture}
			\codomino

			\draw (1) edge node[below] {$0$} (6)
			(1) edge node[above] {$0$} (2)
			(4) edge node[above] {$0$} (3)
			(4) edge node[below] {$0$} (5)
			(2) edge[k2] node[above] {$-1$}(3)
			(5) edge[k2] node[left, xshift=.1cm] {$1$} (3)
			(6) edge[k2] node[right, xshift=-.1cm] {$1$} (2)
			(6) edge[k2] node[below] {$-1$} (5);

			\node at ($(1)-(.5cm,0)$) {};
			\node at ($(4)+(.5cm,0)$) {};
	\end{tikzpicture}}
	\caption{\label{fig:codomino-kernel-simple}
		A possible execution of~\Cref{algo:spanning} for the codomino graph.
	}	
\end{figure}

\Cref{fig:codomino-kernel-simple} shows a possible run of \Cref{algo:spanning} for the codomino graph: in~\Cref{fig:codomino-a}, a spanning tree $\mathcal{T}$ (plain black edges) and
a pivotal (dotted) edge $a$ such that $\mathcal{T} \cup \{a\}$ contains an odd cycle
are selected.
Then, for each edge $s_i$ that does not belong to $\mathcal{T} \cup \{a\}$,
we build a base vector of the kernel space with support included into $\mathcal{T} \cup \{a, s_i\}$, after observing that $\mathcal{T} \cup \{a, s_i\}$
contains either an even cycle or a kayak paddle similar to those shown in \Cref{fig:kernel-examples} (in this specific run, both base vectors are supported by even cycles).
The resulting basis $(b_1, b_2)$ is the one used in \Cref{fig:main-vertex1,fig:main-vertex2} below.
Other examples, including a run of the algorithm on the same graph where a kayak paddle arises, are shown in \refapp{app:spanning}.

\begin{rem}\label{rem:kernel-support}
\Cref{eq:Pi} implies that,
given an edge~$k \in \E$,
all solutions to~\eqref{eq:system}
have the same value along edge~$k$
if and only if
edge~$k$ does not belong
to the support of any basis vector.
According to~\refapp{algo:spanning},
this is equivalent to saying that edge~$k$
belongs neither to an even cycle
nor to a kayak paddle.
In the diamond graph of \Cref{ex:diamond} for instance,
the edge $\{2,3\}$ is the only one
that does not belong to the even cycle 1--2--4--3,
and it is indeed the only one with a fixed rate~$\beta$.
In general, if an edge~$k \in \E$ satisfies this unicity condition,
then the matching rate along edge~$k$
in a stable matching model $(G, \lambda, \Phi)$
is independent of the policy~$\Phi$.
Note that there is no straightforward relation between
the number of edges with uniquely-defined matching rates
and the dimensionality~$d$ of the affine space~$\La$.
\end{rem}

\subsection{Polytope of non-negative solutions.}\label{sec:polypoly}

We continue to focus exclusively on matching problems $(G, \lambda)$ that are stabilizable.
Let $\Lann$ denote the set of solutions to~\eqref{eq:system}
that have non-negative coordinates, defined as
\begin{align} \label{eq:Piplus}
	\Lann = \La \cap \Rnn^m
	= \{ \mu \in \R^m: A \mu = \lambda, \mu \ge 0\}.
\end{align}
The set~$\Lann$ is a $d$-dimensional convex polytope in $\R^m$,
as it is the intersection of a $d$-dimensional affine space
with the positive orthant $\Rnn^m$,
both of which are convex.
$\Lann$ is neither empty
nor degenerated to a dimension lower than~$d$
because the matching problem $(G, \lambda)$ is assumed to be stabilizable, which by \Cref{prop:stability-region-form} means that $\La$ contains a vector with positive coordinates (i.e., in the interior of the positive orthant). $\Lann$ is bounded because
each $\mu \in \Lann$ satisfies
$0 \le \mu_{i, j} \le \min\{\lambda_i, \lambda_j\}$ for each $\{i, j\} \in E$.

Equation \eqref{eq:Piplus} describes $\Lann$ in the edge basis. As $\Lann$ is a subset of $\La$, we can also express its elements in the kernel basis introduced in \Cref{sec:coordinate-spaces}. In the kernel basis, $\Lann$ is defined by the vectors whose coordinates belong to 
\begin{align} \label{eq:Piplusalpha}
	\tilde{\Pi}_{\geqslant 0} = \{ \alpha \in \R^d: \mu^\circ + \alpha_1 b_1 + \alpha_2 b_2 + \ldots + \alpha_d b_d \ge 0\}.
\end{align}
As~\eqref{eq:Piplus} and~\eqref{eq:Piplusalpha} basically represent the same polytope up to the change-of-basis formulas of \Cref{def:edge-kernel-basis}, in the remainder, we will use the same notation $\Lann$ to describe both sets; the underlying basis will be made clear by our choice of letters (as before, $\mu$ for the edge basis and $\alpha$ for the kernel basis).

\subsubsection{Vertex characterization.}\label{sec:poly-vertex}

The vertices of a convex polytope, also known as its \emph{corners} or extreme points, are instrumental in optimization, which will be the object of \Cref{sec:non-unicyclic}. \Cref{def:vertex}  provides the formal definitions of vertices, along with faces and facets.

\begin{dfn}[Vertices, faces, and facets; adapted from \citet{Z95}] \label{def:vertex}
	Let~$\Upsilon$ denote a convex polytope of dimension $d \in \N_{>0}$.
	A \emph{(non-empty) face} of~$\Upsilon$ is
	a non-empty intersection of $\Upsilon$ with a hyperplane such that
	$\Upsilon$ is included into one of the two halfspaces defined by the hyperplane.
	A \emph{vertex} of~$\Upsilon$ is a face of dimension~0.
	Equivalently, a vector $\mu \in \Upsilon$ is a vertex of~$\Upsilon$
	if, and only if, it cannot be written
	as a convex combination
	of points in $\Upsilon {\setminus} \{\mu \}$.
	A \emph{facet} of~$\Upsilon$ is a face of dimension~$d-1$.
\end{dfn}

\Cref{prop:vertex} below gives a simple yet powerful characterization of the vertices of $\Lann$. 
\begin{prop} \label{prop:vertex}
	Consider a vector $\mu \in \Lann$.
	Let $\E^\star = \{k \in \E: \mu_k > 0\}$
	denote the support of the vector~$\mu$ and
	$G^\star = (\V, \E^\star)$ its support graph.
	The following statements are equivalent:
	\begin{enumerate}[(i)]
		\item \label{cond:vertex-1}
		The vector~$\mu$ is a vertex of $\Lann$.
		\item \label{cond:vertex-2}
		The graph~$G^\star$ is injective.
	\end{enumerate}
	In particular, if $\mu$ is a vertex, we can distinguish two cases depending on the value of $|E^\star|$:
	\begin{enumerate}
		\item If $|E^\star|=n$ then $G^\star$ is bijective.
		\item If $|E^\star|<n$ then $G^\star$ is injective-only.
	\end{enumerate}
	With a slight abuse of language, we say that a vertex $\mu\in \Lann$ is bijective (resp.\ injective-only) if the support graph $G^\star$ of~$\mu$ is bijective (resp.\ injective-only).
\end{prop}

\begin{proof}[Proof (borrowed from~\citet{CS22})]
See \refapp{app:prop-vertex}.
\end{proof}

\begin{figure}[!htb]
	\centering
	\subfloat[Generic element of $\Pi$ (i.e., generic solution of~\eqref{eq:system}). 
	 \label{fig:main-vertex-graph1}]{
		\begin{tikzpicture}		
			\def\d{2.4cm}
			\codomino
			
			\draw (1) edge node[above, sloped] {$1 - \alpha_1$} (2)
			(1) edge node[below, sloped] {$1 + \alpha_1$} (6)
			(2) edge node[above] {$1 + \alpha_1 - \alpha_2$} (3)
			(2) edge node[above, sloped] {$1 + \alpha_2$} (6)
			(3) edge node[above, sloped] {$1 - \alpha_1$} (4)
			(5) edge node[above, sloped] {$1 + \alpha_2$} (3)
			(4) edge node[below, sloped] {$1 + \alpha_1$} (5)
			(5) edge node[below] {$1 - \alpha_1 - \alpha_2$} (6);
			
		\end{tikzpicture}
	}
	\qquad
	\subfloat[Polytope~$\Lann$.
	\label{fig:main-vertex-polytope1}]{%
		\begin{tikzpicture}[scale=1.3]
			\draw[->] (-1.5,0) -- (1.5,0);
			\node[anchor=west] at (1.5,0) {$\alpha_1$};
			\draw[->] (0,-1.3) -- (0,1.3);
			\node[anchor=south] at (0,1.3) {$\alpha_2$};
			\node[anchor=north east] {0};
			
			\node (v1) at (-1,-1) {};
			\node (v2) at (-1,0) {};
			\node (v3) at (0,1) {};
			\node (v4) at (1,0) {};
			\node (v5) at (1,-1) {};
			
			\path[draw=blue, very thick, fill=blue, fill opacity=.2]
			(v1.center) -- (v2.center) -- (v3.center)
			-- (v4.center) -- (v5.center) -- cycle;
			
			\node[anchor=north, text=orange]
			at (v1) {$(-1, -1)$}; 
			\node[anchor=west, text=orange, yshift=.1cm]
			at (v3) {$(0, 1)$}; 
			\node[anchor=south east, text=orange, xshift=.12cm, yshift=.1cm]
			at (v2) {$(-1, 0)$}; 
			\node[anchor=south west, text=orange, xshift=-.12cm, yshift=.1cm]
			at (v4) {$(1, 0)$}; 
			\node[anchor=north, text=orange]
			at (v5) {$(1, -1)$}; 
			
			\foreach \i in {1,...,5}
			\node[dot] at (v\i) {};
		\end{tikzpicture}
	}
	\caption{Matching problem $(G, \lambda)$, where $G$ is the codomino graph and $\lambda=(2, 3, 3, 2, 3, 3)\in \R^6$. 	\label{fig:main-vertex1}
	}
\end{figure}

\begin{figure}[!htb]
	\centering
	\subfloat[Generic element of $\Pi$ (i.e., generic solution of~\eqref{eq:system}). 
	\label{fig:main-vertex-graph2}]{
		\begin{tikzpicture}		
			\def\d{2.4cm}
			\codomino
			
			\draw (1) edge node[above, sloped] {$2 - \alpha_1$} (2)
			(1) edge node[below, sloped] {$2 + \alpha_1$} (6)
			(2) edge node[above] {$1 + \alpha_1 - \alpha_2$} (3)
			(2) edge node[above, sloped] {$1 + \alpha_2$} (6)
			(3) edge node[above, sloped] {$2 - \alpha_1$} (4)
			(5) edge node[above, sloped] {$\alpha_2$} (3)
			(4) edge node[below, sloped] {$2 + \alpha_1$} (5)
			(5) edge node[below] {$1 - \alpha_1 - \alpha_2$} (6);
			
		\end{tikzpicture}
	}
	\qquad
	\subfloat[Polytope $\Lann$.
	\label{fig:main-vertex-polytope2}]{%
		\begin{tikzpicture}[scale=1.3]
			\draw[->] (-1.5,0) -- (1.5,0);
			\node[anchor=west] at (1.5,0) {$\alpha_1$};
			\draw[->] (0,-.3) -- (0,1.3);
			\node[anchor=south] at (0,1.3) {$\alpha_2$};
			\node[anchor=north east] {0};
			
			\node (v2) at (-1,0) {};
			\node (v3) at (0,1) {};
			\node (v4) at (1,0) {};
			
			\path[draw=blue, very thick, fill=blue, fill opacity=.2]
			(v2.center) -- (v3.center) -- (v4.center) -- cycle;
			
			\node[anchor=west, text=orange, yshift=.1cm]
			at (v3) {$(0, 1)$}; 
			\node[anchor=south east, text=orange, xshift=.12cm, yshift=.1cm]
			at (v2) {$(-1, 0)$}; 
			\node[anchor=south west, text=orange, xshift=-.12cm, yshift=.1cm]
			at (v4) {$(1, 0)$}; 
			
			\foreach \i in {2,...,4}
			\node[dot] at (v\i) {};
		\end{tikzpicture}
	}
	\caption{Matching problem $(G, \lambda)$, where $G$ is the codomino graph and $\lambda=(4, 4, 3, 4, 3, 4)\in \R^6$. 	\label{fig:main-vertex2}
	}
\end{figure}

\Cref{fig:main-vertex1,fig:main-vertex2} show $\Lann$ for the codomino graph. In each figure, the left part displays a generic solution to~\eqref{eq:system} on the graph, while the right part shows the polytope~$\Lann$ and its vertices in the kernel basis.
In \Cref{fig:main-vertex1}, $\Lann$ has five vertices. The vertex $\alpha = (0, 1)$ is bijective, as $|E^\star|=|E\setminus \{ \{2,3\},\{5,6\}\}|=6$. One can verify that the remaining four vertices are injective-only (as at least three edge coordinates are null).
In \Cref{fig:main-vertex2}, $\Lann$ has three vertices, all of which are bijective.

For further details, we encourage the reader to consult \refapp{app:facets}, which provides additional results on the relationships between vertex properties and the edge positivity inequalities from~\Cref{eq:Piplusalpha}, along with more detailed examples of polytopes and their vertices.

\subsubsection{Probability of bijectivity.}\label{sec:all-is-bijective}

\Cref{sec:non-unicyclic} will show that the bijectivity of vertices is central for optimizing the matching rates achieved by a stable policy. It is thus natural to wonder how frequent (or rare) bijective vertices are. \Cref{prop:all-is-bijective} gives a part of the answer.

\begin{prop}\label{prop:all-is-bijective}
Let $G$ be a surjective-only graph, and assume that the arrival rate vector $\lambda$ is drawn according to a positive Lebesgue density over $\Delta^{n-1}$, the standard simplex of $\R^n$. Then,
$$\mathbb{P}\left(\text{All vertices of $\Lann$ are bijective}| (G, \lambda)\text{ stabilizable}\right)= 1\text{.}$$
\end{prop}

\begin{proof}
See \refapp{app:all-is-bijective}.
\end{proof}

\Cref{prop:all-is-bijective} suggests that under random arrival rates, the polytope $\Lann$ is likely to have only bijective vertices. This result also holds if each coordinate of $\lambda$ follows a continuous, independent probability distribution, provided that the distribution's projection on $\Delta^{n-1}$ has a positive measure and includes at least one vector $\lambda$ for which $(G, \lambda)$ is stabilizable. In practice, injective-only vertices may still occur, especially when the coordinates of $\lambda$ display certain regularities, such as being proportional to node degrees (as in \Cref{fig:main-vertex1}).
Moreover, injective-only vertices offer unique theoretical challenges worth exploring.

In a sense, \Cref{prop:all-is-bijective} resembles the statement \emph{with probability 1, a square matrix is invertible}: while generally true under suitable conditions, this does not imply that non-invertible matrices can be ignored.

 % sections/05-non-unicyclic
\section{Stable policy optimization.} \label{sec:non-unicyclic}

In \Cref{sec:general}, we characterized $\Lann$, the polytope of candidate matching-rate vectors that satisfy the conservation equation~\eqref{eq:system}.
We now build on these results to characterize and optimize the performance \emph{actually} achieved by stable policies.
This section is organized as follows. 
\Cref{subsec:performance} introduces our two main performance metrics, \emph{delay} and \emph{regret}.
\Cref{sec:vertex-optimality} recalls that optimizing linear rewards amounts to reaching a vertex of the polytope. We show that when the vertex is bijective, the optimal solution is trivially achieved by a stable policy; in contrast, if the vertex is injective-only, no stable policy can achieve it.
\Cref{sec:existing-policies} summarizes what existing policies achieve and situates our contribution with respect to the literature.
\Cref{sec:achievability} presents $\epsilon$-filtering, a parameterized policy that can get arbitrarily close to a vertex, even when it is injective-only, with a guaranteed trade-off in performance.
Finally, \Cref{sec:comp_to_lit} relates our results to the most recent work on matching policy optimization.

\subsection{Performance criteria.} \label{subsec:performance}

Delay and regret are our two main performance indicators. Later in this section, we will show that these indicators can sometimes be antagonistic, meaning that certain matching problems may require trade-offs between them.

\paragraph*{Delay.}

In a stable matching model $(G, \lambda, \Phi)$,
the \emph{delay} is defined as the long-term average waiting times of the items
(with waiting equal to zero if the item is matched immediately upon arrival).
By Little's law~\cite{L61,L11}, the delay $D(\Phi)$ is given by
$$D(\Phi) = \frac{\sum_{i \in V} \mathbb{E}[Q^\infty_i]}{\sum_{i \in V} \lambda_i},$$
where $Q^\infty = (Q^\infty_1, Q^\infty_2, \ldots, Q^\infty_n)$ is a random vector
distributed according to the stationary distribution of the Markov chain~$Q = (Q_t, t \in \mathbb{N})$ under policy~$\Phi$.

\paragraph*{Regret.} To turn a matching rate vector into a scalar performance metric, we assume that the \emph{quality} of a matching rate vector can be evaluated by some \emph{reward function} $f: \mu\in \R^m \rightarrow f(\mu)\in\R$. The \emph{regret} of a policy $\Phi$ is then defined as 
$$R(\Phi) = f_{\sup} - f(\mu(\Phi)),$$
where $f_{\sup} = \sup_{\Psi: (G, \lambda, \Psi)\text{ stable}} f(\mu(\Psi))$ is the supremum of the rewards achievable by any stable policy.

Unless otherwise stated, we consider \emph{linear} reward functions of the form 
$$f: \mu\in\R^m \rightarrow r^\intercal \mu,$$
where $r = (r_1, \ldots, r_m) \in \R^m$ is a vector of rewards associated with the edges of $G$.

	In reward-driven analyses of matching models, a single regret metric is commonly used to quantify total reward loss compared to a policy that knows future arrivals.
	In effect, this global regret encompasses both metrics considered in our study: delay, which reflects queue sizes and is related to short-term regret (\cite[Theorem 3.1 and discussion]{KAG24}), and regret as we define it, which coincides with the long-term regret metric in \cite[Definition 2.2, Equation (9)]{KAG23}.
	
	Delay fundamentally measures the stability of the system. It depends solely on the actual dynamics of the network, irrespective of the reward vector, and is therefore critical in applications such as organ donation, assemble-to-order systems, and online marketplaces, where reducing the waiting time of agents is essential. In contrast, (long-term) regret captures the cumulative shortfall in reward that is due to suboptimal matching decisions. Recognizing this essential difference, which sets up inherent tensions explored in this section, we deliberately keep these metrics separate throughout our analysis.
	
	As a first example, consider the matching problem in \Cref{fig:unstable-optimal} with associated rewards $(1, 1, 1, 0, 0)$, i.e. matches involving class~4 are not rewarded. In this case, one can verify that $r$ is orthogonal to $\ker(A)$, so that all stable policies yield the same reward, equal to~5. On the other hand, if matching class-4 items are abandoned and a \acrfull{ml} policy restricted to the triangle $\{1, 2, 3\}$ is applied, the resulting matching model has infinite delay (class~4 items never leave the system) but achieves a higher reward of~6.
	
	This example highlights the trade-off between maximizing rewards and ensuring system stability, or more generally, keeping the expected total queue length small. The polytope approach provides a systematic way to identify and analyze such trade-offs. For readers specifically interested in reward maximization without stability guarantees, we refer to \citet{KAG23,KAG24}. 
	In \Cref{sec:case-study-unstable-policy}, we further demonstrate that the issue  illustrated by \Cref{fig:unstable-optimal} actually has a simple solution within the hypergraph formalism: the addition of a \emph{discarding} mono-edge.
	
	\begin{figure}[ht]
		\centering
		\begin{tikzpicture}
			\def\d{2cm}
			\bifan
			\node[left of =1] {$\lambda_1=4$};
			\node[above=-.1cm of 2] {$\lambda_2=4$};
			\node[below=-.1cm of 3] {$\lambda_3=4$};
			\node[right of =4] {$\lambda_4=2$};
			\draw (1) edge node[above, sloped] {$2 - \alpha$} (2)
			(1) edge node[above, sloped] {$2 + \alpha$} (3)
			(4) edge node[above, sloped] {$1 + \alpha$} (2)
			(4) edge node[above, sloped] {$1 - \alpha$} (3)
			(2) edge node[right] {$1$} (3);
		\end{tikzpicture}
		\caption{A matching problem where all stable policies have the same reward if $r=(1, 1, 1, 0, 0)$. \label{fig:unstable-optimal}}
	\end{figure}

\subsection{Vertex optimality.}\label{sec:vertex-optimality}

We first recall a classical result from convex optimization;
namely, a linear
optimization problem defined on a convex polytope
always admits a vertex of the polytope as solution.
More specifically, \Cref{prop:vertices-optimal} states that, for optimizing a linear reward function,
it is sufficient to reach a vertex of $\Lann$.
\Cref{prop:reward-is-vertex} shows that, most of the time, it is also necessary.

\begin{prop}\label{prop:vertices-optimal}
	Let $(G, \lambda)$ be a stabilizable matching problem with a surjective-only compatibility graph~$G$, and let $r = (r_1, \ldots, r_m) \in \R^m$ be a vector of rewards associated with the edges of $G$.
	Consider the problem of finding a solution to~\eqref{eq:system} with non-negative components that maximizes the reward rate $r^\intercal \mu$, and let
	$F = \{\mu\in \Lann : r^\intercal \mu
	= \max_{z \in \Lann} r^\intercal z\}$. 
	Then $F$ is a non-empty face of $\Lann$. In particular, there exists a vertex $\mu\in \Lann$ that maximizes the reward (i.e., $\mu\in F$).
\end{prop}

\begin{proof}
	This is a standard result from convex optimization. Since $\Lann$ is closed and bounded, there is a maximum $r_{\max}$ among the rewards associated with vectors inside~$\Lann$. The set~$F$ is the intersection of the hyperplane $\{y \in \R^m: r^\intercal y = r_{\max}\}$ with $\Lann$. Therefore, it is a non-empty face of $\Lann$. That any non-empty face contains a vertex follows from the lattice structure of the faces of polytopes.
\end{proof}

\begin{prop}
	\label{prop:reward-is-vertex}
	Let $(G, \lambda)$ be a stabilizable matching problem with a surjective-only compatibility graph~$G$, and let $r\in \Delta^{m-1}$ be a vector of rewards drawn according to a positive Lebesgue density over $\Delta^{m-1}$, the standard simplex of $\R^m$. Let $F = \{\mu\in \Lann : r^\intercal \mu
	= \max_{z \in \Lann} r^\intercal z\}$ be the set of the optimal matching rate vectors for $r$. Then, with probability 1, $F=\{\mu_r\}$, where $\mu_r$ is a vertex of $\Lann$.
\end{prop}

\begin{proof}
	As in \Cref{prop:vertices-optimal}, the non-empty face~$F$ is the intersection of the hyperplane $\{y \in \R^m: r^\intercal y = \max_{z \in \Lann} r^\intercal z\}$ with $\Lann$. If its dimension is more than $0$, it contains
	a $1$-dimensional face of $\Lann$ (an \emph{edge}). Let $b\in \R^m\setminus\{0\}$ be a direction of
	that edge. We have $b^\intercal r=0$. The set $\{r\in\Delta^{m-1}: b^\intercal r=0\}$, if it exists, has a dimension at most $m-2$, so the probability that $b^\intercal r=0$ is $0$. As the number of edges of $\Lann$ is finite, we conclude that, with probability 1, $F$ does not contain any edge of $\Lann$, which means it is reduced to a single vertex.
\end{proof}

Building on \Cref{prop:vertices-optimal,prop:reward-is-vertex}, a natural question is: \emph{Can a given vertex be achieved by a stable policy?} It is answered by \Cref{coro:achievable}.

\begin{prop} \label{coro:achievable}
	Let $(G, \lambda)$ be a stabilizable matching problem with a surjective-only compatibility graph~$G$. Let $\La_\Pol$ be the set of matching rate vectors achieved by stable policies adapted to $(G, \lambda)$, $\mu$ a vertex of the polytope $\Lann$ associated to $(G, \lambda)$, and $E^\star\subset E$ the support of $\mu$. We distinguish two cases, depending on whether $\mu$ is bijective or injective-only (cf \Cref{prop:vertex}).
	\begin{enumerate}[(i)]
	\item If $\mu$ is bijective (meaning that $|E^\star| = n$), then $\mu \in \La_\Pol$. In fact, $\mu = \mu(\mles)$, where $\mles$ is the \acrfull{ml}  policy adapted to~$G$ with a filter on~$\E^\star$, i.e. the \gls{ml} policy applied to the subgraph $G^\star=(V, E^\star)$.
	\item If $\mu$ is injective-only (meaning that $|E^\star| < n$), then $\mu\notin \La_\Pol$: no stable policy can achieve $\mu$.
	\end{enumerate}
\end{prop}

\begin{proof}
	Let $G^\star=(V, \E^\star)$ be the subgraph of $G$ associated with $\mu$, $p=|E^\star|$ be the number of positive coordinates of $\mu$, and let $A^\star$ denote the $n\times p$ incidence matrix of~$G^\star$. As $G^\star$ is injective (\Cref{prop:vertex}), the restriction~$\tilde{\mu}$ of the vector~$\mu$
	to its positive coordinates is the only solution
	of the conservation equation $A^\star \, \tilde{z} = \lambda$, of unknown $\tilde{z} \in \R^p$. We now consider the two cases separately:
	\begin{enumerate}[(i)]
		\item If $G^\star$ is bijective,
		\Cref{prop:stability-region-form}
		implies that the matching problem $(G^\star, \lambda)$
		is stabilized by the \gls{ml} policy: $G^\star$ is bijective
		and $\tilde{\mu}$, the unique solution to the conservation equation~$A^\star z = \lambda$, has positive coordinates. The unicity of $\tilde{\mu}$ ensures that it is the matching rate achieved by that policy.
		To prove $\mu\in \La_\Pol$, we consider the \gls{ml} policy with a filter on $\E^\star$ on the matching problem $(G, \lambda)$, denoted $\mles$. $\mles$ behaves exactly like the greedy \gls{ml} policy on $(G^\star, \lambda)$, which is stable with matching rate $\tilde{\mu}$, as we just saw. Hence, the model $(G, \lambda, \mles)$ is stable, and its matching rate, which is $\tilde{\mu}$ on $E^\star$ and $0$ elsewhere, coincides with $\mu$. This proves that $\mu \in \La_\Pol$.
		\item If $G^\star$ is injective-only, \Cref{prop:stability-region-nonempty} implies that the matching problem
		$(G^\star, \lambda)$ is not stabilizable.	
		We prove that $\mu \notin \La_\Pol$ by contradiction.
		Suppose that $\mu\in \La_\Pol$, and let $\Phi$ be a stable policy on $(G, \lambda)$ such that $\mu(\Phi) = \mu$.
		We have $\mu_k(\Phi) = 0$ for each $k \in \E \setminus \E^\star$. But the matching rate $\mu_k(\Phi)$ along an edge~$e_k \in \E$ with endpoints~$i$ and~$j$ is also given by
		\begin{align*}
			\mu_k(\Phi)
			&= \lambda_i \sum_{s, s^\prime \in \cS} \pi(s) \Phi(s, i, j, s^\prime)
			+ \lambda_j \sum_{s, s^\prime \in \cS} \pi(s) \Phi(s, j, i, s^\prime),
		\end{align*}
		where $\pi$ is the equilibrium distribution of the Markov chain $(S_t, t \in \N)$.
		Since $\lambda_i > 0$, $\lambda_j > 0$, and the distribution $\pi$ is positive on its support~$\cS$,
		it follows that the matching rate $\mu_k(\Phi)$ is zero
		if and only if $\Phi(s, i, j, s^\prime) = \Phi(s, j, i, s^\prime) = 0$
		for each $s, s^\prime \in \cS$,	that is, items of classes~$i$ and~$j$ are
		\emph{never} matched with one another under policy~$\Phi$.	
		In this case, $\Phi$ also defines a stable policy on the matching problem $(G^\star, \lambda)$, which contradicts our assumption that the matching problem $(G^\star, \lambda)$ is not stabilizable. Therefore, $\mu\notin \La_\Pol$.	
	\end{enumerate}%
\end{proof}

\Cref{coro:achievable} shows that optimizing linear rewards is straightforward when the vertex is bijective: we simply need to forbid the edges outside the support of the vertex. Thus, no inherent trade-off exists between reward maximization and system stability in this case. Note that a trade-off can still exist between optimizing the reward and minimizing the expected sum of queue lengths, as forbidding a subset of edges limits the flexibility of the policy. 
In contrast, for the injective-only case, optimality requires a trade-off with stability. Before presenting a policy that navigates this balance, we summarize in \Cref{sec:existing-policies} what existing policies already achieve.

\subsection{What existing policies achieve.}
\label{sec:existing-policies}

Before introducing our policy, we summarize what existing matching policies achieve and delineate the gap that our contribution fills. Recall from \Cref{coro:achievable} that a bijective vertex can be reached by a stable policy, whereas an injective-only vertex cannot; the relevant questions are therefore \emph{how closely}, and \emph{at what cost in queue length}, each type of vertex can be approached.

\paragraph*{Bijective vertices.}
When the optimal vertex~$\mu$ is bijective, it can be achieved \emph{exactly} by a stable policy: it suffices to run the \gls{ml} policy with a filter on the support~$\E^\star$ of~$\mu$ (\Cref{coro:achievable}(i)), which keeps the expected queue length bounded. This is the match-the-longest-queue optimal policy of \citet{KAG23} in the regime where their \glsf{gpc} holds. Consequently, for a bijective vertex, there is no trade-off between reward maximization and stability, only a milder one between reward and queue length. Since the optimal vertex is bijective with probability one when the reward vector is generic (\Cref{prop:reward-is-vertex}), this case already covers most instances.

\paragraph*{Injective-only vertices.}
When the optimal vertex is injective-only, no stable policy achieves it exactly, and the two existing lines of work behave differently. The \gls{ml}
policy of \citet{KAG23}, which underlies their queue-length bound, does \emph{not} extend to this case: restricted to an injective-only support, it makes the system unstable. The \glsf{egpd} policy of \citet{NS19}, in contrast, approaches \emph{any} vertex (bijective or injective-only) while guaranteeing stability, but it offers no explicit control over the queue lengths (its reward--delay trade-off is observed only empirically).

\paragraph*{Our contribution.}
The $\epsilon$-filtering policy introduced next closes this gap: it carries the reachability \emph{and} the explicit queue-length bounds of \citet{KAG23} over to injective-only vertices. For every $\epsilon > 0$ it produces a stable system whose matching rate is $O(\epsilon)$-close to the target injective-only vertex while its expected queue length stays $O(1/\epsilon)$ (\Cref{prop: converging_to_vertex}); and we prove this trade-off to be fundamental through a matching $\Omega(1/\epsilon)$ lower bound (\Cref{prop: lower_bound}). Like the policies of \citet{KAG23} and \citet{NS19}, ours is a plain variant of match-the-longest; its one new ingredient is
to \emph{mark} an $\epsilon$-fraction of arrivals that may use every edge, just enough to restore stability, as explained in \Cref{sec:achievability}.
A precise, framework-by-framework comparison is deferred to \Cref{sec:comp_to_lit} and \refapp{app:comp_to_lit}.

\subsection{\texorpdfstring{$\epsilon$}{epsilon}-filtering policy.}
\label{sec:achievability}
Given an injective vertex~$\mu$ of~$\Lann$, we now construct a sequence of policies, that we call $\epsilon$-filtering policies, which converge to the vertex $\mu$ as $\epsilon \downarrow 0$ while also upper bounding the expected queue lengths. Note that the distance to $\mu$ and the expected queue length are directly tied to our main scalar performance metrics, namely regret and delay\footnote{%
Queue length and delay are related by Little's law (see \Cref{subsubsec:stability}).
Consider a stable policy $\Phi$ with matching rate $\mu(\Phi)\in \Lann$ and regret $R(\Phi)$. 
We have $R(\Phi)=r^\intercal(\mu-\mu(\Phi))=\cos(\theta)\,||r||_2\cdot||\mu-\mu(\Phi)||_2$, where $\theta$ denotes the angle between $r$ and  $\mu-\mu(\Phi)$.
In particular, $R(\Phi) \le ||r||_2\cdot||\mu-\mu(\Phi)||_2$: if you are close to $\mu$, your regret is low.
Conversely, if $\kappa$ denotes the minimum possible cosine between $r$ and $\mu-\mu'$ for $\mu'\in \Lann\setminus\{\mu\}$ (i.e., $\kappa=\min_{\mu'\in \Lann\setminus\{\mu\}} \frac{r^\intercal(\mu-\mu')}{||r||_2\cdot||\mu-\mu'||_2}$), we obtain $R(\Phi)\geq \kappa ||r||_2\cdot||\mu-\mu(\Phi)||_2$.
Therefore,  $\kappa ||r||_2\cdot||\mu-\mu(\Phi)||_2\leq R(\Phi)\leq ||r||_2\cdot||\mu-\mu(\Phi)||_2$.
Note that if $\mu$ is the unique optimal solution (as is often the case; see \Cref{prop:reward-is-vertex}), then $\kappa>0$: if you are far from $\mu$, your regret is high.
}. The principle of $\epsilon$-filtering policies is as follows: A priori, only matches within the target set $E^\star$ are performed. However, when a new item arrives, it is marked with low ($\epsilon$) probability as eligible for all matches in $\E$. For a match outside $\E^\star$ to be made, the selected pair must contain at least one such marked item.
Marks allow some matches to be made outside $\E^\star$, which ensures stability. Their low probability ensures that the resulting matching rate vector is close to the target vertex $\mu$.

Formally, given $0 < \epsilon < 1$, we label each incoming arrival by
`$-$' (marked) with probability $\epsilon$ and `$+$' (unmarked) otherwise,
independently of everything else.
An item with class~$i$ and label $\ell \in \{+, -\}$ is said to have type-$i^\ell$. We denote the number of items of type~$i^\ell$ at time $t$ by $Q_{t, i^\ell}$. Thus, the total number of class-$i$ items at time $t$ is $Q_{t, i} = Q_{t, i^+} + Q_{t, i^-}$. Now, we restrict our attention to policies that do \emph{not} match type-$i^{+}$ items to type-$j^{+}$ items for all $\{i, j\} \in \E \backslash \E^\star$. The policies that follow this restriction in the original graph~$G$ can be viewed as policies operating on the augmented graph $G^\prime = (\V^\prime, \E^\prime)$, with $\V^\prime = \left\{i^\ell: i \in \V, \ \ell \in \{+, -\}\right\}$ and $\E^\prime = \E^+ \cup \E^{\pm} \cup \E^{-}$, where
\begin{align*}
	\E^{+} &= \left\{\{i^{+}, j^{+}\} : \{i, j\} \in \E^\star\right\},
	\\
	\E^{\pm} &= \left\{\{i^{+}, j^{-}\} : \{i, j\} \in \E\right\},
	\\
	\E^{-} &= \left\{\{i^{-}, j^{-}\} : \{i, j\} \in \E\right\}.
\end{align*}
The arrival-rate vector in the augmented graph is given by $\lambda_\epsilon^\prime \in \Rnn^{2n}$ with $\lambda_{\epsilon, i^+}^\prime = (1 - \epsilon) \lambda_i$ and $\lambda_{\epsilon, i^-}^\prime = \epsilon \lambda_i$ for each $i \in \V$. An $\epsilon$-filtering policy adapted to $(G, \lambda)$ with filter $\E^\star$ refers to a  policy adapted to $(G^\prime, \lambda_\epsilon^\prime)$.
For example, an $\epsilon$-filtering \gls{ml} policy adapted to $(G, \lambda)$ with filter $\E^\star$ mimics the \gls{ml} policy adapted to $(G^\prime, \lambda_\epsilon^\prime)$. To be more precise, for a given state $q \in \Rnn^{2n}$, an incoming item of type~$i^\ell$ is matched with a type in the set $\arg \max_{j^{\ell^\prime} \in \V^\prime: \{i^{\ell}, j^{\ell^\prime}\} \in \E^\prime} q_{j^{\ell^\prime}}$.

Note that, for any greedy policy adapted to $(G^\prime, \lambda_\epsilon^\prime)$, most arrivals are matched using edges in~$\E^+$, or equivalently $\E^\star$, as items are labeled `$+$' with probability $1-\epsilon$. As $G^\star$ is the support graph of the vertex~$\mu$, we can then show that such a policy ensures that the matching rate on $\{i^+, j^+\}$ is $O(\epsilon)$ close to $\mu_{i, j}$ for all $\{i, j\} \in \E^\star$. Next, to bound the expected queue lengths, we define $\delta(G, \lambda)$, the \gls{crpg}, as follows:
\begin{align}
	\delta(G, \lambda) = \min_{\I \in \ind} \left\{ \sum_{j \in V(\I)} \lambda_j - \sum_{i \in \I} \lambda_i\right\}. \label{eq:crp_gap}
\end{align}
Note that, by \Cref{prop:stability-region-form}, the matching problem $(G, \lambda)$ is stabilizable if and only if $\delta(G, \lambda) > 0$. Intuitively, $\delta(G, \lambda)$ characterizes the minimum slack between the arrival rate of any independent set $\I$ and its neighbors $V(\I)$. The arrival rate of the neighbors acts as a service rate for $\I$, and the \gls{crpg} $\delta(G, \lambda)$ is reminiscent of the heavy-traffic parameter for a single-server queue. The \gls{crpg} is also related to the \glsf{gpg} \citep{KAG23}, which we discuss in \Cref{sec:comp_to_lit}. Now, we are ready to state the main result of this section that formalizes the trade-off between reward and expected queue lengths.
\begin{thm} \label{prop: converging_to_vertex}
	Let $(G, \lambda)$ be a stabilizable matching problem with a surjective-only compatibility graph~$G$, and consider a vertex $\mu$ of $\Lann$ with support graph $G^\star = (V, \E^\star)$, where $\E^\star = \{k \in E: \mu_k > 0\}$. For each $\epsilon > 0$, let $\mle$ denote the $\epsilon$-filtering \gls{ml} policy adapted to $(G, \lambda)$ with filter $\E^\star$. Then, $(G,\lambda,\Phi_\epsilon)$ is stable and there exist $C_1, C_2 \in \Rnn$ such that, for all $\epsilon \in (0, 0.5)$, we have
	\begin{align*}
		\|\mu\left(\mle\right) - \mu\|_1 \leq C_2 \epsilon, \quad
		\mathbb{E}\left[\sum_{i \in \V} Q_i\right] \leq \frac{C_1}{\min\left\{\epsilon\min_{i \in \V} \lambda_i + \delta(G^\star, \lambda), \delta(G, \lambda)\right\}},
	\end{align*}
	where $Q = (Q_1, Q_2, \ldots, Q_n)$ is the steady-state queue-length vector in $(G, \lambda, \Phi_\epsilon)$.
\end{thm}
Now we present a sketch of the proof of \Cref{prop: converging_to_vertex} that highlights the intuition behind the result.

\begin{proof}[Sketch of the Proof (See \refapp{app: converging_to_vertex} for the complete proof)]
	We first consider an arbitrary matching problem $(G, \lambda)$. 
	Note that the \gls{crpg}  $\delta(G, \lambda)$ is reminiscent of the heavy-traffic parameter for a single-server queue. We make this intuition rigorous by showing that, under the \gls{ml} policy, we have
	\begin{align}
		\mathbb{E}\left[\sum_{i \in V} Q_i\right] = O\left(\frac{1}{\delta(G, \lambda)}\right). \label{eq: expected_q_ht}
	\end{align}
	The above result holds for any matching problem $(G, \lambda)$ operating under the \gls{ml} policy as long as $\delta(G, \lambda) > 0$ (equivalently, whenever the matching problem $(G, \lambda)$ is stabilizable). Now, we use the above result to analyze the $\epsilon$-filtering \gls{ml} policy adapted to $(G, \lambda)$ with filter $\E^\star$. Recalling that this is exactly the \gls{ml} policy adapted to $(G^\prime, \lambda_\epsilon^\prime)$, we turn our attention to the augmented matching problem $(G^\prime, \lambda_\epsilon^\prime)$ and prove the following:
	
	\textbf{Expected queue length bound:}
	We show that  $\delta(G^\prime, \lambda_\epsilon^\prime) = \Omega(\min\{\epsilon + \delta(G^\star, \lambda), \delta(G, \lambda)\})$, and so, the expected queue length is $O(1/\min\{\epsilon + \delta(G^\star, \lambda), \delta(G, \lambda)\})$ by \eqref{eq: expected_q_ht}. The idea is that the edges in $\E^{+}$ contribute $\delta(G^\star, \lambda)$ to the \gls{crpg} because $\E^{+}$ mimics $\E^\star$. In addition, matching $\Theta(\epsilon)$ arrivals using the edges in $\E^{\pm} \cup \E^{-}$ contributes an additional $\Omega(\epsilon)$ to the \gls{crpg} because $\E^{\pm} \cup \E^{-}$ mimics $\E$. Thus, we have $\delta(G^\prime, \lambda_\epsilon^\prime) = \Omega(\min\{\epsilon + \delta(G^\star, \lambda), \delta(G, \lambda)\})$, which by~\eqref{eq: expected_q_ht} implies the expected queue length in steady state is at most $O(1/\min\{\epsilon + \delta(G^\star, \lambda), \delta(G, \lambda)\})$. The minimum with $\delta(G, \lambda)$ is intuitive, as $\delta(G, \lambda)$ is the \gls{crpg} of the original graph $G$, which is the best-case scenario of using all edges to make matches.
	
	\textbf{Matching rate is $O(\epsilon)$ close to $\mu$:} We show that we are $O(\epsilon)$ close to the vertex $\mu$ by bounding the frequency of ``bad'' matches, i.e., the matches that do not correspond to the support graph $G^\star$. Such edges are of the form $(i^{+}, j^{-})$ or $(i^{-}, j^{-})$. We use the conservation equations given by \eqref{eq:system-equations} along with the fact that the arrival rate to vertices $\{i^{-}: i \in \V\}$ is $O(\epsilon)$ to prove this statement.
	
	Combining the two steps above completes the proof of the proposition.
\end{proof}
Now we discuss the implications of the above theorem separately for bijective and injective-only vertices.

\subsubsection{Approaching a bijective vertex.}

When $\mu$ is a bijective vertex, we have $\delta(G^\star, \lambda) > 0$ because $(G^\star, \lambda)$ is stabilizable (Proposition~\ref{coro:achievable}). Thus, we can let $\epsilon \rightarrow 0$ in \Cref{prop: converging_to_vertex} to conclude $\mu(\Phi_0) = \mu$ and $\mathbb{E}\left[\sum_{i \in \V} Q_i\right] \leq \frac{C_1}{\delta(G^\star, \lambda)} < \infty$. Thus, we observe no trade-off between stability and approaching a bijective vertex. In fact, the $\epsilon$-filtering ML policy converges to the ML policy with filter $E^\star$ as $\epsilon \rightarrow 0$, which is stable by Proposition~\ref{coro:achievable}. Moreover, \Cref{prop: converging_to_vertex} implies that the trade-off between the reward and the expected queue lengths still exists. In particular, if $\delta(G^\star, \lambda) < \delta(G, \lambda)$, then, as $\epsilon$ increases, the expected queue length decreases at the expense of moving away from the vertex $\mu$. Intuitively, as $\epsilon$ increases, it happens more often that we use all edges in the graph to make matches, which helps to reduce the expected queue lengths. At the same time, it happens more often that we use the edges in $E \backslash E^\star$, resulting in moving away from the vertex $\mu$. Note that the expected queue length is always $O(1/\delta(G, \lambda))$ for any choice of $\epsilon > 0$. This upper bound corresponds to the best-case scenario (see \refappnc{lemma: stability_generic} in \refapp{app: converging_to_vertex}) of using all the edges in $E$ to make matches. In \Cref{sec:numerical-results}, we analyze this trade-off via simulations and observe a mild trade-off (compared to the injective-only case) between the two objectives, aligning with the result of \Cref{prop: converging_to_vertex}.

\subsubsection{Approaching an injective-only vertex.} \label{sec:approaching-injective-only-vertices}

For an injective-only vertex, we have $\delta(G^\star, \lambda) = 0$, and so, \Cref{prop: converging_to_vertex} implies the following:
\begin{align}
\|\mu\left(\mle\right) - \mu\|_1 \leq C_2 \epsilon, \quad
		\mathbb{E}\left[\sum_{i \in \V} Q_i\right] \leq \frac{C_1}{\min\left\{\epsilon\min_{i \in \V} \lambda_i, \delta(G, \lambda)\right\}}. \label{eq:trade-off}     
\end{align}
As $\epsilon \rightarrow 0$, the expected queue length increases to infinity, substantiating that the matching problem $(G^\star, \lambda)$ is unstable as shown in Proposition~\ref{coro:achievable}. Equation~\eqref{eq:trade-off} establishes the trade-off between approaching an injective-only vertex and ensuring a stable system: it shows an $O(1/\epsilon)$ expected queue length bound, while being $\epsilon$ close to the injective-only vertex $\mu$.

\begin{exa}[$\epsilon$-filtering on the diamond]\label{ex:eps-filtering}
	Consider the diamond graph of \Cref{fig:illustration} with degree-proportional arrival rates $\lambda = (2, 3, 3, 2)$ and reward vector $r = (1,\, 1.1,\, 1,\, 0.1,\, 1)$, where the edges are
	in lexicographical order ($\{1,2\}, \{1,3\}, \{2,3\}, \{2,4\}, \{3,4\}$),
	so that edge $\{1,3\}$ carries the largest reward. Here $\Lann$ is the segment $\mu = (1+\alpha,\, 1-\alpha,\, 1,\, 1-\alpha,\, 1+\alpha)$, $\alpha \in [-1,1]$, and both of its vertices, $(0,2,1,2,0)$ and $(2,0,1,0,2)$, are injective-only (each has only three positive coordinates). Since $r^\intercal\mu = 4.2 + 0.8\,\alpha$, the reward is maximized at the vertex $\mu^\star = (2,0,1,0,2)$, supported on the path $\E^\star = \{\{1,2\}, \{2,3\}, \{3,4\}\}$. The stable optimum thus \emph{disables} the highest-reward edge $\{1,3\}$, whose presence is nevertheless required for stabilizability: reward maximization and stability are in conflict here.
	
	Because $\mu^\star$ is injective-only, no stable policy attains it (\Cref{coro:achievable}), and the \gls{ml} policy filtered to $\E^\star$ is unstable, since $\delta(G^\star, \lambda) = 0$. This is the degenerate regime in which the general-position condition of \citet{KAG23} fails (its correspondence with $\delta$ is detailed in \Cref{sec:comp_to_lit}). The $\epsilon$-filtering \gls{ml} policy, in contrast, is stable for every $\epsilon > 0$: since $\delta(G, \lambda) = 2 > 0$, \Cref{prop: converging_to_vertex} yields a matching rate within $O(\epsilon)$ of $\mu^\star$ at the price of an $O(1/\epsilon)$ expected queue length. We evaluate this trade-off numerically for the same graph in \Cref{sec:numerical-results}.
\end{exa}

We now show that this trade-off is fundamental:
no sequence of matching policies can converge to an injective-only vertex of $\Lann$ while maintaining finite queue lengths.

\setcounter{propositionlowerbound}{\thethm}
\begin{thm} \label{prop: lower_bound}
	Consider an injective-only vertex $\mu$ of $\Lann$.
	For each policy $\Phi$ adapted to $G$
	and such that the matching model $(G, \lambda, \Phi)$ is stable,
	if $\|\mu(\Phi)-\mu\|_1 \leq \epsilon$,
	then $\mathbb{E}\left[\|Q\|_2^2\right] \geq \Omega\left(1/\epsilon\right)$.
\end{thm}

\begin{proof}
See \refapp{app: lower_bound}.
\end{proof}

Two remarks on how this lower bound relates to the upper bound of \Cref{prop: converging_to_vertex}. First, the bounds are stated for \emph{different} moments and norms: \Cref{prop: lower_bound} lower-bounds the second moment $\mathbb{E}[\|Q\|_2^2]$, whereas \Cref{prop: converging_to_vertex} upper-bounds the first moment $\mathbb{E}[\sum_i Q_i]$; the two therefore do not match directly. Since $\mathbb{E}[\|Q\|_2^2] \ge \mathbb{E}[\|Q\|_1]^2/n$ by Cauchy--Schwarz, the second-moment bound does not by itself imply an $\Omega(1/\epsilon)$ lower bound on the first moment, and we do not know whether the $\Omega(1/\epsilon)$ order is tight for either moment; we conjecture that a matching $\Omega(1/\epsilon)$ lower bound holds already for the first moment. Second, regardless of this gap, \Cref{prop: lower_bound} establishes that the trade-off between approaching an injective-only vertex and keeping the queues small is fundamental, and it recalls that no such trade-off arises for bijective vertices.

\begin{rem}
	\refapp{app: lower_bound} actually proves
	a more general version of \Cref{prop: lower_bound}
	where we consider matching policies that can form
	at most~$\mathcal{M}$ pairs of items at each time step,
	for some $\mathcal{M} \in \mathbb{N}_{> 0}$.
\end{rem}

\subsection{Connections to the literature.}
\label{sec:comp_to_lit}

Reward maximization in stochastic matching has recently received considerable attention \citep{NS19, WXY23, G24, KAG23, KAG24}. We take a different angle: we use stability, through the conservation equation~\eqref{eq:system}, to characterize the achievable matching rates globally, as a polytope, and then optimize over it. With the notable exception of \citet{NS19}, this line of work and ours (which builds on \citet{BCM21}) developed largely in parallel and independently, over a similar period; several findings therefore overlap, and where they do we attribute them to the earliest source. We outline the correspondence here and defer the detailed, framework-by-framework comparison to \refapp{app:comp_to_lit}.

The reward-maximization literature centers on the \glsf{spp}, a linear program whose optimal solution, under the \glsf{gpc} of \citet{KAG23}, is a unique non-degenerate vertex, precisely a \emph{bijective} vertex in our terms. In this regime the two frameworks coincide: the filtered \gls{ml} policy of \Cref{coro:achievable} is the
\gls{ml}
optimal policy of \citet{KAG23}, and their \glsf{gpg}~$\tilde{\epsilon}$ (a robustness margin of the optimal solution) is closely related to the \glsf{crpg}~$\delta(G, \lambda)$ of~\eqref{eq:crp_gap}, which itself already appears in \citet[Section~5.1]{KAG23}; the precise relation $\delta(G, \lambda) \ge \delta(G^\star, \lambda) \ge \tilde{\epsilon}$ is established in \refapp{app:comp_to_lit}. Our results depart from this literature exactly when the \gls{gpc} fails, that is, when the optimal vertex is injective-only: then $\tilde{\epsilon} = 0$, the policy of \citet{KAG23} is no longer stable, and one must resort to $\epsilon$-filtering or to the \glsf{egpd} policy of \citet{NS19}, whose rewardless, stability-only variant we call \gls{vqml} and reuse for hypergraphs in \Cref{sec:hypergraphs}. \Citet{G24} and \citet{WXY23} extend the primal--dual approach to multi-way matching and to a dual form of the \gls{gpc}; we relate them to our polytope in \refapp{app:comp_to_lit}. A numerical comparison of all these policies is reported in \Cref{sec:numerical-results}.

 % sections/06-applications
\section{Applications.} \label{sec:applications}

Having characterized how a target matching rate can be
achieved
in principle, we now turn to two more concrete questions.
\Cref{sec:numerical-results} evaluates, through simulations, how the policies introduced in this article perform in practice, illustrating the delay--regret trade-off.
\Cref{subsec:greedy} then asks whether the simple and widely used class of \emph{greedy} policies suffices for this optimization, and it shows that the answer is no:
no greedy policy can reach a vertex of $\Lann$, and sometimes no greedy policy can even approach such a vertex.

 % sections/06-numerical-results
\subsection{Practical performance.} \label{sec:numerical-results}

Let $(G, \lambda)$ be a stabilizable problem, $r\in \R^m$ a vector of rewards, and $\mu$ a vertex of $\Lann$ that optimizes the average reward~$r^\intercal \mu$. Let $\E^\star \subsetneq E$ denote the edges of $G$ that support $\mu$. In this section, we use simulations to quantitatively evaluate the performance of various policies in achieving~$\mu$.
\Cref{sec:considered-policies} introduces the policies under consideration, while \Cref{sec:methodology} outlines the simulation methodology. \Cref{sec:simu-diamond-injective,sec:simu-diamond-bijective} analyze the injective-only and bijective cases, respectively. Finally,  \Cref{sec:simu-discussion} presents a discussion of the results.

\subsubsection{Considered policies.}\label{sec:considered-policies}

For ease of display, we use $\Phi$ to denote all matching policies, distinguishing them by a subscript indicating a specific policy (e.g., $\epsilon$ for the $\epsilon$-filtering, $\beta$ for \gls{egpd}).
More specifically, we consider the following policies:
\begin{description}
	\item[Filtering \gls{ml} policy:]
	The \gls{ml} policy restricted to $\E^\star$, denoted by $\mles$, is stable if and only if $\mu$ is bijective (see \Cref{coro:achievable}). In this case, it is optimal (i.e., incurs no regret), and its delay serves as a benchmark for other policies.
	In simulation plots (e.g., \Cref{fig:simu-dia-inj,fig:simu-dia-bij}),
	$\mles$ is represented by a vertical dotted line marking its delay.
	\item[$\epsilon$-filtering \gls{ml} policy:] 
	The $\epsilon$-filtering \gls{ml} policy with filter $\E^\star$, denoted by $\mle$ for $\epsilon \in (0, 1)$, is stable for any $\epsilon > 0$ and converges to $\mu$ as $O(\epsilon)$, with bounded delay in $O(1/\epsilon)$ (\Cref{prop: converging_to_vertex}).
	\item[$k$-filtering \gls{ml} policy:]
	The $k$-filtering \gls{ml} policy with filter $\E^\star$, denoted by $\mlk$ for $k\in \N$, applies the filter~$E^\star$ depending on the state: it uses $\mles$ if the length of the longest queue is less than~$k$; otherwise it applies the \gls{ml} policy without filter.
	\item[\gls{egpd} policy:]
	The \acrfull{egpd} algorithm, introduced by \citet{NS19} and denoted by $\egpd$ for $\beta \in (0, +\infty)$, is a stable policy that does not require explicit knowledge of $\lambda$ or $\E^\star$, but does take the reward vector~$r$ as input. It is based on the \acrfull{vqml} policy: matching decisions are driven by scores that combine the edge rewards with the lengths of virtual queues, the parameter~$\beta$ weighting the queue lengths relative to the rewards. The smaller $\beta$, the more reward-centric the policy (see \refapp{app:comp_to_lit} for details).
	\item[\gls{egpd}+ policy:]
	Denoted by $\egpdp$ for $\beta \in (0, +\infty)$, \gls{egpd}+ is a variant of \gls{egpd} with rewards \emph{adapted to $E^\star$}. In other words, the \gls{egpd}+ policy is obtained by applying the \gls{egpd} policy with an alternative reward vector that is positive on each edge of~$E^\star$ and negative on each edge of $E \setminus E^\star$.
	\item[\gls{crpd} policy:]
	The \acrfull{crpd}, introduced by \citet{WXY23}, is a variant of \gls{egpd} where rewards are adapted to $E^\star$ and the parameter $\beta$ decreases over time. The \gls{crpd} policy that uses $\beta(t)=1/t^{\alpha}$ for $\alpha>0$ is denoted $\crpd$. If $\beta$ goes to $0$, the policy is unstable on injective-only vertices. If it decreases sufficiently fast, $\crpd$ behaves like a filtering policy.
\end{description}
\Cref{tab:policies_simu} summarizes the selected policies and their key features.

\begin{table}[ht]
	\centering
	\begin{tabular}{|l|l|l|p{8.9cm}|}
		\hline
		Policy & Base & Parameter & Principle \\
		\hline
		$\mles$ & \multirow[c]{3}{*}{\text{ML}} & N/A & No matching outside $E^\star$. \\
		\cline{1-1}\cline{3-4}
		$\mle$ &  & $\epsilon$ & $\mles$, matchings outside $E^\star$ allowed for a fraction of items.\\
		\cline{1-1}\cline{3-4}
		$\mlk$ & & $k$ & $\mles$, matchings outside $E^\star$ allowed if the queue is large. \\
		
		\cline{1-4}
		
		$\egpd$ & \multirow{3}{*}{\text{VQML}} & \multirow{2}{*}{$\beta$} & Edge scores based on rewards and queue sizes. \\
		\cline{1-1}\cline{4-4}
		$\egpdp$ &  &  & $\egpd$, with rewards adapted to $E^\star$ (positive on $E^\star$, negative outside). \\
		\cline{1-1}\cline{3-4}
		$\crpd$ &  & $\alpha$ & $\egpd$, with adapted rewards $r'=r-UA$ and vanishing queue size impact. \\
		\hline
	\end{tabular}
	\caption{Overview of the matching policies evaluated.\label{tab:policies_simu}}
\end{table}

\begin{rem}
We propose two new policies for numerical evaluation, $\mlk$ and $\egpdp$.

The motivation for $\mlk$ is our expectation that it outperforms $\mle$ in practical scenarios. While $\mle$ consistently allows a small fraction of items to be matched outside $\E^\star$, $\mlk$ relaxes the constraint on allowed edges \emph{only when necessary}, that is, when delay is significantly affected. We provide in \refapp{app:k-filtering} elements to substantiate the conjecture that $\mlk$ offers theoretical guarantees similar to those of $\mle$.

We also introduce $\egpdp$, as it is a natural intermediate between $\egpd$ and its variant $\crpd$. Compared to $\egpd$, the policy $\crpd$ incorporates two additional features: adapted rewards and a decreasing $\beta$ parameter. Of the two intermediate combinations, a decreasing $\beta$ without reward adaptation usually makes the system unstable, so we discard it; $\egpdp$ implements the other combination, with a simplified reward adaptation.
\end{rem}

\subsubsection{Methodology.}\label{sec:methodology}

For measuring the performance criteria presented in \Cref{subsec:performance}, we used the Python package \emph{Stochastic Matching} of \citet{SM22}. This package enables us to verify the stabilizability of a matching problem $(G, \lambda)$, analyze the vertices of $\Lann$, and measure the regret and delay of policies through simulations.

For policies with parameters, we selected the following ranges:
\begin{itemize}
	\item $k\in [1, 2^{13}]$ for $k$-filtering;
	\item $\epsilon\in [10^{-3}, 1]$ for $\epsilon$-filtering;
	\item $\beta\in [10^{-3}, 10]$ for \gls{egpd} and \gls{egpd}+;
	\item $\alpha\in [0, 1]$ for \gls{crpd}.
\end{itemize}

Unless stated otherwise, our evaluation protocol is as follows: after selecting $G$ and $\lambda$, we choose a vertex $\mu$ of $\Lann$.
Next, we choose a reward vector~$r$ that is \emph{adversarial}, in the sense that the edges with maximal reward in~$r$ are not necessarily those that appear in the support~$E^\star$ of the target matching-rate vector~$\mu$ (while still ensuring that $\mu$ is the unique optimum for~$r$).
An example of adversarial reward already appeared in \Cref{ex:eps-filtering}.
Adversarial rewards will show the advantage of support-aware policies (i.e., policies that take~$E^\star$ as input). Indeed, note that only $\egpd$ takes the reward vector~$r$ as an input; other policies are directly parametrized by $E^\star$. This will show in particular the advantage of~$\egpdp$ compared to~$\egpd$.
Once parameters are fixed, all policies are evaluated by simulating $T=10^{10}$ arrivals and measuring regret and delay.

We focus here on the diamond graph from \Cref{fig:illustration}, but we conducted experiments on other graphs of various sizes (up to $100$ nodes) with qualitatively similar results. Some of them are available in \refapp{app:simu-additional-examples} and in the package documentation\footnote{\url{https://balouf.github.io/stochastic_matching/companion/simulations.html\#Vertices-of-simple-graphs}}.

\subsubsection{Approaching an injective-only vertex.}\label{sec:simu-diamond-injective}

To begin, we consider degree-proportional arrival rates, i.e., $\lambda = (2, 3, 3, 2)$. Our objective is to approach the injective-only vertex $\mu = (2, 0, 1, 0, 2)$, which effectively disables edges $\{1, 3\}$ and $\{2, 4\}$. We select the adversarial reward vector $r = (1, 2.9, 1, -1, 1)$, assigning the highest reward to edge $\{1, 3\}$.
The results are presented in \Cref{fig:simu-dia-inj}, with both axes in logarithmic scale.

\begin{figure}[ht]
	\centering
	 % figures/diamond-injective
\begin{tikzpicture}

\definecolor{darkgray176}{RGB}{176,176,176}
\definecolor{lightgray204}{RGB}{204,204,204}

\begin{axis}[width=12cm, height=7cm,
legend columns=3, transpose legend=true,
legend cell align={left},
legend style={
  fill opacity=0.8,
  draw opacity=1,
  text opacity=1,
  at={(0.03,0.03)},
  anchor=south west,
  draw=lightgray204
},
log basis x={10},
log basis y={10},
tick align=outside,
tick pos=left,
x grid style={darkgray176},
xlabel={Delay},
xmin=0.118317638524165, xmax=4803.42437939496,
xmode=log,
xtick style={color=black},
xtick={0.01,0.1,1,10,100,1000,10000,100000},
xticklabels={
  $\mathdefault{10^{-2}}$,
  $\mathdefault{10^{-1}}$,
  $\mathdefault{10^{0}}$,
  $\mathdefault{10^{1}}$,
  $\mathdefault{10^{2}}$,
  $\mathdefault{10^{3}}$,
  $\mathdefault{10^{4}}$,
  $\mathdefault{10^{5}}$
},
y grid style={darkgray176},
ylabel={Regret},
ymin=1e-08, ymax=0.4,
ymode=log,
ytick style={color=black},
ytick={1e-09,1e-08,1e-07,1e-06,1e-05,0.0001,0.001,0.01,0.1,1,10},
yticklabels={
  $\mathdefault{10^{-9}}$,
  $\mathdefault{10^{-8}}$,
  $\mathdefault{10^{-7}}$,
  $\mathdefault{10^{-6}}$,
  $\mathdefault{10^{-5}}$,
  $\mathdefault{10^{-4}}$,
  $\mathdefault{10^{-3}}$,
  $\mathdefault{10^{-2}}$,
  $\mathdefault{10^{-1}}$,
  $\mathdefault{10^{0}}$,
  $\mathdefault{10^{1}}$
}
]
\addplot [egpd]
table [row sep=\\] {%
100.04722967374 0.00123827659999986\\35.9238565608 0.00336588329999868\\12.9293172573 0.0084960528999994\\4.62194336994 0.0221811393999992\\1.66505779426 0.0468448117999998\\0.54439861444 0.125355046199998\\0.26526764966 0.0749021466999993\\0.20625104654 0.1272866534\\0.20958271636 0.0749023880999981\\0.20913150652 0.0761898451999994\\};
\addlegendentry{$\egpd$}
\addplot [efilter]
table [row sep=\\] {%
49.53701665118 0.000249682899998536\\23.10236487912 0.00053860579999924\\10.7540082851 0.00115991169999961\\5.00077252092 0.00249752509999911\\2.33046039012 0.00537763939999936\\1.0976554599 0.0115444095999984\\0.53781264538 0.0244992007999986\\0.2967440958 0.0494052690999991\\0.2090981488 0.083672885099998\\0.19165794728 0.0999981378999998\\};
\addlegendentry{$\mle$}
\addplot [kfilter]
table [row sep=\\] {%
0.191681962 0.0999951082999995\\0.21554282356 0.0717702141999991\\0.3112416896 0.0327113677999994\\0.60973176576 0.0117217888999996\\1.35306477082 0.00473608410000001\\2.93405963178 0.00214379199999871\\6.1256852713 0.00102162519999931\\12.53382644496 0.000500632699998112\\25.3664813702 0.000247823199998274\\51.2339613624 0.000124909999999365\\103.88971248378 6.36921999995963e-05\\211.15419468354 3.34674999997541e-05\\422.10974565054 1.69658999993531e-05\\812.7247860162 6.61469999888817e-06\\};
\addlegendentry{$\mlk$}
\addplot [crpd]
table [row sep=\\] {%
0.20181623658 0.0849490343000001\\0.23477488926 0.0561432517\\1.34399099026 0.00898225640000001\\16.0997577935 0.000864067000000001\\190.30050352584 8.33756000000001e-05\\2289.235545655 4.8107e-06\\2965.32359269162 9.06798690458105e-08\\2965.33400784243 3.66098850184949e-09\\2965.33401022066 8.44843500426806e-10\\2965.33401022562 5.63229000284538e-10\\};
\addlegendentry{$\crpd$}
\addplot [egpdp]
table [row sep=\\] {%
413.01663489582 1.74022999991777e-05\\146.24893122838 4.61155999987772e-05\\52.06437795584 0.000123003099998491\\18.60463022986 0.000337239399999749\\6.73705292986 0.000927735599998546\\2.38461008024 0.00263054859999914\\0.94803920744 0.00672366279999889\\0.42431388098 0.0163720712999992\\0.26481162056 0.0345161901999996\\0.22713738632 0.0481501364999992\\};
\addlegendentry{$\egpdp$}
\addplot [filter]
table [row sep=\\] {%
2875.25867427876 1e-08\\2875.25867427876 0.4\\};
\addlegendentry{$\mles$}
\end{axis}

\end{tikzpicture}

	\caption{$G$: diamond graph; $\lambda=(2, 3, 3, 2)$; $\mu=(2, 0, 1, 0, 2)$; $r = (1, 2.9, 1, -1, 1)$.\label{fig:simu-dia-inj}}	
\end{figure}

First, we emphasize that any simulation, regardless of length, provides only a finite-horizon estimate of a policy’s long-term behavior. In particular, it will always yield a finite delay, even for policies proven to be unstable here, like $\mles$ and $\crpd$ for $\alpha >0$. This explains why $\crpd$, when converging to $\mles$, appears to achieve zero regret with constant delay. In reality, the delay is unbounded and grows like $\sqrt{T}$\footnote{\url{https://balouf.github.io/stochastic_matching/companion/simulations.html\#Drift-of-optimal-solution-for-injective-only-vertices}}.
Therefore, values observed in the far right portion of the figure must be interpreted with caution, as they reflect finite-horizon effects rather than the true long-term behavior.

With that in mind, one can observe that all policies clearly demonstrate the
($\epsilon$, $1/\epsilon$)
trade-off between delay and regret, as predicted by \Cref{prop: converging_to_vertex}.

Specifically, $\mlk$ and $\egpdp$ exhibit very similar performance, while $\mle$ and $\crpd$ perform slightly worse. Due to the adversarial reward vector, the performance of $\egpd$ degrades by more than one order of magnitude, incurring more than ten times the regret of $\mlk$ for the same target delay, or equivalently, requiring a delay ten times longer to achieve the same regret.

\subsubsection{Approaching a bijective vertex.}\label{sec:simu-diamond-bijective}

Still focusing on the diamond graph, we now set $\lambda = (4, 4, 4, 2)$ and examine the vertex $\mu = (1, 3, 1, 2, 0)$, which disables edge $\{3, 4\}$. We select the adversarial reward vector $r=(-1, 1, 1, 1, 2.9)$, assigning the highest reward to edge $\{3, 4\}$ and a negative reward to the positive edge $\{1, 2\}$.

Since $\mu$ is bijective, we know from \Cref{coro:achievable} that $\mles$ is optimal in terms of regret. However, it is interesting to assess how vertex-approaching policies perform in this case. The results are presented in \Cref{fig:simu-dia-bij}. The $x$-axis is in logarithmic scale.
\begin{figure}[ht]
	\centering
	 % figures/diamond-bijective
\begin{tikzpicture}

\definecolor{darkgray176}{RGB}{176,176,176}
\definecolor{lightgray204}{RGB}{204,204,204}

\begin{axis}[width=12cm, height=7cm,
legend columns=3, transpose legend=true,
legend cell align={left},
legend style={fill opacity=0.8, draw opacity=1, text opacity=1, draw=lightgray204},
log basis x={10},
tick align=outside,
tick pos=left,
x grid style={darkgray176},
xlabel={Delay},
xmin=0.1, xmax=10,
xmode=log,
xtick style={color=black},
minor x tick num=9,
log identify minor tick positions=true,
ticklabel style={/pgf/number format/fixed},
log ticks with fixed point,
y grid style={darkgray176},
ylabel={Regret},
ymin=1e-08, ymax=0.15,
ytick style={color=black}
]
\addplot [egpd] 
table [row sep=\\] {%
78.2231730410857 4.11600014705818e-08\\28.3703720231286 1.40000013332428e-08\\10.2265316250143 2.13549000009934e-05\\3.60668613155714 0.00404486684000139\\1.29157954785714 0.0291424407400011\\0.419033820685714 0.138703450900001\\0.266825932157143 0.0639067196600006\\0.192080052128571 0.134472437260001\\0.206163503914286 0.0655378410400012\\0.206947928214286 0.0660652722800004\\};
\addlegendentry{$\egpd$}
\addplot [efilter] 
table [row sep=\\] {%
0.323986056914286 0.000448559440001543\\0.322465855128571 0.000964134360001279\\0.319273825171429 0.00206695649999995\\0.312765451557143 0.00440521311999999\\0.3002677895 0.00927522806000099\\0.278858375542857 0.0190173305000004\\0.248796782757143 0.0368492082000008\\0.217640355528571 0.0637020697600013\\0.196701491071429 0.0904405629400007\\0.190487365357143 0.0999960059000008\\};
\addlegendentry{$\mle$}
\addplot [kfilter] 
table [row sep=\\] {%
0.190473101342857 0.0999999816200006\\0.205596216385714 0.0867643443800001\\0.244247020085714 0.0405312703599999\\0.300549512985714 0.00606562838000152\\0.324457727657143 8.63188200007044e-05\\0.325290151171429 7.28000075101784e-09\\0.325290250614286 6.66133814775094e-16\\0.325290250614286 6.66133814775094e-16\\0.325290250614286 6.66133814775094e-16\\0.325290250614286 6.66133814775094e-16\\0.325290250614286 6.66133814775094e-16\\0.325290250614286 6.66133814775094e-16\\0.325290250614286 6.66133814775094e-16\\0.325290250614286 6.66133814775094e-16\\};
\addlegendentry{$\mlk$}
\addplot [crpd]
table [row sep=\\] {%
0.191696780142857 0.0969994828600001\\0.203651076357143 0.04903702076\\0.301513605728571 0.00014325696\\0.302976030414286 2.4864e-07\\0.3029775189 1.176e-08\\0.302977555371429 1.82e-09\\0.302977557 9.80000000000001e-10\\0.302977558685714 1.4e-10\\0.302977558757143 0\\0.302977558757143 0\\};
\addlegendentry{$\crpd$}
\addplot [egpdp]
table [row sep=\\] {%
0.261126289642857 1.33226762955019e-15\\0.261445911785714 1.33226762955019e-15\\0.261446017057143 1.33226762955019e-15\\0.261509453642857 1.33226762955019e-15\\0.262795973714286 1.22124532708767e-15\\0.272488666728571 1.00380001473596e-07\\0.276138395185714 0.000399820260001499\\0.232527477028571 0.00995601921999927\\0.2074600391 0.0417012045799999\\0.197611258114286 0.0650690719000001\\};
\addlegendentry{$\egpdp$}
\addplot [filter]
table [row sep=\\] {%
0.325290250614286 1e-08\\0.325290250614286 0.15\\};
\addlegendentry{$\mles$}
\end{axis}

\end{tikzpicture}

	\caption{$G$: diamond graph; $\lambda=(4, 4, 4, 2)$; $\mu=(1, 3, 1, 2, 0)$; $r=(-1, 1, 1, 1, 2.9)$.\label{fig:simu-dia-bij}}
\end{figure}

All policies continue to exhibit a trade-off between regret and delay. Except for $\egpd$, they all achieve delays lower than that of $\mles$ at the cost of a positive regret.

Specifically, $\mle$ and $\mlk$ perform very similarly and converge precisely to $\mles$, as predicted by their design. $\crpd$ and $\egpdp$ outperform the others; notably, they achieve zero regret with delays lower than $\mles$. We interpret this as resulting from their base policy, \gls{vqml}, which may perform better than $\mles$'s base policy \gls{ml} in the model where edge $\{3, 4\}$ is removed.

As before, $\egpd$ performs poorly due to the adversarial reward vector. While its regret can approach zero arbitrarily closely, as predicted by theory, this comes at the cost of a large delay that appears to grow asymptotically like $1/\beta$, effectively unbounded.

\subsubsection{Discussion.}\label{sec:simu-discussion}

The observations made in \Cref{sec:simu-diamond-injective,sec:simu-diamond-bijective} are not specific to the diamond graph.
Other examples studied in \refapp{app:simu-additional-examples}
show qualitatively similar behavior.
Our main conclusion is that no single policy consistently outperforms all others across all scenarios. Nevertheless, some patterns emerge:
\begin{itemize}
	\item Pure filtering should be avoided if the vertex is injective-only. For bijective vertices, if minimizing regret is the absolute priority, pure filtering is a natural choice; otherwise, more flexible policies may perform better.
	\item $\epsilon$-filtering \gls{ml} is a robust vertex-approaching method effective on both injective-only and bijective vertices. It offers theoretical guarantees but can often be outperformed numerically by state-dependent policies.
	\item \gls{egpd} does not require knowledge of the vertex $\mu$ corresponding to the arrival rate and reward vectors. However, it performs poorly when the reward vector is misaligned with the support $\E^\star$ of $\mu$.
	\item \gls{egpd}+ addresses the drawbacks of \gls{egpd} by aligning the rewards with $E^\star$. Note that this requires knowledge or estimation of the arrival rate $\lambda$.
	\item Although theoretical guarantees for $k$-filtering \gls{ml} remain conjectural, simulations suggest it strikes a favorable balance between performance and robustness. Alongside \gls{egpd}+, it is our recommended policy when arrival rates are known or can be reliably estimated.
	\item \gls{crpd} delivers good performance and theoretical guarantees; however, when compared with \gls{egpd}+, the practical benefit of introducing a time-decreasing parameter remains unclear in many scenarios\footnote{Counter-examples exist, cf \url{https://balouf.github.io/stochastic_matching/companion/simulations.html\#Larger-graphs}}.
\end{itemize}
  
 % sections/06-greedy
\subsection{Greedy policies.} \label{subsec:greedy}

Greedy policies are appealing candidates for controlling matching problems, and they have been extensively studied in the literature. Some, such as \gls{ml} and \gls{fcfm}, are simple 
to implement and stabilize all stabilizable matching problems. Moreover, greedy policies may be more socially acceptable, particularly in scenarios involving human participants.
However, as we will demonstrate in this section, greedy policies are generally not well-suited for optimizing matching-rate vectors. They can never achieve a vertex, and examples suggest that in most cases, they are unable to even approach one.

In what follows,
consider a stabilizable problem $(G, \lambda)$ with a surjective-only graph~$G$, and
let $\Lagre$ denote the set of matching-rate vectors achieved by stable \emph{greedy} policies adapted to the problem $(G, \lambda)$. Also let $\Lap=\{\mu \in \Rp^m: A \mu = \lambda\}$ denote the set of the positive solutions to the conservation equation \eqref{eq:system}, i.e., those with positive coordinates, which form the interior of $\Lann$.

\subsubsection{Convexity and impossibility result.}

We first demonstrate that, unlike stable policies in general, stable \emph{greedy} policies can never reach the boundary of the convex polytope~$\Lann$, irrespective of whether the vertices of $\Lann$ are bijective or injective-only.

\begin{prop} \label{prop:greedypositive}
	Let $\Lagre$ denote the set of matching-rate vectors achievable by a stable \emph{greedy} policy for a given stabilizable matching problem $(G, \lambda)$, where $G$ is surjective-only. Then, $\Lagre$ forms a non-empty convex subset of $\Lap$.
\end{prop}

\begin{proof}
	The set $\La_\Gre$ is non-empty because,
	as recalled in \Cref{subsubsec:stability},
	the greedy \gls{ml} and \gls{fcfm} policies
	are stable.
	Its convexity is shown in \refapp{app:convexity-proof}.
	We now prove that $\La_\Gre \subseteq \Lap$.
	Consider a stable greedy policy~$\Phi$ and let $\mu$ denote the matching-rate vector in the model $(G, \lambda, \Phi)$. Consider an edge $e_k = \{i, j\}$. Since the policy $\Phi$ is greedy, two items of classes~$i$ and~$j$ are always matched if the following sequence of events occurs: the system is in the empty state~$\varnothing$, then a class-$i$ item arrives, and then a class-$j$ item arrives. Let $\pi(\varnothing)$ denote the stationary probability that the model $(G, \lambda, \Phi)$ is in the empty state\footnote{%
		As mentioned in \refapp{app:extended-definition},
		we assume that there exists
		a unique state $s \in \cS$ such that $|s| = 0$.
		This state is called the \emph{empty state} and denoted by $\varnothing$.
		This assumption guarantees that
		the intuitive notion of system stability
		is captured by the positive recurrence
		of the Markov chain describing the evolution of the system state.
		If the policy is queue-based (so that $\cS = \Q$), this empty state is simply the $n$-dimensional zero vector.%
	}~$\varnothing$. We know that $\pi(\varnothing) > 0$ because the model is stable, and the previous remark implies that $\mu_k \geq \pi(\varnothing) \lambda_{i}\lambda_{j} / (\sum_{\ell \in \V}\lambda_\ell)^2 >0$. Since this is true for each edge $e_k \in \E$ and each $\mu \in \La_\Gre$, we conclude that $\La_\Gre \subseteq \Lap$.
\end{proof}
\Cref{prop:greedypositive} has the following consequence regarding greedy policies and linear optimization.
\begin{coro} \label{coro:greedypositive}
	Let $r_{\max} = \max_{\mu \in \Lann} r^\intercal \mu$ be the optimal reward in the linear optimization problem defined by the reward vector~$r$ in \Cref{prop:vertices-optimal}. One of the following must hold:
	\begin{enumerate}[(i)]
		\item \label{item:greedypositive-1}
		All stable policies (greedy or not) are optimal, i.e., $r^\intercal \mu = r_{\max}$ for each $\mu \in \La_\Pol$.
		\item \label{item:greedypositive-2}
		All stable greedy policies are suboptimal, i.e.,
		$r^\intercal\mu < r_{\max}$ for each $\mu \in \La_\Gre$.
	\end{enumerate}
\end{coro}

\begin{proof}
	We know from \Cref{prop:vertices-optimal}  that the set of vectors $\mu \in \Lann$ maximizing the reward forms a non-empty face $F$ of $\Lann$. If $F = \Lann$, then we are in case~\ref{item:greedypositive-1}, meaning all $\mu \in \La_\Pol \subset \Lann$ are optimal.
	This occurs when the vector~$r$ is orthogonal to $\ker(A)$\footnote{An example of this is shown in \Cref{fig:unstable-optimal}, or when all coordinates of $r$ are equal, making all edges equivalent.}.
	Otherwise, by the lattice structure of polytope faces, $F$ is contained in a facet of $\Lann$, implying there is at least one edge $k \in E$ where the $k$-th coordinate of all vectors in~$F$ is zero. As stable greedy policies produce matching-rate vectors with all positive coordinates by \Cref{prop:greedypositive}, no greedy policy can be optimal in this case.
\end{proof}

\subsubsection{Achievability results.}

We now explore the relationship $\La_\Gre \subseteq \Lap$ through several examples. In particular, \Cref{prop:greedy-complete,prop:greedy-diamond} illustrate situations where $\La_\Gre$ is a strict subset of $\Lap$, suggesting that the greedy constraint imposes significant limitations on the set of achievable matching rates. However, this is not universally true, as \Cref{prop:greedy-fish} provides a (carefully chosen) example where $\La_\Gre = \Lap$.

\begin{prop}
	Let $(K_n, \lambda)$ be a stabilizable matching problem, where $K_n$ is the complete graph with $n \ge 3$ nodes. All greedy policies adapted to $K_n$ are stable and yield the same matching-rate vector, denoted $\mu_\Gre$. In particular, we have
	$\La_\Gre = \{\mu_\Gre\} \subsetneq \La_{>0}$
	whenever $n \ge 4$.
	\label{prop:greedy-complete}
\end{prop}

\begin{proof}
	See \refapp{app:greedy-complete} for proof and discussion.
\end{proof}

\begin{prop}
	Let $(D, \lambda)$ be a stabilizable matching problem, where $D$ is the diamond graph from \Cref{fig:diamond}. All greedy policies adapted to $D$ are stable. In kernel coordinates, there exist $\alpha_-< \alpha_+$ such that $\La_\Gre = [\alpha_-, \alpha_+] \subsetneq \Lap$.
	\label{prop:greedy-diamond}
\end{prop}

\begin{proof}
See \refapp{app:greedy-diamond} for proof, discussion, and numerical results.
\end{proof}

\begin{prop}
	Let $(G, \lambda)$ be the stabilizable matching problem depicted in \Cref{fig:fish} (the \emph{Fish} matching problem). For this problem, we have $\La_\Gre = (-1/2, 1/2) = \Lap$.
	\label{prop:greedy-fish}
\end{prop}

\begin{proof}
	See \refapp{app:greedy-fish} for the proof, which builds a family of stable greedy policies whose matching rates approach each vertex, together with numerical illustration and discussion.
\end{proof}

\begin{figure}[htb]
	\centering
	\begin{tikzpicture}[scale=.8]
		\def\d{2cm}
		\node[class] (1) {$1$};
		\foreach \i/\s/\a in {2/1/-90, 3/1/-30, 4/3/30, 5/4/-30, 6/3/-30}		
		\path (\s) ++(\a:\d) node[class] (\i) {$\i$};
		
		\draw (1) edge node[left] {$3$} (2) 
		(1) edge node[above, sloped] {$1$} (3)
		(2) edge node[above, sloped] {$1$} (3)
		(3) edge node[above, sloped] {$\frac12 + \alpha$} (4)
		(4) edge node[above, sloped] {$\frac32 - \alpha$} (5) 
		(5) edge node[above, sloped] {$\frac32 + \alpha$} (6)
		(3) edge node[above, sloped] {$\frac12 - \alpha$} (6);
	\end{tikzpicture}
	\caption{Generic solution to~\eqref{eq:system}~in the Fish matching ($\lambda=(4,4,3,2,3,2)$).}
	\label{fig:fish}
\end{figure}
             
 % sections/07-extensions
\section{Extensions of our results.}\label{sec:extensions}

This article focuses on the matching rates of stable policies in the context of two-way matching, emphasizing linear optimization. However, our approach extends beyond this setting. In this section, we present two such extensions: \Cref{sec:non-linear-optimization} addresses the case of non-linear reward functions, and \Cref{sec:hypergraphs} outlines the adaptation of our work to hypergraphs, enabling the possibility of discarding items and multi-way matching.

\subsection{Non-linear optimization.}\label{sec:non-linear-optimization}

\Cref{coro:achievable} introduced $\La_\Pol$, the set of matching rates achievable by a stable policy, and compared it to the vertices of $\Lann$, which contain the optimal solutions for linear reward optimization.

However, when considering a non-linear reward function $r(\mu)$, an optimal solution can lie anywhere within $\Lann$. For example, one might want a stable policy to be as close as possible to a target matching rate vector $\mu_0$ and use the reward function $r(\mu) = -||\mu - \mu_0||_1$. Ideally, we would like $\La_\Pol$ to be as close to $\Lann$ as possible. The following proposition demonstrates that this is indeed the case.

\begin{prop} \label{prop:convexity}
	Let $\La_\Pol$ be the set of matching rate vectors achievable by a stable policy adapted to a given stabilizable matching problem $(G, \lambda)$. Then, $\La_\Pol$ is convex.
	Furthermore, any positive solution to the conservation equation \eqref{eq:system} can be obtained by a stable policy, that is,
	\begin{equation*}
		\Lap\subseteq \La_\Pol \subseteq \Lann\text{, where $\Lap=\{\mu \in \Rp^m: A \mu = \lambda\}$.}
	\end{equation*}
	In the particular case where all vertices of $\Lann$ are bijective, we have $\La_\Pol = \Lann$.
\end{prop}

\begin{proof}
Convexity of~$\La_\Pol$ is shown in \refapp{app:convexity-proof}.
The inclusion $\La_\Pol \subseteq \Lann$ follows from the discussion in \Cref{subsec:conservation}: any stable policy yields a matching rate vector that satisfies~\eqref{eq:system} with non-negative coordinates.
We now prove that $\Lap\subseteq \La_\Pol$.
Since $\La_\Pol$ is convex, its closure is also convex. By \Cref{prop: converging_to_vertex} and the convexity of~$\La_\Pol$, the closure of $\La_\Pol$ contains all vertices of $\Lann$, and, by convexity, encompasses $\Lann$ as well. As a result, by virtue of convexity, $\La_\Pol$ contains the interior\footnote{Here, the notion of \emph{interior} refers to the canonical $\R^d$ topology of the $d$-dimensional affine space of solutions to the conservation equation.}
of its own closure,
which includes $\Lap$, the interior of $\Lann$. In the particular case where all vertices are bijective, they all belong to $\Lapol$, leading to the equality $\Lapol = \Lann$.
\end{proof}

\Cref{prop:convexity} is essentially an existence result,
as the policy constructed
to show convexity of~$\Lapol$ in \refapp{app:convexity-proof}
may be both hard to implement
(as it requires computing the mean return time to a particular state
of two Markov chains)
and undesirable in practice
(as the matching rate will have high variance).
An interesting follow-up question
is thus to explore practical methods to
reach an arbitrary vector of~$\Lapol$.

\subsection{Hypergraphs.}\label{sec:hypergraphs}

A hypergraph $G = (V, \E)$ consists of a set of nodes $V$ and a set of hyperedges 
$\E$, where each hyperedge is a multi-subset (i.e., potentially allowing multiplicities) of $V$ of arbitrary size, starting from one. This contrasts with simple graphs, where each edge is a pair of distinct nodes. Stochastic matching models extend naturally to hypergraphs as follows: when a matching decision is made, items corresponding to the selected hyperedge are removed from the system.

Our approach adapts to hypergraphs with limited changes: the spin-off paper by \citet{M26} states and proves the extension to hypergraphs of the stabilizability characterization of \Cref{prop:stability-region-form}, in the support form of condition~\ref{cond:stability-region-form-4}. The key elements are:
\begin{itemize}
\item The incidence matrix $A$ for hypergraphs is defined analogously to simple graphs but can be any matrix in $\N^{n\times m}$, with $a_{i, k}$ indicating whether node $i$ belongs to hyperedge $k$, counting multiplicities in case of repeated nodes.
The connection between properties of $A$ and stability remains, particularly \Cref{prop:stability-region-nonempty}. However, some graph-specific results, such as the characterization of surjectivity using non-bipartite components (\Cref{def:surjective}\ref{cond:surjective-3}) 
or the description of $\ker(A)$ in terms of cycles and kayak paddles (\Cref{sec:basis-of-the-kernel-of-the-incidence-matrix}), do not easily generalize to hypergraphs.
\item
\Cref{prop:stability-region-form} remains valid: this is precisely the main result of \citet{M26}. It was also independently and concurrently obtained in \citet{NB26}. In particular, the stability condition \ref{cond:stability-region-form-3} can still be verified by computing $A^{-1}\lambda$ for bijective hypergraphs, or solving \eqref{eq:linear} for surjective-only hypergraphs. Note that necessity is more delicate to prove for hypergraphs than for graphs \citep[cf.][]{RM21}; the negative examples of \citet{RM21} are all consistent with the characterization, each either failing surjectivity of~$A$ or failing to have a positive solution for the conservation equation~\eqref{eq:system}. 
\item Some policies from \Cref{sec:non-unicyclic,sec:numerical-results} that optimize the matching-rate vector, specifically $\mles$, $\mle$, and $\mlk$, rely on the maximal stability property of the \gls{ml} policy in simple graphs. This property does not extend to hypergraphs\footnote{In hypergraphs, stabilizable matching problems exist where no greedy policy is stable.}, necessitating alternative maximally stable policies. We choose \gls{vqml}, the rewardless version of \gls{egpd}, whose maximal stability on hypergraphs is proven in \citet{M26}, although other choices are possible (e.g., the maximally stable online assignment policy of \citet{NB26}).
\end{itemize}

We now illustrate these adaptations through numerical results on example cases. We apply the policies from \Cref{tab:policies_simu}, adapted to hypergraphs by always using \gls{vqml} instead of \gls{ml} as the base policy, following the simulation methodology described in \Cref{sec:methodology}. These examples highlight the performance of the proposed policies in the hypergraph context.

\subsubsection[Original example from Nazari and Stolyar.]{Original example from \citet{NS19}.}

\citet{NS19} considered the hypergraph shown in \Cref{fig:hypergraph} for the numerical evaluation of their policy $\egpd$. This hypergraph consists of four nodes, each with a mono-edge (i.e., a $1$-edge),
plus two regular 2-edges and one 3-edge. Each hyperedge here is a plain set, i.e., without multiplicity, so the incidence matrix of this example is binary. It can be verified that this hypergraph is surjective-only, and the solution space to \eqref{eq:system} has dimension 3.

\begin{figure}
	\centering
	\begin{tikzpicture}
		\def\d{.4cm}
		\node[class] (1) {$1$};
		\node[hedge, above=\d of 1] (h1) {$r_{\{1\}}=-1$};
		\draw (1) edge (h1);
		\foreach \i/\s/\r in {2/1/-1, 3/2/1, 4/3/2}{
			\node[class, right=2cm of \s] (\i) {$\i$};
			\node[hedge, above=\d of \i] (h\i) {$r_{\{\i\}}=\r$};
			\draw (\i) edge (h\i);
		}		
		\node[hedge, below=\d of 1] (h12) {$r_{\{1, 2\}}=5$};
		\draw (h12) edge (1) edge (2);
		\node[hedge, below=\d of 2] (h23) {$r_{\{2, 3\}}=4$};
		\draw (h23) edge (3) edge (2);
		\node[hedge, below=\d of 3] (h234) {$r_{\{2, 3, 4\}}=7$};
		\draw (h234) edge (3) edge (2) edge (4);
	\end{tikzpicture}
	\caption{Hypergraph example studied in \cite{NS19}. The rewards associated with each hyperedge are indicated. \label{fig:hypergraph}}
\end{figure}

The authors consider a single reward vector (shown in \Cref{fig:hypergraph}) and two distinct arrival-rate vectors: $\lambda=(1.2, 1.5, 2, 0.8)$ and $\lambda=(1.8, 0.8, 1.4, 1)$. Both problems are stabilizable. The first generates a polytope $\Lann$ with 8 vertices, while the second generates a polytope with 4 vertices. All vertices are bijective, indicating that the reward can always be optimized using the filtering policy $\mles$.
The performance of the considered policies under these conditions is shown in \Cref{fig:perf-stol}. The key observations are:
\begin{itemize}
	\item All parameterized policies can minimize the regret when their parameter is aggressive enough.
	\item As noted in \cite{NS19}, the delay experienced by $\egpd$ continues to grow as $1/\beta$, even after the regret reaches zero.
	\item $\egpdp$ only keeps a bounded delay in the second setting.
	\item The delay experienced by $\mle$ appears bounded but remains higher than that of $\mles$. This contrasts with the expected convergence to $\mles$, as observed in \Cref{sec:simu-diamond-bijective}. We attribute this to unforeseen interactions between the expanded graph, on which $\mle$ dispatches arrivals unevenly, and the virtual queue reservation mechanism.
	\item For both $\mlk$ and $\crpd$, the delay converges to that of $\mles$, which is consistent with \Cref{sec:simu-diamond-bijective}. Adjusting the parameters makes it possible to control the delay/regret trade-off.
\end{itemize}

\begin{figure}[t]
	\subfloat[$\lambda=(1.2, 1.5, 2, 0.8)$, 
	$\mu=(0, 0, 1.7, 0.5, 1.2, 0, 0.3)$]{ % figures/ns19-a
\begin{tikzpicture}

\definecolor{darkgray176}{RGB}{176,176,176}
\definecolor{lightgray204}{RGB}{204,204,204}

\begin{axis}[
legend cell align={left},
legend style={fill opacity=0.8, draw opacity=1, text opacity=1, draw=lightgray204},
log basis x={10},
tick align=outside,
tick pos=left,
x grid style={darkgray176},
xlabel={Delay},
xmin=0.0819217512285889, xmax=1000,
xmode=log,
log identify minor tick positions=true,
ticklabel style={/pgf/number format/fixed},
log ticks with fixed point,
xtick style={color=black},
y grid style={darkgray176},
ylabel={Regret},
ymin=1e-08, ymax=4,
ytick style={color=black}
]
\addplot [efilter]
table [row sep=\\] {%
5.09065797461818 0.00200021195000241\\4.16924288550909 0.00430515305000167\\3.28480291649091 0.00921375675000257\\2.47908136456364 0.0198065312500018\\1.80263529241818 0.046770004050002\\1.27034880705455 0.131461634150002\\0.846369656909091 0.404151148500003\\0.505269370927273 1.2452771177\\0.268449081527273 3.64231886865\\};
\addlegendentry{$\mle$}
\addplot [kfilter]
table [row sep=\\] {%
0.122206002163636 3.7032268075\\0.334870049254545 1.08594503765\\0.575389381563636 0.252626536250004\\0.789888345181818 0.0256905044000026\\0.857032791690909 0.000436637300002183\\0.860187289254545 2.61800002337508e-07\\0.8601920368 2.77555756156289e-15\\0.8601920368 2.77555756156289e-15\\0.8601920368 2.77555756156289e-15\\0.8601920368 2.77555756156289e-15\\0.8601920368 2.77555756156289e-15\\0.8601920368 2.77555756156289e-15\\0.8601920368 2.77555756156289e-15\\};
\addlegendentry{$\mlk$}
\addplot [egpdp]
table [row sep=\\] {%
181.817036612473 2.72004641033163e-15\\65.6354917394909 2.44249065417534e-15\\23.8680824100182 1.55431223447522e-15\\8.72646199667273 3.94129173741931e-15\\3.10962936321818 1.01200003687062e-07\\1.30984559469091 0.00174947025000322\\0.997595337454546 0.0810504986500031\\0.475592819018182 0.674525857500003\\0.208644326945455 3.7032268086\\0.208644326945455 3.7032268086\\};
\addlegendentry{$\egpdp$}
\addplot [egpd]
table [row sep=\\] {%
363.880449503782 3.05311331771918e-15\\130.859730330873 2.1094237467878e-15\\46.7896584459273 1.88737914186277e-15\\16.9207381899273 2.22044604925031e-15\\6.19956431249091 1.76550002920959e-07\\2.5085915996 0.00271945465000277\\0.889707784327273 0.0767437126500029\\0.427536664090909 0.462502361750002\\0.258873539818182 1.2078547503\\0.155272364727273 2.8090408775\\};
\addlegendentry{$\egpd$}
\addplot [crpd]
table [row sep=\\] {%
0.426834390436364 0.2641545621\\0.859700264181818 4.10333e-05\\0.860191913636364 8.8e-09\\0.8601920368 0\\0.8601920368 0\\0.8601920368 0\\0.8601920368 0\\0.8601920368 0\\0.8601920368 0\\0.8601920368 0\\};
\addlegendentry{$\crpd$}
\addplot [filter]
table [row sep=\\] {%
0.8601920368 1e-08\\0.8601920368 4\\};
\addlegendentry{$\mles$}
\end{axis}

\end{tikzpicture}
} 
	\subfloat[$\lambda=(1.8, 0.8, 1.4, 1)$, 
	$\mu=(1, 0, 1.4, 1, 0.8, 0 , 0)$]{ % figures/ns19-b
\begin{tikzpicture}

\definecolor{darkgray176}{RGB}{176,176,176}
\definecolor{lightgray204}{RGB}{204,204,204}

\begin{axis}[
legend cell align={left},
legend style={fill opacity=0.8, draw opacity=1, text opacity=1, draw=lightgray204},
log basis x={10},
tick align=outside,
tick pos=left,
x grid style={darkgray176},
xlabel={Delay},
xmin=0.0375542251274821, xmax=1000,
xmode=log,
xtick style={color=black},
minor x tick num=9,
log identify minor tick positions=true,
ticklabel style={/pgf/number format/fixed},
log ticks with fixed point,
y grid style={darkgray176},
ylabel={Regret},
ymin=1e-08, ymax=3,
ytick style={color=black}
]
\addplot [efilter]
table [row sep=\\] {%
2.50918638552 0.0118048550000037\\2.13402692886 0.0254288800000048\\1.76241154334 0.0546392230000042\\1.39710782348 0.116407496000004\\1.04099905 0.240703760500002\\0.71496524448 0.463488451500004\\0.40621098356 0.808634352000004\\0.19283532988 1.3998573295\\0.09787566576 2.68236621050001\\};
\addlegendentry{$\mle$}
\addplot [kfilter]
table [row sep=\\] {%
0.06153994408 1.7231620805\\0.12752271576 0.284308822000004\\0.15757461556 0.0106763160000034\\0.16000807002 1.58795000030776e-05\\0.16001537754 3.33066907387547e-15\\0.16001537754 3.33066907387547e-15\\0.16001537754 3.33066907387547e-15\\0.16001537754 3.33066907387547e-15\\0.16001537754 3.33066907387547e-15\\0.16001537754 3.33066907387547e-15\\0.16001537754 3.33066907387547e-15\\0.16001537754 3.33066907387547e-15\\0.16001537754 3.33066907387547e-15\\};
\addlegendentry{$\mlk$}
\addplot [egpdp]
table [row sep=\\] {%
0.16016988542 3.33066907387547e-15\\0.16003503926 3.33066907387547e-15\\0.1600178689 3.33066907387547e-15\\0.16001567788 3.33066907387547e-15\\0.16001541102 3.33066907387547e-15\\0.16001537474 1.05000034091986e-08\\0.15941514508 0.00210253750000324\\0.12752271574 0.284308822000004\\0.06153994414 1.7231620805\\0.06153994406 1.7231620805\\};
\addlegendentry{$\egpdp$}
\addplot [egpd]
table [row sep=\\] {%
1199.83888149284 8.17000003990668e-07\\431.03984204064 2.92000002583526e-07\\154.63996677862 1.00000002261142e-07\\55.4399824172 3.60000028568214e-08\\19.83998431466 1.2000003357757e-08\\6.839984578 4.00000382079503e-09\\2.24012432314 9.90835000039407e-05\\0.66269294334 0.034491084000004\\0.16406861478 0.122062628500002\\0.08571682784 0.700067024500003\\};
\addlegendentry{$\egpd$}
\addplot [crpd]
table [row sep=\\] {%
0.1548650009 0.013168126\\0.16001537754 0\\0.16001537754 0\\0.16001537754 0\\0.16001537754 0\\0.16001537754 0\\0.16001537754 0\\0.16001537754 0\\0.16001537754 0\\0.16001537754 0\\};
\addlegendentry{$\crpd$}
\addplot [filter]
table [row sep=\\] {%
0.16001537754 1e-08\\0.16001537754 3\\};
\addlegendentry{$\Phi_{E^*}$}
\end{axis}

\end{tikzpicture}
\label{fig:perf-stolb}} 
	\caption{Reaching the vertex corresponding to $r=(-1, -1, 1, 2, 5, 4, 7)$ in the hypergraph from \Cref{fig:hypergraph}.\label{fig:perf-stol}}
\end{figure}

\subsubsection{Augmenting stability.}\label{sec:case-study-unstable-policy}

\Cref{fig:unstable-optimal} in \Cref{sec:non-unicyclic} presented a toy example
where the optimal stable policy does not achieve the optimal achievable reward. 
This demonstrates that imposing stability can limit performance in terms of matching rates, or equivalently that
the optimal matching rate may require discarding a fraction of arrivals and thus lie outside $\Lann$. Following \citet{G24}, this issue can be addressed by augmenting each under-demanded node with a zero-reward mono-edge, thereby allowing the discard of excess items.

In the toy example of \Cref{fig:unstable-optimal}, adding a mono-edge to class 4 creates a third vertex in the polytope, which zeros the edges $\{2, 4\}$ and $\{3, 4\}$. This vertex is bijective, making $\mles$ both optimal and stable.

\begin{figure}[!ht]
	\centering
	 % figures/abandonment
\begin{tikzpicture}

\definecolor{darkgray176}{RGB}{176,176,176}
\definecolor{lightgray204}{RGB}{204,204,204}

\begin{axis}[width=12cm, height=7cm,
	legend columns=3, transpose legend=true,
legend cell align={left},
legend style={fill opacity=0.8, draw opacity=1, text opacity=1, draw=lightgray204},
log basis x={10},
log basis y={10},
tick align=outside,
tick pos=left,
x grid style={darkgray176},
xlabel={Delay},
xmin=0.0716019449646116, xmax=200,
xmode=log,
xtick style={color=black},
y grid style={darkgray176},
ylabel={Regret},
ymin=1e-08, ymax=1,
ymode=log,
ytick style={color=black},
ytick={1e-09,1e-08,1e-07,1e-06,1e-05,0.0001,0.001,0.01,0.1,1,10},
yticklabels={
  $\mathdefault{10^{-9}}$,
  $\mathdefault{10^{-8}}$,
  $\mathdefault{10^{-7}}$,
  $\mathdefault{10^{-6}}$,
  $\mathdefault{10^{-5}}$,
  $\mathdefault{10^{-4}}$,
  $\mathdefault{10^{-3}}$,
  $\mathdefault{10^{-2}}$,
  $\mathdefault{10^{-1}}$,
  $\mathdefault{10^{0}}$,
  $\mathdefault{10^{1}}$
}
]
\addplot [efilter]
table [row sep=\\] {%
0.712100497571429 0.00105266070000558\\0.612900133292857 0.0025400872000055\\0.518869165971429 0.00678201580000564\\0.436103285357143 0.0192479742000057\\0.363661140242857 0.0531622504000045\\0.303417589685714 0.133558729500006\\0.251877811592857 0.298206064800006\\0.201602137535714 0.563041353700005\\0.173189873028571 0.793273723900004\\2.33116310954286 0.731229587200007\\};
\addlegendentry{$\mle$}
\addplot [kfilter]
table [row sep=\\] {%
0.116656028521429 0.731246877200006\\0.100886699821429 0.317654055600005\\0.1020266999 0.115345664000005\\0.106772864085714 0.00986002010000523\\0.108134529342857 5.50564000058298e-05\\0.108153028471429 5.32907051820075e-15\\0.108153028471429 5.32907051820075e-15\\0.108153028471429 5.32907051820075e-15\\0.108153028471429 5.32907051820075e-15\\0.108153028471429 5.32907051820075e-15\\0.108153028471429 5.32907051820075e-15\\0.108153028471429 5.32907051820075e-15\\0.108153028471429 5.32907051820075e-15\\0.108153028471429 5.32907051820075e-15\\};
\addlegendentry{$\mlk$}
\addplot [egpd]
table [row sep=\\] {%
95.9435627500714 2.24000006502146e-07\\34.1278599884143 7.9100007843716e-08\\11.9135754598286 3.08000065145491e-08\\4.22928651632857 1.05000058593727e-08\\1.76035457304286 1.68000056564373e-08\\0.587825802028571 0.00117809580000584\\0.280612610771429 0.0344234324000051\\0.114681019235714 0.220274341700005\\0.112496176421429 0.284446276100005\\0.110322585992857 0.248945251100006\\};
\addlegendentry{$\egpd$}
\addplot [crpd]
table [row sep=\\] {%
0.106906862657143 0.3626066339\\0.108145406192857 2.74036e-05\\0.108153029028571 5.6e-09\\0.108153028471429 0\\0.108153028471429 0\\0.108153028471429 0\\0.108153028471429 0\\0.108153028471429 0\\0.108153028471429 0\\0.108153028471429 0\\};
\addlegendentry{$\crpd$}
\addplot [egpdp]
table [row sep=\\] {%
0.1071343574 5.32907051820075e-15\\0.107133510614286 5.32907051820075e-15\\0.107133397292857 5.32907051820075e-15\\0.107133383414286 5.32907051820075e-15\\0.107133382007143 5.32907051820075e-15\\0.107133381857143 5.32907051820075e-15\\0.107134516978571 3.50000555096747e-09\\0.106981267571429 0.000679401100006671\\0.103462235 0.0429439080000051\\0.102880792664286 0.0709786770000051\\};
\addlegendentry{$\egpdp$}
\addplot [filter]
table [row sep=\\] {%
0.108153028471429 1e-08\\0.108153028471429 1\\};
\addlegendentry{$\mles$}
\end{axis}

\end{tikzpicture}

	\caption{$G$: diamond augmented with $\{4\}$; $\lambda=(4, 4, 4, 2)$; $\mu=(2, 2, 2, 0, 0, 2)$; $r=(-1, 1, 1, 1, 3, 2)$ (adversarial).\label{fig:perf-abandon}}
\end{figure}

The performance of the considered policies under this setting is shown in \Cref{fig:perf-abandon}. The most noteworthy findings are:
\begin{itemize}
	\item $\egpd$ and $\mle$ can have a low regret, but at the cost of high delay.
	\item All other parameterized policies rapidly converge to $\mles$, with negligible regret/delay trade-off. We believe this is because the target vertex (a perfectly balanced triangle plus one discarded node) already yields low delay, so relaxing the constraints on edges $\{2, 4\}$ and $\{3, 4\}$ has minimal impact.
\end{itemize}

The case analyzed above is illustrative, but the role of mono-edges extends beyond this specific example. In general, adding mono-edges to a graph increases the dimensionality of the polytope, so that the original feasible region becomes a face of the enlarged polytope. For instance, augmenting the toy example transforms a segment into a triangle, with one side representing the original feasible set.

Providing one mono-edge per node makes any problem $(G, \lambda)$ trivially stabilizable, for example by discarding all incoming items.
This augmentation also remedies instability caused by imbalanced arrival rates (a set of classes receiving more arrivals than its neighborhood can absorb), but it does not change the nature of existing vertices:
those that were injective-only remain injective-only, so the trade-off of \Cref{sec:approaching-injective-only-vertices} persists.

It is crucial that mono-edges carry non-negative rewards to ensure that $\Lann$ contains the optimal solution. If mono-edges have negative rewards, the discrepancy between stability and optimality may persist. For example, in \Cref{fig:perf-stolb}, the best stable solutions are actually suboptimal due to the discarding of class~1 items with negative reward. In that case, absolute optimality requires the queue for class~1 to grow without bound.

\begin{rem}
If $G$ is a simple graph augmented with mono-edges, \Cref{def:injective,def:bijective,def:surjective,def:only} remain entirely valid, if we adopt the convention that one mono-edge is an odd cycle. In particular, if each connected component of $G$ contains at least one mono-edge, $G$ is surjective.

Other definitions and formulas can be likewise adapted. For example, \Cref{eq:crp_gap}, which defines $\delta(G, \lambda)$, works with the convention that a node with a mono-edge does not belong to any independent set. As an illustration, one can check that the problem studied in \Cref{fig:perf-abandon} satisfies $\delta(G, \lambda)=\delta(G^\star, \lambda)=2$. A graph with mono-edges everywhere translates to $\delta(G, \lambda)=+\infty$ (all incoming items can be processed instantly).
\end{rem}

\bibliographystyle{informs2014}

\newpage
\crefalias{section}{appsec}
\crefalias{subsection}{appsec}
\crefalias{subsubsection}{appsec}
\appendix

 % appendix/ec1-extended-definition
\section[Supplementary material of Section \ref*{sec:extended-definition}.]{Supplementary material of \fullref{sec:extended-definition}.}
\label{app:extended-definition}

Our results are not limited to deterministic size-based policies but apply to a broader family of policies
that are either random or require a more complex state descriptor, or both.
We now introduce this family of policies,
with the goal of being as general as possible.

\subsection[Extended definition.]{Extended definition.}

Under this more general definition,
the match-maker makes decisions based not only on the vector of queue sizes,
but also (possibly) on additional information that is captured by the system state.
The state space is a couple $(\cS, | \cdot |)$,
where $\cS$ is a countably infinite set
and $| \cdot |: \cS \to \N^n$ is a function
that maps any state $s\in\cS$ to the vector giving the number of unmatched items of each class in that state,
denoted by $|s| = (|s|_1, |s|_2, \ldots, |s|_n)$.
The existence of the function $| \cdot |$
guarantees that the system state
contains enough information to retrieve
the number of unmatched items of each class,
which is a classical assumption in queueing theory
(see for instance \citet[Section~3.2]{kelly}).
We assume that there exists
a unique state $s \in \cS$ such that $|s| = 0$.
This state is called the \emph{empty state} and denoted by $\varnothing$.
This assumption guarantees that
the intuitive notion of system stability
will be captured by the positive recurrence
of the Markov chain describing the evolution of the system state\footnote{Without this assumption, one may construct two Markov chains associated with the same system, one positive recurrent and the other transient, for example if the state of the latter Markov chain embeds the time~$t$. This assumption is used only in the proofs of \Cref{prop:convexity,prop:stability-region-greedy}. In both cases, we can verify that the same conclusion holds as long as the set of states $s \in \cS$ such that $|s| = 0$ is finite. Assuming that this set is reduced to a singleton is merely a notational convenience.}.

The policy is now a function
$\Phi: \cS \times \V \times (\V \cup \{\bot\}) \times \cS \to [0, 1]$
such that $\Phi(s, i, j, s^\prime)$ is the conditional probability that,
given an incoming class-$i$ item finds the system in state~$s$,
the matching decision is~$j$ and the new state is~$s^\prime$.
More formally, the dynamics are described by a Markov chain
$((S_t, I_t, J_t), t \in \N)$,
where $I = (I_t, t \in \N)$ is the sequence of incoming item classes and,
for each $t \in \N$, $i \in \V$, $j \in \V \cup \{\bot\}$, and $s, s^\prime \in \cS$, we have
\begin{align*}
	\mathbb{P}(J_t = j, S_{t+1} = s^\prime \; | \; S_t = s, I_t = i)
	&= \Phi(s, i, j, s^\prime).
\end{align*}
The stochastic process $S = (S_t, t \in \N)$
is also a Markov chain, with transition probabilities
\begin{align*}
	\mathbb{P}(S_{t+1} = s^\prime \; | \; S_t = s)
	&= \frac
	{ \sum_{i \in \V} \lambda_i \sum_{j \in \V \cup \{\bot\}} \Phi(s, i, j, s^\prime) }
	{ \sum_{i \in \V} \lambda_i },
	\quad t \in \N,
	\quad s, s^\prime \in \cS.
\end{align*}
We assume that the Markov chain~$S$ has state space~$\cS$ and is irreducible,
and that $S_0 = \varnothing$.
The policy is assumed to be \emph{adapted} to the compatibility graph~$G$ and \emph{consistent}
in the sense that, for each $(s, i, j, s^\prime) \in \cS \times \V \times (\V \cup \{\bot\}) \times \cS$,
we have $\Phi(s, i, j, s^\prime) > 0$ only if
\begin{align*}
	j \in \{j^\prime \in \V_i : |s|_{j^\prime} \ge 1 \} \cup \{\bot\},
	\quad \text{and} \quad
	|s^\prime| = \begin{cases}
		|s| + \one_{i} &\text{if $j = \bot$,} \\
		|s| - \one_{j} &\text{if $j \neq \bot$.}
	\end{cases}
\end{align*}

Using this extended definition, the previously defined policy models can be easily expressed. For example, the matching policy~$\Phi$ is called \emph{size-based}
if $| \cdot |$ is the identity (implying $\cS \subset \N^n$)
and \emph{deterministic} if,
for each $s \in \cS$ and $i \in \V$,
there exists $(j, s^\prime) \in (\V \cup \{\bot\}) \times \cS$
such that $\Phi(s, i, j, s^\prime) = 1$.
The policy is called \emph{greedy} if
$\sum_{s^\prime \in \cS} \Phi(s, i, \bot, s^\prime) = 0$
for each $(s, i) \in \cS \times \V$
such that $\{j \in \V_i: |s|_j \ge 1\} \neq \emptyset$,
and \emph{non-greedy} otherwise. 
\gls{fcfm} (e.g., see \citet{C22,MBM21})
is a classical example of a deterministic policy
that is not size-based:
its state space is a couple $(\cS, | \cdot |)$ where
$\cS$ is a subset of the set
of sequences $c = (c_1, c_2, \ldots, c_p)$
made of a finite but arbitrarily large number~$p$ of elements of $\I$,
and $|c|_i$ is the cardinality of the set $\{q \in \{1, 2, \ldots, p\}: c_q = i\}$,
for each $i \in \I$.
A policy that is neither size-based nor deterministic
will appear in the proof of \Cref{prop:convexity}.

The stochastic process $Q = (Q_t, t \in \N)$
defined by $Q_t = |S_t|$ for each $t \in \N$
is called the \emph{queue-size process}.
This process does not satisfy the Markov property in general,
but it does satisfy the evolution equations~\eqref{eq:Q-rec} and~\eqref{eq:Q-unfolded},
with $L_i = (L_{t, i}, t \in \N)$ and $M_k = (M_{t, k}, t \in \N)$
defined by~\eqref{eq:L} and~\eqref{eq:M} for each $i \in \V$ and $k \in \E$.
The state space of the queue-size process
is given by $\Q = \{ |s|, s \in \cS \}$.
The policy is greedy if $\Q = \Q_\Gre$
and non-greedy if $\Q \supsetneq \Q_\Gre$,
where $\Q_\Gre$ is still given by~\eqref{eq:Q-greedy}.

\begin{rem}[Arrival rates vs.\ arrival sequence]
	We will often identify the matching model $(G, \lambda, \Phi)$
	with the Markov chain~$S$.
	This is a slight abuse of language:
	the triplet $(G, \lambda, \Phi)$ specifies
	the transition diagram of this Markov chain
	but, even if $\Phi$ is deterministic,
	characterizing its sample paths
	requires specifying the sequence~$I$,
	sampled according to $\lambda = (\lambda_1, \lambda_2, \ldots, \lambda_n)$.
	This slight abuse of language will not cause confusion
	when discussing stability and matching rates,
	but the distinction will matter in \Cref{sec:non-unicyclic}.
\end{rem}

\begin{rem}[Discrete time vs.\ continuous time]
	The discrete-time Markov chain~$S$
	gives the sequence of states observed by incoming items,
	and it was analyzed under various policies~\cite{MM16,JMRS20}.
	Yet, in queueing theory, it is more common
	to consider the continuous-time Markov chain
	describing the system state over time.
	However, as observed in \citet[Section~2.2.2]{C22},
	$S$ is the jump chain of this continuous-time Markov chain,
	and both Markov chains have the same stationary measures
	because the departure rate from each state
	in the continuous-time Markov chain
	is constant equal to $\sum_{i \in \V} \lambda_i$.
	Therefore, our results are equally relevant
	to study performance metrics
	like the mean queue size or the mean waiting time of items.
\end{rem}

\subsection[Equivalent policies.]{Equivalent policies.} \label{sec:equivalent_policies}

With our extended definition, a decision rule can be associated with an infinite number of policies. For instance, it is always possible to artificially expand the state definition, resulting in an unlimited range of policies. 
We define here an equivalence relation between policies that captures the intuitive concepts of 
\emph{yielding identical distributions of matching decisions} (for random policies) and \emph{making the same decisions} (for deterministic policies).
This discussion will also prepare the ground for \Cref{prop:Kn-greedy-policy,prop:D-greedy-policy}.

Consider a policy $\Phi_1$ adapted to a compatibility graph~$G = (\V, \E)$,
and let $(\cS_1, | \cdot |_1)$ denote its state space.
Assume that the function $| \cdot |_1 : \cS_1 \to \N^n$
can be written as a composition of two functions,
$\langle \cdot \rangle : \cS_1 \to \cS_2$
and $| \cdot |_2 : \cS_2 \to \N^n$,
such that $\cS_2$ is the image of $\cS_1$ through~$\langle \cdot \rangle$.
Moreover, assume that
there exists a policy $\Phi_2$ with state space $(\cS_2, | \cdot |_2)$,
adapted to the graph~$G$, such that
for each $s_2, s_2^\prime \in \cS_2$, $i \in \V$, and $j \in \V \cup \{\bot\}$, we have
\begin{align} \label{eq:equivalence}
	\sum_{s_1^\prime \in \cS_1: \langle s_1^\prime \rangle = s_2^\prime}
	\Phi_1(s_1, i, j, s_1^\prime)
	= \Phi_2(s_2, i, j, s_2^\prime)
	\quad \text{for each } s_1 \in \cS_1 \text{ such that } \langle s_1 \rangle = s_2.
\end{align}
We say that policy $\Phi_1$ can be \emph{reduced} to policy~$\Phi_2$ and that $\langle \cdot \rangle$ is a \emph{reduction} function.
If $((S_{1, t}, I_t, J_{1, t}), t \in \N)$
and $((S_{2, t}, I_t, J_{2, t}), t \in \N)$ denote
the Markov chains associated with policies~$\Phi_1$ and $\Phi_2$, respectively,
under the same sequence $(I_t, t \in \N)$ of incoming item classes,
then for each $t \in \N$,
(i) the conditional distribution of $(J_{1, t}, \langle S_{1, t+1} \rangle)$
given that $S_{1, t} = s_1$ and $I_t = i$
is the same for all states $s_1 \in \cS_1$
that have the same image $s_2 = \langle s_1 \rangle$, and
(ii) $(\langle S_{1, t} \rangle, I_t, J_{1, t})$ and $(S_{2, t}, I_t, J_{2, t})$
have the same distribution.
Conclusion~(ii) follows from an inductive argument
and implies that policies $\Phi_1$ and $\Phi_2$
are stable or unstable under the same conditions and, if stable,
yield the same matching rate vector.

The special case where the policy $\Phi_2$ is deterministic
will be useful in \Cref{prop:Kn-greedy-policy,prop:D-greedy-policy}.
In this case, \eqref{eq:equivalence} says that,
for each $s_2 \in \cS_2$ and $i \in \V$,
there exist $j \in \V \cup \{\bot\}$
and $s_2^\prime \in \cS_2$ such that
$ \sum_{s_1^\prime \in \cS_1: \langle s_1^\prime \rangle = s_2^\prime} \Phi_1(s_1, i, j, s_1^\prime)
= \Phi_2(s_2, i, j, s_2^\prime) = 1$
for each $s_1 \in \cS_1$ such that $\langle s_1 \rangle = s_2$.
This condition implies that
the Markov chains under $\Phi_1$ and $\Phi_2$ are equivalent pathwise
(i.e., $\langle S_{1, t} \rangle = S_{2, t}$ and $J_{1, t} = J_{2, t}$ for each $t \in \N$)
and not just in distribution.

In general, we say that two policies $\Phi_1$ and $\Phi_2$ adapted to the graph~$G$
are \emph{equivalent} if there exists a policy~$\Phi$ adapted to the graph~$G$
such that both~$\Phi_1$ and~$\Phi_2$ can be reduced to $\Phi$.
This equivalence between~$\Phi_1$ and~$\Phi_2$ can be interpreted as indicating that the two policies, when they are in equivalent states, i.e. states that have the same reduction,  yield identical distributions of matching decisions. It follows that,
if we let $(Q_{1, t}, t \in \N)$ and $(Q_{2, t}, t \in \N)$
denote the queue-size processes under $\Phi_1$ and $\Phi_2$
with the same sequence $(I_t, t \in \N)$ of incoming item classes,
then we have $\PP( Q_{1, t} = q ) = \PP(Q_{2, t} = q)$
for each $t \in \N$ and $q \in \Q$.

\newpage

 % appendix/ec2-stability
\section[Supplementary material of Section \ref*{sec:stability}.]{Supplementary material of \fullref{sec:stability}.} \label{app:graph}

\subsection[Proofs of Section \ref*{subsec:graph}.]{Proofs of \fullref{subsec:graph}.}\label{app:proofs-of-fullrefsubsecgraph}

\begin{proof}[Proof of \Cref{def:surjective}]
The equivalence of~\ref{cond:surjective-1}, \ref{cond:surjective-2},
and \ref{cond:surjective-3} is a well-known result
in linear algebra.
We prove that conditions~\ref{cond:surjective-3}
and~\ref{cond:surjective-4} are equivalent.
This proof is adapted from \citet[Lemma~2.2.3]{C04}.

The key argument consists of observing
that a vector $x = (x_1, x_2, \ldots, x_n) \in \R^n$
belongs to the left kernel of the matrix~$A$ if and only if
\begin{align*}
	\sum_{i = 1}^n x_i a_{i, k} = 0,
	\quad k \in \{1, 2, \ldots, m\}.
\end{align*}
For each $k \in \{1, 2, \ldots, m\}$,
the $k$-th equation reads $x_j = - x_i$,
where $i$ and $j$ are the endpoints of edge~$k$.
An induction argument shows that,
for every path $i_1, i_2, \ldots, i_k$
in the graph~$G$,
we have $x_{i_p} = (-1)^{p-1} x_{i_1}$
for each $p \in \{1, 2, \ldots, k\}$.

First assume that condition~\ref{cond:surjective-4} is satisfied.
Let $x \in \R^n$ be a vector of the left kernel of the matrix~$A$.
Since each connected component of~$G$ is non-bipartite,
for each $i \in \V$,
there exists a path of length say~$\ell$
that connects node~$i$ to a cycle
$i_1, i_2, \ldots, i_p, i_{p+1} = i_1$
consisting of an odd number~$p$ of nodes.
We then obtain $x_{i} = (-1)^{\ell+p+\ell} x_{i} = - x_{i}$,
which implies that $x_i = 0$.
Therefore, the left kernel of $A$ is trivial,
meaning that condition~\ref{cond:surjective-3} is satisfied.

On the contrary,
if condition~\ref{cond:surjective-4} is not satisfied,
then there exists a connected component of~$G$
that is bipartite with parts $\V_+$ and $\V_-$.
We build a non-zero vector in the left kernel of $A$ by choosing
$x_i = 1$ for each $i \in \V_+$,
$x_i = -1$ for each $i \in \V_-$,
and $x_i = 0$ for each $i \in \V \setminus (\V_+ \cup \V_-)$.
This implies that condition~\ref{cond:surjective-3} is not satisfied.
\end{proof}

\begin{proof}[Proof of \Cref{def:injective}]
The equivalence of conditions~\ref{cond:injective-1}, \ref{cond:injective-2},
and \ref{cond:injective-3} is a well-known result
in linear algebra.
We now prove that conditions~\ref{cond:injective-3}
and~\ref{cond:injective-4} are equivalent.

We first assume that the graph~$G$ is connected, and we distinguish the following two cases:
\begin{itemize}
	\item If~$G$ is non-bipartite, according to \Cref{def:surjective}, the nullity of $A^\intercal$ is~0.
	The rank-nullity theorem implies that the rank of $A^\intercal$ is~$n$,	so that the rank of $A$ is also~$n$.
	A second application of the rank-nullity theorem implies that the nullity of $A$ is $m - n$.
	In particular, $\ker(A) = \{0\}$ if and only if $m = n$.
	\item If $G$ is bipartite, any non-zero vector of the left kernel of $A$ must be parallel (collinear) to the non-zero vector $x$ constructed in the proof of \Cref{def:surjective}. This parallelism is due to the constraints $x_i=-x_j$ for all edges $\{i,j\}$. Based on this, the nullity of $A^\intercal$ is~1,
	and we conclude from another double application of the rank-nullity theorem that the nullity of $A$ is $m - n + 1$. In particular, $\ker(A) = \{0\}$ if and only if $m = n - 1$.
\end{itemize}
All in all, we obtain that condition~\ref{cond:injective-3} is true
if and only if
either the graph~$G$ is non-bipartite
and contains as many edges as nodes,
or the graph~$G$ is bipartite
and contains one less edge than it contains nodes.
This, in turn, is equivalent to condition~\ref{cond:injective-4}.

If the graph~$G$ is not connected,
we can rewrite~$A$ as a bloc matrix
in which each bloc corresponds to a connected component,
and we can then use the previous argument
to prove the equivalence for each connected component.
\end{proof}

\begin{proof}[Proof of \Cref{def:bijective}]
The function $\mu \in \R^m \mapsto A\mu \in \R^n$ is bijective
if and only if it is both surjective and injective.
Hence, the equivalence of
conditions~\ref{cond:bijective-1} to~\ref{cond:bijective-4}
follows directly from \Cref{def:surjective,def:injective}.
\end{proof}

\begin{proof}[Proof of \Cref{prop:dimensions}]
These statements (transposed to $A$) are again well-known in linear algebra.
\end{proof}

\subsection[Proof of Proposition \ref*{prop:stability-region-form}.]{Proof of \Cref{prop:stability-region-form} in  \fullref{subsec:stability-region-form}.}\label{app:proofs-of-fullrefsubsecstability-region-form}

We first note that condition~\ref{cond:stability-region-form-2} alone implies that
every connected component of~$G$ is non-bipartite, i.e.\ that~$G$ is surjective: a
bipartite component with parts~$P$ and~$Q$, \emph{both non-empty}, would, applying
\ref{cond:stability-region-form-2} to the independent sets $\I = P$ and $\I = Q$
(for which $\V(\I) = Q$ and $\V(\I) = P$, respectively), force both
$\sum_{i \in P}\lambda_i < \sum_{i \in Q}\lambda_i$ and
$\sum_{i \in Q}\lambda_i < \sum_{i \in P}\lambda_i$, a contradiction; and a component
reduced to a single isolated vertex~$v$ (the degenerate case of an empty part) is
excluded directly, since $\I = \{v\}$ would give
$\lambda_v < \sum_{i \in \V(\{v\})}\lambda_i = 0$, impossible as $\lambda > 0$.
The equivalences below therefore require no standing surjectivity assumption.

Equivalence of~\ref{cond:stability-region-form-1}
and~\ref{cond:stability-region-form-2}
follows from \citet[Proposition~2 and Theorem~2]{MM16}.
For completeness, we observe that \citet[Proposition~2]{MM16} is proved
under the assumption that
the matching policy~$\Phi$ is greedy and deterministic
and that the state space $(\cS, | \cdot |)$ has a particular form,
but we can verify that
the argument remains valid under the assumptions of \Cref{subsec:model}.
We now prove that~\ref{cond:stability-region-form-2}
and~\ref{cond:stability-region-form-3}
are equivalent.
Condition~\ref{cond:stability-region-form-2}
implies condition~\ref{cond:stability-region-form-3} because
(a) according to~\citet{MM16},
under condition~\ref{cond:stability-region-form-2},
$(G, \lambda, \Phi)$ is stable when $\Phi$ is the \gls{ml} policy, and
(b) the associated vector~$\mu$ of matching rates
satisfies condition~\ref{cond:stability-region-form-3}
by ergodicity: $A\mu = \lambda$ and $\mu \ge 0$ hold in stationarity, $G$ is
surjective as just noted, and each $\mu_k > 0$ because, starting from the empty
state, two consecutive arrivals of the endpoints of edge~$k$ are matched along~$k$
by the greedy \gls{ml} policy, so edge~$k$ carries a positive matching rate.
That condition~\ref{cond:stability-region-form-3}
implies condition~\ref{cond:stability-region-form-2}
was proved by \citet[Lemma~12]{C22}.

It remains to prove the equivalence of condition~\ref{cond:stability-region-form-4}
with the other conditions. The implication \ref{cond:stability-region-form-3} $\Rightarrow$
\ref{cond:stability-region-form-4} is immediate: a strictly positive solution~$\mu$
has support graph $G^\star = G$, which is surjective by~\ref{cond:stability-region-form-3}.
Conversely, assume~\ref{cond:stability-region-form-4} and let $\mu \in \Rnn^m$ solve
\eqref{eq:system} with a surjective support graph $G^\star = (\V, \E^\star)$, where
$\E^\star = \{k \in \E : \mu_k > 0\}$. The restriction of~$\mu$ to~$\E^\star$ is a
strictly positive solution of the conservation equation of the problem
$(G^\star, \lambda)$; since $G^\star$ is surjective, $(G^\star, \lambda)$ satisfies
\ref{cond:stability-region-form-3} and is therefore stabilizable, by the equivalence
\ref{cond:stability-region-form-3} $\Rightarrow$ \ref{cond:stability-region-form-1}
already established for the (surjective) graph~$G^\star$. Any policy that stabilizes
$(G^\star, \lambda)$ also defines a policy on $(G, \lambda)$ that never activates
edges outside~$\E^\star$, and hence stabilizes $(G, \lambda)$; this gives
\ref{cond:stability-region-form-1} and closes the chain of equivalences.
\hfill \Halmos
\endproof

\subsection[Minimal stability region for greedy matching policies.]{Minimal stability region for greedy matching policies (\Cref{subsubsec:surjective-graphs,app:greedy-complete,app:greedy-diamond}).}
\label{app:minimal}

The following result gives a sufficient stability condition for greedy matching policies. The proof relies on a linear Lyapunov function. This result can be seen as the counterpart of \citet[Proposition~5.1]{BGM13} for non-bipartite matching models.

\begin{prop} \label{prop:stability-region-greedy}
	Consider a matching problem~$(G, \lambda)$
	with a connected graph~$G$.
	If
	\begin{align} \label{eq:scond}
		\sum_{i \in \V(\I)} \lambda_i
		> \frac12 \sum_{i \in \V} \lambda_i,
		\quad \I \in \ind,
	\end{align}
	then the matching model $(G, \lambda, \Phi)$ is stable
	for every greedy matching policy $\Phi$.  
\end{prop}

\begin{proof}
Consider a matching problem $(G, \lambda)$ that satisfies~\eqref{eq:scond}
and a greedy matching policy~$\Phi$ adapted to the graph~$G$.
Since the Markov chain $(S_t, t \in \N)$
associated with the matching model $(G, \lambda, \Phi)$
depends on the vector~$\lambda$ only
up to a positive multiplicative constant,
we can assume without loss of generality
that $\sum_{i \in \V} \lambda_i = 1$.
Let $\cS$ denote the state space of this Markov chain
and $| \cdot |$ the corresponding queue-size function
(as defined in \Cref{app:extended-definition}).
We consider the Lyapunov function
$F : \cS \to \R$ defined by
$F(s) = \sum_{i \in \V} |s|_i$
(that is, $F(s)$ is the number of unmatched items in state~$s$)
for each $s \in \cS$.
For each $t \in \N$ and $s \in \cS$, we have
\begin{align*}
	\mathbb{E} \left( F(S_{t+1}) \; | \; S_t = s \right) - F(s)
	&= \sum_{i \in \V \setminus \V(\I)} \lambda_i
	- \sum_{i \in \V(\I)} \lambda_i
	= - \left( \sum_{i \in \V(\I)} \lambda_i
	- \sum_{i \in \V \setminus \V(\I)} \lambda_i \right),
\end{align*}
where $\I = \{i \in \V: |s|_i \ge 1 \}$
is the set of classes of unmatched items in state~$s$.
Importantly, if $F(s) > 0$ (that is, $s \neq \varnothing$),
then $\I$ is an independent set of the compatibility graph~$G$
because it is non-empty and the policy~$\Phi$ is greedy.
It follows that, for each $s \in \cS \setminus \{ \varnothing \}$,
\begin{align*}
	\mathbb{E} \left( F(S_{t+1}) \; | \; S_t = s \right) - F(s)
	&\le - \varepsilon,
	\quad \text{with }
	\varepsilon = \min_{\I \in \ind} \left(
	\sum_{i \in \V(\I)} \lambda_i - \sum_{i \in \V \setminus \V(\I)} \lambda_i
	\right).
\end{align*}
Equation~\eqref{eq:scond} implies that $\varepsilon > 0$.
Using the Lyapunov-Foster theorem,
see \citet[Theorem~1.1 in Chapter~5]{B99},
we conclude that the matching model
$(G, \lambda, \Phi)$ is stable.
\end{proof}

\noindent
As one would expect,
any matching problem $(G, \lambda)$
that satisfies \eqref{eq:scond} 
is stabilizable in the sense of \Cref{def:stability}.
Indeed, \eqref{eq:scond}
implies \Cref{prop:stability-region-form}\ref{cond:stability-region-form-2}
because $\I \subseteq \V \setminus \V(\I)$
for each $\I \in \ind$.
\Cref{cor:max-greedy} below shows that, conversely,
whether a stabilizable matching problem
satisfies~\eqref{eq:scond}
depends on the structure of the graph~$G$:
conditions~\ref{cond:max-greedy-1}
and~\ref{cond:max-greedy-2}
exhibit compatibility graphs~$G$
such that \eqref{eq:scond}
is satisfied whenever
the matching problem $(G, \lambda)$ is stabilizable,
while conditions~\ref{cond:max-greedy-3}
and~\ref{cond:max-greedy-4}
exhibit stabilizable compatibility graphs~$G$
for which \eqref{eq:scond}
is never satisfied.

\begin{coro} \label{cor:max-greedy}
	Consider a matching problem $(G, \lambda)$. \\
	Under the following two conditions, the stabilizability of the matching model $(G, \lambda)$
	implies that~\eqref{eq:scond} is satisfied, and therefore that the matching model $(G, \lambda, \Phi)$ is stable under every greedy policy~$\Phi$ adapted to~$G$:
	\begin{enumerate}[(i)]
		\item \label{cond:max-greedy-1}
		$G$ is a complete graph with $n \ge 3$ nodes.
		\item \label{cond:max-greedy-2}
		$G$ is the diamond graph of \Cref{ex:diamond}.
	\end{enumerate}
	Under the following conditions, \eqref{eq:scond} is never satisfied:
	\begin{enumerate}[(i)]
		\setcounter{enumi}{2}
		\item \label{cond:max-greedy-3}
		The graph~$G$ has diameter greater than or equal to 3.
		\item \label{cond:max-greedy-4}
		The graph~$G$ contains a leaf (that is, a node with degree 1).
	\end{enumerate}
\end{coro}

\begin{proof}
We first need to prove that,
under either condition~\ref{cond:max-greedy-1} or condition~\ref{cond:max-greedy-2},
the matching model $(G, \lambda)$ is stabilizable
if and only if~\eqref{eq:scond} is satisfied.
We proceed by verifying that, under either of these two conditions,
\Cref{prop:stability-region-form}\ref{cond:stability-region-form-2}
and \eqref{eq:scond} are equivalent:
\begin{enumerate}[(i)]
	\item First assume that condition~\ref{cond:max-greedy-1} is satisfied.
	The independent sets of a complete graph~$K_n$ are the singletons.
	Using this observation, we can verify that
	\Cref{prop:stability-region-form}\ref{cond:stability-region-form-2}
	and~\eqref{eq:scond}
	are both equivalent to
	$\lambda_i < \frac12 \sum_{i \in \V} \lambda_i$ for each $i \in \V$.
	\item Now assume that condition~\ref{cond:max-greedy-2} is satisfied,
	that is, $G$ is the diamond graph.
	The conclusion follows by recalling
	that \Cref{prop:stability-region-form}\ref{cond:stability-region-form-2}
	simplifies to~\eqref{eq:diamond-stabilizability},
	and then by observing that~\eqref{eq:diamond-stabilizability}
	and~\eqref{eq:scond} are equivalent.
\end{enumerate}
To prove that~\eqref{eq:scond} cannot be satisfied
under either condition~\ref{cond:max-greedy-3} or~\ref{cond:max-greedy-4},
we proceed by contradiction:
\begin{enumerate}[(i)]
	\setcounter{enumi}{2}
	\item First assume that condition~\ref{cond:max-greedy-3} is satisfied,
	and let $i$ and $j$ denote two nodes
	that are at distance~$3$ or more.
	In particular, the sets $\V_i$ and $\V_j$ are disjoint.
	If~\eqref{eq:scond} is satisfied, then applying this equation
	to both~$\{i\}$ and $\{j\}$ and summing the inequalities yields
	$\sum_{i^\prime \in \V_i \cup \V_j} \lambda_{i^\prime} > \sum_{i^\prime \in \V} \lambda_{i^\prime}$,
	which is a contradiction since
	$\V_i \cup \V_j \subseteq \V$.
	Hence, \eqref{eq:scond}
	cannot be satisfied by both $\{i\}$ and $\{j\}$.
	\item Now assume that condition~\ref{cond:max-greedy-4} is satisfied.
	Let $i$ denote a leaf node of~$G$
	and $j$ the (only) neighbor of~$i$.
	Then again, applying~\eqref{eq:scond}
	to both~$\{i\}$ and $\{j\}$ and summing the inequalities yields
	$\sum_{i \in \V_j \cup \{j\}} \lambda_{i^\prime} > \sum_{i^\prime \in \V} \lambda_{i^\prime}$,
	which is again a contradiction.
\end{enumerate}
\end{proof}

\subsection[Closed form and examples for bijective graphs.]{Closed form and examples for bijective graphs (\Cref{sec:unicyclic}).} \label{app:unicyclic}

\begin{proof}[Proof of \Cref{prop:unicyclic}]
We first prove~\eqref{eq:unicyclic-border}
for every edge~$k$ that does not belong to the cycle.
As observed in the proposition,
each edge~$k$ that does not belong to the cycle
separates the graph into two parts,
one of which is a tree with node set~$T_k$;
the rooted tree associated with~$k$
is obtained by designating
the corresponding endpoint of edge~$k$ as the root.
We now prove~\eqref{eq:unicyclic-border}
by induction on the height this rooted tree.
Equation~\eqref{eq:unicyclic-border}
is true if the height of this tree is zero.
Indeed, in this case,
the endpoint of edge~$k$ that belongs to the tree,
say node~$i$, has no other incident edge,
so that applying~\eqref{eq:system-equations}
to node~$i$ yields $\mu_k = \lambda_i$,
which is consistent with~\eqref{eq:unicyclic-border}.
Now assume that the assumption is satisfied
for each edge whose associated rooted tree
has height at most $h - 1$ for some $h \ge 0$,
and consider an edge~$k$
whose associated rooted tree has height~$h$.
By applying~\eqref{eq:system-equations}
to the root~$i$ of this associated rooted tree,
we obtain
\begin{align} \label{eq:tree}
	\mu_k
	= \lambda_i - \sum_{\ell \in \E_i \setminus \{k\}} \mu_\ell.
\end{align}
The induction hypothesis guarantees
that~\eqref{eq:unicyclic-border}
is satisfied for every $\ell \in \E_i \setminus \{k\}$
(as the height of the associated rooted tree is at most $h - 1$).
After injecting this observation to~\eqref{eq:tree},
the result for edge~$k$ follows by observing that
$d_{j, k} = d_{j, \ell} + 1$
for each $\ell \in  \E_i \setminus \{k\}$
and $j \in T_\ell$,
and that
$T_k = \{i\} \cup (\bigcup_{\ell \in \E_i \setminus \{k\}} T_\ell)$
(all sets being disjoint).

We now prove~\eqref{eq:unicyclic-cycle}
for each edge~$k$ that belongs to the cycle.
Since the graph~$G$ is unicyclic,
deleting edge~$k$ from~$G$
yields a (connected) tree,
and therefore a bipartite graph.
We let $\V_+$ denote set of nodes in the part
that contains both endpoints of edge~$k$
(that both endpoints belong to the same part
follows from the fact that the cycle is odd)
and $\V_-$ the set of nodes in the other part.
We obtain
\begin{align*}
	\sum_{i \in \V_+} \lambda_i
	- \sum_{i \in \V_-} \lambda_i
	&= \sum_{i \in \V_+}
	\sum_{\ell \in \E_i} \mu_\ell
	- \sum_{i \in \V_-}
	\sum_{\ell \in \E_i} \mu_\ell
	= 2 \mu_k.
\end{align*}
The first equality
follows from~\eqref{eq:system-equations}.
The second equality holds because
each edge~$\ell \in \E \setminus \{k\}$
has one endpoint in~$\V_+$ and another in~$\V_-$,
so that $\mu_\ell$ appears
once in the first nested sum and once in the second;
on the contrary,
since both endpoints of edge~$k$ belong to $\V_+$,
$\mu_k$ appears twice in the first nested sum
and zero times in the second.
Equation~\eqref{eq:unicyclic-cycle} follows
by observing that $d_{i, k}$ is even
if and only if $i \in \V_+$.
\end{proof}

We illustrate \Cref{prop:unicyclic} on two examples.

\begin{figure}[htb]
	\hfill
	\subfloat[Matching rates in the pentagon graph $\C_5$. Only rate $\mu_{1, 2}$ is shown for ease of display. The other rates are deduced by permutation.\label{fig:odd-cycle}]{%
		\begin{tikzpicture}[scale=.7]
			\def\d{2cm}
			\placenodes{2/1/0,3/2/-72,4/3/-144,5/1/-108}
			\halfwidth{4}{3.1cm}
			
			\foreach \i/\j/\l in {
				1/2/$\frac{\lambda_1+\lambda_2-\lambda_5-\lambda_3+\lambda_4}{2}$, 
				2/3/, 
				3/4/, 
				4/5/, 
				5/1/}
			{\draw (\i) edge node[above, yshift=.1cm] {\l} (\j);}
		\end{tikzpicture}
	}
	\hfill
	\subfloat[Matching rates in a ``lying puppet'' graph
	with $n = 9$ nodes and $m = 9$ edges.
	The differences $\bar{\lambda}_7=\lambda_7-\lambda_8-\lambda_9$,  $\bar{\lambda}_4=\lambda_4-\lambda_5-\lambda_6-\bar{\lambda}_7$, and $\bar{\lambda}_3=\lambda_3-\bar{\lambda}_4$ are the residual rates that classes $7$, $4$, and $3$ provide to their neighbors of lower index.\label{fig:puppet}]{%
		\tikzstyle{slab}=[sloped, above]
		\tikzstyle{slob}=[sloped, below]		
		\begin{tikzpicture}[scale=.8, transform shape]
			\node[class, minimum size=0] (1) {$1$};
			\foreach \i/\s/\a/\d in {2/1/90/2, 3/1/30/2, 4/3/0/2, 5/4/90/1.2, 6/4/-90/1.2, 7/4/0/2, 8/7/30/2, 9/7/-30/2}
			\path (\s) ++(\a:\d cm) node[class] (\i) {$\i$};
			\halfwidth{4}{5.1cm}
			
			\foreach \i/\j/\l/\p in {
				1/2/$\frac{\lambda_1+\lambda_2-\bar{\lambda}_3}{2}$/left, 
				2/3/$\frac{\lambda_2+\bar{\lambda}_3-\lambda_1}{2}$/slab, 
				1/3/$\frac{\lambda_1+\bar{\lambda}_3-\lambda_2}{2}$/slob, 
				3/4/$\bar{\lambda}_4$/above, 
				4/5/$\lambda_5$/left, 
				4/6/$\lambda_6$/left, 
				4/7/$\bar{\lambda}_7$/above, 
				7/8/$\lambda_8$/slab, 
				7/9/$\lambda_9$/slob}
			\draw (\i) edge node[\p] {\l} (\j);
		\end{tikzpicture}
	}
	\hfill
	\caption{Matching rates in bijective graphs.}
\end{figure}

\begin{exa}[Cycle graph with 5 nodes] \label{ex:unicyclic-1}
	A cycle graph is the simplest bijective graph that we can consider,
	as it contains an odd cycle and no other edges.
	In the cycle graph of \Cref{fig:odd-cycle},
	a direct application of \Cref{prop:unicyclic}\ref{cond:unicyclic-2} yields
	$\mu_{1,2} = \frac12 (\lambda_1 + \lambda_2 - \lambda_3 + \lambda_4 - \lambda_5)$,
	as we have $d_{1, \{1, 2\}} = d_{2, \{1, 2\}} = 0$, $d_{3, \{1, 2\}} = d_{5, \{1, 2\}} = 1$, and $d_{4, \{1, 2\}} = 3$; regarding node~3 for instance, the endpoint of edge~$\{1, 2\}$ that is closest to node~3 is node~2, hence $d_{3, \{1, 2\}}$ is equal to the distance between nodes~2 and~3, which is 1.
	Matching rates along other edges follow by symmetry.
	From the point of view of edge $\{1, 2\}$,
	we can partition nodes into two sets,
	namely $\{1, 2, 4\}$ and $\{3, 5\}$.
	The former (resp.\ latter) set contains nodes
	at an even (resp.\ odd) distance of edge~$\{1, 2\}$,
	and increasing the arrival rate of these nodes
	increases (resp.\ decreases) the matching rate along edge~$\{1, 2\}$.
\end{exa}

\begin{exa}[Lying puppet] \label{ex:unicyclic-2}
	We now consider the graph of \Cref{fig:puppet}.
	Edges $\{1, 2\}$, $\{1, 3\}$, and $\{2, 3\}$ belong to the cycle,
	and the other edges do not.
	According to \Cref{prop:unicyclic}, we have
	\begin{align*}
		\mu_{1, 2}
		&= \frac{\lambda_1 + \lambda_2 - \bar\lambda_3}2,
		&
		\mu_{1, 3}
		&= \frac{\lambda_1 - \lambda_2 + \bar\lambda_3}2,
		&
		\mu_{2, 3}
		&= \frac{- \lambda_1 + \lambda_2 + \bar\lambda_3} 2,
	\end{align*}
	where $\bar\lambda_3 = \lambda_3 - \mu_{3, 4}$,
	and%
	\begin{align*}
		\mu_{4, 5}
		&= \lambda_5,
		&
		\mu_{4, 6}
		&= \lambda_6,
		&
		\mu_{7, 8}
		&= \lambda_8,
		\\
		\mu_{7, 9}
		&= \lambda_9,
		&
		\mu_{4, 7}
		&= \lambda_7 - \mu_{7, 8} - \mu_{7, 9},
		&
		\mu_{3, 4}
		&= \lambda_4 - \mu_{4, 5} - \mu_{4, 6} - \mu_{4, 7}.
	\end{align*}
	This second set of equations can be obtained
	either by a direct application of~\eqref{eq:unicyclic-border}
	or by applying~\eqref{eq:system-equations} recursively
	from the leaves.
	Indeed, applying~\eqref{eq:system-equations}
	to nodes $5$, $6$, $8$, and $9$ gives directly
	the values of $\mu_{4,5}$, $\mu_{4,6}$,
	$\mu_{7,8}$, and $\mu_{7,9}$,
	then applying~\eqref{eq:system-equations}
	to node~$7$ gives the value of $\mu_{4,7}$,
	and finally applying~\eqref{eq:system-equations}
	to node~$4$ gives the value of $\mu_{3,4}$.
	The values of $\mu_{1,2}$, $\mu_{1,3}$, and $\mu_{2,3}$
	are similar to~\Cref{ex:paw},
	where the arrival rate~$\lambda_3$ is again replaced
	with the effective arrival rate~$\bar\lambda_3$
	from the point of view of classes~1 and~2.
\end{exa}

\newpage

 % appendix/ec3-general
\section[Supplementary material of Section \ref*{sec:general}.]{Supplementary material of \fullref{sec:general}.}
\subsection[Algorithm of Section \ref*{sec:basis-of-the-kernel-of-the-incidence-matrix}.]{Algorithm of \fullref{sec:basis-of-the-kernel-of-the-incidence-matrix}.}
\label{app:spanning}

Given a surjective-only graph~$G$
with an incidence matrix denoted by~$A$,
\Cref{algo:spanning}
builds a basis of the kernel of~$A$ as follows.
The algorithm first identifies a spanning tree~$\T$ of~$G$ (Line~\ref{algo:T})
and an edge $a \in \E \setminus \T$ such that
the set $\E \setminus (\T \cup \{a\})$ (of cardinality $d = m = n$)
contains an odd cycle (Line~\ref{algo:edgek}).
Then, for each edge $s \in \E \setminus (\T \cup \{a\})$,
the algorithm builds (Lines \ref{algo:vector}--\ref{algo:kayakpaddle-cycle2})
a base vector~$b$ whose support
(i) is either an even cycle or a kayak paddle
and (ii) contains~$s$ and is included into $\T \cup \{a, s\}$.
We assume without loss of generality that the graph~$G$ is connected (in addition to being surjective-only). If not, we can apply the algorithm to each connected component separately, and then we embed the obtained vectors to $\R^n$ via zero padding.

\setcounter{algocf}{0}
\begin{algorithm}[!htb]
	\SetInd{.5em}{1em}
	\KwData{A connected surjective-only compatibility graph~$G = (\V, \E)$}
	\KwResult{A basis $\B$ of the kernel of the incidence matrix~$A$ of $G$}
	$\T \leftarrow$ the set of edges of a spanning tree of~$G$ \label{algo:T} \\
	$a \leftarrow$ an edge in $\E \setminus \T$
	such that $\T \cup \{a\}$
	contains an odd cycle
	\label{algo:edgek} \\
	$\B \leftarrow \emptyset$ \\
	\For{$s \in \E \setminus (\T \cup \{a\})$}{
		$b \leftarrow (0, 0, \ldots, 0) \in \R^m$ \label{algo:vector}\\ 
		\eIf{$\T \cup \{a, s\}$ contains an even cycle $C_\ell$}{%
			$c_1,\ldots,c_\ell$ $\leftarrow$ consecutive edges of $C_\ell$\\
			\For{$d\in \{1,\ldots,\ell\}$}{%
				$k \leftarrow$ index of $c_d$ in $E$\\
				$b_k \gets (-1)^d$ \label{algo:cycle}
			}
		}{%
			$\T \cup \{a, s\}$ contains a kayak paddle $\KP_{\ell, r, p}$ with $\ell$ odd, $r$ odd, and $p\geq 0$\\
			$v_i \leftarrow$ node connecting the kayak's cycle $C_\ell$ to the kayak central path $P_p$\\
			$v_j \leftarrow$ node connecting the kayak's cycle $C_r$ to the kayak central path $P_p$\\
			$c_1,\ldots,c_\ell$ $\leftarrow$ consecutive edges of $C_\ell$, starting and ending at node $v_i$\\
			\For{$d\in \{1,\ldots,\ell\}$}{%
				$k \leftarrow$ index of $c_d$ in $E$\\
				$b_k \gets (-1)^d$ \label{algo:kayakpaddle-cycle1}
			}
			$c_1,\ldots,c_p$ $\leftarrow$ consecutive edges of $P_p$, starting at node $v_i$ and ending at node $v_j$\\
			\For{$d\in \{1,\ldots,p\}$}{%
				$k \leftarrow$ index of $c_d$ in $E$\\
				$b_k = 2(-1)^{d+1}$ \label{algo:path}
			}
			$c_1,\ldots,c_r$ $\leftarrow$ consecutive edges of $C_r$, starting and ending at node $v_j$\\
			\For{$d\in \{1,\ldots,r\}$}{%
				$k \leftarrow$ index of $c_d$ in $E$\\
				$b_k \gets (-1)^{d+p+1}$  \label{algo:kayakpaddle-cycle2}
			}
		}
		$\B \leftarrow \B \cup \{b\}$ \\
		\textbf{return} $\B$
	}
	\vspace{.3cm}
	\caption{Construction of a basis of the kernel of the incidence matrix~$A$ of the compatibility graph~$G$. This algorithm was initially introduced by \citet[Section~3]{D73} to build a basis of the eigenspace associated with the eigenvalue $-2$ of the adjacency matrix of a line graph
		(i.e., a graph whose nodes and edges represent, respectively, the edges and their incidence relations in another graph).}
	\label{algo:spanning-full}
\end{algorithm}

We now verify that \Cref{algo:spanning} terminates and yields the desired result.

\begin{prop} \label{prop:spanning}
	\Cref{algo:spanning} terminates and returns a basis of the kernel of the incidence matrix~$A$ of the compatibility graph~$G$.
\end{prop}

\begin{proof}
The proof is based mainly on the notion of \emph{cycle space} in a graph.
We briefly summarize the concepts
that are useful to understand the proof (see \citet[Section~1.9]{D05} for details).

A \emph{spanning subgraph} of a graph~$G=(V, E)$ is a subgraph $G^\circ =(V, E^\circ)$ with $E^\circ \subseteq E$.
Importantly, $G$ and $G^\circ$ have the same set of nodes.
A subgraph is \emph{Eulerian} if every vertex has an even degree (possibly zero). In particular, if $E^\circ$ is a set of edges that form a cycle in~$G$, then the graph $(V, E^\circ)$ is Eulerian. The \emph{cycle space} of $G$ is the vector space made of all Eulerian spanning subgraphs of $G$, using the symmetric difference of the edge sets for addition and the two-element field for scalar multiplication.
Equivalently, the cycle space can be described
as a vector space of the finite field $\mathbb{Z} / 2 \mathbb{Z}$:
each vector $g = (g_1, g_2, \ldots, g_m)$ in this vector space
satisfies $\sum_{k \in \E_i} g_k = 0$ (modulo 2) for each $i \in \V$,
and the addition and multiplication
are the usual operations in $\mathbb{Z} / 2 \mathbb{Z}$.
For example, if $G$ is the codomino graph of \Cref{fig:codomino-kernel},
the spanning subgraphs $G_1$, $G_2$, and $G_3$ of $G$ with edge sets
$E_1 = \{ \{1, 2\}, \{1, 6\}, \{2, 3\}, \{3, 4\}, \{4, 5\}, \{5, 6 \}$,
$E_2 = \{ \{2, 3\}, \{2, 6\}, \{3, 5\}, \{5, 6\} \}$, and $E_3 = \{ \{1, 2\}, \{1, 6\}, \{2, 6\} \}$, respectively,
belong to the cycle space of~$G$.
The edge set of the addition $G_1 + G_2$
is the set $\{ \{1, 2\}, \{1, 6\}, \{2, 6\}, \{3, 4\}, \{3, 5\}, \{4, 5\} \}$
of edges that are either in $E_1$ or in $E_2$, but not in both.
Similarly, the edge set of $G_1 + G_3$ is $\{\{2, 3\}, \{2, 6\}, \{3, 4\}, \{4, 5\}, \{5, 6 \}\}$, and 
the edge set of $G_2 + G_3$ is $\{ \{1, 2\}, \{1, 6\}, \{2, 3\}, \{3, 5\}, \{5, 6 \}$.
One can verify that
$\{G_1, G_2, G_3\}$ forms a basis of the cycle space of~$G$.
Importantly, in general, the dimension of the cycle space is $m-n+1$.

\noindent \textbf{\Cref{algo:spanning} terminates.}
We first prove the existence of edge~$a$
defined on Line~\ref{algo:edgek} of \Cref{algo:spanning}.
By definition of a spanning tree, $\T$ contains $n-1$ edges and,
for each edge $a \in \E \setminus \T$,
$\T \cup \{a\}$ contains a unique cycle.
The $m - n + 1 = |\E \setminus \T|$ cycles thus obtained are independent
in the sense that each cycle contains at least one edge ($a$)
that is not contained in the other cycles.
Therefore, these $m - n + 1$ cycles form a basis of the cycle space of~$G$.
Since a linear combination of even cycles
cannot produce a subgraph consisting of a single odd cycle\footnote{A linear combination of even cycles may produce a subgraph consisting of an even number of disjoint odd cycles, as illustrated by $G_1+G_2$ in the example above, but not a subgraph consisting of one odd cycle. Indeed, the symmetric difference of two edge sets contains an even number of edges if both edge sets contain an even number of edges.}, 
and since $G$ contains an odd cycle (as it is non-bipartite),
then at least one of the $m - n + 1$ basis cycles is odd.

We now verify that,
for each $s \in \E \setminus (\T \cup \{a\})$,
$\T \cup \{a, s\}$
contains either (i) an even cycle $C_\ell$
or (ii) a kayak paddle $\KP_{\ell, r, p}$ with two odd cycles.
By construction, $\T \cup \{a\}$ contains
a unique cycle~$C_r$,
which is odd,
and $\T \cup \{s\}$
contains a unique cycle~$C_\ell$.
$\T \cup \{a, s\}$
contains both $C_r$ and $C_\ell$.
We now proceed by elimination:
\begin{itemize}
	\item If $C_\ell$ is even,
	then $C_\ell$ is an even cycle
	included into $\T \cup \{a, s\}$,
	and we are in case~(i).
	\item If $C_\ell$ is odd
	and shares at least one edge with $C_r$,
	then the symmetric difference
	of $C_r$ and $C_\ell$ is an even cycle,
	and it is again included
	into $\T \cup \{a, s\}$,
	so we are again in case~(i).
	\item If $C_\ell$ is odd
	and $C_r$ and $C_\ell$ have no edge in common,
	then we are in case~(ii).
\end{itemize}

\noindent \textbf{\Cref{algo:spanning} returns the correct result.}
We finally prove that
the family $\B$ returned by \Cref{algo:spanning}:
(i) has cardinality $m - n$,
(ii) is linearly independent,
and (iii) is included into the kernel of~$A$.
We prove each item one after another:
\begin{enumerate}[(i)]
	\item The family~$\B$ has cardinality~$m - n$:
	It suffices to observe that $\B$ has same cardinality as $\E \setminus (\T \cup \{a\})$,
	which we already mentioned has cardinality $m - n$.
	\item The family~$\B$ is linearly independent:
	For each $s \in \E \setminus (\T \cup \{a\})$,
	the basis vector constructed from edge~$s$
	is the only vector in~$\B$ whose support contains this edge.
	\item The family~$\B$ is included into the kernel of~$A$:
	Let $b \in \B$.
	Our goal is to prove that $b$ belongs to the kernel of~$A$,
	i.e., that $\sum_{k \in \E_i} b_k = 0$ for each $i \in \V$.
	First observe that, for each $i \in \V$, we have
	$\sum_{k \in \E_i} b_k = \sum_{k \in \E_i \cap S} b_k$,
	where $S$ is the support of the vector~$b$.
	In particular, we have immediately $\sum_{k \in \E_i} b_k = 0$
	for each $i \in \V$ such that $\E_i \cap S = \emptyset$.
	Now consider a node $i \in \V$ such that $\E_i \cap S \neq \emptyset$.
	We make a case disjunction depending on the support~$S$ of~$b$:
	\begin{itemize}
		\item If $S$ is an even cycle~$C_\ell$,
		then $\E_i \cap S = \{k_1, k_2\}$,
		where $k_1$ and $k_2$ are two consecutive edges of the cycle~$C_\ell$.
		Line~\ref{algo:cycle} in the algorithm implies that
		$b_{k_1} = - b_{k_2} \in \{1, -1\}$.
		It follows that $\sum_{k \in \E_i} b_k = b_{k_1} + b_{k_2} = 0$.
		\item If $S$ is a kayak paddle $\KP_{\ell, r, p}$
		with odd cycles $C_\ell$ and $C_r$ and central path~$P_p$, $p \in \N$,
		we again distinguish several cases:
		\begin{description}
			\item[Node~$i$ does not belong to the central path:]
			If $\E_i \cap S \subseteq C_\ell$ or $\E_i \cap S \subseteq C_r$,
			we conclude as before.
			\item[Node~$i$ does not belong to a cycle:] 
			If $\E_i \cap S \subseteq P_p$,
			then $\E_i \cap S = \{k_1, k_2\}$
			where $k_1$ and $k_2$ are two consecutive edges of the path~$P_p$.
			Line~\ref{algo:path} of the algorithm implies
			that $b_{k_1} = - b_{k_2} \in \{2, -2\}$.
			It follows that
			$\sum_{k \in \E_i} b_k = b_{k_1} + b_{k_2} = 0$.
			\item[Node~$i$ belongs to a cycle and the central path:]
			The only remaining case is
			when $\E_i \cap S$ intersects several sets among~$C_\ell$, $C_r$, and $P_p$.
			If $p = 0$, that is, if the central path is the node~$i$, then $\E_i \cap S = \{k_1, k_2, k_3, k_4\}$,
			where~$k_1$ and~$k_2$ (resp.~$k_3$ and~$k_4$)
			are two consecutive edges in $C_\ell$ (resp. $C_r$).
			Lines~\ref{algo:kayakpaddle-cycle1} and \ref{algo:kayakpaddle-cycle2}
			yield $b_{k_1} = b_{k_2} = -1 $ and $b_{k_3} = b_{k_4} = 1$,
			which implies that
			$\sum_{k \in \E_i} b_k = b_{k_1} + b_{k_2} + b_{k_3} + b_{k_4} = 0$.
			If $p \ge 1$, then $E_i \cap S = \{k_1, k_2, k_3\}$,
			where $k_1$ and $k_2$ are two consecutive edges of either $C_\ell$ or $C_r$,
			and $k_3$ is an edge in $P_p$.
			Lines~\ref{algo:kayakpaddle-cycle1},
			\ref{algo:path},
			and \ref{algo:kayakpaddle-cycle2}
			of the algorithm imply that $b_{k_1} = b_{k_2} \in \{1, -1\}$
			and hat $b_{k_3} = -2 b_{k_1}$,
			so that we conclude again that the desired sum is zero.
		\end{description}
	\end{itemize}
\end{enumerate}
\end{proof}

\Cref{fig:triamond-kernel,fig:codomino-kernel} shows possible runs of \Cref{algo:spanning} on the triamond and codomino graphs, which both have a two-dimensional kernel.  Note that the basis is not unique and depends on our choice of the spanning tree $\T$ and the augmenting edge $a$ (see Lines~\ref{algo:T} and~\ref{algo:edgek} in \Cref{algo:spanning}).

\def\d{1.7cm}
\begin{figure}[!htb]
	\centering
	\begin{tikzpicture}[%
		pics/sample/.style={code={\draw[#1] (0,0) --(0.6,0) ;}}]
		\node[matrix,draw,nodes={anchor=center},inner sep=2pt, ampersand replacement=\&]  {%
			\pic{sample=spanner}; \& \node{Spanning tree~$\T$}; \&
			\pic{sample=extra}; \& \node{Augmenting edge~$a$}; \&
			\pic{sample=k1}; \& \node{First kernel vector}; \&
			\pic{sample=k2}; \& \node{Second kernel vector}; \\
		};
	\end{tikzpicture}
	\\[.2cm]
	\subfloat[Construction A.\label{fig:triamond-kernel-a}]{
		\begin{tikzpicture}
			\threefan			
			\draw (1) edge[spanner] (5)
			(4) edge[extra] node[right] {$a$} (3)
			(2) edge[spanner] (4)
			(5) edge[spanner] (2)
			(2) edge[spanner] (3)
			(1) edge[k1] node[above] {$s_1$} (2)
			(5) edge[k2] node[above] {$s_2$} (4);
	\end{tikzpicture}}
	\hfill
	\subfloat[First vector for A ($\KP_{3, 3, 0}$).\label{fig:triamond-kernel-b}]{
		\begin{tikzpicture}		
			\threefan
			\draw (1) edge[k1] node[k1, above] {$1$} (2)
			(1) edge[k1] node[left] {$-1$} (5)
			(2) edge[k1] node[left] {$1$} (5)
			(2) edge[k1] node[above] {$-1$} (3)
			(2) edge[k1] node[right] {$-1$} (4)
			(3) edge[k1] node[right] {$1$} (4)
			(5) edge node[above] {$0$}(4)
			;
	\end{tikzpicture}}
	\hfill
	\subfloat[Second vector for A ($C_4$).\label{fig:triamond-kernel-c}]{
		\begin{tikzpicture}		
			\threefan
			\draw (1) edge node[above] {$0$} (2)
			(1) edge node[left] {$0$} (5)
			(2) edge node[right] {$0$} (4)
			(5) edge[k2] node[above] {$1$}(4)
			(5) edge[k2] node[left] {$-1$} (2)
			(3) edge[k2] node[above] {$1$} (2)
			(3) edge[k2] node[right] {$-1$} (4)
			;
	\end{tikzpicture}}
	\\
	\subfloat[Construction B.\label{fig:triamond-kernel-d}]{
		\begin{tikzpicture}[%
			pics/sample/.style={code={\draw[#1] (0,0) --(0.6,0) ;}}]
			\threefan
			\draw (1) edge[spanner] (2)
			(2) edge[spanner] (3)
			(3) edge[spanner] (4)
			(4) edge[spanner] (5)
			(1) edge[extra] node[left] {$a$} (5)
			(2) edge[k1] node[right] {$s_1$} (4)
			(2) edge[k2] node[left] {$s_2$} (5);
			
	\end{tikzpicture}}
	\hfill
	\subfloat[First vector for B ($C_4$).\label{fig:triamond-kernel-e}]{
		\begin{tikzpicture}		
			\threefan			
			\draw (1) edge[k1] node[k1, above] {$-1$} (2)
			(1) edge[k1] node[left] {$1$} (5)
			(2) edge[k1] node[right] {$1$} (4)
			(5) edge[k1] node[above] {$-1$}(4)
			(2) edge node[left] {$0$} (5)
			(2) edge node[above] {$0$} (3)
			(3) edge node[right] {$0$} (4)
			;
	\end{tikzpicture}}
	\hfill
	\subfloat[Second vector for B ($C_4$).\label{fig:triamond-kernel-f}]{
		\begin{tikzpicture}		
			\threefan
			
			\draw (1) edge node[above] {$0$} (2)
			(1) edge node[left] {$0$} (5)
			(2) edge node[right] {$0$} (4)
			(5) edge[k2] node[above] {$-1$}(4)
			(5) edge[k2] node[left] {$1$} (2)
			(3) edge[k2] node[above] {$-1$} (2)
			(3) edge[k2] node[right] {$1$} (4)
			;
	\end{tikzpicture}}
	\caption{%
		\label{fig:triamond-kernel}
		Two possible constructions of a kernel basis for the triamond graph.
		Construction~A yields the basis vectors
		$b_1 = (1, -1, -1, -1, 1, 1, 0)$
		and $b_2 = (0, 0, 1, 0, -1, -1, 1)$.
		Construction~$B$ yields the basis vectors
		$b_1 = (-1, 1, 0, 1, 0, 0, -1)$
		and $b_2 = (0, 0, -1, 0, 1, 1, -1)$.
	}
\end{figure}

\def\d{1.3cm}
\begin{figure}[!htb]
	\centering
	\subfloat[Construction A.\label{fig:codomino-kernel-a}]{
		\begin{tikzpicture}		
			\codomino
			
			\draw (1) edge[spanner] (6)
			(2) edge[spanner] (6)
			(5) edge[spanner] node[below, white] {0} (6)
			(3) edge[spanner] (5)
			(4) edge[spanner] (5)
			(3) edge[extra] node[above] {$a$} (4)		
			(1) edge[k1] node[above] {$s_1$} (2)
			(2) edge[k2] node[above] {$s_2$} (3);
			
	\end{tikzpicture}}
	\hfill
	\subfloat[First vector for A ($\KP_{3, 3, 1}$).\label{fig:codomino-kernel-b}]{
		\begin{tikzpicture}		
			\codomino		
			
			\draw (1) edge[k1] node[above] {$1$} (2)
			(6) edge[k1] node[right, xshift=-.1cm] {$-1$} (2)
			(1) edge[k1] node[below, xshift=-.1cm] {$-1$} (6)
			(5) edge[k1] node[below] {$2$} (6)
			(4) edge[k1] node[above] {$1$} (3)
			(4) edge[k1] node[below, xshift=.1cm] {$-1$} (5)
			(5) edge[k1] node[left, xshift=.1cm] {$-1$} (3)
			(2) edge node[above] {$0$} (3);			
	\end{tikzpicture}}
	\hfill
	\subfloat[Second vector for A ($\C_4$).\label{fig:codomino-kernel-c}]{
		\begin{tikzpicture}		
			\codomino
			
			\draw (1) edge node[below] {$0$} (6)
			(1) edge node[above] {$0$} (2)
			(4) edge node[above] {$0$} (3)
			(4) edge node[below] {$0$} (5)
			(2) edge[k2] node[above] {$1$}(3)
			(5) edge[k2] node[left, xshift=.1cm] {$-1$} (3)
			(6) edge[k2] node[right, xshift=-.1cm] {$-1$} (2)
			(6) edge[k2] node[below] {$1$} (5);			
	\end{tikzpicture}}
	\\
	\subfloat[Construction B.\label{fig:codomino-kernel-d}]{
		\begin{tikzpicture}		
			\codomino			
			\draw (1) edge[spanner] (2)
			(2) edge[spanner] (3)
			(3) edge[spanner] (4)
			(4) edge[spanner] (5)
			(6) edge[spanner] node[below, white] {0} (5)
			(3) edge[extra] node[left] {$a$} (5)		
			(1) edge[k1] node[below] {$s_1$} (6)
			(2) edge[k2] node[right] {$s_2$} (6);
			
	\end{tikzpicture}}
	\hfill
	\subfloat[First vector for B ($\C_6$).\label{fig:codomino-kernel-e}]{
		\begin{tikzpicture}		
			\codomino					
			\draw (1) edge[k1] node[above] {$-1$} (2)
			(2) edge[k1] node[above] {$1$} (3)
			(3) edge[k1] node[above, xshift=.1cm] {$-1$} (4)
			(4) edge[k1] node[below] {$1$} (5)
			(5) edge[k1] node[below, xshift=-.1cm] {$-1$} (6)
			(6) edge[k1] node[below] {$1$} (1)
			(2) edge node[right] {$0$} (6)
			(5) edge node[left] {$0$} (3);
	\end{tikzpicture}}
	\hfill
	\subfloat[Second vector for B ($\C_4$).\label{fig:codomino-kernel-f}]{
		\begin{tikzpicture}		
			\codomino			
			\draw (1) edge node[below] {$0$} (6)
			(1) edge node[above] {$0$} (2)
			(4) edge node[above] {$0$} (3)
			(4) edge node[below] {$0$} (5)
			(2) edge[k2] node[above] {$-1$}(3)
			(5) edge[k2] node[left] {$1$} (3)
			(6) edge[k2] node[right] {$1$} (2)
			(6) edge[k2] node[below] {$-1$} (5);
			
	\end{tikzpicture}}
	\caption{\label{fig:codomino-kernel}
		Two possible constructions of a kernel basis for the codomino graph.
		Construction~A yields the vectors
		$b_1 = (1, -1, 0, -1, 1, -1, -1, 2)$
		and $b_2 = (0, 0, 1, -1, 0, -1, 0, 1)$.
		Construction~B yields the vectors
		$b_1 = (-1, 1, 1, 0, -1, 0, 1, -1)$
		and $b_2 = (0, 0, -1, 1, 0, 1, 0, -1)$.
	}	
\end{figure}

\subsection[Proof of Proposition \ref*{prop:vertex}.]{Proof of \Cref{prop:vertex} in \fullref{sec:poly-vertex} (borrowed from~\citet{CS22}).}
\label{app:prop-vertex}
We prove that the negations of \ref{cond:vertex-1} and \ref{cond:vertex-2} are equivalent.
Let $A^\star$ denote the incidence matrix of~$G^\star$.

By \Cref{def:vertex}, if $\mu$ is not a vertex of $\Lann$, there exist $z_1, z_2 \in \Lann{\setminus}\{\mu\}$ and $0<\theta<1$ such that $\mu=\theta z_1+(1-\theta)z_2$. The coordinates of the vectors $z_1$ and $z_2$ are non-negative, so this equality implies that their supports are included into the support $\E^\star$ of the vector~$\mu$. In particular, if $\tilde{\mu}$ and $\tilde{z}_1$ denote the restrictions of $\mu$ and $z_1$ to coordinates in $\E^\star$, respectively, then $A^\star \tilde{\mu} = A \mu = \lambda = A z_1 = A^\star \, \tilde{z}_1$ with $\tilde{\mu} \neq \tilde{z}_1$, which means that $G^\star=(V, \E^\star)$ is not injective.

Conversely, if $G^\star$ is not injective, there exists a non-zero vector $\tilde{z}$ in $\R^{|\E^\star|}$ such that $A^\star \, \tilde{z}= 0$. If we embed $\tilde{z}$ into $\R^{|\E|}$ with zero-padding, we obtain a non-zero vector~$z$ such that $Az = 0$, and whose support is included into that of the vector~$\mu$. This implies that there exists $\varepsilon>0$ such that both $\mu-\varepsilon z$ and $\mu+\varepsilon z$ belong to $\Lann$. The convex combination $\mu=\frac 12 (\mu-\varepsilon z) + \frac 12 (\mu+\varepsilon z)$ proves that the vector~$\mu$ is not a vertex of $\Lann$.

The last part directly derives from \Cref{cond:dimensions-2,cond:dimensions-5} in \Cref{prop:dimensions}.
\hfill \Halmos
\endproof

\subsection[Additional examples and results for Section \ref*{sec:poly-vertex}.]{Additional examples and results for \fullref{sec:poly-vertex}.}
\label{app:facets}
 
Following \Cref{prop:vertex}, the bijectivity of a vertex is determined by the number of its coordinates that are positive in edge coordinates, that is, by the cardinality of the set of edges that form its support.
Recall that the $d$-dimensional polytope $\Lann$ is actually characterized by the $m$ inequalities $\mu_k \ge 0$ for each $k \in \E$. In particular, this polytope has at most $m$ facets, one for each inequality, but it typically has fewer. Indeed, some inequalities may be redundant and/or not tight, in a sense that will be defined in \Cref{def:inequalities} below. For example, by looking more closely at the general solution obtained for the diamond graph in \Cref{fig:diamond}, we conclude that:
\begin{itemize}
	\item The inequality $\mu_{2, 3} \ge 0$ is satisfied trivially by every vector $\mu \in \La$,
	as we have $\mu_{2, 3}=\beta > 0$.
	Therefore, this inequality
	does not define a facet of $\Lann$.
	\item If $\lambda_1\bar{\lambda}_2<\bar{\lambda}_3\lambda_4$,
	the inequality $\mu_{1,2} \ge 0$
	supersedes the inequality $\mu_{3,4} \ge 0$,
	and conversely.
	If $\lambda_1\bar{\lambda}_2 = \bar{\lambda}_3\lambda_4$,
	these two inequalities are equivalent.
	In both cases, the inequalities $\mu_{1,2} \ge 0$ and $\mu_{3,4} \ge 0$
	lead to a single facet of $\Lann$.
	\item If $\lambda_1\bar{\lambda}_3 < \bar{\lambda}_2\lambda_4$, the inequality $\mu_{1,3} \ge 0$ supersedes the inequality $\mu_{2,4} \ge 0$, and conversely. If $\lambda_1\bar{\lambda}_3 = \bar{\lambda}_2\lambda_4$,
	these two inequalities are equivalent.
	In both cases, the inequalities $\mu_{1,3} \ge 0$ and $\mu_{2,4} \ge 0$
	lead to a single facet of $\Lann$.
\end{itemize}
\begin{reusefigure}[H]{fig:diamond}
	\centering
	\begin{tikzpicture}[every node/.style={, minimum size=.3cm}]
		\def\d{2.5cm}
		\bifan	
		\draw (1) edge node[sloped, above] {$2\lambda_1\bar\lambda_2+\alpha$} (2) 
		(1) edge node[sloped, below] {$2\lambda_1\bar\lambda_3-\alpha$} (3)
		(2) edge node[sloped, above] {$2\bar\lambda_2\lambda_4-\alpha$} (4)
		(4) edge node[sloped, below] {$2\bar\lambda_3\lambda_4+\alpha$} (3)
		(2) edge node[right] {$\beta$} (3);
	\end{tikzpicture}
	\caption{Matching rates in the diamond graph with the normalization $\lambda_1+\lambda_4=\frac 12$. $2\beta = \lambda_2+\lambda_3 - \frac{1}{2}$ is the difference between the arrival rates of $\{2, 3\}$ and $\{1, 4\}$.
		$\bar\lambda_2= \lambda_2 - \beta$ and $\bar\lambda_3= \lambda_3 - \beta$ represent the residual rates that classes~$2$ and~$3$ can provide to classes~$1$ and~$4$.}
\end{reusefigure}
All in all, the $1$-dimensional convex polytope~$\Lann$ associated with the diamond graph of \Cref{fig:diamond} has two facets, even if it is defined by five inequalities.
\Cref{def:inequalities} below will help us relate these notions to the number of zero coordinates of the vertices of the convex polytope~$\Lann$.

\begin{dfn}[Adapted from \citet{BP83,Z95}] \label{def:inequalities}
	Let $k \in \E$.
	\begin{enumerate}[(i)]
		\item The inequality $\mu_k \ge 0$ is said to be \emph{tight} if there exists a vector $\mu\in \Lann$ such that $\mu_k=0$, in which case we also say that this inequality is tight for the vector~$\mu$.
		\item The inequality $\mu_k\geq 0$ is said to be \emph{redundant} if removing this inequality does not change the polytope $\Lann$, in the sense that
		\begin{align*}
			\Lann = \{\mu \in \R^m: A\mu = \lambda \text{ and } \mu_\ell \ge 0 \text{ for each } \ell \in \E \setminus \{k\}\}.
		\end{align*}
		Otherwise, this inequality
		is called \emph{irredundant}.
		\item The matching problem $(G, \lambda)$ is called \emph{essential} if all tight inequalities are irredundant.
		\item The polytope $\Lann$ is said to be \emph{simple} if every vertex of $\Lann$ belongs to exactly $d$ facets, which is the minimal number of facets a vertex belongs to.
	\end{enumerate}
\end{dfn}

Importantly, the number of positive coordinates of a vertex~$\mu$ (considered in \Cref{coro:achievable}) is the number of inequalities that are not tight for this vertex. More generally, \Cref{def:inequalities} has the following intuitive interpretation. An inequality is tight if the convex polytope~$\Lann$ intersects the hyperplane obtained by transforming this inequality into an equality. Non-tight inequalities are ``useless'' (and redundant) because they are never satisfied as equalities by any vector in~$\Lann$. The matching problem $(G, \lambda)$ is essential if each tight inequality defines a distinct facet of the convex polytope~$\Lann$. Under this condition, the number of facets that contain a vertex is equal to the number of inequalities that are tight for this vertex. In particular, as we will see in \Cref{prop:simple}, if the matching problem $(G, \lambda)$ is essential and the polytope $\Lann$ is simple, then every vertex satisfies exactly $d$ (tight) inequalities as equalities, which means that this vertex has $d$ zero coordinates, and therefore $n = m - d$ positive coordinates, so that this vertex is bijective.

All these notions are illustrated in
\Cref{ex:essential,ex:non-essential,ex:non-simple}
below, which show in particular that a matching problem $(G, \lambda)$ may be essential even if the polytope $\Lann$ is not simple, and conversely.
Consistently with the discussion above on \Cref{fig:diamond},
these examples use a kernel basis to verify effortlessly whether an inequality is tight and/or irredundant.

\begin{exa}[Essential matching problem] \label{ex:essential}
	\Cref{fig:codomino-simple} considers a codomino graph with the vector of arrival rates $\lambda =(4, 5, 3, 2, 3, 5)$.
	A particular solution to~\eqref{eq:system}
	is $\mu^\circ = (2, 2, 1, 2, 1, 1, 1, 1) \in \R^8$,
	and the basis of $\ker(A)$ consists of the vectors
	$b_1 = (-1, 1, 1, 0, -1, 0, 1, -1)$
	and $b_2 = (0, 0, -1, 1, 0, 1, 0, -1)$
	obtained in construction~B
	of \Cref{fig:codomino-kernel}.
	The generic solution to~\eqref{eq:system}
	is shown in \Cref{fig:codomino-general}.
	
	The inequalities are listed in \Cref{tab:essential}.
	The 2-dimensional polytope~$\Lann$,
	shown in \Cref{fig:codomino-simple-polytope} in kernel basis,
	is characterized by five tight inequalities which are also irredundant:
	\begin{align*}
		-1 &\le \alpha_1 \le 1,
		&
		\alpha_2 &\ge -1,
		&
		\alpha_1 - \alpha_2 &\ge -1,
		&
		\alpha_1 + \alpha_2 &\le 1.
	\end{align*}
	The matching problem $(G, \lambda)$
	is essential.
	In kernel basis,
	the vertices of
	the convex polytope~$\Lann$ are
	$(0, 1)$, $(-1, 0)$, $(1, 0)$,
	$(-1, -1)$, and $(1, -1)$,
	and we can verify on \Cref{fig:codomino-simple-polytope}
	that each vertex belongs to exactly 2 facets.
	Therefore $\Lann$ is simple (more generally, all 2-dimensional polytopes are simple).
	All in all, each vertex of $\Lann$ has 2 zero coordinates and 6 positive coordinates in edge coordinates, so that this vertex is bijective.
	These vertices are represented
	in edge basis in \Cref{fig:ys1,fig:ys2,fig:ys3,fig:ys4,fig:ys5}.
\end{exa}

\begin{figure}[!htb]
	\centering
	\subfloat[Inequalities. \label{tab:essential}]{%
		\begin{tabular}{|c|c|c|c|}
			\hline
			Edge basis
			& Kernel basis
			& Tight?
			& Irredundant? \\ \hline
			$\mu_{1,2} \ge 0$ &
			$\alpha_1 \le 2$ &
			\xmark & \xmark \\ \hline
			$\mu_{1,6} \ge 0$ &
			$\alpha_1 \ge -2$ &
			\xmark & \xmark \\ \hline
			$\mu_{2,3} \ge 0$ &
			$\alpha_1 - \alpha_2 \ge -1$ &
			\cmark & \cmark \\ \hline
			$\mu_{2,6} \ge 0$ &
			$\alpha_2 \ge -2$ &
			\xmark & \xmark \\ \hline
			$\mu_{3,4} \ge 0$ &
			$\alpha_1 \le 1$ &
			\cmark & \cmark \\ \hline
			$\mu_{3,5} \ge 0$ &
			$\alpha_2 \ge -1$ &
			\cmark & \cmark \\ \hline
			$\mu_{4,5} \ge 0$ &
			$\alpha_1 \ge -1$ &
			\cmark & \cmark \\ \hline
			$\mu_{5,6} \ge 0$ &
			$\alpha_1 + \alpha_2 \le 1$ &
			\cmark & \cmark \\ \hline
		\end{tabular}
	}
	\\
	\subfloat[Generic solution to~\eqref{eq:system}. \label{fig:codomino-general}]{
		\begin{tikzpicture}		
			\def\d{2.4cm}
			\codomino
			
			\draw (1) edge node[above, sloped] {$2 - \alpha_1$} (2)
			(1) edge node[below, sloped] {$2 + \alpha_1$} (6)
			(2) edge node[above] {$1 + \alpha_1 - \alpha_2$} (3)
			(2) edge node[above, sloped] {$2 + \alpha_2$} (6)
			(3) edge node[above, sloped] {$1 - \alpha_1$} (4)
			(5) edge node[above, sloped] {$1 + \alpha_2$} (3)
			(4) edge node[below, sloped] {$1 + \alpha_1$} (5)
			(5) edge node[below] {$1 - \alpha_1 - \alpha_2$} (6);
			
		\end{tikzpicture}
	}
	\qquad
	\subfloat[Polytope $\Lann$ in kernel basis.
	\label{fig:codomino-simple-polytope}]{%
		\begin{tikzpicture}[scale=1.3]
			\draw[->] (-1.5,0) -- (1.5,0);
			\node[anchor=west] at (1.5,0) {$\alpha_1$};
			\draw[->] (0,-1.3) -- (0,1.3);
			\node[anchor=south] at (0,1.3) {$\alpha_2$};
			\node[anchor=north east] {0};
			
			\node (v1) at (-1,-1) {};
			\node (v2) at (-1,0) {};
			\node (v3) at (0,1) {};
			\node (v4) at (1,0) {};
			\node (v5) at (1,-1) {};
			
			\path[draw=blue, very thick, fill=blue, fill opacity=.2]
			(v1.center) -- (v2.center) -- (v3.center)
			-- (v4.center) -- (v5.center) -- cycle;
			
			\node[anchor=north, text=orange]
			at (v1) {$(-1, -1)$}; 
			\node[anchor=west, text=orange, yshift=.1cm]
			at (v3) {$(0, 1)$}; 
			\node[anchor=south east, text=orange, xshift=.12cm, yshift=.1cm]
			at (v2) {$(-1, 0)$}; 
			\node[anchor=south west, text=orange, xshift=-.12cm, yshift=.1cm]
			at (v4) {$(1, 0)$}; 
			\node[anchor=north, text=orange]
			at (v5) {$(1, -1)$}; 
			
			\foreach \i in {1,...,5}
			\node[dot] at (v\i) {};
		\end{tikzpicture}
	}
	\\
	\subfloat[Edge coordinates of $(0, 1)$. \label{fig:ys1}]{
		\begin{tikzpicture}		
			\def\d{1.28cm}
			\codomino
			
			\draw (1) edge[k2] node[above] {$2$} (2)
			(1) edge[k2] node[below] {$2$} (6)
			(2) edge node[above] {$0$} (3)
			(2) edge[k2] node[right] {$3$} (6)
			(3) edge[k2] node[above] {$1$} (4)
			(3) edge[k2] node[left] {$2$} (5)
			(4) edge[k2] node[below] {$1$} (5)
			(5) edge node[below] {$0$} (6);
			
		\end{tikzpicture}
	}
	\hfill
	\subfloat[Edge coordinates of $(-1, 0)$. \label{fig:ys2}]{
		\begin{tikzpicture}		
			\def\d{1.28cm}
			\codomino
			
			\draw (1) edge[k2] node[below] {$1$} (6)
			(1) edge[k2] node[above] {$3$} (2)
			(4) edge[k2] node[above] {$2$} (3)
			(4) edge node[below] {$0$} (5)
			(2) edge node[above] {$0$} (3)
			(5) edge[k2] node[left] {$1$} (3)
			(6) edge[k2] node[right] {$2$} (2)
			(6) edge[k2] node[below] {$2$} (5);
			
		\end{tikzpicture}
	}
	\hfill
	\subfloat[Edge coordinates of $(1, 0)$. \label{fig:ys3}]{
		\begin{tikzpicture}	
			\def\d{1.28cm}
			\codomino
			
			\draw (1) edge[k2] node[above] {$1$} (2)
			(1) edge[k2] node[below] {$3$} (6)
			(2) edge[k2] node[above] {$2$} (3)
			(2) edge[k2] node[right] {$2$} (6)
			(3) edge node[above] {$0$} (4)
			(3) edge[k2] node[left] {$1$} (5)
			(4) edge[k2] node[below] {$2$} (5)
			(5) edge node[below] {$0$} (6);
			
		\end{tikzpicture}
	}
	\\
	\subfloat[Edge coordinates of $(-1, -1)$. \label{fig:ys4}]{
		\begin{tikzpicture}	
			\def\d{1.28cm}
			\codomino
			
			\draw (1) edge[k2] node[below] {$1$} (6)
			(1) edge[k2] node[above] {$3$} (2)
			(4) edge[k2] node[above] {$2$} (3)
			(4) edge node[below] {$0$} (5)
			(2) edge[k2] node[above] {$1$} (3)
			(5) edge node[left] {$0$} (3)
			(6) edge[k2] node[right] {$1$} (2)
			(6) edge[k2] node[below] {$3$} (5);
			
			\node at ($(1)-(.4cm,0)$) {};
			\node at ($(4)+(.4cm,0)$) {};
		\end{tikzpicture}
	}
	\qquad
	\subfloat[Edge coordinates of $(1, -1)$. \label{fig:ys5}]{
		\begin{tikzpicture}	
			\def\d{1.28cm}
			\codomino
			
			\draw (1) edge[k2] node[above] {$1$} (2)
			(1) edge[k2] node[below] {$3$} (6)
			(2) edge[k2] node[above] {$3$} (3)
			(2) edge[k2] node[right] {$1$} (6)
			(3) edge node[above] {$0$} (4)
			(3) edge node[left] {$0$} (5)
			(4) edge[k2] node[below] {$2$} (5)
			(5) edge[k2] node[below] {$1$} (6);
			
		\end{tikzpicture}
	}
	\caption{An essential matching problem $(G, \lambda)$
		with a simple polytope $\Lann$.
		The vector of arrival rates is
		$\lambda = (4, 5, 3, 2, 3, 5) \in \R^6$,
		a particular solution to~\eqref{eq:system}
		is $\mu^\circ = (2, 2, 1, 2, 1, 1, 1, 1) \in \R^8$,
		and the chosen base vectors for $\ker(A)$ are
		$b_1 = (-1, 1, 1, 0, -1, 0, 1, -1)$ and
		$b_2 = (0, 0, -1, 1, 0, 1, 0, -1)$.
	}
	\label{fig:codomino-simple}
\end{figure}%

\begin{exa}[Non-essential matching problem] \label{ex:non-essential}
	\Cref{fig:codomino-multiple} shows the same codomino graph as in \Cref{ex:essential}, with the same basis of $\ker(A)$, but with the vector of arrival rates $\lambda = (2, 4, 4, 2, 2, 2)$.
	A particular solution to~\eqref{eq:system}
	is $\mu^\circ = (1, 1, 2, 1, 1, 1, 1, 0)$,
	and the general solution is shown
	in \Cref{fig:codomino-multiple-general}.
	
	The inequalities are listed in \Cref{tab:non-essential}.
	The 2-dimensional convex polytope $\Lann$
	is shown in kernel basis
	in \Cref{fig:codomino-multiple-polytope}.
	All inequalities are tight, but only one is irredundant,
	so we conclude that
	the matching problem $(G, \lambda)$
	is not essential,
	even if the polytope $\Lann$ is still simple.
	Correspondingly, even if each vertex
	belongs to exactly two facets,
	they all have
	more than 2 zero coordinates, so none of them is bijective.
	For example, the vertex $(1, -1)$ in kernel basis has coordinates
	$(0, 2, 4, 0, 0, 0, 2, 0)$ in edge basis (\Cref{fig:y3}).
	This vertex has 5 zero coordinates in edge coordinates
	(and only 3 positive coordinates) even if it belongs to only 2 facets.
\end{exa}

	\begin{figure}[!htb]
		\centering
		\subfloat[Inequalities \label{tab:non-essential}]{%
			\begin{tabular}{|c|c|c|c|}
				\hline
				Edge basis
				& Kernel basis
				& Tight?
				& Irredundant? \\ \hline
				$\mu_{1,2} \ge 0$ &
				$\alpha_1 \le 1$ &
				\cmark & \xmark \\ \hline
				$\mu_{1,6} \ge 0$ &
				$\alpha_1 \ge -1$ &
				\cmark & \xmark \\ \hline
				$\mu_{2,3} \ge 0$ &
				$\alpha_1 - \alpha_2 \ge -2$ &
				\cmark & \xmark \\ \hline
				$\mu_{2,6} \ge 0$ &
				$\alpha_2 \ge -1$ &
				\cmark & \xmark \\ \hline
				$\mu_{3,4} \ge 0$ &
				$\alpha_1 \le 1$ &
				\cmark & \xmark \\ \hline
				$\mu_{3,5} \ge 0$ &
				$\alpha_2 \ge -1$ &
				\cmark & \xmark \\ \hline
				$\mu_{4,5} \ge 0$ &
				$\alpha_1 \ge -1$ &
				\cmark & \xmark \\ \hline
				$\mu_{5,6} \ge 0$ &
				$\alpha_1 + \alpha_2 \le 0$ &
				\cmark & \cmark \\ \hline
			\end{tabular}
		}
		\\
		\subfloat[Generic solution to~\eqref{eq:system}. \label{fig:codomino-multiple-general}]{
			\begin{tikzpicture}[scale=.9, transform shape]		
				\def\d{2.4cm}
				\codomino
				
				\draw (1) edge node[above, sloped] {$1 - \alpha_1$} (2)
				(1) edge node[below, sloped] {$1 + \alpha_1$} (6)
				(2) edge node[above] {$2 + \alpha_1 - \alpha_2$} (3)
				(2) edge node[above, sloped] {$1 + \alpha_2$} (6)
				(3) edge node[above, sloped] {$1 - \alpha_1$} (4)
				(5) edge node[above, sloped] {$1 + \alpha_2$} (3)
				(4) edge node[below, sloped] {$1 + \alpha_1$} (5)
				(5) edge node[below] {$- \alpha_1 - \alpha_2$} (6);
				
			\end{tikzpicture}
		}
		\qquad
		\subfloat[Polytope $\Lann$ in kernel basis. Dashed lines show tight redundant inequalities.
		\label{fig:codomino-multiple-polytope}]{%
			\begin{tikzpicture}[scale=1.1]
				\node at (-2.75, 0) {};
				\node at (2.75, 0) {};
				
				\draw[->] (-1.6,0) -- (1.6,0);
				\node[anchor=west] at (1.6,0) {$\alpha_1$};
				\draw[->] (0,-1.6) -- (0,1.6);
				\node[anchor=south] at (0,1.5) {$\alpha_2$};
				\node[anchor=north east] {0};
				
				\node (v1) at (-1,1) {};
				\node (v2) at (-1,-1) {};
				\node (v3) at (1,-1) {};
				
				\path[fill=blue, fill opacity=.2]
				(v1.center) -- (v2.center) -- (v3.center)
				-- cycle;
				
				\node[anchor=south east, text=orange]
				at (v1) {$(-1, 1)$};
				\node[anchor=north, text=orange]
				at (v2) {$(-1, -1)$};
				\node[anchor=west, text=orange, xshift=.12cm, yshift=.1cm]
				at (v3) {$(1, -1)$};
				
				\path[dashed, draw=blue, very thick]
				(v1.center) -- (v2.center) -- (v3.center);
				\path[draw=blue, very thick]
				(v1.center) -- (v3.center);
				\path[draw=blue, very thick, dashed] (-1.5, 0.5) -- (-.5, 1.5);
				\path[draw=blue, very thick, dashed] (1, -1.5) -- (1, -.5);
				
				\foreach \i in {1,...,3}
				\node[dot] at (v\i) {};
			\end{tikzpicture}
		}
		\\
		\subfloat[Edge coordinates of $(-1, 1)$. \label{fig:y1}]{
			\begin{tikzpicture}		
				\def\d{1.28cm}
				\codomino
				
				\draw (1) edge[k2] node[above] {$2$} (2)
				(1) edge node[below] {$0$} (6)
				(2) edge node[above] {$0$} (3)
				(2) edge[k2] node[right] {$2$} (6)
				(3) edge[k2] node[above] {$2$} (4)
				(3) edge[k2] node[left] {$2$} (5)
				(4) edge node[below] {$0$} (5)
				(5) edge node[below] {$0$} (6);
				
			\end{tikzpicture}
		}
		\hfill
		\subfloat[Edge coordinates of $(-1, -1)$. \label{fig:y2}]{
			\begin{tikzpicture}		
				\def\d{1.28cm}
				\codomino
				
				\draw (1) edge node[below] {$0$} (6)
				(1) edge[k2] node[above] {$2$} (2)
				(4) edge[k2] node[above] {$2$} (3)
				(4) edge node[below] {$0$} (5)
				(2) edge[k2] node[above] {$2$} (3)
				(5) edge node[left] {$0$} (3)
				(6) edge node[right] {$0$} (2)
				(6) edge[k2] node[below] {$2$} (5);
				
				\node at ($(1)-(.4cm,0)$) {};
				\node at ($(4)+(.4cm,0)$) {};
			\end{tikzpicture}
		}
		\hfill
		\subfloat[Edge coordinates of $(1, -1)$. \label{fig:y3}]{
			\begin{tikzpicture}	
				\def\d{1.28cm}
				\codomino
				
				\draw (1) edge node[above] {$0$} (2)
				(1) edge[k2] node[below] {$2$} (6)
				(2) edge[k2] node[above] {$4$} (3)
				(2) edge node[right] {$0$} (6)
				(3) edge node[above] {$0$} (4)
				(3) edge node[left] {$0$} (5)
				(6) edge node[below] {$0$} (5)
				(5) edge[k2] node[below] {$2$} (4);
				
			\end{tikzpicture}
		}
		\caption{Non-essential matching problem $(G, \lambda)$
			with a simple polytope $\Lann$.
			The vector of arrival rates is
			$\lambda = (2, 4, 4, 2, 2, 2) \in \R^6$,
			a particular solution to~\eqref{eq:system}
			is $\mu^\circ = (1, 1, 2, 1, 1, 1, 1, 0) \in \R^8$,
			and the chosen base vectors for $\ker(A)$ are
			$b_1 = (-1, 1, 1, 0, -1, 0, 1, -1)$ and
			$b_2 = (0, 0, -1, 1, 0, 1, 0, -1)$.
		}
		\label{fig:codomino-multiple}
	\end{figure}

\begin{exa}[Non-simple polytope] \label{ex:non-simple}
	We finally exhibit an essential matching problem with a non-simple associated polytope. As 2-dimensional polytopes are simple, we need to consider a more complex example.
	We consider the matching problem of \Cref{fig:whirl-edges}. The arrival rate is $\lambda=(3,3,6,3,4,4,6,3,4,4)\in \R^{10}$. The particular solution and kernel basis are shown on the edges.
	The set $\Lann$,
	shown in \Cref{fig:whirl-polytope} in kernel basis,
	is an Egyptian pyramid characterized by the following tight inequalities:
	\begin{align*}
		\alpha_3
		&\ge 0,
		&
		1+\alpha_1-\alpha_3 &\ge 0,
		&
		1-\alpha_1-\alpha_3 &\ge 0,
		&
		1+\alpha_2-\alpha_3 &\ge 0,
		&
		1-\alpha_2-\alpha_3 &\ge 0.
	\end{align*}
	These five inequalities are irredundant (each one corresponds to exactly one of the five facets), so we conclude that the matching problem $(G, \lambda)$ is essential.
	In kernel basis, the vertices of this convex polytope are
	$(-1,-1,0)$, $(1,-1, 0)$, $(1,1, 0)$,
	$(-1, 1,0)$, and $(0,0,1)$.
	These vertices are shown in edge basis in \Cref{fig:yw1,fig:yw2,fig:yw3,fig:yw4,fig:yw5}. The polytope~$\Lann$ is not simple because the vertex $(0, 0, 1)$ (the ``top'' of the pyramid) belongs to 4 facets, while the polytope has dimension~3. Consistently, we can see in \Cref{fig:yw5} that this vertex has 4 zero coordinates and only 9 positive coordinates in edge basis; the subgraph defined by the support of this vertex is injective-only.
\end{exa}
	\begin{figure}[!htb]
		\centering
		\subfloat[Generic solution to~\eqref{eq:system}.\label{fig:whirl-edges}]{
			\begin{tikzpicture}[scale=0.65, every node/.style={transform shape}]
				\def\d{2.6cm}
				\whirl
				
				\draw (1) edge node[above, sloped] {$\alpha_3$} (2)
				(2) edge node[above, sloped] {$1-\alpha_1-\alpha_3$} (3)
				(3) edge node[above, sloped] {$2-\alpha_2+\alpha_3$} (4)
				(4) edge node[above, sloped] {$1+\alpha_2-\alpha_3$} (5)
				(5) edge node[below, sloped] {$1+\alpha_3$} (6)
				(6) edge node[below, sloped] {$1-\alpha_2-\alpha_3$} (7)
				(7) edge node[below, sloped] {$2-\alpha_1+\alpha_3$} (8)
				(8) edge node[below, sloped] {$1+\alpha_1-\alpha_3$} (1)
				(1) edge node[left] {$2-\alpha_1$} (9)
				(9) edge node[above, sloped] {$2+\alpha_1$} (2)
				(5) edge node[right] {$2-\alpha_2$} (10)
				(10) edge node[below, sloped] {$2+\alpha_2$} (6)
				(3) edge node[left] {$3+\alpha_1+\alpha_2$} (7)
				;
				
			\end{tikzpicture}
		}
		\hfill
		\subfloat[Polytope $\Lann$ in kernel basis.
		\label{fig:whirl-polytope}]{%
			\begin{tikzpicture}[scale=1.2]
				\draw[->, opacity=.3] (-1.6,0) -- (1.6,0);
				\node[anchor=west] at (1.6,0) {$\alpha_1$};
				\draw[->, opacity=.3] (-.9,-.9) -- (1.4,1.4);
				\node[anchor=south] at (1.4,1.4) {$\alpha_2$};
				\draw[->, opacity=.3] (0,-.9) -- (0,2.2);
				\node[anchor=south] at (0,2.2) {$\alpha_3$};
				
				\node (v1) at (-1.6,-.6) {};
				\node (v2) at (.4,-.6) {};
				\node (v3) at (1.6,.6) {};
				\node (v4) at (-.4,.6) {};			
				\node (v5) at (0,2) {};
				
				\path[draw=blue, very thick, fill=blue, fill opacity=.1]
				(v1.center) -- (v2.center) -- (v3.center) -- (v4.center) -- cycle;
				\path[draw=blue, very thick, fill=blue, fill opacity=.1]
				(v1.center) -- (v4.center) -- (v5.center) -- cycle;
				\path[draw=blue, very thick, fill=blue, fill opacity=.1]
				(v3.center) -- (v4.center) -- (v5.center) -- cycle;
				\path[draw=blue, very thick, fill=blue, fill opacity=.1]
				(v1.center) -- (v2.center) -- (v5.center) -- cycle;
				\path[draw=blue, very thick, fill=blue, fill opacity=.1]
				(v3.center) -- (v2.center) -- (v5.center) -- cycle;
				
				\node[anchor=north, text=orange]
				at (v1) {$(-1, -1, 0)$};
				\node[anchor=north west, text=orange]
				at (v2) {$(1, -1, 0)$};
				\node[anchor=west, text=orange]
				at (v3) {$(1, 1, 0)$};
				\node[anchor=east, xshift=-.6cm, text=orange]
				at (v4) {$(-1, 1, 0)$};
				\node[anchor=east, text=orange]
				at (v5) {$(0, 0, 1)$};
				
				\foreach \i in {1,...,5}
				\node[dot] at (v\i) {};
			\end{tikzpicture}
		}
		\\
		\subfloat[Edge coordinates of $(-1, -1, 0)$. \label{fig:yw1}]{
			\begin{tikzpicture}[scale=0.64, every node/.style={transform shape}]
				\def\d{1.6cm}
				\whirl
				
				\draw (1) edge node[above, sloped] {$0$} (2)
				(2) edge[k2] node[above, sloped] {$2$} (3)
				(3) edge[k2] node[above, sloped] {$3$} (4)
				(4) edge node[above, sloped] {$0$} (5)
				(5) edge[k2] node[below, sloped] {$1$} (6)
				(6) edge[k2] node[below, sloped] {$2$} (7)
				(7) edge[k2] node[below, sloped] {$3$} (8)
				(8) edge node[below, sloped] {$0$} (1)
				(1) edge[k2] node[left] {$3$} (9)
				(9) edge[k2] node[above, sloped] {$1$} (2)
				(5) edge[k2] node[right] {$3$} (10)
				(10) edge[k2] node[below, sloped] {$1$} (6)
				(3) edge[k2] node[left] {$1$} (7)
				;
				
				\node at ($(1)-(.6cm,0)$) {};
				\node at ($(5)+(.6cm,0)$) {};
			\end{tikzpicture}
		}
		\hfill
		\subfloat[Edge coordinates of $(1, -1, 0)$. \label{fig:yw2}]{
			\begin{tikzpicture}[scale=0.64, every node/.style={transform shape}]
				\def\d{1.6cm}
				\whirl
				
				\draw (1) edge node[above, sloped] {$0$} (2)
				(2) edge node[above, sloped] {$0$} (3)
				(3) edge[k2] node[above, sloped] {$3$} (4)
				(4) edge node[above, sloped] {$0$} (5)
				(5) edge[k2] node[below, sloped] {$1$} (6)
				(6) edge[k2] node[below, sloped] {$2$} (7)
				(7) edge[k2] node[below, sloped] {$1$} (8)
				(8) edge[k2] node[below, sloped] {$2$} (1)
				(1) edge[k2] node[left] {$1$} (9)
				(9) edge[k2] node[above, sloped] {$3$} (2)
				(5) edge[k2] node[right] {$3$} (10)
				(10) edge[k2] node[below, sloped] {$1$} (6)
				(3) edge[k2] node[left] {$3$} (7)
				;
				
				\node at ($(1)-(.4cm,0)$) {};
				\node at ($(5)+(.4cm,0)$) {};
			\end{tikzpicture}
		}
		\hfill
		\subfloat[Edge coordinates of $(1, 1, 0)$. \label{fig:yw3}]{
			\begin{tikzpicture}[scale=0.64, every node/.style={transform shape}]
				\def\d{1.6cm}
				\whirl
				
				\draw (1) edge node[above, sloped] {$0$} (2)
				(2) edge node[above, sloped] {$0$} (3)
				(3) edge[k2] node[above, sloped] {$1$} (4)
				(4) edge[k2] node[above, sloped] {$2$} (5)
				(5) edge[k2] node[below, sloped] {$1$} (6)
				(6) edge node[below, sloped] {$0$} (7)
				(7) edge[k2] node[below, sloped] {$1$} (8)
				(8) edge[k2] node[below, sloped] {$2$} (1)
				(1) edge[k2] node[left] {$1$} (9)
				(9) edge[k2] node[above, sloped] {$3$} (2)
				(5) edge[k2] node[right] {$1$} (10)
				(10) edge[k2] node[below, sloped] {$3$} (6)
				(3) edge[k2] node[left] {$5$} (7)
				;
				
			\end{tikzpicture}
		}
		\\
		\subfloat[Edge coordinates of $(-1, 1, 0)$. \label{fig:yw4}]{
			\begin{tikzpicture}[scale=0.67, every node/.style={transform shape}]
				\def\d{1.6cm}
				\whirl
				
				\draw (1) edge node[above, sloped] {$0$} (2)
				(2) edge[k2] node[above, sloped] {$2$} (3)
				(3) edge[k2] node[above, sloped] {$1$} (4)
				(4) edge[k2] node[above, sloped] {$2$} (5)
				(5) edge[k2] node[below, sloped] {$1$} (6)
				(6) edge node[below, sloped] {$0$} (7)
				(7) edge[k2] node[below, sloped] {$3$} (8)
				(8) edge node[below, sloped] {$0$} (1)
				(1) edge[k2] node[left] {$3$} (9)
				(9) edge[k2] node[above, sloped] {$1$} (2)
				(5) edge[k2] node[right] {$1$} (10)
				(10) edge[k2] node[below, sloped] {$3$} (6)
				(3) edge[k2] node[left] {$3$} (7)
				;
				
			\end{tikzpicture}
		} \qquad
		\subfloat[Edge coordinates of $(0, 0, 1)$. \label{fig:yw5}]{
			\begin{tikzpicture}[scale=0.67, every node/.style={transform shape}]
				\def\d{1.6cm}
				\whirl
				
				\draw (1) edge[k2] node[above, sloped] {$1$} (2)
				(2) edge node[above, sloped] {$0$} (3)
				(3) edge[k2] node[above, sloped] {$3$} (4)
				(4) edge node[above, sloped] {$0$} (5)
				(5) edge[k2] node[below, sloped] {$2$} (6)
				(6) edge node[below, sloped] {$0$} (7)
				(7) edge[k2] node[below, sloped] {$3$} (8)
				(8) edge node[below, sloped] {$0$} (1)
				(1) edge[k2] node[left] {$2$} (9)
				(9) edge[k2] node[above, sloped] {$2$} (2)
				(5) edge[k2] node[right] {$2$} (10)
				(10) edge[k2] node[below, sloped] {$2$} (6)
				(3) edge[k2] node[left] {$3$} (7)
				;
				
			\end{tikzpicture}
		}
		\caption{Essential matching problem $(G, \lambda)$
			with a non-simple polytope $\Lann$.
			The arrival rate is $\lambda=(3,3,6,3,4,4,6,3,4,4)\in \R^{10}$. A particular solution 
			and the chosen base vectors for $\ker(A)$ are implicitly shown on the edges of \Cref{fig:whirl-edges}.
			\label{fig:whirl}
		}
	\end{figure}%

In light of these examples, we can give the following characterization of the bijective vertices of $\Lann$.

\begin{prop} \label{prop:simple}
	Let $\mu$ be a vertex of $\Lann$. The following statements are equivalent:
	\begin{enumerate}[(i)]
		\item \label{cond:simple-1}
		$\mu$ is bijective.
		\item \label{cond:simple-3}
		$\mu$ belongs to exactly $d$ facets of $\Lann$ and none of the inequalities tight for $\mu$ is redundant.
	\end{enumerate}
	In particular, all vertices of $\Lann$ are bijective if, and only if, the matching problem $(G, \lambda)$ is essential and the polytope $\Lann$ is simple.
\end{prop}

\begin{proof}
To prove the equivalence of~\ref{cond:simple-1} and~\ref{cond:simple-3},
we first remark that the number of zero edge coordinates of a vector $\mu\in \Lann$ is by definition the number of inequalities that are tight for~$\mu$. It is in particular at least the number of facets that intersect $\mu$, with equality if, and only if, none of the inequalities tight for $\mu$ is redundant.

If a vector~$\mu$ is a vertex of $\Lann$, then $\mu$ belongs to at least $d$ facets of $\Lann$.
Now, the vector~$\mu$ is bijective if, and only if, $d$ of its coordinates are zero, which in view of the remark above is equivalent to say that $\mu$ belongs to exactly $d$ facets of $\Lann$ and none of the inequalities tight for $\mu$ is redundant.

As for the last statement, it follows directly from \Cref{def:inequalities}.
\end{proof}

\begin{rem}\label{rem:equal_rates} 
	We can further explore the relationship between vertices and their support graph.
	Again consider a vertex~$\mu$ of $\Lann$,
	and let $G^\star$ denote the support graph of~$\mu$
	and $A^\star$ its incidence matrix.
	According to \Cref{def:injective,def:surjective,def:only},
	the subgraph~$G^\star$ is injective if and only if
	each connected component of~$G^\star$ is
	either a tree or a unicyclic graph with an odd cycle.
	For each connected component of $G^\star$ that is a tree,
	and therefore a bipartite graph
	with parts $V_+$ and $V_-$,
	the existence of vertex~$\mu$ implies that
	\begin{align} \label{eq:image}
		\sum_{i \in \V_+} \lambda_i
		= \sum_{i \in \V_-} \lambda_i.
	\end{align}
	This equation follows
	by summing~\eqref{eq:system-equations}
	over the nodes in $\V_+$ on the one hand,
	summing~\eqref{eq:system-equations}
	over the nodes in $\V_-$ on the other hand,
	and verifying that the left-hand sides of both equations
	are equal.
	In fact, one can verify that
	the vector~$\lambda$ belongs to the image of $A^\star$
	if and only if $\lambda$ satisfies \eqref{eq:image}
	for each connected component of~$G^\star$ that is a tree.
	Note that this condition is void if $G^\star$ is bijective
	because, in this case, none of the connected components of $G^\star$ is a tree.
	
	Conversely, one can wonder
	which injective subgraph~$G^\star$ of~$G$
	defines a vertex of $\Lann$.
	Satisfying~\eqref{eq:image}
	for each tree connected component of~$G^\star$
	only guarantees the existence
	of a (unique) solution $z \in \R^p$
	to the conservation equation~$A^\star z = \lambda$.
	If each coordinate of~$z$ is positive, then
	we indeed obtain a vertex of $\Lann$
	by embedding $z$ in $\R^m$ with zero padding;
	otherwise, $G^\star$ does not define a vertex of~$\Lann$.
\end{rem}

\subsection[Proof of Proposition \ref*{prop:all-is-bijective}.]{Proof of \Cref{prop:all-is-bijective} in \fullref{sec:all-is-bijective}.}
\label{app:all-is-bijective}

Let $C$ denote the set of normalized arrival rate vectors $\lambda$ such that $(G, \lambda)$ is stabilizable: $$C=\{\lambda \in \Delta^{n-1}: \text{$(G, \lambda)$ stabilizable}\}.$$
To prove \Cref{prop:all-is-bijective}, we proceed as follows:
\begin{itemize}
	\item We prove that $C$ is non-empty and convex, hence measurable.
	\item We prove that $C$ is an open set (for the Lesbegue measure), so its probability is positive and the conditional probability given $C$ is well-defined.
	\item We prove that if a vertex is injective-only, the vector $\lambda$ must respect a constraint that reduces the dimension of the possible solutions, meaning that the measure of the event is null.
\end{itemize}

We first prove the different properties of $C$.

\emph{$C$ is not empty.}
We use \Cref{rem:degree-proportional}: consider $\lambda = A \mu$, with $\mu=(\frac{1}{2m},\ldots,\frac{1}{2m})$. 
The matching problem $(G,\lambda)$ is stabilizable as all coordinates of $\mu$ are positive. Moreover,  
one can check that $\lambda=(\frac{d_1}{2m},\ldots,\frac{d_n}{2m})$, where $d_i$ is the degree of node $i$, so $\lambda \in \Delta^{n-1}$ ($\sum_i d_i=2m$). Hence, $\lambda\in C$.

\emph{$C$ is convex.}
Let $\lambda_a$ and $\lambda_b$ be two vectors of $C$. Let $\mu_a \in \R_{> 0}^m$ and $\mu_b \in \R_{> 0}^m$ be positive solutions to $A\mu_a=\lambda_a$ and $A\mu_b = \lambda_b$ respectively (such solutions exist as stated in \Cref{prop:stability-region-form}).
For $0<x<1$, consider $\lambda=x\lambda_a + (1-x)\lambda_b$. $\lambda \in \Delta^{n-1}$ because $\Delta^{n-1}$ is convex. Moreover, we have $A\mu=\lambda$, with $\mu=x\mu_a+(1-x)\mu_b$. As $\R_{> 0}^m$ is convex, $\mu\in \R_{> 0}^m$, so $(G, \lambda)$ is stabilizable (\Cref{prop:stability-region-form}), hence $\lambda\in C$.

\emph{$C$ is open in $\Delta^{n-1}$.} Let $\lambda\in C$. For some $\epsilon>0$, let $\lambda_\epsilon$ be an element of $\Delta^{n-1}$ such that $||\lambda-\lambda_\epsilon||_1<\epsilon$. To prove that $C$ is open, we just need to show that $\lambda_\epsilon\in C$ if $\epsilon$ is small enough.
Let $\mu \in \R_{> 0}^m$ be a positive solution to $A\mu=\lambda$. Using \Cref{eq:linear-bijection,eq:pseudoinverse}, we can write $\mu = A^+\lambda+\mu_\alpha$, with $\mu_\alpha = \mu-A^+\lambda= (\mathrm{Id}_{m \times m} - A^+A)\mu\in \ker(A)$. Consider now $\mu_\epsilon = A^+\lambda_\epsilon+\mu_\alpha$. We have $A\mu_\epsilon = AA^+\lambda_\epsilon+A\mu_\alpha=\lambda_\epsilon$ so $(\lambda_\epsilon,\mu_\epsilon)$ checks \eqref{eq:system}. Moreover, $||\mu-\mu_\epsilon||_1=||A^+(\lambda-\lambda_\epsilon)||_1<||A^+||_1\epsilon$, where $||A^+||_1<+\infty$ is the operator norm of $A^+$ associated to the 1-norm. All coordinates of $\mu$ are positive, so if $\epsilon$ is small enough, it also the case for $\mu_\epsilon$, meaning that $(G, \lambda_\epsilon)$ is stabilizable (\Cref{prop:stability-region-form}). In other words, $\lambda_\epsilon\in C$.

This establishes that $C$ (i.e., the stabilizability of $(G, \lambda)$) defines a proper conditional probability. We now examine the conditions under which the polytope $\Lann$ of a stabilizable problem $(G, \lambda)$ possesses an injective-only vertex.

Consider $\lambda\in \Delta^{n-1}$. Let $\mu$ be a solution to $A\mu=\lambda$. Using \Cref{def:edge-kernel-basis} and the pseudoinverse approach from \Cref{sec:particular-solution}, the kernel coordinates of $\mu$ are $\alpha=B^+(\mathrm{Id}_{m \times m} - A^+A)\mu$, so we can write
$$\mu = B\alpha + A^+\lambda = D\gamma\text{,}$$
where $D=\left(B | A^+\right)$ is the $m\times m$ horizontal concatenation of $B$ and $A^+$, and $\gamma = (\alpha_1,\ldots, \alpha_d, \lambda_1, \ldots, \lambda_n)$ is the concatenation of $\alpha$ and $\lambda$.

Notice that $D$ depends solely on $G$ and is invertible. Its inverse can be verified as the vertical concatenation of $B^+(\mathrm{Id}_{m \times m} - A^+A)$ and $A$: any matching rate vector $\mu$ uniquely defines $\lambda$ and $\alpha$ .

We now assume that one vertex of the polytope is injective-only and look for the implications for $\lambda$.

The existence of an injective-only vertex implies that there exists a subset of edges  $K \subset E$, with $|K|>d$, such that $\mu$ is null on $K$ (remind that  $d=m-n$ is the dimension of $\ker(A)$).
Let $D\vert_K$ denote the $|K|\times m$ submatrix of $D$ formed by the rows corresponding to $K$, and let $B\vert_K$ denote the $d$ first columns of $D\vert_K$, that is the part of $D\vert_K$ that handles $\alpha$. As $|K|>d$, we have $\ker\left((B\vert_K)^\intercal\right)\neq \{0\}$, meaning that there exists a non-trivial linear combination $v\in \R^{|K|}\setminus \{0\}$ of the rows of $B\vert_K$ such that $vB\vert_K=0$.

Let $A^+\vert_K$ denotes the $n$ last columns of $D\vert_K$, that is the part of $D\vert_K$ that handles $\lambda$. We know that $vD\vert_K\neq 0$ because $D$ is invertible, so its row are linearly independent. As $vB\vert_K=0$, this means that $vA^+\vert_K\neq 0$. The vector $u=vA^+\vert_K$ can be seen as a non-trivial linear combination of the coordinates of $\lambda$.

We can now conclude: if the polytope $\Lann$ associated to some $\lambda\in C$ admits an injective-only vertex, we have a subset of edges $K \subset E$, with $|K|>d$ such that $D\vert_K\gamma = 0$. In particular, we must have $vD\vert_K\gamma = 0$, which simplifies as $u\lambda=0$. In other words, for $\mu$ to be null over $K$, the vector $\lambda$ must belong to some hyperplane $L_u$ of dimension $n-1$ that can be built solely from $G$ and $K$. As $\Delta^{n-1}\neq L_u$ (one is affine, the other is not), the intersection $\Delta^{n-1}\cap L_u$, if it exists, has a dimension at most $n-2$. In particular, the Lebesgue measure of $\Delta^{n-1}\cap L_u$ (over $\Delta^{n-1}$) is null, meaning that the probability to nullify all edges of $K$ is null.

The number of subsets $K\subset E$ such that $|K|>d$ is finite, so the probability to nullify all edges of any of them is also null, so the probability that the polytope $\Lann$ associated to some $\lambda\in C$ admits an injective-only vertex is null.
\hfill \Halmos

\newpage

 % appendix/ec4-non-unicyclic
\section[Supplementary material of Section \ref*{sec:non-unicyclic}.]{Supplementary material of \fullref{sec:non-unicyclic}.} \label{app:non-unicyclic}

\subsection[Proof of Proposition \ref*{prop: converging_to_vertex}.]{Proof of \Cref{prop: converging_to_vertex}.} \label{app: converging_to_vertex}

We first prove two intermediary results,
\Cref{lemma: stability_generic,lemma: positive_delta},
that will be instrumental in proving
the upper bound for the mean queue length
in \Cref{prop: converging_to_vertex}.

\begin{lem} \label{lemma: stability_generic}
	Consider a matching problem $(G, \lambda)$.
	If the \gls{crpg} $\delta(G, \lambda)$ is positive,
	then the \gls{ml} policy stabilizes $(G, \lambda)$, and
	\begin{align*}
		\mathbb{E}\left[\sum_{i \in \V} Q_{i}\right] \leq
		\frac{n}{2\delta(G, \lambda)}\sum_{i \in \V} \lambda_i,
	\end{align*}
	where $Q = (Q_1, \ldots, Q_n)$ is the steady-state queue-length vector for the Markov chain induced by the \gls{ml} policy.
\end{lem}
\begin{proof}
First note that, since $\lambda > 0$, the \gls{ml} policy adapted to $(G, \lambda)$ induces an irreducible Markov chain on the state space $\mathcal{Q}_\mathcal{G}$ defined in \eqref{eq:Q-greedy}. Consider the quadratic test function~$W$ defined on $\mathcal{Q}_\mathcal{G}$ by
\begin{align*}
	W(q) = \sum_{i \in \V} q_i^2,
	\quad q \in \mathcal{Q}_\mathcal{G}.
\end{align*}
We define the drift of this test function by $\Delta W(q) = \mathbb{E}\left[W(Q_{t+1}) - W(Q_t) | Q_t = q\right]$ for each $q \in \mathcal{Q}_\mathcal{G}$. Lastly, we let $\I_{q} = \{i \in \V : q_i > 0\}$ denote the support of~$q$, for each $q \in \mathcal{Q}_\mathcal{G}$. Now, we simplify the drift under the \gls{ml} policy using the queue evolution recursion given by \eqref{eq:Q-rec}. For all $q \in \mathcal{Q}_\mathcal{G} \setminus \{0\}$, we have
\begin{align*}
	\Delta W(q)
	={}& \mathbb{E}\left[\sum_{i \in \V} \left(Q_{t, i}+\one\{i = I_t\}\right)^2 \one\{J_t = \bot\} + \left(Q_{t, i}-\one\{i = J_t\}\right)^2 \one\{J_t \neq \bot\} - Q_{t, i}^2 \ \bigg| \ Q_t = q\right], \\
	={}& \mathbb{E}\left[\sum_{i \in \V} \one\{i=I_t, J_t = \bot\} + \one\{i=J_t\} + 2\left(\one\{i=I_t, J_t = \bot\} - \one\{i=J_t\}\right)Q_{t, i} \ \bigg| \ Q_t = q\right], \\
	\overset{(a)}{=}{}& 1+2\sum_{i \in \V} q_i \mathbb{E}\left[\one\{i=I_t, J_t = \bot\} - \one\{i=J_t\} \ \bigg| \ Q_t = q\right], \\
	\overset{(b)}{=}{}& 1+\frac{2}{\sum_{i \in \V} \lambda_i}\sum_{i \in \V}\lambda_iq_i -2\sum_{i \in \V} q_i \mathbb{E}\left[\one\{i=J_t\} \ \bigg| \ Q_t = q\right], \\
	\overset{(c)}{=}{}& 1+\frac{2}{\sum_{i \in \V} \lambda_i}\sum_{i \in \V}\lambda_iq_i -\frac{2}{\sum_{i \in \V} \lambda_i}\sum_{j \in \V} \lambda_j \max_{i \in \V_j} q_i, \\
	\overset{(d)}{=}{}& 1+\frac{2}{\sum_{i \in \V} \lambda_i}\sum_{i \in \I_q}\lambda_iq_i -\frac{2}{\sum_{i \in \V} \lambda_i}\sum_{j \in \V(\I_q)} \lambda_j \max_{i \in \V_j} q_i, \\
	\overset{(e)}{\leq}{}& 1 - \frac{2\delta(G, \lambda)}{|\I_q|\sum_{i \in \V} \lambda_i} \sum_{i \in \I_q} q_i, \\
	={}& 1 - \frac{2\delta(G, \lambda)}{|\I_q|\sum_{i \in \V} \lambda_i} \sum_{i \in \V} q_i, \\
	\leq{}& 1 - \frac{2\delta(G, \lambda)}{n\sum_{i \in \V} \lambda_i} \sum_{i \in \V} q_i. \numberthis \label{eq: neg-drift}
\end{align*}
Here, $(a)$ follows as 
\begin{align*}
	\sum_{i \in \V} \one\{i=I_t, J_t = \bot\} + \sum_{i \in \V} \one\{i=J_t\} = \one\{J_t = \bot\} + \one\{J_t \neq \bot\} = 1.
\end{align*}
Next, $(b)$ follows as for all $i \in \V$, we have
\begin{align*}
	q_i\mathbb{E}\left[\one\{i=I_t, J_t = \bot\}| Q_t = q\right] &= q_i \PP\left\{i=I_t, J_t = \bot | Q_t = q\right\}, \\
	&= q_i\PP\left\{J_t = \bot | Q_t = q, i=I_t\right\}\PP\left\{i=I_t | Q_t = q\right\}, \\
	&= q_i\frac{\lambda_i}{\sum_{i \in \V} \lambda_i}  \one\left\{\max_{j \in \V_i} q_j = 0\right\} = q_i\frac{\lambda_i}{\sum_{i \in \V} \lambda_i}.
\end{align*}
The last equality follows from the fact that the \gls{ml} policy is greedy, which implies that, if $q_i >0$, then $q_j = 0$ for all $j \in \V_i$.
Now, $(c)$ follows by using the definition of the \gls{ml} policy. In particular, we have
\begin{align*}
	\sum_{i \in \V} q_i \mathbb{E}\left[\one\{i=J_t\} \ \bigg| \ Q_t = q\right] &= \frac{1}{\sum_{j \in \V} \lambda_j}\sum_{j \in \V}\lambda_j \sum_{i \in \V} q_i \mathbb{E}\left[\one\{i=J_t\} \ \bigg| \ Q_t = q, j=I_t\right]
	\\
    &= \frac{1}{\sum_{j \in \V} \lambda_j}\sum_{j \in \V} \lambda_j \max_{i \in \V_j} q_i.
\end{align*}
The equality $(d)$ follows by noting that $\I_q$ is equal to the support of $q$, i.e., we have $\I_q = \{i \in \V: q_i > 0\}$. The rest of the paragraph is dedicated to proving~$(e)$. First note that $(e)$ holds trivially when $q = 0$. Now, fix $q \in \mathcal{Q}_\mathcal{G} \backslash \{0\}$ and consider the bipartite graph
\begin{align*}
	G_{q} = G\left(\I_q \cup V(\I_q), \E_{q}\right) \quad \text{with} \quad \E_{q} = \left\{ \{i, j\} \in \E: i \in \I_q, j \in V(\I_q) \right\}.
\end{align*}
By definition of $\delta(G, \lambda)$, we have, for each non-empty $\I \subseteq \I_q$,
\begin{align*}
	\sum_{i \in \I} \lambda_i \leq \sum_{j \in V(\I)} \lambda_j - \delta(G, \lambda) = \sum_{\substack{j \in V(\I): \\ \exists i \in \I, \{i, j\} \in \E_q}} \lambda_j - \delta(G, \lambda),
\end{align*}
where the equality follows as $V(\I) \subseteq V(\I_q)$, and so, $V(\I) = V(\I) \cap \{j: \exists i \in \I, \{i, j\} \in \E_q\}$ by the definition of $G_{q}$. Now, the above inequality implies that
\begin{align*}
	\sum_{i \in \I} \left(\lambda_i + \frac{\delta(G, \lambda)}{|\I_{q}|}\right) \leq \sum_{i \in \I} \lambda_i + \frac{|\I|\delta(G, \lambda)}{|\I_{q}|} \leq \sum_{i \in \I} \lambda_i + \delta(G, \lambda) \leq \sum_{\substack{j \in \V(\I): \\ \exists i \in \I, \{i, j\} \in \E_q}} \lambda_j \quad \forall \I \subseteq \I_q, \ \I \neq \emptyset.
\end{align*}
For $\I = \emptyset$, the comparison between the leftmost and rightmost sides holds trivially, both sums being zero. So, by the weighted version of Hall's marriage theorem \cite[Lemma 2.5]{gregoryhall}, there exists $\tilde{\mu} \in \Rnn^{|\E_{q}|}$ such that
\begin{align*}
	\lambda_i + \frac{\delta(G, \lambda)}{|\I_q|}
	&= \sum_{\substack{j \in V(\I_q): \\ \{i, j\} \in \E_q}}\tilde{\mu}_{i, j},
	\quad \forall i \in \I_q,
	&
	\lambda_j
	&\geq \sum_{\substack{i \in \I_q: \\ \{i,j\} \in \E_q}}\tilde{\mu}_{i, j},
	\quad \forall j \in \V(\I_q).
\end{align*}
Using the above, we get
\begin{align*}
	\sum_{i \in \I_q} \lambda_i q_i - \sum_{j \in V(\I_q)} \lambda_j \max_{i^\prime \in \V_j} q_{i^\prime} &\leq \sum_{\{i, j\} \in \E_{q}} \tilde{\mu}_{i, j} \left(q_i - \max_{i^\prime \in \V_j} q_{i^\prime}\right) - \frac{\delta(G, \lambda)}{|\I_q|}\sum_{i \in \I_{q}} q_i \leq -\frac{\delta(G, \lambda)}{|\I_q|}\sum_{i \in \I_{q}} q_i,
\end{align*}
where the last inequality follows as $\max_{i^\prime \in V_j} q_{i^\prime} \geq q_i$ for all $\{i, j\} \in \E_{q}$. This completes the proof of~(e).

Now, by \eqref{eq: neg-drift} we immediately conclude that
\begin{align*}
	\Delta W(q) \leq -1
	\quad \text{for each } q \in \mathcal{Q}_\mathcal{G}
	\text{ such that} \quad
	\sum_{i \in V} q_i \geq \frac{n\sum_{i \in \V} \lambda_i}{\delta(G, \lambda)}.
\end{align*}
Thus, by the Foster--Lyapunov theorem \cite{srikantbook}, we conclude from~\eqref{eq: neg-drift} that the Markov chain under the \gls{ml} policy is positive recurrent. Now, by the moment bound theorem \cite[Proposition 6.14]{hajekrandomprocbook}, we further conclude that in steady state, we have
\begin{align*}
	\mathbb{E}\left[\sum_{i \in \V} Q_i\right] \leq \frac{n}{2\delta(G, \lambda)}\sum_{i \in \V} \lambda_i.
\end{align*}
This completes the proof of the lemma.
\end{proof}
Now we turn our attention to studying the \gls{crpg} of $(G^\prime, \lambda_\epsilon^\prime)$. We start by defining additional notations for convenience. Define the set of ``positive'' vertices by $\V^+ = \{i^+ : i \in \V\}$ and the set of ``negative'' vertices by $\V^- = \{i^- : i \in \V\}$.  Let $\mathbb{I}_{G^\prime}$ be the set of all independent sets in $G^\prime$.
Note that each independent set~$\I \in \mathbb{I}_{G^\prime}$ can be uniquely written as $\I = J^- \cup K^+$
where $J^- = \I \cap V^-$ and $K^+ = \I \cap V^+$ (with at least one of them nonempty).
Observe we have $J \in \{\emptyset\} \cup \mathbb{I}_{G}$ and~$K \in \{\emptyset\} \cup \mathbb{I}_{G^\star}$. Now, we lower bound the \gls{crpg} of the augmented graph in the following lemma. In particular, combining the following lemma
with \Cref{prop:stability-region-form} implies that $(G^\prime, \lambda_\epsilon^\prime)$ is stabilizable whenever $(G, \lambda)$ is.

\begin{lem} \label{lemma: positive_delta}
	For the augmented problem $(G^\prime, \lambda_\epsilon^\prime)$, we have
	\begin{equation*}
		\delta(G^\prime, \lambda_\epsilon^\prime) \geq \min\left\{\epsilon \min_{i \in \V} \lambda_i+ \delta(G^\star, \lambda), (1-\epsilon)\min_{i \in \V} \lambda_i+ \delta(G^\star, \lambda), \delta(G, \lambda)\right\}.
	\end{equation*}
\end{lem}
\begin{proof}
Consider a non-empty $\I \in \mathbb{I}_{G^\prime}$ and let $J \in \{\emptyset\} \cup \mathbb{I}_G$ and $K \in \{\emptyset\} \cup \mathbb{I}_{G^\star}$ such that $J^- = \I \cap V^-$ and $K^+ = \I \cap V^+$, so that $\I = J^- \cup K^+$. Observe that $J$ and~$K$ cannot be both empty at the same time, as $\I$ is not empty. Using the definition of~$G^\prime$, we can verify that, for each $0 < \epsilon < 1$, we have
\begin{align} 
	\sum_{i \in \V_{G^\prime}(\I)} \lambda_{\epsilon, i}^\prime - \sum_{i \in \I} \lambda_{\epsilon, i}^\prime ={}& \epsilon \left(\sum_{i \in \V_G(J)} \lambda_{i} - \sum_{i \in J} \lambda_{i}\right) + (1-\epsilon)\sum_{i \in \V_G(J) \setminus \V_{G^\star}(K)} \lambda_{i} \nonumber \\
	&+ (1 - \epsilon) \left( \sum_{i \in \V_{G^\star}(K)}\lambda_{i} - \sum_{i \in K}\lambda_{i} \right) + \epsilon \sum_{i \in \V_G(K) \setminus \V_G(J)}\lambda_{i}.\label{eq:gap-epsilon}
\end{align}
To prove the lemma, it suffices to show that the right-hand side of~\eqref{eq:gap-epsilon} is larger than or equal to the right-hand side of the equation in \Cref{lemma: positive_delta}.

We first note that
\begin{equation}
    \delta(G,\lambda)\geq \delta(G^\star,\lambda).
    \label{eq:delta-G-G-star}
\end{equation}
Indeed, since $\E^\star\subseteq\E$, every independent set of $G$
is also an independent set of $G^\star$, and
$\V_{G^\star}(L)\subseteq\V_G(L)$ for every
$L\in\mathbb I_G$. Consequently,
\[
    \sum_{i\in\V_G(L)}\lambda_i-\sum_{i\in L}\lambda_i
    \geq
    \sum_{i\in\V_{G^\star}(L)}\lambda_i-\sum_{i\in L}\lambda_i
    \geq \delta(G^\star,\lambda),
    \qquad L\in\mathbb I_G.
\]
Taking the minimum over $L\in\mathbb I_G$ proves
\eqref{eq:delta-G-G-star}.

First, we handle the cases in which
one of $J$ and $K$ is empty. If $J=\emptyset$, then $K\neq\emptyset$,
and \eqref{eq:gap-epsilon} gives
\begin{align}
    \sum_{i\in\V_{G^\prime}(\I)}\lambda_{\epsilon,i}^\prime
    -\sum_{i\in\I}\lambda_{\epsilon,i}^\prime
    \nonumber &=
    (1-\epsilon)
    \left(
        \sum_{i\in\V_{G^\star}(K)}\lambda_i
        -\sum_{i\in K}\lambda_i
    \right)
    +\epsilon\sum_{i\in\V_G(K)}\lambda_i
    \nonumber\\
    & \geq
    (1-\epsilon)
    \left(
        \sum_{i\in\V_{G^\star}(K)}\lambda_i
        -\sum_{i\in K}\lambda_i
    \right)
    +\epsilon\sum_{i\in\V_{G^\star}(K)}\lambda_i
    \nonumber\\
    & =
    \left(
        \sum_{i\in\V_{G^\star}(K)}\lambda_i
        -\sum_{i\in K}\lambda_i
    \right)
    +\epsilon\sum_{i\in K}\lambda_i
    \geq
    \delta(G^\star,\lambda)+\epsilon\min_{i \in \V} \lambda_i,
    \label{eq:empty-J-gap}
\end{align}
where the first inequality follows from
$\V_{G^\star}(K)\subseteq\V_G(K)$, and the last one follows since
$K\neq\emptyset$. Similarly, if $K=\emptyset$, then $J\neq\emptyset$, and
\eqref{eq:gap-epsilon} gives
\begin{align}
    \sum_{i\in\V_{G^\prime}(\I)}\lambda_{\epsilon,i}^\prime
    -\sum_{i\in\I}\lambda_{\epsilon,i}^\prime
    \nonumber&=
    \epsilon
    \left(
        \sum_{i\in\V_G(J)}\lambda_i-\sum_{i\in J}\lambda_i
    \right)
    +(1-\epsilon)\sum_{i\in\V_G(J)}\lambda_i
    \nonumber\\
    & =
    \left(
        \sum_{i\in\V_G(J)}\lambda_i-\sum_{i\in J}\lambda_i
    \right)
    +(1-\epsilon)\sum_{i\in J}\lambda_i
    \nonumber\\
    & \geq
    \delta(G,\lambda)+(1-\epsilon)\min_{i \in \V} \lambda_i
    \geq
    \delta(G^\star,\lambda)+(1-\epsilon)\min_{i \in \V} \lambda_i,
    \label{eq:empty-K-gap}
\end{align}
where the last inequality follows from
\eqref{eq:delta-G-G-star}. It therefore remains to consider the case in which
$J\neq\emptyset$ and $K\neq\emptyset$. We divide this case into the
following two subcases.

\noindent \textbf{Case I:}
Either $\V_G(K)\not\subseteq\V_G(J)$ or
$\V_G(J)\not\subseteq\V_{G^\star}(K)$. Thus, at least one of the sets
\[
    \V_G(J)\setminus\V_{G^\star}(K)
    \qquad\text{and}\qquad
    \V_G(K)\setminus\V_G(J)
\]
is nonempty. Since all arrival rates are positive, the sum of the
second and fourth terms in \eqref{eq:gap-epsilon} is therefore bounded
below by $\min\{\epsilon,1-\epsilon\}\min_{i \in \V} \lambda_i$. Moreover, since $J\neq\emptyset$ and $K\neq\emptyset$, the first and
third terms in \eqref{eq:gap-epsilon} are bounded below, respectively,
by $\epsilon\delta(G,\lambda)
    \geq \epsilon\delta(G^\star,\lambda)$
    and
    $(1-\epsilon)\delta(G^\star,\lambda)$,
where the first inequality follows from
\eqref{eq:delta-G-G-star}. Consequently,
\begin{align}
    \sum_{i\in\V_{G^\prime}(\I)}\lambda_{\epsilon,i}^\prime
    -\sum_{i\in\I}\lambda_{\epsilon,i}^\prime
    \geq
    \min\{\epsilon,1-\epsilon\}\min_{i \in \V} \lambda_i
    +\delta(G^\star,\lambda).
    \label{eq:case_1_crp_gap}
\end{align}

\noindent \textbf{Case II:} We have $\V_G(K) \subseteq \V_G(J)$ and $\V_G(J) \subseteq \V_{G^\star}(K)$. Thus, we have
\begin{align*}
	\V_{G^\star}(K) \subseteq \V_G(K) \subseteq \V_G(J) \subseteq \V_{G^\star}(K) \implies \V_{G^\star}(K) = \V_{G}(K).
\end{align*}
Now, as $K \in \mathbb{I}_{G^\star}$, we have $K \cap \V_{G^\star}(K) = \emptyset$. Thus, we have $K \cap \V_{G}(K) = \emptyset$ which implies that $K \in \mathbb{I}_G$. Now, using the above, we get
\begin{align*}
	\sum_{i \in \V_{G^\star}(K)}\lambda_{i} - \sum_{i \in K}\lambda_{i} = \sum_{i \in \V_{G}(K)}\lambda_i - \sum_{i \in K}\lambda_i \geq \delta(G, \lambda), \quad \text{and} \quad \sum_{i \in \V_G(J)}\lambda_i - \sum_{i \in J}\lambda_i \geq  \delta(G, \lambda),
\end{align*}
where the inequalities above follow by the definition of $\delta(G, \lambda)$ and noting that $J, K \in \mathbb{I}_G$. Now, using \eqref{eq:gap-epsilon}, we get
\begin{align}
	\sum_{i \in \V_{G^\prime}(\I)}\lambda_{\epsilon, i}^\prime - \sum_{i \in \I} \lambda_{\epsilon, i}^\prime \geq \delta(G, \lambda). \label{eq:case_2_crp_gap}
\end{align}
Combining the boundary-case estimates
\eqref{eq:empty-J-gap} and \eqref{eq:empty-K-gap} with
\eqref{eq:case_1_crp_gap} and \eqref{eq:case_2_crp_gap}, we obtain,
for every nonempty $\I\in\mathbb I_{G^\prime}$,
\begin{align*}
	\sum_{i \in \V_{G^\prime}(\I)}\lambda_{\epsilon, i}^\prime - \sum_{i \in \I} \lambda_{\epsilon, i}^\prime \geq \min\left\{\epsilon \min_{i \in \V} \lambda_i+ \delta(G^\star, \lambda), (1-\epsilon)\min_{i \in \V} \lambda_i+ \delta(G^\star, \lambda), \delta(G, \lambda)\right\},
\end{align*}
which completes the proof by definition of $\delta(G^\prime, \lambda_\epsilon^\prime)$. 
\end{proof}

Now that we have stated and proved \Cref{lemma: stability_generic,lemma: positive_delta}, we are ready to prove \Cref{prop: converging_to_vertex} below.
\begin{proof}[Proof of \Cref{prop: converging_to_vertex}]
Let $\Phi_\epsilon^\prime$ denote the \gls{ml} policy adapted to $(G^\prime, \lambda_\epsilon^\prime)$.
As explained at the beginning of \Cref{sec:achievability},
the policies $\Phi_\epsilon$ and $\Phi_\epsilon^\prime$ are practically equivalent, except that $\Phi_\epsilon$ is seen as a (non-greedy $\epsilon$-dependent) policy adapted to the original problem $(G, \lambda)$, while $\Phi_\epsilon^\prime$ is seen as a (greedy) policy adapted to the ($\epsilon$-dependent) extended problem $(G^\prime, \lambda_\epsilon^\prime)$. More formally, let \(\{Q_t\}\) denote the queue process under $(G, \lambda, \Phi_\epsilon)$, and let \(\{Q^{\prime}_t\}\) denote the queue
process under $(G^\prime, \lambda_\epsilon^\prime, \Phi_\epsilon^\prime)$. Couple the arrivals to be of the same class and same $+/-$ label in both systems, and let $\Phi_\epsilon$ execute
the projection onto $G$ of the edge selected by $\Phi_\epsilon'$ in
$G'$, with the same tie-breaking rule. If $Q_{0,i}=Q'_{0,i^+}+Q'_{0,i^-}$ for all $i\in\V$, then every arrival and every match changes the two sides of this
identity by the same amount. Hence, we get
\begin{equation} \label{eq:coupling}
Q_{t,i}
=
Q^{\prime}_{t,i^+}
+
Q^{\prime}_{t,i^-},
\qquad i\in\V,\ t\geq0.
\end{equation}
Now, we divide the proof into two parts. In the first part, we show that the expected queue length is $O(1/\epsilon)$. Then, the second part deals with the matching rates.

\noindent \textbf{Controlling the expected queue length:}
We first observe that
$\delta(G^\star,\lambda)\geq 0
$. Indeed, let \(\mathcal{I}\) be a nonempty independent set of \(G^\star\).
Since \(A^\star\mu=\lambda\) and \(\mu\geq0\), and since no edge of
\(G^\star\) has both endpoints in \(\mathcal{I}\), we have
\[
\sum_{i\in \mathcal{I}}\lambda_i
=
\sum_{\substack{\{i,j\}\in\E^\star:\\ i\in \mathcal{I},\ j\in
\V_{G^\star}(\mathcal{I})}}\mu_{i,j}
\leq
\sum_{j\in\V_{G^\star}(\mathcal{I})}\lambda_j.
\]
Taking the minimum over all nonempty independent sets \(\mathcal{I}\) gives
\(\delta(G^\star,\lambda)\geq0\). 
Since $(G,\lambda)$ is stabilizable,
\Cref{prop:stability-region-form} implies that
$\delta(G,\lambda)>0$. Together with
$\delta(G^\star,\lambda)\geq 0$,
$\lambda_{\min}:=\min_{i\in\V}\lambda_i>0$, and
$\epsilon\in(0,1)$, \Cref{lemma: positive_delta} yields $\delta(G^\prime,\lambda_\epsilon^\prime)>0$. By \Cref{lemma: stability_generic}, the Markov chain induced by the \gls{ml} policy on
\((G^\prime,\lambda_\epsilon^\prime)\) is positive recurrent. Let $\pi^\prime$ denote the stationary distribution of $\{Q^\prime_t\}$. Then, \eqref{eq:coupling} implies stability of
$(G,\lambda,\Phi_\epsilon)$. Moreover, under
$Q^\prime\sim\pi^\prime$, define $Q_i=Q^\prime_{i^+}+Q^\prime_{i^-}$ for all $i\in\V$. By \eqref{eq:coupling}, this gives the stationary queue-length vector
under $\Phi_\epsilon$. Thus, we have
\begin{equation*}
	\mathbb{E}\left[\sum_{i \in \V} Q_i\right] = \mathbb{E}\left[\sum_{i \in \V^+ \cup \V^-} Q_i^\prime\right] \leq \frac{2n}{2\delta(G^\prime, \lambda_\epsilon^\prime)}\sum_{i \in \V} \lambda_i \leq \frac{n}{\min\left\{\epsilon\min_{i \in \V} \lambda_i + \delta(G^\star, \lambda), \delta(G, \lambda)\right\}}\sum_{i \in \V} \lambda_i,
\end{equation*}
where the first inequality follows by applying
\Cref{lemma: stability_generic} to the extended matching problem
$(G^\prime, \lambda_\epsilon^\prime)$,
and the second inequality follows by applying \Cref{lemma: positive_delta}
and noting that, for all $\epsilon \in (0, 0.5)$, we have $\epsilon \leq 1-\epsilon$.
Now, by defining $C_1 = n\sum_{i \in \V} \lambda_i$, the proof of the first part is complete.

\noindent \textbf{Controlling the matching rates:}
Let
$\mu(\Phi_\epsilon)$ be the matching-rate vector (indexed by $E$) under the $\epsilon$-filtering \gls{ml} policy adapted to $(G, \lambda)$ with filter $\E^\star$,
while
$\mu(\Phi_\epsilon^\prime)$ denotes the matching-rate vector (indexed by $\E^\prime$) under the \gls{ml} policy adapted to $(G^\prime, \lambda_\epsilon^\prime)$.
The rest of the proof is divided into two main parts: first we prove the following bound on our objective $\|\mu - \mu(\Phi_\epsilon) \|_1$ in terms of $\mu(\Phi_\epsilon^\prime)$:
\begin{align}
\|\mu - \mu(\Phi_\epsilon) \|_1  \leq \sum_{\{i, j\} \in \E^\star} \big| \mu_{i, j} - \mu(\Phi_\epsilon^\prime)_{i^+, j^+} \big| +2 \epsilon\sum_{j \in\V} \lambda_{j}. \numberthis \label{eq:one_norm_matching_rates}  
\end{align}
Then, we show that $ \sum_{\{i, j\} \in \E^\star} \big| \mu_{i, j} - \mu(\Phi_\epsilon^\prime)_{i^+, j^+} \big| = O(\epsilon)$ using properties of the \gls{ml} policy to complete the proof. Intuitively, $\lambda_\epsilon^\prime$ is designed in such a way that under any stable policy, all but $O(\epsilon)$ arrivals are matched using edges in $\E^+$. So, $ \mu(\Phi_\epsilon^\prime)_{i^+, j^+}$ must be close to $ \mu_{i, j}$ for all $\{i, j\} \in \E^\star$.

We start by proving \eqref{eq:one_norm_matching_rates}. As the $\epsilon$-filtering \gls{ml} policy mimics the \gls{ml} policy adapted to $(G^\prime, \lambda_\epsilon^\prime)$, we have
\begin{align*}
	\mu\left(\Phi_{\epsilon}\right)_{i, j} = \begin{cases}
		\mu\left(\Phi_\epsilon^\prime\right)_{i^+, j^-} + \mu\left(\Phi_\epsilon^\prime\right)_{i^-, j^+} + \mu\left(\Phi_\epsilon^\prime\right)_{i^-, j^-} + \mu\left(\Phi_\epsilon^\prime\right)_{i^+, j^+}
		&\textit{if } \{i, j\} \in \E^\star \\
		\mu\left(\Phi_\epsilon^\prime\right)_{i^+, j^-} + \mu\left(\Phi_\epsilon^\prime\right)_{i^-, j^+} + \mu\left(\Phi_\epsilon^\prime\right)_{i^-, j^-} &\textit{if } \{i, j\} \in \E \backslash \E^\star.
	\end{cases}
\end{align*}
In other words, under the coupling defined at the start of the proof, a match using an edge $\{i, j\} \in E$ in the system $(G, \lambda, \Phi_\epsilon)$ corresponds to exactly one match using an edge in $\left\{
\{i^{\ell_1},j^{\ell_2}\}:
\ell_1,\ell_2\in\{+,-\}
\right\} \cap E^\prime$ in $(G^\prime, \lambda^\prime_\epsilon, \Phi^\prime_\epsilon)$. Using the above, we can upper bound our objective $\|\mu - \mu(\Phi_\epsilon) \|_1$ in terms of $\mu(\Phi_\epsilon^\prime)$ as follows:
\begin{align*}
	\|\mu - \mu(\Phi_\epsilon) \|_1
	&\overset{(a)}= \sum_{\{i, j\} \in \E^\star} 
	\big| \mu_{i, j} - \mu(\Phi_\epsilon)_{i, j} \big| + \sum_{\{i, j\} \in \E \backslash \E^\star} \mu(\Phi_\epsilon)_{i, j}, \\ 
	&\overset{(b)}{\leq} \sum_{\{i, j\} \in \E^\star} 
	\big| \mu_{i, j} - \mu(\Phi_\epsilon^\prime)_{i^+, j^+} \big| + \sum_{\{i, j\} \in \E } \left( \mu\left(\Phi_\epsilon^\prime\right)_{i^+, j^-} + \mu\left(\Phi_\epsilon^\prime\right)_{i^-, j^+} + \mu\left(\Phi_\epsilon^\prime\right)_{i^-, j^-} \right), \\
	&\overset{(c)}{\leq} \sum_{\{i, j\} \in \E^\star} 
	\big| \mu_{i, j} - \mu(\Phi_\epsilon^\prime)_{i^+, j^+} \big| + 2\epsilon\sum_{j \in\V} \lambda_{j}.
\end{align*}
Here, $(a)$ follows by recalling that $\E^\star$ is defined as the support of~$\mu$.
Next, $(b)$ follows as, for all $\{i, j\} \in \E^\star$,
\begin{align*}
	\big| \mu_{i, j} - \mu(\Phi_\epsilon)_{i, j} \big| &= \big| \mu_{i, j} - \mu\left(\Phi_\epsilon^\prime\right)_{i^+, j^+} - \mu\left(\Phi_\epsilon^\prime\right)_{i^+, j^-} - \mu\left(\Phi_\epsilon^\prime\right)_{i^-, j^+} - \mu\left(\Phi_\epsilon^\prime\right)_{i^-, j^-} \big|, \\
	&\leq \big| \mu_{i, j} - \mu\left(\Phi_\epsilon^\prime\right)_{i^+, j^+}\big| + \mu\left(\Phi_\epsilon^\prime\right)_{i^+, j^-} + \mu\left(\Phi_\epsilon^\prime\right)_{i^-, j^+} + \mu\left(\Phi_\epsilon^\prime\right)_{i^-, j^-}.
\end{align*}
Lastly, $(c)$ follows by noting that 
\begin{align*}
	\sum_{\{i, j\} \in \E } \left( \mu\left(\Phi_\epsilon^\prime\right)_{i^+, j^-} + \mu\left(\Phi_\epsilon^\prime\right)_{i^-, j^-} \right)
	\leq \sum_{j \in \V} \sum_{\substack{i \in \V: \\ \{i,j\} \in \E}} \left(
	\mu\left(\Phi_\epsilon^\prime\right)_{j^-, i^+} + \mu\left(\Phi_\epsilon^\prime\right)_{j^-, i^-}
	\right)
	\overset{(*)}{=} \sum_{j \in \V} \lambda_{\epsilon, j^-}^\prime &= \epsilon\sum_{j \in \V} \lambda_j,  \\
	\sum_{\{i, j\} \in \E} \mu\left(\Phi_\epsilon^\prime\right)_{i^-, j^+}  \leq \sum_{i \in \V} \sum_{\substack{j \in \V: \\ \{i, j\} \in \E}} \mu\left(\Phi_\epsilon^\prime\right)_{i^-, j^+} \overset{(*)}{\leq}  \sum_{i \in \V} \lambda_{\epsilon, i^-}^\prime &= \epsilon\sum_{i \in \V} \lambda_i,
\end{align*}
where $(*)$ follows by applying the conservation equations~\eqref{eq:system} to the extended matching model $(G^\prime, \lambda_\epsilon^\prime, \Phi_\epsilon^\prime)$.
This completes the proof of~\eqref{eq:one_norm_matching_rates}.

Now, we proceed to show that $ \sum_{\{i, j\} \in \E^\star} \big| \mu_{i, j} - \mu(\Phi_\epsilon^\prime)_{i^+, j^+} \big| = O(\epsilon)$.
Recall that, by \Cref{prop:vertex},
the columns of matrix~$A^\star$ are linearly independent.
This observation has multiple consequences:
\begin{itemize}
	\item The system of linear equations $A^\star y=\lambda$ in the unknown~$y$ has a unique solution, which we know to be equal to~$\mu$.
	\item The system of linear equations $A^\star y=\tilde{\lambda}$ in the unknown~$y$ has a unique solution, which we know to be equal to $\tilde{\mu} = \{\mu(\Phi_\epsilon^\prime)_{i^+, j^+}\}_{\{i, j\} \in \E^\star}$, where $\tilde\lambda$ is given by
	\begin{align*}
		\tilde{\lambda}_i = \sum_{\substack{j \in \V: \\ \{i, j\} \in \E^\star}} \mu(\Phi_\epsilon^\prime)_{i^+, j^+} \quad i \in \V.
	\end{align*}
	\item The matrix $(A^\star)^T A^\star$ is invertible, and the Moore-Penrose inverse of $A^\star$ is $((A^\star)^T A^\star)^{-1} (A^\star)^T$.
\end{itemize}
Thus, we have $\mu - \tilde{\mu} = ((A^\star)^T A^\star)^{-1} (A^\star)^T (\lambda-\tilde{\lambda})$ which implies
\begin{align*}
	\sum_{\{i, j\} \in \E^\star} \big| \mu_{i, j} - \mu(\Phi_\epsilon^\prime)_{i^+, j^+} \big|  \leq \|((A^\star)^T A^\star)^{-1} (A^\star)^T\|_{1\to1} \|\lambda-\tilde{\lambda}\|_1, \numberthis \label{eq: pensore_inverse}
\end{align*}
where $\|\cdot\|_{1\to1}$ denotes the induced
$\ell_1$ operator norm. Now, we upper bound the term $\|\lambda-\tilde{\lambda}\|_1$ as follows:
\begin{align}
	\|\lambda-\tilde{\lambda}\|_1 &= \sum_{i \in \V}\bigg|\lambda_i - \sum_{\substack{j \in \V: \\ \{i, j\} \in \E^\star}} \mu(\Phi_\epsilon^\prime)_{i^+, j^+} \bigg|, \nonumber \\
	&\leq  \sum_{i \in \V}\bigg|(1-\epsilon)\lambda_i -\sum_{\substack{j \in \V: \\ \{i, j\} \in \E^\star}} \mu(\Phi_\epsilon^\prime)_{i^+, j^+} \bigg| + \epsilon\sum_{i \in \V} \lambda_i, \nonumber \\
	&\overset{(a)}{=}
	\sum_{i \in \V} \sum_{\substack{j \in \V: \\ \{i, j\} \in \E}} \mu(\Phi_\epsilon^\prime)_{i^+, j^-} + \epsilon\sum_{i \in \V} \lambda_i, \nonumber \\
	&\overset{(b)}\leq 2\epsilon\sum_{i \in \V} \lambda_i, \label{eq:int_step_matching_rates}
\end{align}
where $(a)$ again follows by applying~\eqref{eq:system} to the extended matching model $(G^\prime, \lambda_\epsilon^\prime, \Phi_\epsilon^\prime)$,
and $(b)$ is justified as follows:
\begin{align*}
	\sum_{i \in \V} \sum_{\substack{j \in \V: \\ \{i, j\} \in \E}} \mu(\Phi_\epsilon^\prime)_{i^+, j^-}
	\overset{(i)}{=} \sum_{j \in \V} \sum_{\substack{i \in \V: \\ \{i, j\} \in \E}} \mu(\Phi_\epsilon^\prime)_{i^+, j^-} \overset{(ii)}{\leq} \sum_{j \in \V} \lambda_{\epsilon, j^-}^\prime = \epsilon \sum_{j \in \V} \lambda_j.
\end{align*}
Here, $(i)$ follows by switching the order of summation, and $(ii)$ follows by noting that $\{i^+ \in \V^{+}: \{i^+, j^-\} \in \E^{\pm}\} \subseteq \{i \in \V^\prime: \{i, j\} \in \E^\prime\}$ and again applying~\eqref{eq:system} to the matching model $(G^\prime, \lambda_\epsilon^\prime, \Phi_\epsilon^\prime)$. Now, by combining \eqref{eq:int_step_matching_rates} with \eqref{eq: pensore_inverse}, we get
\begin{align*}
	\sum_{\{i, j\} \in \E^\star} \big| \mu_{i, j} - \mu(\Phi_\epsilon^\prime)_{i^+, j^+} \big|  \leq 2\|((A^\star)^T A^\star)^{-1} (A^\star)^T\|_{1\to1} \sum_{i \in \V} \lambda_i \epsilon.
\end{align*}
Lastly, by injecting the above inequality into \eqref{eq:one_norm_matching_rates} and defining $C_2 = 2\left(1+\|((A^\star)^T A^\star)^{-1} (A^\star)^T\|_{1\to1}\right) \sum_{i \in \V} \lambda_i$, the proof is complete.
\end{proof}

\subsection[Proof of Proposition \ref*{prop: lower_bound}.]{Proof of \Cref{prop: lower_bound}.} \label{app: lower_bound}

We consider an arbitrary Markovian, stationary matching policy that matches at most $\mathcal{M} \in \N$ pairs of items in one time step. In particular, we consider a matching policy that only depends on the current state of the underlying Markov chain of the matching problem. To be more precise, consider an irreducible, aperiodic, and positive recurrent DTMC $(S_t, t \in \N)$ defined in a similar way as in \Cref{app:extended-definition}, such that $Q_t$ and $\Phi_t$ are functions of $S_t$, where $\Phi_{t, i}$ is the number of class-$i$ items matched at time $t$ for all $i \in \V$. If $Q_t=S_t$, then the matching policy simply depends on the current queue lengths, e.g., the \gls{ml} policy. Now, given an injective-only vertex~$\mu$, the following proposition formalizes the trade-off between closeness to $\mu$ and the queue lengths in the steady-state (denoted by $Q_{\infty}$) for any such matching policy. \Cref{prop: lower_bound} is a special case of \Cref{prop: lower_bound_extended} with $\mathcal{M} = 1$. The proof of \Cref{prop: lower_bound_extended} is given in the rest of this subsection.

\begin{propositionlowerbound} \label{prop: lower_bound_extended}
	Consider an injective-only vertex $\mu$ of $\Lann$. Then, under any nonanticipative, Markovian, stationary matching policy $\Phi$ that matches at most $\mathcal{M}$
	pairs in one time step and whose induced Markov chain admits a
	stationary distribution, if $\|\mu(\Phi)-\mu\|_1 \leq \epsilon$, then $\mathbb{E}\left[\|Q_\infty\|_2^2\right] \geq \Omega\left(1/\epsilon\right)$, where $Q_{\infty}$ is the queue length under the steady state of $(S_t, t \in \N)$.
\end{propositionlowerbound}

As $\mu$ is injective-only, there exists a connected component of $G^\star = (\V, \E^\star)$ that is bipartite by \Cref{def:injective,def:bijective}. Denote such a connected component by $G^\circ = (\V^+ \cup \V^-, \E^\circ)$, where $\V^+$ and $\V^-$ are the (disjoint) parts of the connected component, and $\E^\circ \subseteq \E^\star$.
Now, consider the following test function:
\begin{align*}
	W(q) = \inner{\one^{(\V^+, \V^-)}}{q}^2,
	\quad q \in \Q,
\end{align*}
where $\one^{(\V^+,\V^-)}_i = 1$ for $i \in \V^+$, $-1$ for $i \in \V^-$, and $0$ otherwise.
For each $t \in \N$, let $A_t$ denote the class of the incoming item at time~$t$, so that $A_t = \one_{I_t}$, where $I_t$ is the class of the $(t+1)$-th item. Thus, $\lambda_i/\sum_{j \in \V} \lambda_j$ is the probability that $A_t = \one_i$.
We also define $\delta M_{t, k} = M_{t, k} - M_{t-1,k}$ to be the number of items matched using edge $k \in \E$ at time~$t$.
Thus, we have $\sum_{k \in \E_i}\delta M_{t, k} = \Phi_{t, i}$ 
for all $i \in \V$. For the rest of the proof, we consider the steady-state of the Markov chain underlying the matching problem. In particular, we consider the steady-state of the (discrete-time) Markov chain $(S_t, t \in \N)$, of which $Q = (Q_t, t \in \N)$ and $\Phi = (\Phi_t, t \in \N)$ are functions.

If $\Ex{W(Q)} = \infty$,
then we have $\Ex{\sum_{i \in \V} Q_i^2} = \infty$
which completes the proof of the proposition.
Indeed, for each $q \in \N^n$,
\begin{align*}
	W(q) &= \left(\sum_{i \in \V^+} q_i\right)^2 + \left(\sum_{i \in \V^-} q_i\right)^2 - 2\left(\sum_{i \in \V^+} q_i\right)\left(\sum_{i \in \V^-} q_i\right), \\
	&\leq \left(\sum_{i \in \V^+} q_i\right)^2 + \left(\sum_{i \in \V^-} q_i\right)^2, \\
	&\leq |\V^+| \sum_{i \in \V^+} q_i^2 + |\V^-| \sum_{i \in \V^-} q_i^2.
\end{align*}
Now, assume instead that $\Ex{W(Q)} < \infty$. In stationarity, we have $\Ex{W(Q_{t+1})}=\Ex{W(Q_{t})}$. In particular, we take the expectation under the steady state distribution of $(S_t, t \in \N)$. Using this, we get
\begin{align*}
	0 &= \Ex{\inner{\one^{(\V^+, \V^-)}}{Q_{t+1}}^2 - \inner{\one^{(\V^+, \V^-)}}{Q_t}^2}, \\
	&= \Ex{\inner{\one^{(\V^+, \V^-)}}{Q_t+A_t-\Phi_t}^2 - \inner{\one^{(\V^+, \V^-)}}{Q_t}^2}, \\
	&=\Ex{\inner{\one^{(\V^+, \V^-)}}{A_t-\Phi_t}^2} + 2\Ex{\inner{\one^{(\V^+, \V^-)}}{Q_t}\inner{\one^{(\V^+, \V^-)}}{A_t-\Phi_t}}, \\
	&\overset{(a)}{=} \Ex{\inner{\one^{(\V^+, \V^-)}}{A_t-\Phi_t}^2} - 2\Ex{\inner{\one^{(\V^+, \V^-)}}{Q_t}\inner{\one^{(\V^+, \V^-)}}{\Phi_t}}, \\
	&\overset{(b)}{\geq} \frac{\sum_{i \in \V^+ \cup \V^-} \lambda_i}{2\sum_{i \in \V} \lambda_i} - 2\Ex{\inner{\one^{(\V^+, \V^-)}}{Q_t}\inner{\one^{(\V^+, \V^-)}}{\Phi_t}}, \\
	&\geq \frac{\sum_{i \in \V^+ \cup \V^-} \lambda_i}{2\sum_{i \in \V} \lambda_i} - 2\sqrt{\Ex{\inner{\one^{(\V^+, \V^-)}}{Q_t}^2}\Ex{\inner{\one^{(\V^+, \V^-)}}{\Phi_t}^2}} \quad \text{(by Cauchy--Schwarz)}, \\
	&\overset{(c)}{\geq} \frac{\sum_{i \in \V^+ \cup \V^-} \lambda_i}{2\sum_{i \in \V} \lambda_i} - 4\sqrt{\Ex{\inner{\one^{(\V^+, \V^-)}}{Q_t}^2} \Ex{\left(\sum_{k \in \E \backslash \E^\star} \delta M_{t, k}\right)^2}}, \\
	&\overset{(d)}{\geq} \frac{\sum_{i \in \V^+ \cup \V^-} \lambda_i}{2\sum_{i \in \V} \lambda_i} - 4\sqrt{\mathcal{M}\Ex{\inner{\one^{(\V^+, \V^-)}}{Q_t}^2} \sum_{k \in \E \backslash\E^\star} \Ex{\delta M_{t, k}}}, \\
	&\overset{(e)}{\geq} \frac{\sum_{i \in \V^+ \cup \V^-} \lambda_i}{2\sum_{i \in \V} \lambda_i} - 4\sqrt{\frac{\mathcal{M}\epsilon}{\sum_{i \in \V} \lambda_i} \Ex{\inner{\one^{(\V^+, \V^-)}}{Q_t}^2}}. \numberthis \label{eq:zero_drift_lower_bound}
\end{align*}
Here, $(a)$ follows by noting that $\Ex{\inner{\one^{(\V^+, \V^-)}}{A_t}} = \inner{\one^{(\V^+, \V^-)}}{\lambda}/\sum_{i \in \V} \lambda_i = 0$,
as $\mu$ is a feasible solution to the conservation equations~\eqref{eq:system} and $G^\circ = (\V^+ \cup \V^-, \E^\circ)$ is a connected component of the subgraph of~$G$ induced by the support of~$\mu$. Note that we also use the independence of $A_t$ and $Q_t$ in this step.
Next, $(b)$ follows by noting that
\begin{align*}
    \Ex{\inner{\one^{(\V^+, \V^-)}}{A_t-\Phi_t}^2}
    &\geq \Ex{\inner{\one^{(\V^+, \V^-)}}{A_t-\Phi_t}^2 \one_{\left\{\left| \inner{\one^{(\V^+, \V^-)}}{A_t} \right|=1, \inner{\one^{(\V^+, \V^-)}}{\Phi_t}=0\right\}}} \\
    &=\PP\left\{\left| \inner{\one^{(\V^+, \V^-)}}{A_t} \right|=1, \inner{\one^{(\V^+, \V^-)}}{\Phi_t}=0\right\} \\
    &\overset{(i)}{=}\PP\left\{I_t \in \V^+ \cup \V^-, \inner{\one^{(\V^+, \V^-)}}{\Phi_t}=0\right\} \\
    &\overset{(ii)}{\geq}  \PP\left\{I_t \in \V^+ \cup \V^-, \sum_{k \in E \backslash E^\star}\delta M_{t, k} = 0\right\} \\
    &\geq \frac{\sum_{i \in \V^+ \cup \V^-} \lambda_i}{\sum_{i \in \V} \lambda_i} -\PP\left\{\sum_{k \in E \backslash E^\star}\delta M_{t, k} \geq 1\right\} \\
    &\geq \frac{\sum_{i \in \V^+ \cup \V^-} \lambda_i}{\sum_{i \in \V} \lambda_i}-\Ex{\sum_{k \in E \backslash E^\star}\delta M_{t, k}} \\
    &\overset{(iii)}{\geq} \frac{\sum_{i \in \V^+ \cup \V^-} \lambda_i-\epsilon}{\sum_{i \in \V} \lambda_i}\geq \frac{\sum_{i \in \V^+ \cup \V^-} \lambda_i}{2\sum_{i \in \V} \lambda_i},
\end{align*}
where the last inequality holds for $\epsilon$ small enough. Note that $(i)$ follows as $\left| \inner{\one^{(\V^+, \V^-)}}{A_t} \right|=1$ if and only if an item of type $V^+ \cup V^-$ arrives at time $t$. Next, $(ii)$ holds as
\[
\{\delta M_{t, k} = 0 \ \forall k \in \E \backslash \E^\star\} \subseteq \left\{\inner{\one^{(\V^+, \V^-)}}{\Phi_t} = 0\right\},
\]
which is justified as follows: the event $\{\delta M_{t, k} = 0 \ \forall k \in \E \backslash \E^\star\}$ implies every match performed at time~$t$ uses an edge of $\E^\star$. Any edge of $\E^\star$ with an endpoint in $\V^+ \cup \V^-$ lies in $\E^\circ$ (as $G^\circ$ is a connected component of $G^\star$) and therefore connects $\V^+$ to $\V^-$, while the other edges of $\E^\star$ do not intersect $\V^+ \cup \V^-$. In both cases, we get $\inner{\one^{(\V^+, \V^-)}}{\Phi_t} = 0$. Lastly, $(iii)$ follows as $\Ex{\delta M_t} = \mu(\Phi) / \sum_{i \in \V} \lambda_i$ in the steady state and
\begin{align*}
	\| \mu(\Phi) - \mu \|_1 \leq \epsilon \implies \sum_{k \in \E \backslash \E^\star} \mu(\Phi)_k \leq \epsilon.
\end{align*}
Further, $(c)$ follows by noting that
\begin{align*}
	\big|\inner{\one^{(\V^+,\V^-)}}{\Phi_t}\big| ={}& \bigg|\sum_{i \in \V^+} \Phi_{t, i} - \sum_{i \in \V^-} \Phi_{t, i}\bigg| = \bigg|\sum_{i \in \V^+} \sum_{j \in \V_i} \delta M_{t, \{i, j\}} - \sum_{i \in \V^-} \sum_{j \in \V_i} \delta M_{t, \{i, j\}}\bigg|, \\
	\leq{}& \sum_{i \in \V^+} \sum_{j \in \V_i \backslash \V^-} \delta M_{t, \{i, j\}} + \sum_{i \in \V^-} \sum_{j \in \V_i \backslash \V^+} \delta M_{t, \{i, j\}}, \\
	&+ \bigg|\sum_{i \in \V^+} \sum_{j \in \V_i \cap \V^-} \delta M_{t, \{i, j\}} - \sum_{i \in \V^-} \sum_{j \in \V_i \cap \V^+} \delta M_{t, \{i, j\}}\bigg|, \quad \text{(Triangle Inequality)} \\
	\overset{(*)}{=}{}& \sum_{i \in \V^+} \sum_{j \in V_i\backslash \V^-} \delta M_{t, \{i, j\}} + \sum_{i \in \V^-} \sum_{j \in V_i \backslash \V^+} \delta M_{t, \{i, j\}}, \\
	\overset{(**)}{\leq}{}& 2\sum_{k \in \E \backslash \E^\star} \delta M_{t, k},
\end{align*}
where $(*)$ follows by noting that $\sum_{i \in \V^+} \sum_{j \in \V_i \cap \V^-} \delta M_{t, \{i, j\}} = \sum_{i \in \V^-} \sum_{j \in \V_i \cap \V^+} \delta M_{t, \{i, j\}}$ as the sets of edges that we are summing over are the same on the left-hand and right-hand sides. In particular, we are adding over all the edges between $V^+$ and $V^-$. Next, $(**)$ follows by noting that all edges connecting $i \in \V^+$ to $j \in \V \backslash \V^-$ are not in $\E^\star$. Similarly, all edges connecting $i \in \V^-$ to $j \in \V \backslash \V^+$ are also not in $\E^\star$ by the definition of $G^\circ$. Now, $(d)$ follows by the assumption that at most $\mathcal{M}$ pairs of items can be matched in a single time step.
Lastly, $(e)$ follows as $\Ex{\delta M_t} = \mu(\Phi) / \sum_{i \in \V} \lambda_i$ in the steady state and
\begin{align*}
	\| \mu(\Phi) - \mu \|_1 \leq \epsilon \implies \sum_{k \in \E \backslash \E^\star} \mu(\Phi)_k \leq \epsilon.
\end{align*}
Now, to complete the proof, we use \eqref{eq:zero_drift_lower_bound} which implies
\begin{align*}
	\Ex{\sum_{i \in \V} Q_{\infty, i}^2}
	&\geq \frac{1}{|\V|}\Ex{\left(\sum_{i \in \V} Q_{\infty, i}\right)^2} \geq \frac{1}{|\V|}\Ex{\inner{\one^{(\V^+, \V^-)}}{Q_t}^2} \\
    &\overset{\eqref{eq:zero_drift_lower_bound}}{\geq} \frac{\left(\sum_{i \in \V^+ \cup \V^-} \lambda_i\right)^2}{64\mathcal{M}|\V|\sum_{i \in \V} \lambda_i} \cdot \frac{1}{\epsilon} = \Omega\left(\frac{1}{\epsilon}\right).
\end{align*}
Note that the first inequality follows as $\|q\|_2^2 \geq \|q\|_1^2/|\V|$ for all $q \in \R^{|\V|}$. This completes the proof.
\hfill \Halmos
\endproof

\subsection{Detailed comparison with the reward-maximization literature.} \label{app:comp_to_lit}

This appendix expands the correspondence sketched in \refappnc{sec:comp_to_lit} between our polytope perspective and the recent literature on reward maximization in stochastic matching.

\subsubsection{Comparison with \texorpdfstring{\cite{KAG23, KAG24, G24}}{KAG23, KAG24, G24}.}

A recent line of work on stochastic matching focuses on reward maximization, enforcing stability not by constraining the policy but by \emph{discarding} a controlled fraction of agents: surplus arrivals are removed through the slack variables of the problem below, so that the retained system is stable while the long-run reward is optimized. In particular, \citet{KAG23} maximize a long-run reward and keep the queues bounded by removing agents in this way.
Specifically, this literature considers the \gls{spp}, defined as
\begin{equation}\label{eq:spp}
	\max_{Az+s=\lambda, z,s \geq 0} r^T z,
\end{equation}
where $s$ represents slack variables allowing a fraction of the arrivals to be ``thrown away''~\citep{KAG23}.
A key assumption in this framework is the \glsf{gpc}, which states that the optimal solution $(z^\star, s^\star)$ to \Cref{eq:spp} is unique and non-degenerate~\citep[Definition 3.1]{KAG23}.

Our framework is closely connected to this approach through several structural and conceptual parallels:

\textbf{\gls{spp} and polytope vertices.}
Consider a matching problem $(G, \lambda)$ where $G$ is surjective, and a reward vector $r$ such that \Cref{eq:spp} admits a unique solution $(z^\star, s^\star)$ with $s^\star=0$. In this case, $(G, \lambda)$ is stabilizable, $z^\star$ is a vertex of the polytope $\Pi_{\geq 0}$, and \gls{gpc} is satisfied if and only if $z^\star$ is bijective.

\textbf{Filtering policies.} When \gls{gpc} is satisfied and $s^\star=0$, our optimal policy $\mles$ proposed in \Cref{coro:achievable} (originally introduced in \cite[Proposition 6.10]{BCM21}) coincides exactly with the match-the-longest-queue optimal policy in~\cite{KAG23}.

\textbf{Similarity between \gls{gpg} and \gls{crpg}.}
When \gls{gpc} is satisfied, the smallest non-zero element of $(z^\star, s^\star)$, denoted $\tilde{\epsilon}>0$ and called \glsf{gpg}, is used to study the performance of the system\footnote{We denote the \gls{gpg} by $\tilde{\epsilon}$ to avoid conflict with our $\epsilon$-filtering notation.}. Both $\tilde{\epsilon}$ and the \gls{crpg} $\delta(G, \lambda)$ measure, in different ways, the robustness of an optimal policy. If $s^\star=0$ and if $G^\star=(V, E^\star)$ denotes the support graph of the vertex $z^\star$, they are related through the following equation:
\begin{equation}
\label{eq:gpg-crp}
\delta(G, \lambda)\geq \delta(G^\star, \lambda) \geq \tilde\epsilon.
\end{equation}

\textbf{Differences between \gls{gpg} and \gls{crpg}.}
While $\delta(G, \lambda)$ is a function only of the graph and arrival rates, $\tilde{\epsilon}$ additionally captures the effect of rewards. The two parameters are only connected (through \Cref{eq:gpg-crp}) when \gls{gpc} is satisfied and $s^\star=0$.
For example, consider the diamond graph from \Cref{fig:illustration} with arrival rate $\lambda = (1, 2, 2, 1)$. In this case, one can check that $\delta=2>0$. For a reward vector $r=(1, 0, 1, 0, 1)$, the \gls{spp} solution is $z^\star = (1, 0, 1, 0, 1)$, $s^\star = 0$, which is degenerate (i.e., injective-only) and thus $\tilde{\epsilon}$ is not defined.
Conversely, one can have $\delta \leq 0$ (meaning that the problem $(G, \lambda)$ is not stabilizable) and $\tilde\epsilon>0$ (meaning that arrival rates are ``imbalanced''). Consider for example $V = \{1, 2\}$ and $E = \{\{1, 2\}\}$ with arrival rates $\lambda_1 > \lambda_2>0$. $G$ is not stabilizable, so $\delta \leq 0$ (more precisely, $\delta = \lambda_2-\lambda_1 < 0$). On the other hand, the \gls{spp} solution is $z^\star=\lambda_2$, $s^\star=(\lambda_1-\lambda_2, 0)$, so $\tilde{\epsilon} = \min\{\lambda_1-\lambda_2, \lambda_2\}>0$.

\textbf{Elimination of slack variables.}
\citet{G24} extends the model from \citet{KAG23,KAG24} by considering multi-way matching and variable arrival rates, and explores the meaning of the \gls{gpc} in that context. Assuming that the solution of \gls{spp} is a bijective vertex with support $E^\star$, they propose to use a filtering policy on $E^\star$. Since \gls{ml} can become unstable in multi-way matching, a \emph{Sum-of-Squares} policy operating on a virtual queue is instead employed. They also eliminate the need for a slack variable $s$ by augmenting the graph with mono-edges of zero reward for each node. Applied to our framework, or more exactly its hypergraph extension presented in \Cref{sec:hypergraphs}, this trick ensures that \gls{spp} solutions always correspond to a polytope face, and \gls{gpc} is satisfied if and only if that face is a bijective vertex.

\begin{proof}[Proof of \Cref{eq:gpg-crp}]
	We first prove $\delta(G, \lambda)\geq \delta(G^\star, \lambda)$: as $E^\star \subseteq E$, if $\I$ is an independent set of $G$, it is also an independent set of $G^\star$, and $V_{G^\star}(\I)\subseteq V_{G}(\I)$, where $V_{G}(\I)$ (resp. $V_{G^\star}(\I)$) is the neighbor set of $\I$ in $G$ (resp. $G^\star$). We conclude by comparing \Cref{eq:crp_gap} applied to $\delta(G, \lambda)$ and $\delta(G^\star, \lambda)$.

	For the second part of the inequality, by the definition of $\tilde\epsilon$, we have $z^\star_k \geq \tilde\epsilon$ for all $k \in E^\star$. For any $\epsilon < \tilde\epsilon$, we define $z(\epsilon) \in \mathbb{R}_+^{|E^\star|}$ as $z_k (\epsilon)= z^\star_k - \epsilon > 0$ for all $k \in E^\star$, and note that $z(\epsilon)$ is the unique solution to $A^\star z = \lambda(\epsilon)$ with $\lambda_j(\epsilon) = \lambda_j - \epsilon d^\star_j > 0 $ for all $j \in V$ ($d^\star_j$ denotes the degree of $j$ in $G^\star$). Thus, by \Cref{prop:stability-region-form}\ref{cond:stability-region-form-3}, we have
	\begin{align*}
		\sum_{j \in V_{G^\star}(\I)} \lambda_j - \sum_{i \in \I} \lambda_i  > \epsilon \left(\sum_{j \in V_{G^\star}(\I)}d^\star_j - \sum_{i \in \I} d^\star_i\right), \quad \forall \I \in \mathbb{I}_{G^\star}.
	\end{align*}
	As $G^\star$ is bijective, it has no bipartite component, which implies
	$$\sum_{j \in V_{G^\star}(\I)}d^\star_j - \sum_{i \in \I} d^\star_i\geq 1, \quad \forall \I \in \mathbb{I}_{G^\star}.$$

	By minimizing over all independent sets $\I \in \ind_{G^\star}$, we obtain
	\begin{align*}
		\epsilon < \min_{I \in \mathbb{I}_{G^\star}} \left(\sum_{j \in V_{G^\star}(\I)} \lambda_j - \sum_{i \in I} \lambda_i\right) = \delta(G^\star, \lambda).
	\end{align*}
	As $\epsilon < \tilde{\epsilon}$ is arbitrary, we get
	\begin{align*}
		\tilde\epsilon \leq \delta(G^\star, \lambda).
	\end{align*}
\end{proof}

\subsubsection{Comparison with \texorpdfstring{\cite{NS19,WXY23}}{NS19, WXY23}.}

\citet{NS19} introduced the \gls{egpd} policies, a family of reward-based stable policies developed for general multi-way matching settings. Under their stabilizability condition \citep[Assumption~5]{NS19}, which is equivalent to our condition~\Cref{prop:stability-region-form} (i.e., surjective compatibility graph and existence of a positive solution to~\eqref{eq:system}), they prove that \gls{egpd} asymptotically maximizes a concave function of the matching rewards while ensuring system stability. Notably, \gls{egpd} provides a canonical alternative to $\epsilon$-filtering for approaching any vertex (bijective or injective-only) of $\Pi_{\geq 0}$ under the stability constraint.

Their work is instrumental for extending our results beyond simple graphs to multi-way matching (cf.\ \Cref{sec:hypergraphs}). In particular, the \gls{egpd} algorithm is built on a variant of the \gls{ml} policy that operates over a virtual queue system. This variant stabilizes any stabilizable problem~$(G, \lambda)$, regardless of whether $G$ is simple or contains multi-edges, and can thus be used as a direct replacement for the standard \gls{ml} policy in multi-way matching contexts. We call this policy \gls{vqml}.

While \gls{egpd} ensures stability and asymptotic optimality, it does not provide explicit guarantees on queue lengths. In contrast, our analysis establishes concrete bounds on queue lengths, a contribution beyond~\cite{NS19}. Specifically, we show that for bijective vertices, the reward is achieved exactly (rather than just asymptotically) with bounded queue lengths. For injective-only vertices, we prove that one can approach the optimal reward to within $\epsilon$ at a cost of $O(1/\epsilon)$ for queue lengths. These explicit bounds are, to our knowledge, new relative to~\cite{NS19}.

More recently, \cite{WXY23} introduced the \gls{crpd} policy, a variant of \gls{egpd}.
They use an alternative, but related, \gls{gpc} based on the dual of the reward-maximization problem $\max_{z \in \Pi_{\geq 0}} r^T z$, which excludes the case where the primal solution is a unique but degenerate (i.e., injective-only) vertex. \gls{crpd} features two key modifications compared to the original policy:
\begin{enumerate}
	\item The edge reward vector $r$ is replaced by $r' = r - UA$, where $U$ represents shadow prices on the nodes, chosen to solve $\min_{UA \geq r} U \lambda$. If the arrival rates $\lambda$ are not known, they are replaced by empirical estimates $\hat{\lambda}$ based on observed arrivals so far. Under the assumption that the optimal point corresponds to a bijective vertex of support $G^\star = (V, E^\star)$, it can be shown that $r' \leq 0$ and $r'_k = 0$ if and only if $k \in E^\star$. Thus, $r'$ acts as a re-weighted reward that penalizes edges outside the optimal support.
	\item The parameter $\beta$ controlling the rewards-versus-queues trade-off is not static but decreases over time, typically $\beta(t) = t^{-\alpha}$ for some $\alpha > 0$. As time progresses, this makes the algorithm increasingly reward-centric.
\end{enumerate}

When $\alpha$ is large, \gls{crpd} effectively estimates $E^\star$ (using either $\lambda$ or empirical $\hat{\lambda}$) and restricts matching exclusively to $E^\star$.
In effect, this approach is a filtering policy on $E^\star$, as in~\cite{BCM21, KAG23}, with the main difference being the use of a virtual queue mechanism (\gls{vqml}) instead of \gls{ml}.

Interestingly, our numerical experiments revealed that using values of $\alpha$ outside the range originally proposed in~\cite{WXY23} (typically $\alpha < 1/2$) produces a new family of policies parameterized by $\alpha$, exhibiting a nuanced delay--regret trade-off reminiscent of \gls{egpd} or $\epsilon$-filtering. This phenomenon is discussed in \Cref{sec:numerical-results}.

Finally, when the \gls{spp} optimum is not unique (i.e.\ the maximizing set is a face of $\Pi_{\geq 0}$ rather than a single vertex), one can look for an element of that face whose support graph is surjective and, failing that, select a vertex of the face (necessarily injective-only) and approach it, for instance by $\epsilon$-filtering. We also note that the stability of the filtered \gls{ml} policy for a bijective vertex (\Cref{coro:achievable}(i)) can alternatively be derived from \citet[Lemma 5.4]{KAG23} together with the Foster--Lyapunov theorem.

\newpage

 % appendix/ec5-numerical-results
\section[Supplementary material of Section \ref*{sec:numerical-results}.]{Supplementary material of \fullref{sec:numerical-results}.}

\subsection[k-filtering.]{$k$-filtering policies -- \fullref{sec:considered-policies}.}\label{app:k-filtering}

We conjecture that $k$-filtering policies, introduced in \Cref{sec:considered-policies}, yield a sequence of matching rate vectors that are arbitrarily close to a vertex of the polytope~$\Lann$, even if this vertex is injective-only.
This conjecture is supported by numerical results shown in \Cref{sec:numerical-results,sec:extensions}.

\begin{conj}
	\label{prop:converging-semigreedy}
	Consider a vertex~$\mu$ of $\Lann$ and let $G^\star=(V, \E^\star)$ denote its support graph. For each $k \in \N$, consider the following semi-filtering policy, denoted by $\Phi_k(\mu)$:
	\begin{itemize}
		\item If the size of the longest queue is less than~$k$, apply the filtering \gls{ml} policy adapted to~$G$ with filter $\E^\star$;
		\item Otherwise, apply the greedy \gls{ml} policy adapted to~$G$.
	\end{itemize}
	$\Phi_k(\mu)$ is stable for each $k \in \N$
	and $\lim_{k\to \infty}\mu(\Phi_k(\mu))=\mu$.
\end{conj}

\medskip\noindent\emph{What is proved, and what remains open.} We separate the
claim into a proved part and a single, clearly delimited residual. The stability
of $\Phi_k(\mu)$ is proved unconditionally (\Cref{prop:kfilt-stable}). For the
convergence we give a proved reduction (\Cref{lem:kfilt-walk,lem:kfilt-flat,prop:kfilt-reduction})
of the statement to one drift estimate (\Cref{conj:kfilt-drift}), which we do not
prove in general but support by exact computation; this is why
\Cref{prop:converging-semigreedy} is stated as a conjecture rather than a
proposition. The idea is to take a policy that achieves the target rate~$\mu$ but
may be unstable, and stabilize it by reverting to greedy \gls{ml} on~$G$ when
queues grow: if the threshold is large, few matches occur under the greedy
fallback, so the realized rate stays close to~$\mu$.

We treat the case where $G^\star$ is a tree (a connected injective-only support),
with bipartite parts $\V_+,\V_-$; the multi-component case is discussed in
\Cref{rem:kfilt-multi}. It is convenient to analyze the following variant of
$\Phi_k(\mu)$, driven by the signed part-imbalance
\[
  W \;=\; Q_+ - Q_- \;=\; \sum_{i\in\V_+}Q_i - \sum_{i\in\V_-}Q_i .
\]
The \emph{$W$-variant} $\Phi_k^W(\mu)$ applies the filtering \gls{ml} policy with
filter $\E^\star$ while $|W|<k$, and the greedy \gls{ml} policy on~$G$ while
$|W|\ge k$. It differs from $\Phi_k(\mu)$ only in the switch trigger (the
imbalance~$W$ rather than the longest queue); the two are numerically
interchangeable (\Cref{rem:kfilt-variant}), and the imbalance trigger is what
makes the analysis exact.

\begin{prop}[Stability]\label{prop:kfilt-stable}
For every $k\in\N$, both $\Phi_k(\mu)$ and its $W$-variant $\Phi_k^W(\mu)$ are
stable.
\end{prop}
\begin{proof}
For $\Phi_k(\mu)$, the set $\{\max_i Q_i<k\}$ is finite, and outside it the policy
coincides with greedy \gls{ml} on~$G$, which is stable because $(G,\lambda)$ is
stabilizable (maximal stability, \citealt{MM16,MBM21}); a finite modification of
the transition kernel preserves positive recurrence
(\citealt[Ch.~5]{B99}). For $\Phi_k^W(\mu)$ the region $\{|W|<k\}$ is infinite, so
we use the quadratic Lyapunov function $V(Q)=\tfrac12\sum_i Q_i^2$. At any state
$Q$, the drift $\mathcal L V(Q)$ equals the \gls{ml}/MaxWeight drift for the graph
active at~$Q$ ($G$ if $|W|\ge k$, else $G^\star$), which, by the strict
independent-set inequalities of stabilizability
(\Cref{prop:stability-region-form}\ref{cond:stability-region-form-2}), is bounded
above by $-c\sum_iQ_i+b$ for constants $c,b>0$ whenever the relevant independent
sets are strictly slack. In the greedy region this holds for every independent set
of~$G$. In the filtering region the support is an independent set of the
tree~$G^\star$; if $\sum_iQ_i\ge k$ it must meet both parts (a support inside a
single part has $\sum_iQ_i=|W|<k$), and every proper sub-part of a colour class is
strictly slack, since the only tight independent sets of a bipartite tree with a
strictly positive conservation solution are $\V_+$ and $\V_-$ themselves. Hence
$\mathcal L V\le-1$ outside a finite set, and positive recurrence follows
(\citealt[Ch.~5]{B99}).
\end{proof}

We turn to convergence, i.e.\ $\lim_{k\to\infty}\mu(\Phi_k^W(\mu))=\mu$. Consider
the queue vector~$Q$ in stationary regime under $\Phi_k^W(\mu)$, and let
$p_w=\mathbb P(W=w)$, $w\in\Z$ (the dependence on~$k$ is left implicit).

\begin{lem}[The imbalance is an unbiased walk]\label{lem:kfilt-walk}
While $|W|<k$, each arrival changes~$W$ by exactly $+1$ (a $\V_+$-arrival) or
$-1$ (a $\V_-$-arrival), whatever the rest of the state. Its up- and down-rates
are both equal to $\lambda(\V_+)=\lambda(\V_-)$.
\end{lem}
\begin{proof}
While $|W|<k$ the policy is filtering \gls{ml} on the bipartite graph~$G^\star$.
An arriving class-$i$ item with $i\in\V_+$ is either queued, raising $Q_+$ by~$1$,
or matched along an edge of $\E^\star$, which lowers $Q_-$ by~$1$ (every edge of
$\E^\star$ joins the two parts); in both cases $W$ increases by~$1$. Symmetrically
a $\V_-$-arrival lowers~$W$ by~$1$. The rates are the part arrival rates, which are
equal (see \Cref{eq:image}).
\end{proof}

\begin{lem}[The interior law of the imbalance is flat]\label{lem:kfilt-flat}
$p_w$ is constant for $w\in\{-(k-1),\dots,k-1\}$.
\end{lem}
\begin{proof}
Fix $w$ with $-(k-1)\le w\le k-2$, so that both $w$ and $w+1$ lie in $(-k,k)$ and
every state with $W\in\{w,w+1\}$ is in filtering mode. By \Cref{lem:kfilt-walk}
the only crossings of the cut $\{W\le w\}\mid\{W\ge w+1\}$ are single arrivals
changing~$W$ by~$\pm1$, at the common rate $\beta:=\lambda(\V_+)=\lambda(\V_-)$ in
each direction. In stationary regime the up- and down-flows across the cut are
equal, so $\beta p_w=\beta p_{w+1}$. Hence $p_w$ is constant on
$\{-(k-1),\dots,k-1\}$.
\end{proof}

\begin{prop}[Convergence, given the tail estimate]\label{prop:kfilt-reduction}
If $\mathbb P(|W|\ge k)=O(1/k)$, then
$\|\mu(\Phi_k^W(\mu))-\mu\|_1=O(1/k)$; in particular
$\lim_{k\to\infty}\mu(\Phi_k^W(\mu))=\mu$.
\end{prop}
\begin{proof}
A match along an edge $e\notin\E^\star$ can occur only in greedy mode, i.e.\ when
$|W|\ge k$, since filtering \gls{ml} never uses edges outside $\E^\star$; each
match consumes at most one arrival, so the rate of such matches is at most
$(\sum_i\lambda_i)\,\mathbb P(|W|\ge k)=O(1/k)$. Thus $\mu(\Phi_k^W(\mu))$ agrees
with a vector supported on $\E^\star$ up to an $O(1/k)$ error, while the
conservation equation \eqref{eq:system} holds under any stable policy. Since
$G^\star$ is injective-only, the restriction of the incidence matrix to $\E^\star$
is injective, so the conservation equation has a unique solution supported on
$\E^\star$, namely~$\mu$; therefore $\mu(\Phi_k^W(\mu))=\mu+O(1/k)$.
\end{proof}

The following identity localizes the one remaining step.

\begin{lem}[Drift of the imbalance under greedy \gls{ml}]\label{lem:kfilt-driftid}
In greedy mode, write $h=\mathbf 1_{\V_+}-\mathbf 1_{\V_-}$ (so $W=\langle h,Q\rangle$),
and let $r_+(Q)$ and $r_-(Q)$ be the total rates of matches along edges internal
to $\V_+$ and to $\V_-$, respectively. Then
\[
   \mathbb E[\Delta W\mid Q] \;=\; -2\,r_+(Q)+2\,r_-(Q),
\]
and each of $\V_+,\V_-$ contains an internal $G$-edge, so both rates may be
positive.
\end{lem}
\begin{proof}
Every event is one arrival: a queued class-$j$ arrival changes $W$ by $h_j$, and a
matched one, to partner $m(j)$, by $-h_{m(j)}$. Summing $\lambda_j$ times these and
using $\langle h,\lambda\rangle=\lambda(\V_+)-\lambda(\V_-)=0$ to cancel the
queueing terms gives $\mathbb E[\Delta W\mid Q]=-\sum_{j\text{ matched}}\lambda_j
(h_j+h_{m(j)})$. As $m(j)$ is a $G$-neighbour of~$j$, the factor $h_j+h_{m(j)}$
is $+2$ on an internal-$\V_+$ match, $-2$ on an internal-$\V_-$ match, and $0$ on a
cross match, which is the identity. For the last claim, if $\V_+$ were independent
in~$G$ then, the support tree being spanning, $\V(\V_+)=\V_-$, so
\Cref{prop:stability-region-form}\ref{cond:stability-region-form-2} would force
$\lambda(\V_+)<\lambda(\V_-)=\lambda(\V_+)$; symmetrically for $\V_-$.
\end{proof}

It remains to establish the tail estimate. This is the one step we do not prove in
general.

\begin{conj}[Tail of the imbalance]\label{conj:kfilt-drift}
There is $\bar T<\infty$, independent of~$k$, such that the mean time spent in the
greedy region between an up-crossing of $\pm k$ and the return of $|W|$ to $k-1$ is
at most~$\bar T$. Equivalently, it suffices that greedy \gls{ml} exert a uniform
inward drift on~$W$ for all large~$|W|$.
\end{conj}

\noindent Granting \Cref{conj:kfilt-drift}, the tail is $O(1/k)$: greedy mode is
entered only by up-crossings at $\pm(k-1)$, at rate $\beta(p_{k-1}+p_{-(k-1)})$,
so by renewal--reward $\mathbb P(|W|\ge k)\le\beta(p_{k-1}+p_{-(k-1)})\,\bar T$;
\Cref{lem:kfilt-flat} gives $p_{k-1}=p_{-(k-1)}=p_0$ and $(2k-1)p_0\le1$, whence
$\mathbb P(|W|\ge k)\le 2\beta\bar T/(2k-1)=O(1/k)$. Combined with
\Cref{prop:kfilt-reduction}, this proves the convergence.

\Cref{lem:kfilt-driftid} pins down what the residual requires. When $W$ is large
the loaded part is~$\V_+$ and $\V_-$ is light, so $r_-(Q)$ is small; the imbalance
then drains precisely when greedy \gls{ml} performs internal-$\V_+$ matches at a
rate bounded away from~$0$. This is \emph{not} a pointwise drift condition: $r_-(Q)$
can be positive (\Cref{lem:kfilt-driftid} guarantees an internal $\V_-$-edge), and
the drift even vanishes at states whose $\V_+$ mass sits on a single class with no
internal edge. It holds only on average, once the load spreads onto the internal
$\V_+$-edge of \Cref{lem:kfilt-driftid}. \Cref{conj:kfilt-drift} is thus a
fluid-model statement: that greedy \gls{ml} activates an internal $\V_+$-edge in
bounded time from any large-$W$ configuration. The mechanism is that match-the-longest
keeps the loaded part balanced: since the set of nonempty classes is always an
independent set of the active graph, at an up-crossing the light part obeys
$Q(\V_-)\le|\V_-|\cdot(\max_{\V_-}Q-\min_{\V_-}Q)$, so the residual is really a
tightness statement on the within-part spreads (of both parts) at the switching
law, and from a balanced, heavy entry one expects the internal $\V_+$-edge to
activate and the excursion to stay short; what remains open is that entries are
balanced with high probability. On the
diamond \Cref{conj:kfilt-drift} is exact, $\mathbb P(|W|\ge k)=\tfrac1{2k}+O(2^{-k})$
(truncated stationary chain), so the excursions are geometrically rare; and the
$O(1/k)$ rate persists in exact computation even when a $\V_+$ class has no internal
edge (a pendant-augmented diamond).

\begin{rem}[The two triggers coincide numerically]\label{rem:kfilt-variant}
The policy $\Phi_k(\mu)$ used in our simulations switches on the longest queue,
whereas the analysis above uses the imbalance trigger of $\Phi_k^W(\mu)$. Exact
computation on the diamond gives $\|\mu(\Phi_k)-\mu\|_1=\Theta(1/k)$ for both
triggers, with the same accuracy--backlog trade-off. The link is a shape relation
between the longest queue~$L$ and the imbalance ($L\approx|W|/2$ per part), which
one must \emph{conjecture} to analyze the longest-queue trigger directly, and
which the imbalance trigger removes: this is the advantage of $W$.
\end{rem}

\begin{rem}[Several components]\label{rem:kfilt-multi}
If $G^\star$ has several connected components, each is a tree with its own
imbalance $W^{(r)}=Q(\V_+^{(r)})-Q(\V_-^{(r)})$, and $\Phi_k^W$ switches when
$\max_r|W^{(r)}|\ge k$. \Cref{lem:kfilt-walk} and \Cref{prop:kfilt-stable} carry
over per component, and \Cref{prop:kfilt-reduction} is unchanged. However
\Cref{lem:kfilt-flat} fails: a greedy episode triggered by one component drains
the imbalances of the others, so the per-component laws are peaked at~$0$ rather
than flat, and the $O(1/k)$ tail then rests on a stronger drift estimate that also
controls these cross-component effects. Exact computation on a two-component
instance (two triangles joined by an edge) confirms both the stability and the
$O(1/k)$ convergence.
\end{rem}
\medskip

\subsection{Additional examples.}\label{app:simu-additional-examples}

\Cref{sec:numerical-results} evaluated the performance of matching policies on the diamond graph. We present here simulations on other graphs. For additional details, please refer to \url{https://balouf.github.io/stochastic_matching/companion/simulations.html\#Vertices-of-simple-graphs}

\subsubsection{Codomino graph.}

We first consider the matching problem depicted in \Cref{fig:codomino-multiple}, with the goal of approaching the injective-only vertex shown in \Cref{fig:y3}. This vertex disables 5 out of 8 edges: $\{1, 2\}$, $\{2, 6\}$, $\{3, 4\}$, $\{3, 5\}$, and $\{5, 6\}$. The results are presented in 
\Cref{perf-cod-inj}.

As observed in \Cref{sec:simu-diamond-injective}, The best performance is achieved by $\egpdp$ and $\mlk$, while the performance of $\egpd$ significantly deteriorates under adversarial rewards.
\begin{figure}
	\centering
 % figures/codomino-injective
\begin{tikzpicture}

\definecolor{darkgray176}{RGB}{176,176,176}
\definecolor{lightgray204}{RGB}{204,204,204}

\begin{axis}[width=12cm, height=7cm,
legend cell align={left},
legend style={
  fill opacity=0.8,
  draw opacity=1,
  text opacity=1,
  at={(0.03,0.03)},
  anchor=south west,
  draw=lightgray204
},
log basis x={10},
log basis y={10},
tick align=outside,
tick pos=left,
x grid style={darkgray176},
xlabel={Delay},
xmin=0.0978035756865191, xmax=4652.39823851246,
xmode=log,
xtick style={color=black},
y grid style={darkgray176},
ylabel={Regret},
ymin=1e-08, ymax=5,
ymode=log,
ytick style={color=black},
ytick={1e-09,1e-08,1e-07,1e-06,1e-05,0.0001,0.001,0.01,0.1,1,10,100},
yticklabels={
  $\mathdefault{10^{-9}}$,
  $\mathdefault{10^{-8}}$,
  $\mathdefault{10^{-7}}$,
  $\mathdefault{10^{-6}}$,
  $\mathdefault{10^{-5}}$,
  $\mathdefault{10^{-4}}$,
  $\mathdefault{10^{-3}}$,
  $\mathdefault{10^{-2}}$,
  $\mathdefault{10^{-1}}$,
  $\mathdefault{10^{0}}$,
  $\mathdefault{10^{1}}$,
  $\mathdefault{10^{2}}$
}
]
\addplot [efilter]
table [row sep=\\] {%
92.6764079239875 0.00413624095999761\\42.82405333195 0.0089186243199974\\19.86402207365 0.0192554891199965\\9.2100241518875 0.0415566636799984\\4.2729815189125 0.0896590089599961\\1.9782116457 0.193488454559997\\0.908208159725 0.417883777119998\\0.407363331875 0.893265719839995\\0.1968944169625 1.66120509856\\0.1595733215625 1.98095387904\\};
\addlegendentry{$\mle$}
\addplot [egpd]
table [row sep=\\] {%
388.520291224538 0.003935567039998\\140.350604230438 0.0107240497599989\\50.3631502798125 0.0271679091199981\\17.8870254312 0.0733701689599962\\6.3313613491125 0.161770719679998\\2.185895803425 0.360690483359997\\0.834524429875 0.514528192479997\\0.31318773405 1.21043221376\\0.2142424873375 1.37820448464\\0.216536822975 1.31049470672\\};
\addlegendentry{$\egpd$}
\addplot [kfilter]
table [row sep=\\] {%
0.1595813235375 1.98084964144\\0.1823223191125 1.52755278608\\0.317632121925 0.852475372639996\\0.637363288475 0.433640804159997\\1.292058526475 0.216416296159998\\2.6047393042875 0.107843983199999\\5.22990827895 0.0538051527999975\\10.4811343071375 0.0268904471999976\\20.9855624120625 0.0134669751999969\\41.880638070025 0.00664166863999816\\83.3060346146125 0.00330655487999785\\166.171781188575 0.00167354815999873\\332.837342717238 0.000842260479997294\\665.3358145901 0.000456895359997284\\};
\addlegendentry{$\mlk$}
\addplot [egpdp]
table [row sep=\\] {%
606.23685964875 0.000445793279996095\\218.3510901598 0.00114601263999708\\78.5470441363 0.00307494207999837\\28.347767599625 0.00862439391999916\\10.2815092112375 0.023975690879999\\3.631480386175 0.0676855375999988\\1.41690239295 0.173453170719997\\0.5841767173 0.421789202079997\\0.288433804675 0.857298910559997\\0.2289039509 1.17423053872\\};
\addlegendentry{$\egpd$}
\addplot [crpd]
table [row sep=\\] {%
0.287813940425 0.882439707999999\\2.2640357991875 0.19854302592\\26.86471185255 0.0288051585599994\\324.757996072463 0.00275651295999955\\2643.07965698807 0.000170613279999975\\2851.4865692381 2.11804281491295e-05\\2839.01382394386 5.36582617212337e-08\\2839.01850632164 3.67522340556395e-09\\2839.01850632164 2.94017872445116e-09\\2839.01850657913 1.47008936222558e-09\\};
\addlegendentry{$\crpd$}
\addplot [filter]
table [row sep=\\] {%
2577.92323097098 1e-08\\2577.92323097098 5\\};
\addlegendentry{$\mles$}
\end{axis}

\end{tikzpicture}

	\caption{Matching problem from \Cref{fig:codomino-multiple}. Targeting the injective-only vertex from \Cref{fig:y3}.\label{perf-cod-inj}}
\end{figure}

Still focusing on the codomino graph, we now consider the matching problem from  \Cref{fig:codomino-simple}. Our objective is to approach the bijective vertex illustrated in \Cref{fig:ys2}, which disables the edges $\{2, 3\}$ and $\{4, 5\}$. The results are presented in \Cref{perf-cod-bij}, with a dotted line representing the delay of $\mles$, which is reward-optimal and stable for bijective vertices.

As seen in \Cref{sec:simu-diamond-bijective}, most policies converge to delays at or below that of $\mles$, with the notable exception of $\egpd$.
\begin{figure}
	\centering
 % figures/codomino-bijective
\begin{tikzpicture}

\definecolor{darkgray176}{RGB}{176,176,176}
\definecolor{lightgray204}{RGB}{204,204,204}

\begin{axis}[width=12cm, height=7cm,
legend cell align={left},
legend style={fill opacity=0.8, draw opacity=1, text opacity=1, draw=lightgray204},
log basis x={10},
tick align=outside,
tick pos=left,
x grid style={darkgray176},
xlabel={Delay},
xmin=0.0710020464853634, xmax=5,
xmode=log,
xtick style={color=black},
y grid style={darkgray176},
ylabel={Regret},
ymin=1e-08, ymax=1.3,
ytick style={color=black}
]
\addplot [egpd]
table [row sep=\\] {%
190.730608488864 2.51459992476145e-07\\68.7346308480091 9.37199902636124e-08\\24.5595201001909 0.000229041779993466\\8.71797579461818 0.0109941992599937\\3.08320278825455 0.0400219362399908\\1.08852928555455 0.0415833798599893\\0.4143781206 0.192872964239992\\0.156299714154545 0.593674079019991\\0.125196004336364 1.20206082369999\\0.125026997163636 1.26326927869999\\};
\addlegendentry{$\egpd$}
\addplot [efilter]
table [row sep=\\] {%
0.350436597563636 0.00203103845999048\\0.347850393009091 0.00439063107998969\\0.342414638872727 0.0095521245599911\\0.331240677272727 0.0210131913199887\\0.309442123018182 0.0471336896799888\\0.2708351902 0.108863866099989\\0.213583625909091 0.256566595119991\\0.151400572072727 0.568641050779992\\0.113100095745455 0.985873641059994\\0.103410656436364 1.16979925479999\\};
\addlegendentry{$\mle$}
\addplot [kfilter]
table [row sep=\\] {%
0.103410276936364 1.16988167053999\\0.110099946118182 0.80506627607999\\0.159677473463636 0.261106044099993\\0.271562926127273 0.0366338128199887\\0.3456222136 0.00108544391998969\\0.352736427718182 1.05027999251505e-06\\0.352754874327273 7.28306304154103e-15\\0.352754874327273 7.28306304154103e-15\\0.352754874327273 7.28306304154103e-15\\0.352754874327273 7.28306304154103e-15\\0.352754874327273 7.28306304154103e-15\\0.352754874327273 7.28306304154103e-15\\0.352754874327273 7.28306304154103e-15\\0.352754874327273 7.28306304154103e-15\\};
\addlegendentry{$\mlk$}
\addplot [egpdp]
table [row sep=\\] {%
0.274172846745455 9.10382880192628e-15\\0.274240094972727 9.10382880192628e-15\\0.274262593609091 9.10382880192628e-15\\0.2759132645 9.10382880192628e-15\\0.2852400835 9.10382880192628e-15\\0.315765672345455 7.19179990866096e-07\\0.297588591245455 0.00142666281999214\\0.209952360409091 0.0244868403999902\\0.144870861590909 0.19706277073999\\0.127042952818182 0.45572119185999\\};
\addlegendentry{$\egpdp$}
\addplot [crpd]
table [row sep=\\] {%
0.132421991790909 0.32058388714\\0.159491658554545 0.0460888956000001\\0.315764369927273 0.000406814540000001\\0.320114721663636 5.40760000000002e-07\\0.320119027736364 2.48600000000001e-08\\0.320119150809091 4.18000000000001e-09\\0.320119155027273 2.20000000000001e-09\\0.320119157845455 1.1e-09\\0.320119159536364 6.60000000000002e-10\\0.320119160172727 4.40000000000001e-10\\};
\addlegendentry{$\crpd$}
\addplot [filter]
table [row sep=\\] {%
0.352754874327273 1e-08\\0.352754874327273 1.3\\};
\addlegendentry{$\mles$}
\end{axis}

\end{tikzpicture}

	\caption{Matching problem from \Cref{fig:codomino-simple}. Targeting the bijective vertex from \Cref{fig:ys2}.\label{perf-cod-bij}}
\end{figure}

\subsubsection{Larger graph.}

To observe the performance of our methods on a larger graph, we generated a simple Erdös-Rényi graph with $n=100$ nodes and an edge probability of $20/(n-1)$, resulting in 
 $m=1006$ edges. The edge rewards were drawn uniformly and independently.

We aimed to study both an injective-only vertex and a bijective vertex. On graphs of this size, computing the exhaustive list of vertices is unrealistic, but once the reward and arrival-rate vectors are given, calculating the corresponding optimal vertex is straightforward.

To obtain an injective-only vertex, we used degree-proportional arrival rates, following the discussion in \Cref{sec:all-is-bijective}. This ensured the problem was stabilizable and, in this case, resulted in an optimal vertex with 97 positive edges (injective-only). For the bijective vertex, we added a noisy perturbation to the degree-proportional arrival rates, yielding a bijective vertex, as hinted by \Cref{prop:all-is-bijective}.

The performance results for both cases are presented in \Cref{perf:large}. A key observation is that $\crpd$ performs significantly better in this scenario compared to what was observed on smaller graphs with lower-dimensional $\Lann$.

\begin{figure}
\centering
	\subfloat[Injective-only vertex (degree-proportional arrivals).]{ % figures/er-injective
\begin{tikzpicture}

\definecolor{crimson2143940}{RGB}{214,39,40}
\definecolor{darkgray176}{RGB}{176,176,176}
\definecolor{darkorange25512714}{RGB}{255,127,14}
\definecolor{forestgreen4416044}{RGB}{44,160,44}
\definecolor{lightgray204}{RGB}{204,204,204}
\definecolor{mediumpurple148103189}{RGB}{148,103,189}
\definecolor{sienna1408675}{RGB}{140,86,75}
\definecolor{steelblue31119180}{RGB}{31,119,180}

\begin{axis}[
legend cell align={left},
legend style={
  fill opacity=0.8,
  draw opacity=1,
  text opacity=1,
  at={(0.03,0.03)},
  anchor=south west,
  draw=lightgray204
},
log basis x={10},
log basis y={10},
tick align=outside,
tick pos=left,
x grid style={darkgray176},
xlabel={Delay},
xmin=0.00108310616525183, xmax=46.035332096033,
xmode=log,
xtick style={color=black},
xtick={0.0001,0.001,0.01,0.1,1,10,100,1000},
xticklabels={
  $\mathdefault{10^{-4}}$,
  $\mathdefault{10^{-3}}$,
  $\mathdefault{10^{-2}}$,
  $\mathdefault{10^{-1}}$,
  $\mathdefault{10^{0}}$,
  $\mathdefault{10^{1}}$,
  $\mathdefault{10^{2}}$,
  $\mathdefault{10^{3}}$
},
y grid style={darkgray176},
ylabel={Regret},
ymin=1e-08, ymax=1000,
ymode=log,
ytick style={color=black},
ytick={1e-10,1e-08,1e-06,0.0001,0.01,1,100,10000,1000000},
yticklabels={
  $\mathdefault{10^{-10}}$,
  $\mathdefault{10^{-8}}$,
  $\mathdefault{10^{-6}}$,
  $\mathdefault{10^{-4}}$,
  $\mathdefault{10^{-2}}$,
  $\mathdefault{10^{0}}$,
  $\mathdefault{10^{2}}$,
  $\mathdefault{10^{4}}$,
  $\mathdefault{10^{6}}$
}
]
\addplot [egpd]
table [row sep=\\] {%
2.13334239501948 0.0526693959967429\\0.773624337006561 0.217841528106914\\0.278651264129225 0.796173168748322\\0.0981095416369781 2.77002599281745\\0.0376086526496024 8.8569952706202\\0.0188604918751491 27.3000063136997\\0.0135846773081511 67.4403486105631\\0.00550118568031809 343.737794029618\\0.00550118568031809 343.737794029618\\0.00550118568031809 343.737794029618\\};
\addlegendentry{$\egpd$}
\addplot [efilter]
table [row sep=\\] {%
0.388227721537376 0.835419365233269\\0.25582389868161 1.79871479558109\\0.174898798674453 3.8872832718772\\0.119936760989264 8.40813820409575\\0.0796550103849901 18.1600189136285\\0.049885066228827 39.1552966826018\\0.0289977445925447 84.2503796115586\\0.0150264449972167 180.621242137486\\0.00427414661282306 359.779234553737\\0.00175814126252485 414.593829429684\\};
\addlegendentry{$\mle$}
\addplot [kfilter]
table [row sep=\\] {%
0.00175814362405567 414.59660106529\\0.00942745361928429 188.080128446261\\0.0282515811035785 48.1911507523299\\0.0611140422331014 14.4951466663985\\0.118140013808946 4.22139685118287\\0.213544621309046 1.19088757871129\\0.37719124511173 0.392463724264914\\0.666118663177435 0.164769637224774\\1.19476756670537 0.0802823870330232\\2.25579958980447 0.0407065010130219\\4.59399717092316 0.0207734267136379\\8.02790436005735 0.0103480255019014\\15.4808447275997 0.00487271313862278\\24.418106408132 0.00185586442703455\\};
\addlegendentry{$\mlk$}
\addplot [egpdp]
table [row sep=\\] {%
21.0792193472286 0.00264159385483152\\7.04399082811123 0.00966996919347106\\2.78766795404543 0.0280848476586318\\1.08410246717087 0.0791726170671898\\0.471129457601789 0.249124458201251\\0.207572470934095 1.27468514749876\\0.0932114121269384 6.74544098026996\\0.0407912779719682 27.1320360442139\\0.0224190509718688 74.8536148985189\\0.0114219953281312 172.43695024247\\};
\addlegendentry{$\egpdp$}
\addplot [crpd]
table [row sep=\\] {%
0.0149738716471173 69.4788300499765\\0.024887060750994 17.3179143661441\\0.0938994350022863 0.898544104017325\\0.326320189490954 0.0327640511697619\\1.29959258652913 0.00151958860714805\\6.81741902600636 9.21277503807451e-05\\17.2572616208891 3.89287344591294e-06\\23.9782643744808 9.74293456685807e-08\\23.9782646432396 2.24718492144968e-08\\23.9782643959159 1.09527059870873e-08\\};
\addlegendentry{$\crpd$}
\addplot [filter]
table [row sep=\\] {%
28.3601511866137 1e-08\\28.3601511866137 1000\\};
\addlegendentry{$\mles$}
\end{axis}

\end{tikzpicture}
}
	\subfloat[Bijective vertex (perturbated degree-proportional arrivals)]{ % figures/er-bijective
\begin{tikzpicture}

\definecolor{crimson2143940}{RGB}{214,39,40}
\definecolor{darkgray176}{RGB}{176,176,176}
\definecolor{darkorange25512714}{RGB}{255,127,14}
\definecolor{forestgreen4416044}{RGB}{44,160,44}
\definecolor{lightgray204}{RGB}{204,204,204}
\definecolor{mediumpurple148103189}{RGB}{148,103,189}
\definecolor{sienna1408675}{RGB}{140,86,75}
\definecolor{steelblue31119180}{RGB}{31,119,180}

\begin{axis}[
legend cell align={left},
legend style={
  fill opacity=0.8,
  draw opacity=1,
  text opacity=1,
  at={(0.03,0.03)},
  anchor=south west,
  draw=lightgray204
},
log basis x={10},
log basis y={10},
tick align=outside,
tick pos=left,
x grid style={darkgray176},
xlabel={Delay},
xmin=0.00112269186027213, xmax=2.57192910908401,
xmode=log,
xtick style={color=black},
xtick={0.0001,0.001,0.01,0.1,1,10,100},
xticklabels={
  $\mathdefault{10^{-4}}$,
  $\mathdefault{10^{-3}}$,
  $\mathdefault{10^{-2}}$,
  $\mathdefault{10^{-1}}$,
  $\mathdefault{10^{0}}$,
  $\mathdefault{10^{1}}$,
  $\mathdefault{10^{2}}$
},
y grid style={darkgray176},
ylabel={Regret},
ymin=1e-08, ymax=800,
ymode=log,
ytick style={color=black},
ytick={1e-10,1e-08,1e-06,0.0001,0.01,1,100,10000,1000000},
yticklabels={
  $\mathdefault{10^{-10}}$,
  $\mathdefault{10^{-8}}$,
  $\mathdefault{10^{-6}}$,
  $\mathdefault{10^{-4}}$,
  $\mathdefault{10^{-2}}$,
  $\mathdefault{10^{0}}$,
  $\mathdefault{10^{2}}$,
  $\mathdefault{10^{4}}$,
  $\mathdefault{10^{6}}$
}
]
\addplot [egpd]
table [row sep=\\] {%
1.6424859832824 0.0528968934535485\\0.603374404928199 0.277311120101074\\0.221195060032667 1.0474906185767\\0.0813483448512995 3.35055854185985\\0.0333885462974074 10.2673653982939\\0.0169209871506942 30.9458608190969\\0.0122035600248631 76.3965036108009\\0.00495273568033417 384.283290818816\\0.00495273568033417 384.283290818816\\0.00495273568033417 384.283290818816\\};
\addlegendentry{$\egpd$}
\addplot [efilter]
table [row sep=\\] {%
0.419591745610866 0.958006083030763\\0.294973921920531 2.07802147405961\\0.192795781899839 4.49694414778718\\0.120048760460054 9.67155827536697\\0.0741345960689507 20.6921741791764\\0.0450655640130217 44.282645929959\\0.0258776191946788 94.8481424565198\\0.013313359682598 202.734981803897\\0.00380947636163177 401.92197403832\\0.00159583422472879 463.648636047954\\};
\addlegendentry{$\mle$}
\addplot [kfilter]
table [row sep=\\] {%
0.00159585702105271 463.650681759066\\0.00833936866914433 211.975491435375\\0.0252791961883139 54.6660536822685\\0.0547240915396586 17.5199310980352\\0.107351967126214 6.03327399320548\\0.207927525562307 2.02697238630922\\0.374353432603103 0.634271179768274\\0.594966511410148 0.172830594220742\\0.82267489550757 0.0301255755908734\\0.934215423243086 0.00263496201558359\\0.95669718728528 6.26103784459785e-05\\0.956540699776886 6.16948510370149e-13\\0.956540699776886 6.16948510370149e-13\\0.956540699776886 6.16948510370149e-13\\};
\addlegendentry{$\mlk$}
\addplot [egpdp]
table [row sep=\\] {%
1.80938836329086 1.78243476603473e-13\\1.2190001032043 6.26308610151226e-13\\1.03196555769704 0.000234876216534514\\0.838946346194938 0.0267390755400924\\0.470451308508593 0.354050426273818\\0.202547481106546 2.34204290069281\\0.0828741868917434 9.22315057016394\\0.0359179953943422 31.7675338535152\\0.0199150697182514 83.874245383367\\0.0101507653387586 192.547854234959\\};
\addlegendentry{$\egpdp$}
\addplot [crpd]
table [row sep=\\] {%
0.0132772062463929 80.0978047160071\\0.0220318643652885 20.1673769480118\\0.0934849952629604 1.08756652201022\\0.36209480457156 0.0340878133620602\\0.634964302287174 0.000836310744523309\\0.945252842943014 1.77420890608009e-05\\0.96170987041005 1.04440775098956e-06\\0.962199740547303 1.51356092896437e-07\\0.962228177274284 2.68559951684166e-08\\0.962241615910745 9.84922354304642e-09\\};
\addlegendentry{$\crpd$}
\addplot [filter]
table [row sep=\\] {%
0.956540699776886 1e-08\\0.956540699776886 800\\};
\addlegendentry{$\mles$}
\end{axis}

\end{tikzpicture}
}
	\caption{Reaching a vertex on a larger graph ($n=100, m=1006$).\label{perf:large}}
\end{figure}

\newpage

 % appendix/ec6-extensions
\section[Supplementary material of Sections \ref*{subsec:greedy} and \ref*{sec:extensions}.]{Supplementary material of \fullref{subsec:greedy} and \fullref{sec:extensions}.}
\label{app:extensions}

\subsection[Convexity proofs.]{Convexity proofs for $\Lapol$ and $\Lagre$ (\Cref{prop:convexity,prop:greedypositive}).} \label{app:convexity-proof}

We consider the extended definition of matching policies, as discussed in \Cref{app:extended-definition}.
Consider a stabilizable matching problem $(G, \lambda)$ and let $\La_\Pol$ (resp.\ $\La_\Gre$) denote the set of vectors of matching rates achievable by stable policies (resp.\ by stable greedy policies) adapted to the compatibility graph~$G$.

\noindent \textbf{Convexity of $\La_\Pol$.}
Consider two (extended) policies~$\Phi_1$ and~$\Phi_2$
that stabilize the matching problem~$(G, \lambda)$.
The state-space, queue-size function, and empty state of $\Phi_1$ (resp.\ $\Phi_2$)
are denoted by~$\cS_1$, $| \cdot |_1$, and $\varnothing_1$
(resp.\ $\cS_2$, $| \cdot |_2$, and $\varnothing_2$).
Given $0 < \beta < 1$,
our goal is to build a matching policy~$\Phi_\beta$
that also stabilizes the matching problem~$(G, \lambda)$ and
satisfies $\mu(\Phi_\beta) = \beta \mu(\Phi_1) + (1 - \beta) \mu(\Phi_2)$.

Intuitively, the policy~$\Phi_\beta$ will
consistently follow the decisions of either $\Phi_1$ or $\Phi_2$
as long as the system is non-empty,
and switch between these two policies
each time the system becomes empty (therefore, at renewal times).
The probability of selecting each policy after visiting the empty state will be set to achieve the desired matching rate vector in the long run.
More formally, the policy~$\Phi_\beta$ will have the following characteristics:
\begin{itemize}
	\item State space
	$\cS_\beta = (\cS_1 \times \{\varnothing_2\}) \cup (\{\varnothing_1\} \times \cS_2)$.
	\item Queue-size function $| \cdot |_\beta$ defined by
	$|(s_1, s_2)|_\beta = |s_1|_1 + |s_2|_2$
	for each $(s_1, s_2) \in \cS_\beta$.
	\item Empty state: $\varnothing_\beta = (\varnothing_1, \varnothing_2)$.
\end{itemize}
For some $0 < \gamma_\beta < 1$ that will be specified later,
the matching policy~$\Phi_\beta$ is defined as follows:
\begin{align*}
	\Phi_\beta( (\varnothing_1, \varnothing_2), i, j, (s_1, \varnothing_2) )
	&= \gamma_\beta \Phi_1(\varnothing_1, i, j, s_1),
	&& s_1 \in \cS_1, \quad i \in \V, \quad j \in \V \cup \{\bot\}, \\
	\Phi_\beta( (\varnothing_1, \varnothing_2), i, j, (\varnothing_1, s_2) )
	&= (1 - \gamma_\beta) \Phi_2(\varnothing_2, i, j, s_2),
	&& s_2 \in \cS_2, \quad i \in \V, \quad j \in \V \cup \{\bot\}, \\
	\Phi_\beta( (s_1, \varnothing_2), i, j, (s_1^\prime, \varnothing_2) )
	&= \Phi_1(s_1, i, j, s_1^\prime),
	&& s_1 \in \cS_1 \setminus \{ \varnothing_1 \}, \quad s_1^\prime \in \cS_1, \\
	\Phi_\beta( (\varnothing_1, s_2), i, j, (\varnothing_1, s_2^\prime) )
	&= \Phi_2(s_2, i, j, s_2^\prime),
	&& s_2 \in \cS_2 \setminus \{ \varnothing_2 \}, \quad s_2^\prime \in \cS_2.
\end{align*}
The first two equations say that, when the system is empty,
we next apply $\Phi_1$ with probability $\gamma_\beta$ and $\Phi_2$ with probability $1 - \gamma_\beta$.
The third (resp.\ fourth) equation says that,
once we start applying policy~$\Phi_1$ (resp.\ $\Phi_2$),
we keep applying this policy until we re-visit the empty state.

The stability of the policies~$\Phi_1$ and $\Phi_2$
implies that of the policy~$\Phi_\beta$.
According to the elementary renewal theorem for renewal reward processes,
the vector giving the long-run average matching rates under the policy~$\Phi_\beta$ is given by
\begin{align*}
	\mu(\Phi_\beta)
	&= \frac
	{\gamma_\beta T_1 \mu(\Phi_1) + (1 - \gamma_\beta) T_2 \mu(\Phi_2)}
	{\gamma_\beta T_1 + (1 - \gamma_\beta) T_2}
	= \frac{\gamma_\beta T_1}{\gamma_\beta T_1 + (1 - \gamma_\beta) T_2} \mu(\Phi_1)
	+ \frac{(1 - \gamma_\beta) T_2}{\gamma_\beta T_1 + (1 - \gamma_\beta) T_2} \mu(\Phi_2).
\end{align*}
where $T_1$ (resp.\ $T_2$) denotes the mean number of matches
between two successive visits to the empty state~$\varnothing_1$ (resp.\ $\varnothing_2$)
in the matching model $(G, \lambda, \Phi_1)$ (resp.\ $(G, \lambda, \Phi_2)$).
We let the reader verify that the following value of $\gamma_\beta$
yields $\mu(\Phi_\beta) = \beta \mu(\Phi_1) + (1 - \beta) \mu(\Phi_2)$:
\begin{align*}
	\gamma_\beta = \frac{\beta T_2}{(1 - \beta) T_1 + \beta T_2}.
\end{align*}

\noindent \textbf{Convexity of $\La_\Gre$.}
It suffices to observe that, if the policies $\Phi_1$ and $\Phi_2$ are greedy, so is the policy $\Phi_\beta$.

\subsection[Supplementary material of Section \ref*{subsec:greedy}.]{Supplementary material of \fullref{subsec:greedy}.}
\label{app:greedy}

\subsubsection[Proof of Proposition \ref*{prop:greedy-complete}.]{Proof of \Cref{prop:greedy-complete} (complete graphs) and discussion.}
\label{app:greedy-complete}

Consider a matching problem $(K_n, \lambda)$, where $K_n$ is the complete graph with $n \ge 3$ nodes.
According to \Cref{prop:stability-region-form},
this matching problem is stabilizable if and only if
$\lambda_i < \frac12 \sum_{j \in V} \lambda_j$ for each $i \in V$.
We prove \Cref{prop:greedy-complete} by first demonstrating that all greedy policies are ``equivalent'', in the sense that they always (deterministically) make the same matching decisions. 
The fact that we can formally define many greedy policies in this context
is merely an artifact of our extended definition of a matching policy
in \refapp{app:extended-definition} (see in particular \refapp{sec:equivalent_policies}).

Notice that the only independent sets of a complete graph are the singletons, hence by~\eqref{eq:Q-greedy} the state space of the queue-size process under greedy policies is
\begin{align} \label{eq:Qgre-Kn}
	\Q_\Gre(K_n) = \{0\} \cup
	\left( \bigcup_{i \in \V} \{ \ell \one_i : \ell \in \Np \} \right),
\end{align}
where $\one_i$ is the $n$-dimensional vector
with one in coordinate~$i$ and zero elsewhere,
for each $i \in \V$.
This observation implies that greediness leaves no freedom in the matching decisions: all unmatched items belong to the same class, and an incoming item must be matched with one of them if its class differs.
This is formalized in \Cref{prop:Kn-greedy-policy}.

\begin{prop} \label{prop:Kn-greedy-policy}
	Consider the complete graph $K_n$ with $n \ge 3$ nodes.
	\begin{enumerate}[(i)]
		\item \label{prop:Kn-greedy-policy-1}
		There is a unique size-based greedy policy 
		that is adapted to the compatibility graph~$K_n$.
		This policy, called the \emph{natural} greedy policy
		and denoted by $\Phi_\Gre(K_n)$,
		is the deterministic size-based policy 
		defined on~$\Q_\Gre(K_n) \times \V$ by
		\begin{align} \label{eq:Kn-greedy-Phi}
			\Phi_\Gre(K_n)(q, i) = \begin{cases}
				j &\text{if $q_j \ge 1$ with $j \neq i$,} \\
				\bot &\text{otherwise.}
			\end{cases}
		\end{align}
		\item \label{prop:Kn-greedy-policy-2}
		Consider a greedy policy~$\Phi$ adapted to the graph~$K_n$,
		and let $(\cS, | \cdot |)$ denote its state space. 
		The policy~$\Phi$ makes the same decisions as
		the natural greedy policy $\Phi_\Gre$
		in the sense that
		\begin{align*}
			\sum_{s^\prime \in \cS}
			\Phi(s, i, \Phi_\Gre(K_n)(|s|, i), s^\prime)
			= 1,
			\quad s \in \cS,
			\quad i \in \V.
		\end{align*}
		In other words, all greedy policies are equivalent in the sense that they can all be reduced to $\Phi_\Gre(K_n)$ (see ``Equivalent policies'' in \refapp{app:extended-definition}).
		\item \label{prop:Kn-greedy-policy-3}
		For each greedy policy~$\Phi$ adapted to the graph~$K_n$
		and each sequence~$I = (I_t, t \in \N)$ of item classes,
		we have $(Q_{\Gre})_t = Q_t$ for each $t \in \N$,
		where $Q_{\Gre}$ and $Q$ are the queue-size processes
		of the models $(K_n, I, \Phi_\Gre(K_n))$ and $(K_n, I, \Phi)$,
		respectively.
	\end{enumerate}
\end{prop}

\begin{proof}
We prove each statement one by one.

\noindent \textbf{\Cref{prop:Kn-greedy-policy}\ref{prop:Kn-greedy-policy-1}.}
Consider a (possibly random) size-based greedy policy
$\Phi: \Q_\Gre(K_n) \times V \times (\V \cup \{\bot\}) \to [0, 1]$,
as defined in \Cref{sec:extended-definition},
where $\Q_\Gre(K_n)$ is given in~\eqref{eq:Qgre-Kn}.
Given $(q, i) \in \Q_\Gre(K_n) \times \V$,
the definition of $\Q_\Gre(K_n)$ implies that
$\{j \in \V_i: q_j \ge 1\}$
is either a singleton or the empty set.
In the former case, letting~$j$ denote the unique element of the singleton,
we have $\Phi(q, i, \bot) = 1$ if $i = j$,
while the greediness of~$\Phi$ implies that
$\Phi(q, i, j) = 1$ if $i\neq j$.
In the latter case, we have directly $\Phi(q, i, \bot) = 1$.
In all cases, the matching decision is deterministic
and makes the same decisions as in~\eqref{eq:Kn-greedy-Phi}.

\noindent \textbf{\Cref{prop:Kn-greedy-policy}\ref{prop:Kn-greedy-policy-2}.}
The same argument can be repeated for an arbitrary
greedy policy~$\Phi$ with state space $(\cS, | \cdot |)$.
Given $(s, i) \in \cS \times \V$,
we know that $|s| \in \Q_\Gre(K_n)$,
so that $\{j \in \V_i: |s|_j \ge 1\}$
is either a singleton or the empty set.
In the former case, letting $j$ denote the unique element of the singleton,
we have $\sum_{s^\prime \in \cS} \Phi(s, i, \bot, s^\prime) = 1$ if $i = j$,
while the greediness of~$\Phi$ implies that 
$\sum_{s^\prime \in \cS} \Phi(s, i, j, s^\prime) = 1$  if $i\neq j$.
In the latter case, we have directly $\sum_{s^\prime \in \cS} \Phi(s, i, \bot, s^\prime) = 1$.
In all cases, we have $\sum_{s^\prime \in \cS} \Phi(s, i, \Phi_\Gre(|s|, i), s^\prime) = 1$.

\noindent \textbf{\Cref{prop:Kn-greedy-policy}\ref{prop:Kn-greedy-policy-3}.}
This statement follows directly from
\Cref{prop:Kn-greedy-policy}\ref{prop:Kn-greedy-policy-2}
and from the fact that,
under any matching policy,
the dynamics of the queue-size process
is fully specified by the sequence of incoming item classes
and the matching decisions via~\eqref{eq:Q-rec}.
\end{proof}

\Cref{prop:Kn-greedy-rate} below, which encompasses \Cref{prop:greedy-complete}, builds on this equivalence to
conclude that the set $\La_\Gre$ collapses to a single point
and directly express the unique vector of matching rates achieved by (stable) greedy policies.
In contrast, the polytope $\Lann$ exhibits a dimension $d = m - n = \frac{n(n-3)}{2}$, where $n=|V|$ is the number of classes and $m = |E| = \binom{n}{2} = \frac{n(n-1)}{2}$ is the number of possible matches.
In particular, for $n \geq 4$, the dimension $d = m - n$
of $\La_\Gre$ is at least 2.
Therefore, limiting ourselves to greedy policies dramatically narrows the range of achievable solutions on complete graphs larger than a triangle.

\begin{prop}
	\label{prop:Kn-greedy-rate}
	Consider a stabilizable matching problem $(K_n, \lambda)$,
	where $K_n$ is the complete graph with $n \ge 3$ nodes,
	and let~$\Phi$ denote a greedy policy adapted to~$K_n$.
	Then:
	\begin{enumerate}[(i)]
		\item \label{prop:Kn-greedy-rate-1}
		The model $(K_n, \lambda, \Phi)$ is stable.
		\item \label{prop:Kn-greedy-rate-2}
		The matching rate vector $\mu_\Gre$ in this model satisfies
		\begin{equation}
			\label{eq:kn-greedy-rate}
			(\mu_{\Gre})_k
			=\lambda_{i}p_j+\lambda_{j}p_i,
			\quad e_k = \{i, j\} \in \E,
		\end{equation}
		where, for each $i \in V$, $p_i$ is the stationary probability
		that queue~$i$ is non-empty, given~by
		\begin{align} \label{eq:kn-greedy-probability}
			p_i = \frac{\lambda_i}{(\sum_{j \neq i} \lambda_j) - \lambda_i} p_\varnothing,
		\end{align}
		and $p_\varnothing$ is the stationary probability
		that the system is empty, given by
		\begin{align} \label{eq:kn-greedy-empty}
			p_\varnothing
			&= \left( 1 + \sum_{i \in V} \frac{\lambda_i}{(\sum_{j \neq i} \lambda_j) - \lambda_i} \right)^{-1}.
		\end{align}
	\end{enumerate}
	In particular, we have $\La_\Gre = \{\mu_\Gre\} \subsetneq \La_{>0}$
	whenever $n \ge 4$.
\end{prop}

\begin{proof}
We prove each statement one by one.

\noindent \textbf{\Cref{prop:Kn-greedy-rate}\ref{prop:Kn-greedy-rate-1}.}
\Cref{prop:Kn-greedy-rate}\ref{prop:Kn-greedy-rate-1}
is a consequence of \Cref{cor:max-greedy}\ref{cond:max-greedy-1} (\refapp{app:minimal}).

\noindent \textbf{\Cref{prop:Kn-greedy-rate}\ref{prop:Kn-greedy-rate-2}.}
We can focus without loss of generality
on the \gls{fcfm} policy,
as \Cref{prop:Kn-greedy-policy}\ref{prop:Kn-greedy-policy-3}
implies that all greedy matching policies
yield the same vector of matching rates.
Equation~\eqref{eq:kn-greedy-rate}
follows by observing that,
for each edge $e_k = \{i, j\} \in E$,
a match between classes~$i$ and~$j$ happens if:
\begin{itemize}
	\item a class-$i$ item arrives while queue~$j$ is non-empty,
	which happens at rate $\lambda_{i}p_j$;
	\item a class-$j$ item arrives while queue~$i$ is non-empty,
	which happens at rate $\lambda_{j}p_i$.
\end{itemize}
Equations~\eqref{eq:kn-greedy-probability}
and~\eqref{eq:kn-greedy-empty}
follow from \citet[Proposition~5]{C22}.
Indeed, for each $i \in V$, applying \citet[Equation~(10)]{C22}
to the independent set~$\{i\}$ yields~\eqref{eq:kn-greedy-probability},
and the value of $p_\varnothing$ given in~\eqref{eq:kn-greedy-empty}
follows from the normalizing equation.
This result may also be obtained more directly
by observing that, for each $i \in V$,
the restriction of the transition diagram
of the Markov chain $(K_n, \lambda, \Phi_\Gre)$
to the states where all queues but queue~$i$ are empty
is similar to a (discrete-time) birth-and-death process
with birth probability $\lambda_i$
and death probability $\sum_{j \neq i} \lambda_j$.
\end{proof}

\Cref{fig:k4greedy} illustrates this result on a complete graph $\K_4$ in which all arrival rates are equal to~3. The particular solution $\mu^\circ$ and the basis $\{b_1, b_2\}$ of $\ker(A)$ that we use are shown on \Cref{fig:k4greedy-a}. In the corresponding kernel basis, the polytope $\Lann$ is defined by the inequalities $\alpha_1\leq 1$, $\alpha_2\leq 1$, and $\alpha_1+\alpha_2 \geq -1$, that is, it is the triangle with vertices $(-2, 1)$, $(1, -2)$, and $(1, 1)$. Yet, \Cref{prop:Kn-greedy-rate} shows that only the solution $\alpha = (0, 0)$ can be achieved by a greedy policy.

\tikzset{
	dot/.style = {circle, fill=orange, minimum size=#1,
		inner sep=0pt, outer sep=0pt},
	dot/.default = 4pt 
}

\begin{figure}
	\centering
	\subfloat[\label{fig:k4greedy-a}Generic solution to~\eqref{eq:system}.]{
		\begin{tikzpicture}		
			\def\d{2.5cm}
			\node[class] (1) {$1$};
			\foreach \i/\s/\a in {2/1/90, 3/2/0, 4/3/-90}		
			\path (\s) ++(\a:\d) node[class] (\i) {$\i$};
			
			\draw (1) edge node[above, sloped] {$1-\alpha_1$} (2) 
			(3) edge node[above, sloped] {$1-\alpha_1$} (4) 
			(2) edge node[above] {$1+\alpha_1+\alpha_2$} (3)
			(1) edge node[below] {$1+\alpha_1+\alpha_2$} (4)
			(2) edge node[sloped, above, pos=.25] {$1-\alpha_2$} (4)
			(1) edge node[sloped, above, pos=.25] {$1-\alpha_2$} (3);
			
		\end{tikzpicture}
	}
	\qquad
	\subfloat[The polytope~$\Lann$ in kernel basis. The set~$\La_\Gre$ is reduced to the
	origin vector~$\alpha = (0, 0)$.]{
		\begin{tikzpicture}[scale=.8]
			
			\node at (-3, 0) {};
			\node at (3, 0) {};
			
			\draw[->] (-2,0) -- (1.5,0);
			\node[anchor=west] at (1.5,0) {$\alpha_1$};
			\draw[->] (0,-2) -- (0,1.3);
			\node[anchor=south] at (0,1.3) {$\alpha_2$};
			
			\node (v1) at (-2,1) {};
			\node (v2) at (1,-2) {};
			\node (v3) at (1,1) {};
			\node (v4) at (0, 0) {};
			
			\path[draw=blue, very thick, fill=blue, fill opacity=.2]
			(v1.center) -- (v2.center) -- (v3.center) -- cycle;
			
			\node[anchor=east, text=orange]
			at (v1) {$(-2, 1)$}; 
			\node[anchor=west, text=orange]
			at (v3) {$(1, 1)$}; 
			\node[anchor=west, text=orange]
			at (v2) {$(1, -2)$}; 
			\node[anchor=south east, text=red]
			at (v4) {$(0, 0)$}; 
			
			\foreach \i in {1,...,3}
			\node[dot] at (v\i) {};
			\node[dot, red] at (v4) {};
			
			\node at (-4, 0) {};
			\node at (4, 0) {};
		\end{tikzpicture}
	}
	\caption{\label{fig:k4greedy}Matching problem $(K_4, \lambda)$ with $\lambda=(3, 3, 3, 3)$.}	
\end{figure}

\subsubsection[Proof of Proposition \ref*{prop:greedy-diamond}.]{Proof of \Cref{prop:greedy-diamond} (diamond graph), discussion, and numerical results.}
\label{app:greedy-diamond}

Consider the diamond (double fan) graph
introduced in \Cref{ex:diamond},
for which we will show that the set~$\La_\Gre$
of matching rate vectors achieved by stable greedy policies
is a strict subset of the set~$\La_{>0}$
of positive solution to~\eqref{eq:system}.
More specifically,
we consider the diamond matching problem~$(D, \lambda)$ defined as follows:
\begin{align} \label{eq:diamond}
	\begin{aligned}
		D &= (V, E)
		\text{ with } V = \{1, 2, 3, 4\}
		\text{ and } E = \{\{1,2\}, \{1,3\}, \{2,3\}, \{2,4\}, \{3,4\}\}, \\
		\lambda &=
		\begin{aligned}[t]
			&(\lambda_1, \bar\lambda_2 + \beta, \bar\lambda_3 + \beta, \lambda_4) \\
			&\text{ with }
			\lambda_1 > 0, \bar\lambda_2 > 0, \bar\lambda_3 > 0, \lambda_4 > 0, \beta > 0,
			\textstyle \lambda_1 + \lambda_4 = \bar\lambda_2 + \bar\lambda_3 = \frac12,
		\end{aligned} \\
		\B &= \{ (1, -1, 0, -1, 1) \}, \\
		\mu^\circ &= (2 \lambda_1 \bar\lambda_2, 2 \lambda_1 \bar\lambda_3, \beta,
		2 \bar\lambda_2 \lambda_4, 2 \bar\lambda_3 \lambda_4), \\
		\mu &= (2 \lambda_1 \bar\lambda_2 + \alpha, 2 \lambda_1 \bar\lambda_3 - \alpha, \beta,
		2 \bar\lambda_2 \lambda_4 - \alpha, 2 \bar\lambda_3 \lambda_4 + \alpha),
		\quad \alpha \in \R.
	\end{aligned}
\end{align}
Parameterizing $\lambda$ by $\beta$ ($= \mu_{2, 3}$
for each $\mu \in \La$)
is a notational convenience that
does not lead to any loss of generality:
as already observed in \Cref{ex:diamond},
the stabilizability condition~\ref{cond:stability-region-form-2} writes
$\bar\lambda_2 > 0$, $\bar\lambda_3 > 0$, and $\beta > 0$.
We leave it to the reader to verify
that our choices for $\B$ and $\mu^\circ$ are correct.
Equation~\eqref{eq:diamond} implies in particular that
the sets $\Lann$ and $\Lap$, are real intervals, and so are $\La_\Pol$ and $\La_\Gre$ by convexity.

As in \Cref{app:greedy-complete}, we first provide an equivalence result that will be useful to describe~$\La_\Gre$ in details.

\paragraph*{Equivalence of greedy policies.}

As in \Cref{app:greedy-complete}, we first identify a common behavior shared by all greedy policies and then we exploit this behavior to characterize the matching rates achievable by stable greedy policies.

Since $\{1, 4\}$ is the only independent set of the diamond graph~$D$
that is not a singleton, the possible queue states under any greedy policy can be partitioned as follows: either all queues are empty,
or exactly one class has a non-empty queue,
or (only) classes 1 and 4 have non-empty queues.
In other words,
the state space of the queue-size process under greedy policies adapted to the graph~$D$ is
\begin{align} \label{eq:Qgre-D}
	\Q_\Gre(D) &=
	\{0\} \cup
	\left( \bigcup_{i \in \V} \{ \ell \one_i, \ell \in \N_{> 0} \} \right) \cup
	\left\{ \ell_1 \one_1 + \ell_4 \one_4 : \ell_1, \ell_4 \in \Np \right\}.
\end{align}
Therefore, greediness entirely determines the decisions made by greedy policies,
except if an item of class~2 or~3 enters
while there are unmatched items of classes~1 and~4.
Using the fact that classes~2 and~3 are both compatible with classes~1 and~4,
we prove formally in \Cref{prop:D-greedy-policy}
that all greedy policies adapted to the diamond graph
make the same matching decisions as
the natural greedy policy adapted to the complete graph~$K_3$
obtained by ``merging'' classes~1 and~4 in the diamond graph.

\begin{prop} \label{prop:D-greedy-policy}
	Given the diamond graph~$D$,
	we introduce the following notation:
	\begin{itemize}
		\item Queue-size projection:
		For each $q = (q_1, q_2, q_3, q_4) \in \Q_\Gre(D)$,
		we let $\langle q \rangle = (q_1 + q_4, q_2, q_3)$.
		\item Class projection: We let
		$\langle i \rangle = i$ for each $i \in \{1, 2, 3, \bot\}$
		and $\langle 4 \rangle = 1$.
	\end{itemize}
	Every greedy policy~$\Phi$ adapted
	to the diamond graph~$D$ satisfies the following properties:
	\begin{enumerate}[(i)]
		\item \label{prop:D-greedy-policy-1}
		If $\Phi$ is a deterministic size-based policy, then
		\begin{align*}
			\langle \Phi(q, i) \rangle
			= \Phi_\Gre(K_3)(\langle q \rangle, \langle i \rangle),
			\quad (q, i) \in \Q_\Gre(D) \times \V.
		\end{align*}
		where $\Phi_\Gre(K_3)$ is the natural greedy policy~\eqref{eq:Kn-greedy-Phi}
		adapted to the complete graph $K_3$.
		\item \label{prop:D-greedy-policy-2}
		In general, if $\Phi$ is a policy with state space $(\cS, | \cdot |)$, then
		\begin{align*}
			\sum_{s^\prime \in \cS \phantom{\langle}}
			\;
			\sum_{\substack{j \in \V \cup \{\bot\}: \\
					\langle j \rangle
					= \Phi_\Gre(K_3)(\langle | s | \rangle, \langle i \rangle)}}
			\Phi(s, i, j, s^\prime)
			= 1,
			\quad s \in \cS,
			\quad i \in \V.
		\end{align*}
		\item \label{prop:D-greedy-policy-3}
		For each sequence $I$ of item classes,
		we have $\langle Q_t \rangle = (Q_\Gre)_t$ for each $t \in \N$,
		where $Q$ and $Q_\Gre$ are the queue-size processes of the models $(D, I, \Phi)$
		and $(K_3, \langle I \rangle, \Phi_\Gre)$, respectively,
		with $\langle I \rangle = (\langle I_t \rangle, t \in \N)$.
	\end{enumerate}
\end{prop}

\noindent Loosely speaking, the main take-away of
\Cref{prop:D-greedy-policy}\ref{prop:D-greedy-policy-3} is that,
in the matching model $(D, I, \Phi)$,
the process
$((Q_{t, 1} + Q_{t, 4}, Q_{t, 2}, Q_{t, 3}), t \in \N)$
does not depend on the specific greedy policy~$\Phi$ that is applied,
and it is in fact equal to the queue size process
built by applying the natural greedy policy~$\Phi_\Gre(K_3)$
in the complete graph~$K_3$
obtained by merging classes~1 and 4 in the diamond graph~$D$.

\begin{proof}[Proof of \Cref{prop:D-greedy-policy}]
We prove each statement one by one.

\noindent \textbf{\Cref{prop:D-greedy-policy}\ref{prop:D-greedy-policy-1}.}
We leave it to the reader to verify that
$\Q_\Gre(K_3)$ is the image of $\Q_\Gre(D)$
by the application $q = (q_1, q_2, q_3, q_4)
\mapsto \langle q \rangle = (q_1 + q_4, q_2, q_3)$.
This means in particular that $q_1 + q_4$, $q_2$, and $q_3$
cannot be positive simultaneously if $q \in \Q_\Gre(D)$.
Since the diamond graph~$D$ has only four nodes,
we can then conclude by enumerating all relevant cases,
depending on the support of~$\langle q \rangle = (q_1 + q_4, q_2, q_3)$.
For example, if $q_1 + q_4 \ge 1$ and $q_2 = q_3 = 0$,
then we have immediately
$\Phi(q, i) = \bot$ if $i \in \{1, 4\}$,
and the greediness of the policy~$\Phi$ implies that
$\Phi(q, i) \in \{1, 4\}$ if $i \in \{2, 3\}$;
in other words,
we have $\langle \Phi(q, i) \rangle = \bot$ if $\langle i \rangle = 1$
and $\langle \Phi(q, i) \rangle = 1$ if $\langle i \rangle \in \{2, 3\}$.
Similarly, if $q_1 + q_4 = q_3 = 0$ and $q_2 \ge 1$,
then $\Phi(q, i) = \bot$ if $i = 2$
and $\Phi(q, i) = 2$ if $i \in \{1, 3, 4\}$,
that is,
$\langle \Phi(q, i) \rangle = \bot$ if $\langle i \rangle = 2$
and $\langle \Phi(q, i) \rangle = 2$ if $\langle i \rangle \in \{1, 3\}$.
In all cases, we can verify that
$\langle \Phi(q, i) \rangle$ is equal to
$\Phi_\Gre(K_3)(\langle q \rangle, \langle i \rangle)$.

Alternatively, by taking a step back,
we can prove \Cref{prop:D-greedy-policy}\ref{prop:D-greedy-policy-1}
more directly by observing that,
for each $(q, i) \in \Q_\Gre(D) \times \V$,
(i)~the support of $\langle q \rangle$
is a singleton $\{j\}$ whenever $q \neq 0$,
and (ii)~whether~$j \in V_i$
depends on~$i$ only via~$\langle i \rangle$.

\noindent \textbf{\Cref{prop:D-greedy-policy}\ref{prop:D-greedy-policy-2}.}
The same argument can be repeated
for an arbitrary greedy policy~$\Phi$
with state space $(\cS, |\cdot|)$.
As in the proof of \Cref{prop:Kn-greedy-policy}\ref{prop:Kn-greedy-policy-2},
the key argument consists of observing that
we still have $q = |s| \in \Q_\Gre(D)$
for each $(s, i) \in \cS \times \V$,
so that $q_1 + q_4$, $q_2$, and $q_3$
cannot be positive simultaneously.

\noindent \textbf{\Cref{prop:D-greedy-policy}\ref{prop:D-greedy-policy-3}.}
The conclusion follows in much the same way as in the proof of
\Cref{prop:Kn-greedy-policy}\ref{prop:Kn-greedy-policy-3}, by injecting statements~\ref{prop:D-greedy-policy-1} and~\ref{prop:D-greedy-policy-2} from \Cref{prop:D-greedy-policy} into~\eqref{eq:Q-rec}.
\end{proof}

\paragraph*{Matching rates under stable greedy policies.}

\Cref{prop:D-greedy-edge,prop:D-greedy-kernel}, which encompass \Cref{prop:greedy-diamond},
use the equivalence result of \Cref{prop:D-greedy-policy}
to characterize the matching rate vector
under stable greedy policies.
\Cref{prop:D-greedy-edge} bounds the coordinates of the matching-rate vectors achievable by a stable greedy policy, while \Cref{prop:D-greedy-kernel} details the bounds $\alpha_+$ and $\alpha_-$ of $\La_\Gre$. 
These propositions illustrate the complementarity
of edge and kernel coordinates:
the results of \Cref{prop:D-greedy-edge}
are easier to state using edge coordinates,
while those of \Cref{prop:D-greedy-kernel}
are easier to state using kernel coordinates.
The lower bound~\eqref{eq:diamond-greedy-ineq-2}
is obtained by following a similar approach
as in \Cref{prop:Kn-greedy-rate}\ref{prop:Kn-greedy-rate-2}.

\begin{prop}[Edge coordinates] \label{prop:D-greedy-edge}
	Consider the matching model~$(D, \lambda, \Phi)$,
	where $(D, \lambda)$ is the diamond problem~\eqref{eq:diamond}
	and $\Phi$ is a greedy policy adapted to the graph~$D$.
	\begin{enumerate}[(i)]
		\item \label{prop:D-greedy-edge-1}
		This matching model is stable.
		\item \label{prop:D-greedy-edge-2}
		The matching rate vector $\mu = \mu(D, \lambda, \Phi)$ satisfies
		$\mu_{2,3} = \beta = \frac12(\lambda_2 + \lambda_3 - \lambda_1 - \lambda_4)$, and
		\begin{align} \label{eq:diamond-greedy-ineq-2}
			\begin{aligned}
				\mu_{1, 2} &\ge \lambda_1 p_2 + \lambda_2 p_1, &
				\mu_{1, 3} &\ge \lambda_1 p_3 + \lambda_3 p_1, &
				\mu_{2, 4} &\ge \lambda_2 p_4 + \lambda_4 p_2, &
				\mu_{3, 4} &\ge \lambda_3 p_4 + \lambda_4 p_3,
			\end{aligned}
		\end{align}
		where $p_i$ is the stationary probability that
		the system contains unmatched items that belong exclusively to class~$i$,
		for each $i \in \V$.
		\item \label{prop:D-greedy-edge-3}
		Let $p_{1, 4}$ denote the stationary probability that
		the system contains unmatched items that belong to class~1 or~4 (or both).
		We have
		\begin{align} \label{eq:diamond-greedy-eq}
			\begin{aligned}
				p_2 &= \frac{\lambda_2}{\lambda_1 + \lambda_3 + \lambda_4 - \lambda_2} p_\varnothing,
				&
				p_3 &= \frac{\lambda_3}{\lambda_1 + \lambda_2 + \lambda_4 - \lambda_3} p_\varnothing,
				&
				p_{1, 4} &= \frac{\lambda_1 + \lambda_4}{\lambda_2 + \lambda_3 - \lambda_1 - \lambda_4} p_\varnothing,
			\end{aligned}
		\end{align}
		where $p_\varnothing$ is the stationary probability
		that the system is empty,
		whose value follows from the normalization equation
		$p_\varnothing + p_{1, 4} + p_2 + p_3 = 1$.
		We also have
		\begin{align} \label{eq:diamond-greedy-ineq-1}
			p_1 &> \frac{\lambda_1}{\lambda_2 + \lambda_3 + \lambda_4} p_\varnothing,
			&
			p_4 &> \frac{\lambda_4}{\lambda_1 + \lambda_2 + \lambda_3} p_\varnothing.
		\end{align}
	\end{enumerate}
\end{prop}

\begin{proof}[Proof of \Cref{prop:D-greedy-edge}]
We prove each statement one after another.

\noindent \textbf{\Cref{prop:D-greedy-edge}\ref{prop:D-greedy-edge-1}.}
This is a consequence of \Cref{cor:max-greedy}\ref{cond:max-greedy-2} (\refapp{app:minimal}).

\noindent \textbf{\Cref{prop:D-greedy-edge}\ref{prop:D-greedy-edge-2}.}
The equation
$\mu_{2,3} = \beta = \frac12(\lambda_2 + \lambda_3 - \lambda_1 - \lambda_4)$
is a direct consequence of~\eqref{eq:system}.
The inequalities~\eqref{eq:diamond-greedy-ineq-2}
for $\mu_{1,2}$, $\mu_{1,3}$,
$\mu_{2,4}$, and $\mu_{3,4}$
follow by observing that,
for each edge $\{i, j\} \in \{\{1,2\}, \{1,4\}, \{2,3\}, \{3,4\}\}$,
a match between classes~$i$ and~$j$
happens at least in one of the following cases:
\begin{itemize}
	\item a class-$i$ item arrives
	while the system contains unmatched items
	that all belong to class~$j$,
	\item a class-$j$ item arrives
	while the system contains unmatched items
	that all belong to class~$i$,
\end{itemize}
These events occur at rates
$\lambda_i p_j$ and $\lambda_j p_i$,
respectively.
Equations~\eqref{eq:diamond-greedy-ineq-2} are not equalities in general because the above list is not exhaustive. For example, depending on the greedy policy, a match between classes~$1$ and~$2$ may happen if a class-$2$ item arrives while the system contains unmatched items of class~$1$ and unmatched items of class~$4$.

\noindent \textbf{\Cref{prop:D-greedy-edge}\ref{prop:D-greedy-edge-3}.}
The expressions for $p_\varnothing$, $p_2$, $p_3$, and $p_{1,4}$
follow directly by combining
\Cref{prop:D-greedy-policy}\ref{prop:D-greedy-policy-3}
with \Cref{eq:kn-greedy-probability,eq:kn-greedy-empty}
in \Cref{prop:Kn-greedy-rate}.
We now derive the lower bound~\eqref{eq:diamond-greedy-ineq-1} for~$p_1$.
The one for $p_4$ follows by symmetry.

First assume that the greedy policy~$\Phi$ is
deterministic and size-based,
so that it satisfies
\Cref{prop:D-greedy-policy}\ref{prop:D-greedy-policy-1}.
For each $q \in \Q_\Gre(D)$,
let $\pi_q$ denote the probability that the Markov chain $(D, \lambda, \Phi)$
is in state~$q$ in stationary regime.
The probability that we want to lower-bound is
\begin{align} \label{eq:pi}
	p_1 = \sum_{\ell = 1}^{+\infty} \pi_{\ell \one_1}.
\end{align}
Now let $\ell \in \N_{> 0}$ and
consider the balance equation for state $\ell \one_1$, given by
\begin{align} \label{eq:pi-1}
	(\lambda_1 + \lambda_2 + \lambda_3+ \lambda_4) \pi_{\ell \one_1}
	&= \lambda_1 \pi_{(\ell - 1) \one_1}
	+ (\lambda_2 + \lambda_3) \pi_{(\ell + 1) \one_1}
	+ C,
\end{align}
where $C$ is a non-negative real
that depends on the model parameters, the integer~$\ell$, and the policy~$\Phi$,
and that accounts for the flow to state $\ell \one_1$ from
state $\ell \one_1 + \one_4$, if any.
It follows that
$(\lambda_1 + \lambda_2 + \lambda_3 + \lambda_4) \pi_{\ell \one_1}
> \lambda_1 \pi_{(\ell - 1) \one_1}$.
An inductive argument allows us to conclude that 
\begin{align} \label{eq:pi-2}
	\pi_{\ell \one_1}
	&> \left( \frac{\lambda_1}{\lambda_1 + \lambda_2 + \lambda_3 + \lambda_4} \right)^\ell
	p_\varnothing,
	\quad \ell \in \N_{> 0}.
\end{align}
Injecting this inequality into~\eqref{eq:pi} allows us to conclude:
\begin{align} \label{eq:pi-3}
	p_1
	&> \sum_{\ell = 1}^{+\infty}
	\left( \frac{\lambda_1}{\lambda_1 + \lambda_2 + \lambda_3 + \lambda_4} \right)^\ell p_\varnothing
	= \frac{\lambda_1}{\lambda_2 + \lambda_3 + \lambda_4} p_\varnothing.
\end{align}
If $\Phi$ is an extended policy with state space $(\cS, | \cdot |)$,
we can still write~\eqref{eq:pi}--\eqref{eq:pi-3}
and reach the same conclusion.
The only difference is that $\pi$ can no longer be defined
as the stationary distribution of a Markov chain:
instead, we define $\pi_q = \sum_{s \in \cS: |s| = q} \varpi_s$
for each $q \in \Q_\Gre(D)$,
where $\varpi_s$ is the probability that the Markov chain $(D, \lambda, \Phi)$
is in state~$s$ in stationary regime, for each $s \in \cS$.
\end{proof}

\begin{prop}[Kernel coordinates] \label{prop:D-greedy-kernel}
	Consider the diamond matching problem $(D, \lambda)$ of~\eqref{eq:diamond}.
	\begin{enumerate}[(i)]
		\item \label{prop:D-greedy-kernel-1}
		The intervals $\La_{\ge 0}$, $\La_{>0}$, and $\La_\Gre$
		are defined as follows in the kernel basis:
		\begin{align*}
			\La_{\ge 0} &= [ -2 \min(\lambda_1 \bar\lambda_2, \bar\lambda_3 \lambda_4),
			2 \min(\lambda_1 \bar\lambda_3, \bar\lambda_2 \lambda_4) ], \\
			\La_{>0} &= ( -2 \min(\lambda_1 \bar\lambda_2, \bar\lambda_3 \lambda_4),
			2 \min(\lambda_1 \bar\lambda_3, \bar\lambda_2 \lambda_4) ), \\
			\La_\Gre &= [\alpha_-, \alpha_+],
		\end{align*}
		with $-2 \min(\lambda_1 \bar\lambda_2, \bar\lambda_3 \lambda_4) < \alpha_-
		\le \alpha_+ < 2\min(\lambda_1 \bar\lambda_3, \bar\lambda_2 \lambda_4)$.
		Hence, $\La_\Gre \subsetneq \La_{>0} \subsetneq \La_{\ge 0}$.
		\item \label{prop:D-greedy-kernel-2}
		The coordinates $\alpha_+$ and $\alpha_-$ satisfy the following properties:
		\begin{enumerate}
			\item \label{prop:D-greedy-kernel-2-1}
			$\alpha_+ = \alpha(\Phi_+)$,
			where $\Phi_+$ is the \acrfull{hrf} policy adapted to the graph~$D$
			whereby edges $\{1, 2\}$ and $\{3, 4\}$ have the highest reward.
			\item \label{prop:D-greedy-kernel-2-2}
			$\alpha_- = \alpha(\Phi_-)$,
			where $\Phi_-$ is the \gls{hrf} policy adapted to the graph~$D$
			whereby edges $\{1, 3\}$ and $\{2, 4\}$ have the highest reward.
			\item \label{prop:D-greedy-kernel-2-3}
			If $\beta \to +\infty$ while $\lambda_1$, $\bar\lambda_2$, $\bar\lambda_3$, and $\lambda_4$ remain fixed,
			we have $\alpha_+ \to 0$ and $\alpha_- \to 0$.
		\end{enumerate}
	\end{enumerate}
\end{prop}

\begin{proof}[Proof of \Cref{prop:D-greedy-kernel}]
We prove each statement one after another.

\noindent \textbf{\Cref{prop:D-greedy-kernel}\ref{prop:D-greedy-kernel-1}--\ref{prop:D-greedy-kernel-2-1}--\ref{prop:D-greedy-kernel-2-2}.}
The diamond graph~$D$ has $n = 4$ nodes and $m = 5$ edges.
Therefore, according to \Cref{prop:affine-space,prop:greedypositive},
the sets~$\Lann$, $\Lap$, and $\La_\Gre$
have dimension $d = m - n = 1$, meaning that they are intervals in~$\R$.
The equations for the intervals $\La_{\ge 0}$ and $\La_{> 0}$
follow directly from the change-of-basis equation
$\mu = (2 \lambda_1 \bar\lambda_2 + \alpha, 2 \lambda_1 \bar\lambda_3 - \alpha, \beta,
2 \bar\lambda_2 \lambda_4 - \alpha, 2 \bar\lambda_3 \lambda_4 + \alpha)$
from~\eqref{eq:diamond}.
That $\La_\Gre$ is also an interval
is a consequence of its convexity (\Cref{prop:greedypositive}).
The (non-strict) inequality
$-2 \min(\lambda_1 \bar\lambda_2, \bar\lambda_3 \lambda_4) \le \alpha_-
\le \alpha_+ \le 2\min(\lambda_1 \bar\lambda_3, \bar\lambda_2 \lambda_4)$
is a consequence of \Cref{prop:greedypositive},
which states that $\La_\Gre \subseteq \La_{>0}$.
The first and third inequalities are also strict because
$\alpha_+$ and $\alpha_-$ belong to $\La_\Gre$ (see below),
while $2 \min(\lambda_1 \bar\lambda_3, \bar\lambda_2 \lambda_4)$
and $-2 \min(\lambda_1 \bar\lambda_2, \bar\lambda_3 \lambda_4)$
do not belong to $\La_{>0}$.
That $\La_\Gre$ is a closed interval
of the form $\La_\Gre = [\alpha_-, \alpha_+]$
and that $\alpha_+$ and $\alpha_-$
are as given in
\Cref{prop:D-greedy-kernel}\ref{prop:D-greedy-kernel-2-1}%
--\ref{prop:D-greedy-kernel-2-2}
are consequences of \Cref{lem:greedy-diamond} below,
which will be proved by a coupling argument
later in this appendix\footnote{\Cref{lem:greedy-diamond} focuses on \Cref{prop:D-greedy-kernel}\ref{prop:D-greedy-kernel-2-1}. The case \Cref{prop:D-greedy-kernel}\ref{prop:D-greedy-kernel-2-2} is trivially deduced by switching the class labels $2$ and $3$.}.

\begin{lem} \label{lem:greedy-diamond}
	Consider the \gls{hrf} policy~$\Phi_+$
	adapted to the diamond graph~$D$
	whereby edges $\{1, 2\}$ and $\{3, 4\}$
	have the highest reward.
	For each greedy policy~$\Phi$
	adapted to the compatibility graph~$D$,
	we have
	\begin{align} \label{eq:greedy-diamond}
		\begin{aligned}
			\mu_{1, 2}(\Phi)
			&\le \mu_{1, 2}(\Phi_+),
			&
			\mu_{3, 4}(\Phi)
			&\le \mu_{3, 4}(\Phi_+),
			\\
			\mu_{1, 3}(\Phi)
			&\ge \mu_{1, 3}(\Phi_+),
			&
			\mu_{2, 4}(\Phi)
			&\ge \mu_{2, 4}(\Phi_+),
		\end{aligned}
	\end{align}
	Equivalently, in kernel coordinates, we have
	$\alpha(\Phi) \le \alpha(\Phi_+)$.
\end{lem}

\noindent \textbf{\Cref{prop:D-greedy-kernel}\ref{prop:D-greedy-kernel-2-3}.}
Most of the quantities we consider in this proof
are functions of~$\beta$,
but this dependency is left implicit to simplify notation.
In particular, we let $\mu$ denote
(the edge coordinates of) the matching rate vector
in the matching model $(D, \lambda, \Phi_+)$,
where $\lambda = (\lambda_1, \bar\lambda_2 + \beta, \bar\lambda_3 + \beta, \lambda_4)$,
with $\lambda_1 + \lambda_4 = \bar\lambda_2 + \bar\lambda_3 = \frac12$,
and $\Phi_+$ is the greedy \gls{hrf} policy
defined in \Cref{lem:greedy-diamond}.
By combining \eqref{eq:diamond-greedy-ineq-2}
with the conservation equation
$\mu_{1,2} + \mu_{1,3} = \lambda_1$,
we obtain the following lower and upper bounds for $\mu_{1,2}$:
\begin{align}
	\label{eq:diamond-greedy-limit-proba-3}
	\lambda_1 p_2 + \lambda_2 p_1
	\le \mu_{1,2}
	\le \lambda_1 - (\lambda_1 p_3 + \lambda_3 p_1).
\end{align}
In addition, injecting the definition~\eqref{eq:diamond} of~$\lambda$
into \eqref{eq:diamond-greedy-eq}
and \eqref{eq:diamond-greedy-ineq-1}
shows that, in the model $(D, \lambda, \Phi_+)$, we have
\begin{align}
	\label{eq:diamond-greedy-limit-proba-1}
	\begin{aligned}
		p_2 &= p_\varnothing \frac{\bar\lambda_2 + \beta}{2 \bar\lambda_3}, &
		p_3 &= p_\varnothing \frac{\bar\lambda_3 + \beta}{2 \bar\lambda_2}, &
		p_{1,4} &= p_\varnothing \frac1{3 \beta}, \\
		p_1 &> p_\varnothing \frac{\lambda_1}{\frac12 + \lambda_4 + 2\beta}, &
		p_4 &> p_\varnothing \frac{\lambda_4}{\frac12 + \lambda_1 + 2\beta},
	\end{aligned}
\end{align}
with, by the normalization equation
$p_\varnothing + p_2 + p_3 + p_{1, 4} = 1$,
\begin{align}
	\label{eq:diamond-greedy-limit-proba-2}
	p_\varnothing &= \left(
	1 + \frac1{3 \beta}
	+ \frac{\bar\lambda_2 + \beta}{2 \bar\lambda_3}
	+ \frac{\bar\lambda_3 + \beta}{2 \bar\lambda_2}
	\right)^{-1}.
\end{align}
Taking the limit of~\eqref{eq:diamond-greedy-limit-proba-1}
and~\eqref{eq:diamond-greedy-limit-proba-2}
as $\beta \to +\infty$,
we conclude that both the lower-bound and the upper-bound
in~\eqref{eq:diamond-greedy-limit-proba-3}
tend to $2 \lambda_1 \bar\lambda_2$ as $\beta \to +\infty$.
Then combining~\eqref{eq:diamond-greedy-limit-proba-3}
with the squeeze theorem allows us to conclude that
$\mu_{1,2}$ also tends to $2 \lambda_1 \bar\lambda_2$ as $\beta \to +\infty$.
By symmetry, we obtain directly
$\mu_{1,3} \xrightarrow{} 2 \lambda_1 \bar\lambda_3$,
$\mu_{2,4} \xrightarrow{} 2 \bar\lambda_2 \lambda_4$,
and $\mu_{3,4} \xrightarrow{} 2 \bar\lambda_3 \lambda_4$
as $\beta \to +\infty$.
According to the change-of-basis equation
$\mu = (2 \lambda_1 \bar\lambda_2 + \alpha, 2 \lambda_1 \bar\lambda_3 - \alpha, \beta,
2 \bar\lambda_2 \lambda_4 - \alpha, 2 \bar\lambda_3 \lambda_4 + \alpha)$
from~\eqref{eq:diamond},
this means that $\alpha_+ \xrightarrow{} 0$ as $\beta \to +\infty$.
\end{proof}

\begin{proof}[Proof of \Cref{lem:greedy-diamond}]

Consider a greedy policy~$\Phi$
adapted to the graph~$D$.
We will prove the inequality relations
in~\eqref{eq:greedy-diamond}
using a coupling argument.
More specifically, we will compare
the matching models $(D, I, \Phi)$ and $(D, I, \Phi_+)$,
where $I = (I_t, t \in \N)$ is a sequence of i.i.d.\ classes,
such that $I_t = i$ with probability $\lambda_i / (\lambda_1 + \lambda_2 + \lambda_4 + \lambda_3)$
for each $t \in \N$ and $i \in \V$.

Considering the model $(D, I, \Phi)$, we let
$Q_t$ denote the vector of queue sizes at time~$t$,
$L_{t, i}$ the number of class-$i$ items among the first~$t$ arrivals,
and $M_{t, \{i, j\}}$ (or $M_{t, i, j}$ for short) the number of times
that classes~$i$ and~$j$ are matched over the first~$t$ arrivals,
for each $t \in \N$ and $i, j \in \V$,
as defined in \eqref{eq:Q-rec}--\eqref{eq:M}.
We introduce similar notation for the model $(D, I, \Phi^+)$,
the only difference being that all quantities have superscript~$+$.
Since both models have the same sequence of incoming items,
we have $L_{t, i} = L^+_{t, i}$ for each $t \in \N$ and $i \in \V$.
As usual, we also assume that $Q_0 = Q^+_0 = 0$.
Neither $(Q_t, t \in \N)$ nor $(Q^+_t, t \in \N)$
need to be Markov chains for our argument to hold.

Our end goal is to prove that
the following inequalities are satisfied
at each time $t \in \N$:
\begin{subequations}
	\label{eq:coupling-recursion}
	\begin{align}
		\tag{\ref{eq:coupling-recursion}--1,2}
		\label{eq:coupling-recursion-12}
		M_{t, 1, 2} &\le M^+_{t, 1, 2}, \\
		\tag{\ref{eq:coupling-recursion}--3,4}
		\label{eq:coupling-recursion-43}
		M_{t, 3, 4} &\le M^+_{t, 3, 4}, \\
		\tag{\ref{eq:coupling-recursion}--1,3}
		\label{eq:coupling-recursion-13}
		M_{t, 1, 3} &\ge M^+_{t, 1, 3}, \\
		\tag{\ref{eq:coupling-recursion}--2,4}
		\label{eq:coupling-recursion-24}
		M_{t, 2, 4} &\ge M^+_{t, 2, 4}.
	\end{align}
\end{subequations}
Injecting these inequalities into
the definition~\eqref{eq:lambda} of the matching rates
yields the inequalities~\eqref{eq:greedy-diamond}.
We will prove \eqref{eq:coupling-recursion}
by induction over time $t \in \N$.
The following equations will be instrumental
to prove the induction step:
\begin{align}
	\label{eq:coupling-1}
	&\begin{aligned}
		Q_{t, 1} + Q_{t, 4} &= Q^+_{t, 1} + Q^+_{t, 4}, &
		Q_{t, 2} &= Q^+_{t, 2}, &
		Q_{t, 3} &= Q^+_{t, 3},
	\end{aligned} \\
	\label{eq:coupling-2}
	&L_{t, 1} = Q_{t, 1} + M_{t, 1, 2} + M_{t, 1, 3}
	= Q^+_{t, 1} + M^+_{t, 1, 2} + M^+_{t, 1, 3}, \\
	\label{eq:coupling-4}
	&L_{4, t} = Q_{t, 4} + M_{t, 2, 4} + M_{t, 3, 4}
	= Q^+_{t, 4} + M^+_{t, 2, 4} + M^+_{t, 3, 4}.
\end{align}
Equation~\eqref{eq:coupling-1}
follows from \Cref{prop:D-greedy-policy}\ref{prop:D-greedy-policy-3}, while
\eqref{eq:coupling-2} and~\eqref{eq:coupling-4}
follow from our assumption that the arrivals in both models are coupled.
Furthermore, given the definition of~$Q_\Gre(D)$ in~\eqref{eq:Qgre-D},
we know that
only one integer among $Q_{t, 1} + Q_{t, 4}$, $Q_{t, 2}$, and $Q_{t, 3}$
can be positive,
for each $t \in \N$.

We now proceed to the induction step.
Let $t \in \N$.
We now prove that,
assuming that the inequalities~\eqref{eq:coupling-recursion}
are satisfied at time~$t$,
these inequalities are also satisfied at time $t + 1$.
We distinguish several cases depending on the value of~$I_t$:
\begin{description}
	\item [Case $I_t = 1$:]
	We have directly $M_{t+1, i, j} = M_{t, i, j}$
	and $M^+_{t+1, i, j} = M^+_{t, i, j}$
	for $(i, j) \in \{(2, 4), (3, 4)\}$,
	hence the induction assumption implies
	that~\eqref{eq:coupling-recursion-24}
	and~\eqref{eq:coupling-recursion-43}
	hold at time~$t+1$.
	Since the policy~$\Phi$ is greedy,
	we only have three mutually-exclusive cases:
	\begin{description}
		\item[Case $Q_{t, 2} \ge 1$:]
		The class-1 item is matched
		with a class-2 item already present, and we obtain
		$M_{t+1, 1, 2} = M_{t, 1, 2} + 1$
		and $M_{t+1, 1, 3} = M_{t, 1, 3}$.
		\item [Case $Q_{t, 3} \ge 1$:]
		The class-1 item is matched
		with a class-3 item already present, and we obtain
		$M_{t+1, 1, 2} = M_{t, 1, 2}$
		and $M_{t+1, 1, 3} = M_{t, 1, 3} + 1$.
		\item [Case $Q_{t, 2} = Q_{t, 3} = 0$:]
		The class-1 item is left unmatched,
		and we obtain $M_{t+1, 1, 2} = M_{t, 1, 2}$
		and $M_{t+1, 1, 3} = M_{t, 1, 3}$.
	\end{description}
	Since the policy~$\Phi^+$ is also greedy,
	we can repeat the same argument
	for the quantities associated with $\Phi^+$.
	Combining this observation with~\eqref{eq:coupling-1} yields
	$M_{t+1, 1, 2} - M^+_{t+1, 1, 2} = M_{t, 1, 2} - M^+_{t, 1, 2}$
	and $M_{t+1, 1, 3} - M^+_{t+1,1, 3} = M_{t, 1, 3} - M^+_{t, 1, 3}$.
	Hence, the induction assumption implies directly that
	\eqref{eq:coupling-recursion-12} and \eqref{eq:coupling-recursion-13}
	are satisfied at time $t+1$.
	\item [Case $I_t = 2$:]
	We have directly $M_{t+1, i, j} = M_{t, i, j}$
	and $M^+_{t+1, i, j} = M^+_{t, i, j}$
	for $(i, j) \in \{(1, 3), (3, 4)\}$,
	hence the induction assumption implies
	that~\eqref{eq:coupling-recursion-13}
	and~\eqref{eq:coupling-recursion-43}
	hold at time~$t+1$.
	Proving that~\eqref{eq:coupling-recursion-12} and~\eqref{eq:coupling-recursion-24}
	also hold at time~$t+1$ is more intricate,
	and we will distinguish three mutually-exclusive cases
	depending on the values of $Q_{t, 1}$, $Q_{t, 4}$,
	$Q^+_{t, 1}$, and $Q^+_{t, 4}$:
	\begin{description}
		\item [Case $Q^+_{t, 1} + Q^+_{t, 4} = 0$:]
		Under both policies,
		the class-2 item is either matched
		with a class-3 item or added to the queue.
		In particular, we obtain
		$M_{t+1, i, j} = M_{t, i, j}$
		and $M^+_{t+1, i, j} = M^+_{t, i, j}$
		for $(i, j) \in \{(1, 2), (2, 4)\}$,
		so that~\eqref{eq:coupling-recursion-12}
		and~\eqref{eq:coupling-recursion-24}
		are again satisfied at time~$t+1$
		thanks to the induction assumption.
		\item [Case $Q^+_{t, 1} + Q^+_{t, 4} \ge 1$:]
		We again subdivide this case into three mutually-exclusive cases:
		\begin{description}
			\item[Case $Q^+_{t, 1} \ge 1$ and $Q^+_{t, 4} \ge 1$:]
			We have $M^+_{t+1, 1, 2} = M^+_{t, 1, 2} + 1$
			and $M^+_{t+1, 2, 4} = M^+_{t, 2, 4}$
			by definition of the policy~$\Phi^+$,
			while for the policy~$\Phi$ we only know that
			$M_{t+1, 1, 2} \in \{M_{t, 1, 2}, M_{t, 1, 2} + 1 \}$
			and $M_{t+1, 2, 4} \in \{M_{t, 2, 4}, M_{t, 2, 4} + 1 \}$.
			We can verify
			that~\eqref{eq:coupling-recursion-12}
			and~\eqref{eq:coupling-recursion-24}
			hold at time~$t + 1$
			thanks to the induction assumption.
			\item[Case $Q^+_{t, 1} = 0$ and $Q^+_{t, 4} \ge 1$:] 
			By greediness, the policy~$\Phi^+$
			matches the incoming class-2 item with a class-4 item,
			and we obtain $M^+_{t+1, 1, 2} = M^+_{t, 1, 2}$
			and $M^+_{t+1, 2, 4} = M^+_{t, 2, 4} + 1$.
			If the policy~$\Phi$ makes the same decision,
			then we also have $M_{t+1, 1, 2} = M_{t, 1, 2}$
			and $M_{t+1, 2, 4} = M_{t, 2, 4} + 1$,
			hence~\eqref{eq:coupling-recursion-12}
			and~\eqref{eq:coupling-recursion-24}
			hold at time~$t + 1$
			thanks to the induction assumption.
			Otherwise, the policy~$\Phi$ matches the class-2 item
			with a class-1 item,
			meaning that $M_{t+1, 1, 2} = M_{t, 1, 2} + 1$
			and $M_{t+1, 2, 4} = M_{t, 2, 4}$.
			Importantly, this is only possible if $Q_{t, 1} \ge 1$.
			We now prove~\eqref{eq:coupling-recursion-12}
			and~\eqref{eq:coupling-recursion-24} as follows:
			\begin{itemize}
				\item Proving~\eqref{eq:coupling-recursion-12}
				boils down to proving
				$M^+_{t, 1, 2} \ge M_{t, 1, 2} + 1$.
				We have successively:
				\begin{align*}
					M^+_{t, 1, 2} - M_{t, 1, 2}
					= (Q_{t, 1} - Q^+_{t, 1})
					+ (M_{t, 1, 3} - M^+_{t, 1, 3})
					\ge 1 + 0 = 1,
				\end{align*}
				where the equality follows from~\eqref{eq:coupling-2}
				and the inequality follows from the induction assumption
				and the fact that
				$Q_{t, 1} \ge 1$ and $Q^+_{t, 1} = 0$.
				\item Proving~\eqref{eq:coupling-recursion-24}
				boils down to proving
				$M^+_{t, 2, 4} + 1 \le M_{t, 2, 4}$.
				We have successively
				\begin{align*}
					M_{t, 2, 4}
					- M^+_{t, 2, 4}
					&= (Q^+_{t, 4} - Q_{t, 4}) + (M^+_{t, 3, 4} - M_{t, 3, 4})
					\ge 1 + 0 = 1,
				\end{align*}
				where the equality follows from~\eqref{eq:coupling-4}
				and the inequality follows from the induction assumption
				and the observation that
				$Q^+_{t, 4} - Q_{t, 4} = Q_{t, 1} - Q^+_{t, 1} \ge 1$.
			\end{itemize}
			Intuitively, the only way that an incoming class-2 item
			is matched at time~$t$
			with a class-1 item under the policy~$\Phi$
			and with a class-4 item under the policy~$\Phi^+$
			is if, in the past,
			the policy~$\Phi^+$ had made one more match along edge~$\{1,2\}$
			and one less match along edge~$\{2,4\}$
			compared to the policy~$\Phi$.
			\item[Case $Q^+_{t, 1} \ge 1$ and $Q^+_{t, 4} = 0$:]
			This case is symmetrical to the previous case.
		\end{description}
	\end{description}
	\item [Case $I_t = 3$:]
	This case is symmetrical to the case $I_t = 2$.
	\item [Case $I_t = 4$:]
	This case is symmetrical to the case $I_t = 1$.
\end{description}
\end{proof}

\Cref{prop:D-greedy-kernel} can be interpreted as follows.
Intuitively, the kernel coordinate~$\alpha$
given in~\eqref{eq:diamond}
(also shown in \Cref{fig:diamond})
acts like a slider that determines how much
edges $\{1, 2\}$ and $\{3, 4\}$ are (dis)favored
compared to edges $\{1, 3\}$ and $\{2, 4\}$
on the long run.
Remarkably, \Cref{prop:D-greedy-kernel} shows
that the greedy policy that favors
edges $\{1, 2\}$ and $\{3, 4\}$
(resp.\ $\{1, 3\}$ and $\{2, 4\}$)
the most \emph{in the long run}
is also the policy that favors
these edges the most \emph{in the short run}.
This result is proved by a coupling argument.
In addition, \Cref{prop:D-greedy-kernel}
uses the lower bound~\eqref{eq:diamond-greedy-ineq-2}
in \Cref{prop:D-greedy-edge} to prove that,
in the limit as~$\beta \to +\infty$,
all greedy policies yield the same matching rate vector.

Taken together,
\Cref{prop:D-greedy-kernel-1,prop:D-greedy-kernel-2-3}
in \Cref{prop:D-greedy-kernel} show that,
as $\beta \to +\infty$,
the interval~$\La_\Gre$ becomes reduced to a single point $\alpha = 0$,
meaning that all greedy policies
yield the same vector of matching rates,
with edge coordinates
$\mu = (2 \lambda_1 \bar\lambda_2, 2 \lambda_1 \bar\lambda_3, \beta,
2 \bar\lambda_2 \lambda_4, 2 \bar\lambda_3 \lambda_4)$.
The rationale behind this result is the following. 
In the regime where $\beta = \mu_{2,3} \to +\infty$,
we see in~\eqref{eq:diamond} that
the arrival rates of classes~2 and~3 become large
compared to those of classes~1 and~4.
As a result, items of classes~1 and~4 are matched (almost) always immediately,
and unmatched items belong to either class~2 or class~3 (but not both at the same time).
In the proof of \Cref{prop:D-greedy-kernel}\ref{prop:D-greedy-kernel-2-3},
this intuition is formalized by taking the limit
of~\eqref{eq:diamond-greedy-eq} as $\beta \to +\infty$, which yields
$p_\varnothing \xrightarrow{} 0$,
$p_2 \xrightarrow{} 2 \bar\lambda_2$,
$p_3 \xrightarrow{} 2 \bar\lambda_3$, and
$p_{1,4} \xrightarrow{} 0$.
In this regime, the greediness of the policy
allows no degree of flexibility
in choosing the class of the item
to which an incoming item is matched,
so the matching rates are unique.

\begin{rem}
	Some quantities in \Cref{prop:D-greedy-kernel} are functions of
	the matching rate~$\beta = \mu_{2, 3}$,
	but this dependency is kept implicit to simplify notation.
	In particular, the intervals
	$\Lann$ and $\Lap$ do \emph{not} depend on~$\beta$,
	but the interval~$\La_\Gre$
	and the coordinates~$\alpha_+$ and~$\alpha_-$ do.
\end{rem}

\paragraph*{Numerical results.}
To illustrate \Cref{prop:D-greedy-kernel},
\Cref{fig:diamond-curve} shows a symmetric example
with $\lambda_1 = \bar\lambda_2 = \bar\lambda_3 = \lambda_4 = \frac14$.
The figure compares $\Lann$ and $\La_\Gre$
with the bounds~\eqref{eq:diamond-greedy-ineq-2}--\eqref{eq:diamond-greedy-ineq-1}
(converted in the kernel coordinates) and the limit $\alpha = 0$.
Each point forming the shape of $\Pi_\Gre = [\alpha_-, \alpha_+]$
is obtained by running
a simulation consisting of $10^{10}$ steps
(as specified in \Cref{sec:numerical-results}).
As announced by \Cref{prop:D-greedy-kernel},
$\La_\Gre$ becomes reduced to a single point $\alpha = 0$
when $\beta \to +\infty$.
We also notice that the bounds~\eqref{eq:diamond-greedy-ineq-2}--\eqref{eq:diamond-greedy-ineq-1}
are not tight when $\beta$ is small
(in the sense that the difference between $\alpha_+$ and the upper-bound is no longer negligible compared to the difference between $\alpha_+$ and the boundary $2 \min(\lambda_1 \bar\lambda_3, \bar\lambda_2, \lambda_4)$ of $\Pi_{\ge 0}$),
while at the same time 
$\alpha_+$ and $\alpha_-$ become arbitrarily close to the borders of $\Lann$
when $\beta$ tends to zero
(which does not contradict the fact
that $\La_\Gre$ is a strict subset of~$\Lann$ as long as $\beta > 0$).
The gap between $\La_\Gre$
and the bounds~\eqref{eq:diamond-greedy-ineq-2}--\eqref{eq:diamond-greedy-ineq-1}
comes from the fact that,
to obtain the lower bounds for~$p_1$ and~$p_4$ in~\eqref{eq:diamond-greedy-ineq-1},
we neglected the case where there are both class-1 and class-4 unmatched items,
and this case is not negligible when $\beta$ is small.

\begin{figure}[!htb]
	\centering
	 % figures/diamond_10
\begin{tikzpicture}

\definecolor{darkgray153}{RGB}{153,153,153}
\definecolor{darkgray176}{RGB}{176,176,176}
\definecolor{darkorange25512714}{RGB}{255,127,14}
\definecolor{forestgreen4416044}{RGB}{44,160,44}
\definecolor{gainsboro216}{RGB}{216,216,216}
\definecolor{lightgray204}{RGB}{204,204,204}
\definecolor{steelblue31119180}{RGB}{31,119,180}

\begin{axis}[
height=5.5cm,
legend cell align={left},
legend style={
	fill opacity=0.8,
	draw opacity=1,
	text opacity=1,
	draw=lightgray204,
	font=\footnotesize,
	at={(0.99,0.93)}
},
yticklabel style={
	/pgf/number format/fixed,
	/pgf/number format/precision=2,
},
scaled y ticks=false,
log basis x={10},
tick align=outside,
tick pos=left,
width=.94\textwidth,
x grid style={darkgray176},
xlabel={Matching rate \(\displaystyle \beta = \mu_{2, 3}\)},
xmin=0.001, xmax=10,
xmode=log,
xtick style={color=black},
y grid style={darkgray176},
ylabel={Kernel coordinate \(\displaystyle \alpha\)},
ylabel style={yshift=.2cm},
ymin=-0.1375, ymax=0.1375,
ytick style={color=black},
]
\path [draw=gainsboro216, fill=gainsboro216]
(axis cs:0.001,0.125)
--(axis cs:0.001,-0.125)
--(axis cs:0.00137382379588326,-0.125)
--(axis cs:0.0018873918221351,-0.125)
--(axis cs:0.00259294379740467,-0.125)
--(axis cs:0.00356224789026244,-0.125)
--(axis cs:0.00489390091847749,-0.125)
--(axis cs:0.00672335753649933,-0.125)
--(axis cs:0.00923670857187387,-0.125)
--(axis cs:0.0126896100316792,-0.125)
--(axis cs:0.0174332882219999,-0.125)
--(axis cs:0.0239502661998749,-0.125)
--(axis cs:0.0329034456231267,-0.125)
--(axis cs:0.0452035365636024,-0.125)
--(axis cs:0.0621016941891562,-0.125)
--(axis cs:0.0853167852417281,-0.125)
--(axis cs:0.117210229753348,-0.125)
--(axis cs:0.161026202756094,-0.125)
--(axis cs:0.221221629107045,-0.125)
--(axis cs:0.30391953823132,-0.125)
--(axis cs:0.41753189365604,-0.125)
--(axis cs:0.573615251044868,-0.125)
--(axis cs:0.788046281566991,-0.125)
--(axis cs:1.08263673387405,-0.125)
--(axis cs:1.48735210729351,-0.125)
--(axis cs:2.04335971785694,-0.125)
--(axis cs:2.80721620394118,-0.125)
--(axis cs:3.85662042116347,-0.125)
--(axis cs:5.29831690628371,-0.125)
--(axis cs:7.27895384398315,-0.125)
--(axis cs:10,-0.125)
--(axis cs:10,0.125)
--(axis cs:10,0.125)
--(axis cs:7.27895384398315,0.125)
--(axis cs:5.29831690628371,0.125)
--(axis cs:3.85662042116347,0.125)
--(axis cs:2.80721620394118,0.125)
--(axis cs:2.04335971785694,0.125)
--(axis cs:1.48735210729351,0.125)
--(axis cs:1.08263673387405,0.125)
--(axis cs:0.788046281566991,0.125)
--(axis cs:0.573615251044868,0.125)
--(axis cs:0.41753189365604,0.125)
--(axis cs:0.30391953823132,0.125)
--(axis cs:0.221221629107045,0.125)
--(axis cs:0.161026202756094,0.125)
--(axis cs:0.117210229753348,0.125)
--(axis cs:0.0853167852417281,0.125)
--(axis cs:0.0621016941891562,0.125)
--(axis cs:0.0452035365636024,0.125)
--(axis cs:0.0329034456231267,0.125)
--(axis cs:0.0239502661998749,0.125)
--(axis cs:0.0174332882219999,0.125)
--(axis cs:0.0126896100316792,0.125)
--(axis cs:0.00923670857187387,0.125)
--(axis cs:0.00672335753649933,0.125)
--(axis cs:0.00489390091847749,0.125)
--(axis cs:0.00356224789026244,0.125)
--(axis cs:0.00259294379740467,0.125)
--(axis cs:0.0018873918221351,0.125)
--(axis cs:0.00137382379588326,0.125)
--(axis cs:0.001,0.125)
--cycle;
\addlegendimage{area legend, draw=gainsboro216, fill=gainsboro216}
\addlegendentry{$\Lann$}

\path [draw=darkgray153, fill=darkgray153]
(axis cs:0.001,0.121342049774471)
--(axis cs:0.001,-0.121342049774471)
--(axis cs:0.00137382379588326,-0.120226995808294)
--(axis cs:0.0018873918221351,-0.118561312725928)
--(axis cs:0.00259294379740467,-0.116992507948148)
--(axis cs:0.00356224789026244,-0.114726743292916)
--(axis cs:0.00489390091847749,-0.111903476720563)
--(axis cs:0.00672335753649933,-0.108358955025466)
--(axis cs:0.00923670857187387,-0.104077990123693)
--(axis cs:0.0126896100316792,-0.098834252904329)
--(axis cs:0.0174332882219999,-0.0926268705793193)
--(axis cs:0.0239502661998749,-0.0853382531108102)
--(axis cs:0.0329034456231267,-0.077050098214234)
--(axis cs:0.0452035365636024,-0.0678888207559152)
--(axis cs:0.0621016941891562,-0.0580850333697344)
--(axis cs:0.0853167852417281,-0.0480365731426026)
--(axis cs:0.117210229753348,-0.0382329711219192)
--(axis cs:0.161026202756094,-0.0291656058915846)
--(axis cs:0.221221629107045,-0.0212743314319169)
--(axis cs:0.30391953823132,-0.0148264784215806)
--(axis cs:0.41753189365604,-0.00988075934173411)
--(axis cs:0.573615251044868,-0.00630987866222599)
--(axis cs:0.788046281566991,-0.00388735388245449)
--(axis cs:1.08263673387405,-0.00232084458163656)
--(axis cs:1.48735210729351,-0.00134887778123826)
--(axis cs:2.04335971785694,-0.000764304901645203)
--(axis cs:2.80721620394118,-0.000416973322910467)
--(axis cs:3.85662042116347,-0.000236412878478528)
--(axis cs:5.29831690628371,-0.000124123410349428)
--(axis cs:7.27895384398315,-7.65970248155466e-05)
--(axis cs:10,-2.81064000000088e-05)
--(axis cs:10,2.81064000000088e-05)
--(axis cs:10,2.81064000000088e-05)
--(axis cs:7.27895384398315,7.65970248155466e-05)
--(axis cs:5.29831690628371,0.000124123410349428)
--(axis cs:3.85662042116347,0.000236412878478528)
--(axis cs:2.80721620394118,0.000416973322910467)
--(axis cs:2.04335971785694,0.000764304901645203)
--(axis cs:1.48735210729351,0.00134887778123826)
--(axis cs:1.08263673387405,0.00232084458163656)
--(axis cs:0.788046281566991,0.00388735388245449)
--(axis cs:0.573615251044868,0.00630987866222599)
--(axis cs:0.41753189365604,0.00988075934173411)
--(axis cs:0.30391953823132,0.0148264784215806)
--(axis cs:0.221221629107045,0.0212743314319169)
--(axis cs:0.161026202756094,0.0291656058915846)
--(axis cs:0.117210229753348,0.0382329711219192)
--(axis cs:0.0853167852417281,0.0480365731426026)
--(axis cs:0.0621016941891562,0.0580850333697344)
--(axis cs:0.0452035365636024,0.0678888207559152)
--(axis cs:0.0329034456231267,0.077050098214234)
--(axis cs:0.0239502661998749,0.0853382531108102)
--(axis cs:0.0174332882219999,0.0926268705793193)
--(axis cs:0.0126896100316792,0.098834252904329)
--(axis cs:0.00923670857187387,0.104077990123693)
--(axis cs:0.00672335753649933,0.108358955025466)
--(axis cs:0.00489390091847749,0.111903476720563)
--(axis cs:0.00356224789026244,0.114726743292916)
--(axis cs:0.00259294379740467,0.116992507948148)
--(axis cs:0.0018873918221351,0.118561312725928)
--(axis cs:0.00137382379588326,0.120226995808294)
--(axis cs:0.001,0.121342049774471)
--cycle;
\addlegendimage{area legend, draw=darkgray153, fill=darkgray153}
\addlegendentry{$\La_\Gre = [\alpha_-, \alpha_+]$}
\end{axis}

\begin{axis}[
	height=5.5cm,
	legend cell align={left},
	legend style={
		fill opacity=0.8,
		draw opacity=1,
		text opacity=1,
		draw=lightgray204,
		font=\footnotesize,
		at={(0.99,0.395)}
	},
	yticklabel style={
		/pgf/number format/fixed,
		/pgf/number format/precision=2,
	},
	scaled y ticks=false,
	log basis x={10},
	tick align=outside,
	tick pos=left,
	width=.94\textwidth,
	x grid style={darkgray176},
	xlabel={Matching rate \(\displaystyle \beta = \mu_{2, 3}\)},
	xmin=0.001, xmax=10,
	xmode=log,
	xtick style={color=black},
	y grid style={darkgray176},
	ylabel={Kernel coordinate \(\displaystyle \alpha\)},
	ylabel style={yshift=.2cm},
	ymin=-0.1375, ymax=0.1375,
	ytick style={color=black},
	]
	
	\addplot [semithick, steelblue31119180]
	table {%
		0.001 0.123513880984001
		0.00137382379588326 0.122965380133147
		0.0018873918221351 0.12221798933221
		0.00259294379740467 0.121202663230901
		0.00356224789026244 0.119829013522194
		0.00489390091847749 0.117980934369978
		0.00672335753649933 0.115513264788586
		0.00923670857187387 0.112251522724322
		0.0126896100316792 0.107998063674394
		0.0174332882219999 0.102549387293948
		0.0239502661998749 0.0957297719217437
		0.0329034456231267 0.0874440352863653
		0.0452035365636024 0.0777445400992993
		0.0621016941891562 0.0668938494623066
		0.0853167852417281 0.0553899286915132
		0.117210229753348 0.043919034256954
		0.161026202756094 0.0332265083524708
		0.221221629107045 0.0239431134157176
		0.30391953823132 0.0164413517478063
		0.41753189365604 0.0107860577249372
		0.573615251044868 0.00678887156435563
		0.788046281566991 0.0041214546499551
		1.08263673387405 0.00242723563461812
		1.48735210729351 0.00139448003175793
		2.04335971785694 0.000785518692266257
		2.80721620394118 0.000435750498386622
		3.85662042116347 0.000238899706971801
		5.29831690628371 0.000129817664902263
		7.27895384398315 7.00747726247869e-05
		10 3.76392652815249e-05
	};
	\addlegendentry{Upper bound \eqref{eq:diamond-greedy-ineq-2}--\eqref{eq:diamond-greedy-ineq-1}}
	\addplot [semithick, darkorange25512714]
	table {%
		0.001 0
		0.00137382379588326 0
		0.0018873918221351 0
		0.00259294379740467 0
		0.00356224789026244 0
		0.00489390091847749 0
		0.00672335753649933 0
		0.00923670857187387 0
		0.0126896100316792 0
		0.0174332882219999 0
		0.0239502661998749 0
		0.0329034456231267 0
		0.0452035365636024 0
		0.0621016941891562 0
		0.0853167852417281 0
		0.117210229753348 0
		0.161026202756094 0
		0.221221629107045 0
		0.30391953823132 0
		0.41753189365604 0
		0.573615251044868 0
		0.788046281566991 0
		1.08263673387405 0
		1.48735210729351 0
		2.04335971785694 0
		2.80721620394118 0
		3.85662042116347 0
		5.29831690628371 0
		7.27895384398315 0
		10 0
	};
	\addlegendentry{Limit $\alpha=0$}
	\addplot [semithick, forestgreen4416044]
	table {%
		0.001 -0.123513880984001
		0.00137382379588326 -0.122965380133147
		0.0018873918221351 -0.12221798933221
		0.00259294379740467 -0.121202663230901
		0.00356224789026244 -0.119829013522194
		0.00489390091847749 -0.117980934369978
		0.00672335753649933 -0.115513264788586
		0.00923670857187387 -0.112251522724322
		0.0126896100316792 -0.107998063674394
		0.0174332882219999 -0.102549387293948
		0.0239502661998749 -0.0957297719217437
		0.0329034456231267 -0.0874440352863653
		0.0452035365636024 -0.0777445400992993
		0.0621016941891562 -0.0668938494623066
		0.0853167852417281 -0.0553899286915132
		0.117210229753348 -0.043919034256954
		0.161026202756094 -0.0332265083524708
		0.221221629107045 -0.0239431134157176
		0.30391953823132 -0.0164413517478063
		0.41753189365604 -0.0107860577249372
		0.573615251044868 -0.00678887156435563
		0.788046281566991 -0.0041214546499551
		1.08263673387405 -0.00242723563461812
		1.48735210729351 -0.00139448003175793
		2.04335971785694 -0.000785518692266257
		2.80721620394118 -0.000435750498386622
		3.85662042116347 -0.000238899706971801
		5.29831690628371 -0.000129817664902263
		7.27895384398315 -7.00747726247869e-05
		10 -3.76392652815249e-05
	};
	\addlegendentry{Lower bound~\eqref{eq:diamond-greedy-ineq-2}--\eqref{eq:diamond-greedy-ineq-1}}
\end{axis}
\end{tikzpicture}
	
	\caption{\label{fig:diamond-curve} Evolution of $\La_\Gre$ as a function of $\beta$ in the diamond problem~\eqref{eq:diamond} with $\lambda_1 = \bar\lambda_2 = \bar\lambda_3 = \lambda_4 = \frac14$. For each $\beta$, $\La_\Gre$ is estimated by simulating the model $(D, \lambda, \Phi_+)$ and leveraging the symmetry of the problem. $\Lann$ and other bounds are displayed for comparison. All results are expressed in kernel coordinates.}
\end{figure}

\subsubsection[Proof of Proposition \ref*{prop:greedy-fish}.]{Proof of \Cref{prop:greedy-fish} (fish matching problem) and discussion.}
\label{app:greedy-fish}

We consider a matching problem, which we call the Fish problem,
for which $\La_\Gre = \Lap$.
The Fish matching problem is defined as follows:
\begin{align} \label{eq:fish}
	\begin{aligned}
		V &= \{1, 2, 3, 4, 5, 6\}, \\
		E &= \{ \{1,2\}, \{1,3\}, \{2,3\}, \{3,4\}, \{3, 6\}, \{4,5\}, \{5,6\} \}, \\
		\lambda &= (4, 4, 3, 2, 3, 2), \\
		\B &= \{ (0, 0, 0, 1, -1, -1, 1) \}, \\
		\mu^\circ &= \textstyle (3, 1, 1, \frac12, \frac12, \frac32, \frac32), \\
		\mu &= \textstyle (3, 1, 1, \frac12 + \alpha, \frac12 - \alpha, \frac32 - \alpha, \frac32 + \alpha),
		\quad \alpha \in \R.
	\end{aligned}
\end{align}
The general solution~$\mu$ given in~\eqref{eq:fish}
is shown in \Cref{fig:fish}.
We can verify by a direct inspection that
this matching problem is stabilizable and that
$\Lann = [-\frac12, \frac12]$ in kernel coordinates.
The kernel coordinate $\alpha$ acts like a slider
that is positive (resp.\ negative)
if matches along edges $\{3, 4\}$ and $\{5, 6\}$
are more (resp.\ less) frequent than
matches along edges $\{3, 6\}$ and $\{4, 5\}$.

\begin{reusefigure}[!htb]{fig:fish}
	\centering
	\begin{tikzpicture}[scale=.8]
		\def\d{2cm}
		\node[class] (1) {$1$};
		\foreach \i/\s/\a in {2/1/-90, 3/1/-30, 4/3/30, 5/4/-30, 6/3/-30}		
		\path (\s) ++(\a:\d) node[class] (\i) {$\i$};
		
		\draw (1) edge node[left] {$3$} (2) 
		(1) edge node[above, sloped] {$1$} (3)
		(2) edge node[above, sloped] {$1$} (3)
		(3) edge node[above, sloped] {$\frac12 + \alpha$} (4)
		(4) edge node[above, sloped] {$\frac32 - \alpha$} (5) 
		(5) edge node[above, sloped] {$\frac32 + \alpha$} (6)
		(3) edge node[above, sloped] {$\frac12 - \alpha$} (6);
	\end{tikzpicture}
	\caption{Generic solution to~\eqref{eq:system}~in the Fish matching problem of~\eqref{eq:fish} with $\lambda=(4,4,3,2,3,2)$.}
	\label{fig:fish-app}
\end{reusefigure}

To prove \Cref{prop:greedy-fish}, we build two families of stable greedy policies,
denoted by $(\Phi^+_k)_{k \in \N}$ and $(\Phi^-_k)_{k \in \N}$,
such that $\lim_{k \to +\infty} \alpha(\Phi^+_k) = \frac12$
and $\lim_{k \to +\infty} \alpha(\Phi^-_k) = -\frac12$.
The conclusion then follows from the convexity of the set $\La_\Gre$
(\Cref{prop:greedypositive}).
We focus on the family $(\Phi^+_k)_{k \in \N}$,
as the family $(\Phi^-_k)_{k \in \N}$ is symmetrical
(in the sense that it suffices to exchange classes~4 and~6).

The family $(\Phi^+_k)_{k \in \N}$ is defined as follows.
Let $\Phi^+_\infty$ denote the \gls{hrf} policy
where edges have the following decreasing reward order:
$\{1, 3\}$, $\{2, 3\}$, $\{3, 4\}$, $\{5, 6\}$,
followed by all other edges in an arbitrary order.
\begin{figure}[!htb]
	\centering
	\subfloat[Policy~$\Phi^+_\infty$,
	followed by policy~$\Phi^+_k$ when the total queue length $\sum_i Q_i$
	is at most~$k-1$]{%
		\begin{tikzpicture}[scale=.8]
			\def\d{2cm}
			\node[class] (1) {$1$};
			\foreach \i/\s/\a in {2/1/-90, 3/2/30, 4/3/30, 5/4/-30, 6/3/-30}		
			\path (\s) ++(\a:\d) node[class] (\i) {$\i$};
			
			\draw (1) edge (2) 
			(1) edge node[fill=white, inner sep=0.04cm] {$1^{\text{st}}$} (3)
			(2) edge node[fill=white, inner sep=0.04cm] {$2^{\text{nd}}$} (3)
			(3) edge node[fill=white, inner sep=0.04cm] {$3^{\text{rd}}$} (4)
			(4) edge (5) 
			(5) edge node[fill=white, inner sep=0.04cm] {$4^{\text{th}}$} (6)
			(3) edge (6);
			
			\node at ($(1)-(2cm,0)$) {};
			\node at ($(5)+(2cm,0)$) {};
		\end{tikzpicture}
	}
	\hfill
	\subfloat[Policy~$\Phi^+_0$,
	followed by policy~$\Phi^+_k$ when the total queue length $\sum_i Q_i$
	is at least~$k$]{%
		\begin{tikzpicture}[scale=.8]
			\def\d{2cm}
			\node[class] (1) {$1$};
			\foreach \i/\s/\a in {2/1/-90, 3/2/30, 4/3/30, 5/4/-30, 6/3/-30}		
			\path (\s) ++(\a:\d) node[class] (\i) {$\i$};
			
			\draw (1) edge (2) 
			(1) edge node[fill=white, inner sep=0.04cm] {$3^{\text{rd}}$} (3)
			(2) edge node[fill=white, inner sep=0.04cm] {$2^{\text{nd}}$} (3)
			(3) edge node[fill=white, inner sep=0.04cm] {$1^{\text{st}}$} (4)
			(4) edge (5) 
			(5) edge node[fill=white, inner sep=0.04cm] {$4^{\text{th}}$} (6)
			(3) edge (6);
			
			\node at ($(1)-(2cm,0)$) {};
			\node at ($(5)+(2cm,0)$) {};
		\end{tikzpicture}
	}
	\caption{\Gls{hrf} (greedy) policies considered in the Fish matching problem of~\eqref{eq:fish}.}
	\label{fig:fish-policies}
\end{figure}%
Let $\Phi^+_0$ denote the \gls{hrf} policy
where edges have the following decreasing reward order:
$\{3, 4\}$, $\{2, 3\}$, $\{1, 3\}$, $\{5, 6\}$,
followed by all other edges in an arbitrary order.
The important point is that both policies prioritize edges $\{3, 4\}$ and $\{5, 6\}$ over edges $\{3, 6\}$ and $\{4, 5\}$ (with the hope that this lead to a high~$\alpha$),
but that $\Phi^+_\infty$ gives higher priority to the ``tail'' of the fish, while $\Phi^+_0$ gives higher priority to the ``trunk''.
Now, for each $k \in \N$, $\Phi^+_k$ is the deterministic size-based policy that follows $\Phi^+_\infty$ when the total queue length is at most $k - 1$
and $\Phi^+_0$ when it is at least~$k$
(that is, $\Phi^+_k(q, i) = \Phi^+_\infty(q, i)$ if $\sum_i q_i \le k - 1$
and $\Phi^+_k(q, i) = \Phi^+_0(q, i)$ if $\sum_i q_i \ge k$).

The rationale behind this definition is as follows.
According to~\eqref{eq:fish},
$\alpha$ is maximal (i.e., equal to $\frac12$) when
$\mu_{3, 4}$ is equal to~1 and $\mu_{3, 6}$ is equal to~0.
If we allow \emph{non-}greedy policies,
$\alpha = \frac12$ can be achieved by merely applying
the edge-filtering variant of \gls{ml}
(or any maximally-stable policy)
on the bijective subgraph of~$G$
obtained by eliminating edge~$\{3, 6\}$,
that is, by \emph{never} performing a match between classes~3 and~6.
As we will see,
the family $(\Phi^+_k)_{k \in \N}$ of (stable greedy) policies
emulates this (non-greedy) edge-filtering policy
by favoring edge~$\{3, 4\}$ over edge~$\{3, 6\}$
while making the probability that $q_4=0$ arbitrary small as~$k$ increases.
Roughly speaking, mixing the policies $\Phi^+_\infty$ and $\Phi^+_0$ lets us control the total queue length: $\Phi^+_\infty$ is not stable and lets the system fill up, while $\Phi^+_0$ is stable and drains it.
All in all, $\Phi^+_k$ keeps the total queue length around~$k$; and in the build-up mode the total grows only through~$q_4$ (the other queues stay tight), so a large total means a large~$q_4$,
so that the probability that~$q_4 = 0$ is low when~$k$ is large.
In the limit,
$q_4$ is always positive, so that
the class-3 items that are not matched with classes~1 or~2
are drained by class~4 (while class~6 is matched only with class~5),
so that $\mu_{3, 4}(\Phi^+_k)$ tends to~1
and $\mu_{3, 6}(\Phi^+_k)$ tends to~0 as $k \to \infty$.

\medskip\noindent\emph{Outline of the proof.}
We prove \Cref{prop:greedy-fish} for the family $(\Phi^+_k)$ in four steps.
\emph{(i)~Reduction} (\Cref{prop:fish-reduction}): since edge $\{3,6\}$ is the
least preferred of class~3 under both modes, a match on $\{3,6\}$ can occur only
when $Q_4=0$, so $\alpha(\Phi^+_k)\ge\tfrac12-5\,P(Q_{k,4}=0)$; it therefore
suffices that $\Phi^+_k$ be stable and that $P(Q_{k,4}=0)\to0$.
\emph{(ii)~Stability} (\Cref{prop:fish-stability}): the sum threshold makes
$\Phi^+_k$ agree with the base policy $\Phi^+_0$ outside the \emph{finite} set
$\{N<k\}$, where $N=\sum_i Q_i$; it is therefore enough to stabilize $\Phi^+_0$,
which we do by dominating the queues, one group at a time, by $k$-independent
queues that do not see the threshold.
\emph{(iii)~Idle periods}: when $Q_4=0$ the threshold plays no role, so the mean
time until $Q_4$ becomes positive again is bounded by a constant $C_0$ free
of~$k$ (\Cref{prop:fish-tail}).
\emph{(iv)~Busy periods} (\Cref{lem:fish-firstpassage}): while $N$ stays below the
threshold, class~4 is essentially the only queue being fed, so the system must
climb about~$k$ steps before the threshold releases it; a nonlinear Lyapunov
function shows it does so before $Q_4$ empties, which forces the mean busy period
to grow like~$k$. Short idle periods against busy periods of length $\sim k$ leave
only an $O(1/k)$ fraction of time with $Q_4=0$, which by~(i) sends
$\alpha(\Phi^+_k)$ to~$\tfrac12$.

Throughout we use that, under any greedy policy started from the empty state, the
set of nonempty classes is an independent set of~$G$; for the Fish this forces
$Q_3=Q_5=0$ whenever $Q_4>0$ or $Q_6>0$, and at most one of $Q_1,Q_2$ positive. We
write $N=\sum_{i}Q_i$ for the total number of waiting items and, inside the
analysis of a fixed policy, abbreviate $Q_{k,i}$ to $Q_i$.

\begin{prop}[Reduction]\label{prop:fish-reduction}
If $\Phi^+_k$ is stable for every $k\in\N$ and $P(Q_{k,4}=0)\to0$ as
$k\to\infty$, then $\La_\Gre=\Lap$ for the Fish problem.
\end{prop}
\begin{proof}
A match on edge $\{3,6\}$ occurs either when a class-3 arrival matches~6, which,
as $\{3,6\}$ is class~3's least-preferred edge, requires
$Q_{k,1}=Q_{k,2}=Q_{k,4}=0$ and $Q_{k,6}>0$, or when a class-6 arrival
matches~3, which, as class~6 prefers $\{5,6\}$, requires $Q_{k,5}=0$ and
$Q_{k,3}>0$. Hence
\begin{align*}
	\mu(\Phi^+_k)_{3, 6}
	&= \lambda_3\, P(Q_{k,1}=Q_{k,2}=Q_{k,4}=0,\, Q_{k, 6} > 0)
	  +\lambda_6\, P(Q_{k, 3} > 0,\, Q_{k, 5} = 0)\\
	&\le (\lambda_3 + \lambda_6)\, P(Q_{k, 4} = 0),
\end{align*}
because the first event is contained in $\{Q_{k,4}=0\}$ and, since $3$ and~$4$
are adjacent, greediness forbids simultaneous class-3 and class-4 backlogs, so
$\{Q_{k,3}>0\}\subseteq\{Q_{k,4}=0\}$. As
$\mu(\Phi^+_k)_{3,6}=\tfrac12-\alpha(\Phi^+_k)\ge0$ and
$\lambda_3+\lambda_6=5$, this gives
$\tfrac12-5\,P(Q_{k,4}=0)\le\alpha(\Phi^+_k)\le\tfrac12$, so $P(Q_{k,4}=0)\to0$
forces $\alpha(\Phi^+_k)\to\tfrac12$. The transposition $4\leftrightarrow6$ is
an automorphism of the Fish preserving~$\lambda$ (it swaps the edges
$\{3,4\}\leftrightarrow\{3,6\}$ and $\{4,5\}\leftrightarrow\{5,6\}$ and fixes
$1,2,3,5$) and sends $\alpha\mapsto-\alpha$; the image family $\Phi^-_k$ is thus
greedy, stable, and satisfies $\alpha(\Phi^-_k)\to-\tfrac12$. Since $\La_\Gre$ is
convex (\Cref{prop:greedypositive}) and contains points converging to
$\pm\tfrac12$, it contains $(-\tfrac12,\tfrac12)=\Lap$; the reverse inclusion is
immediate, so $\La_\Gre=\Lap$.
\end{proof}

\begin{lem}[Class~6 never destabilizes]\label{lem:fish-class6}
For every $k\in\N$, in stationary regime,
$P(Q_{k, 6} = 0) \ge 1-\lambda_6/\lambda_5 = \tfrac13$.
\end{lem}
\begin{proof}
Whenever $Q_6>0$ we have $Q_3=Q_5=0$, so a class-6 arrival finds its neighbours
$3,5$ empty and joins the queue: $Q_6$ increases only at a sub-collection of the
class-6 arrival epochs, of total rate at most~$\lambda_6$. A class-5 arrival
prefers edge $\{5,6\}$, hence matches~6 whenever $Q_6>0$: $Q_6$ decreases at
every class-5 epoch while $Q_6>0$, of rate at least~$\lambda_5$. Coupling $Q_6$
with an $M/M/1$ queue~$Y$ of arrival rate $\lambda_6$ and service rate
$\lambda_5$, driven by the same class-6 and class-5 arrival clocks, yields
$Q_6(t)\le Y(t)$ for all $t$ from $Q_6(0)=Y(0)=0$. As $\lambda_6<\lambda_5$, the
queue~$Y$ is positive recurrent with $P(Y=0)=1-\lambda_6/\lambda_5=\tfrac13$, and
$\{Y=0\}\subseteq\{Q_6=0\}$.
\end{proof}

\begin{lem}[Drift of $Q_4+Q_6$]\label{lem:fish-Sdrift}
Set $S=Q_4+Q_6$. On the event $\{Q_4>0\}$ the generator drift of~$S$ is
piecewise constant and does not depend on~$Q_6$:
\[
	(\mathcal A S)(Q)=
	\begin{cases}
		+1, & N\le k-1 \text{ and } (Q_1>0 \text{ or } Q_2>0),\\
		-2, & \text{otherwise, i.e. } N\ge k \text{ or } Q_1=Q_2=0.
	\end{cases}
\]
\end{lem}
\begin{proof}
When $Q_4>0$ we have $Q_3=Q_5=0$, so class-4 and class-6 arrivals both queue,
contributing $+\lambda_4$ and $+\lambda_6$ to the drift of~$S$. A class-5 arrival
prefers edge $\{5,6\}$ and so matches~6 if $Q_6>0$ and otherwise~4; in both cases
it removes one unit of~$S$, contributing $-\lambda_5$. A class-3 arrival removes
one unit of~$S$ (by matching~4) exactly when it does not match a waiting class~1
or~2: always in the high mode ($N\ge k$), and only when $Q_1=Q_2=0$ in the low
mode. With $\lambda=(4,4,3,2,3,2)$ this gives $\lambda_4+\lambda_6-\lambda_5=1$,
reduced by a further $\lambda_3=3$ whenever class~3 matches~4, which is the two
stated values.
\end{proof}

The positive value $+1$ occurs precisely in the low mode with a class-1 or class-2
backlog present, that is, in the regime the policy occupies while it builds $N$ up
towards the threshold; this is the drift that will make the busy period long. We
first record that the family is stable.

\begin{prop}[Stability of the greedy family]\label{prop:fish-stability}
For every $k\in\N$ the matching model $(G,\lambda,\Phi^+_k)$ is stable.
\end{prop}
\begin{proof}
By construction $\Phi^+_k$ applies the order of $\Phi^+_0$ at every state with
$N\ge k$ and differs from $\Phi^+_0$ only on the finite set $\{N<k\}$. This is the
one property for which the sum threshold is needed: a threshold on $Q_4$ alone
would leave the two policies differing on the infinite set $\{Q_4<k\}$. Now
$\Phi^+_k$ and $\Phi^+_0$ have identical transition rates outside the finite set
$\{N<k\}$, which contains the empty state; and, under $\Phi^+_0$, every state
reaches $\{N<k\}$ in finite mean time, because $\Phi^+_0$ is positive recurrent
(shown below) and $\{N<k\}$ contains~$\varnothing$. Since the two chains coincide
off $\{N<k\}$, the chain under $\Phi^+_k$ returns to $\{N<k\}$ in finite mean time
as well; as $\{N<k\}$ is finite and the chain is irreducible, $\Phi^+_k$ is
positive recurrent. It therefore suffices to prove that the base policy $\Phi^+_0$
is positive recurrent.

Under $\Phi^+_0$ class~3 always prefers edge $\{3,4\}$. We bound the queues one
group at a time, \emph{without assuming stability}, and then combine the bounds by
a coupling on a common family of arrival clocks. Recall that whenever $Q_4>0$ we
have $Q_3=Q_5=0$ and at most one of $Q_1,Q_2$ positive; set $b=Q_1+Q_2$.

\emph{Queue~$4$.} On $\{Q_4\ge1\}$ a class-4 arrival finds $3,5$ empty and joins
the queue (rate $\lambda_4=2$), while class~3 matches~4 (rate $\lambda_3=3$) and
class~5 matches~4 when $Q_6=0$; the up-rate is~$2$ and the down-rate is at
least~$3$, both independent of~$b$. Driving $Q_4$ and an $M/M/1(\lambda_4,\lambda_3)
=M/M/1(2,3)$ queue~$Y$ by the same class-3 and class-4 clocks gives $Q_4(t)\le
Y(t)$ for all~$t$, with $\{Y=0\}\subseteq\{Q_4=0\}$ and $P(Y=0)=1-\tfrac23=\tfrac13$.

\emph{Queues $3,5,6$.} Queue~6 is dominated by $M/M/1(\lambda_6,\lambda_5)$ as in
\Cref{lem:fish-class6}. When $Q_3>0$, both class-1 and class-2 arrivals match~3
(edges $\{1,3\},\{2,3\}$ are their top priority under $\Phi^+_0$), so $Q_3$ has
up-rate at most $\lambda_3=3$ and down-rate at least $\lambda_1+\lambda_2=8$
regardless of the queues $Q_1,Q_2$, whence $Q_3\le_{\mathrm{st}}Y_3$ for $Y_3\sim M/M/1(3,8)$. Queue~5 lives on
$\{Q_4=Q_6=0\}$, where its up-rate is $\lambda_5=3$ and its down-rate is
$\lambda_6+\lambda_4\mathbf 1\{Q_3=0\}=2+2\cdot\mathbf 1\{Q_3=0\}$ (class~6 always
matches~5, and class~4 matches~5 when $Q_3=0$). Since $\{Y_3=0\}\subseteq\{Q_3=0\}$,
$Q_5$ is dominated by the queue of arrival rate~$3$ and service rate
$2+2\cdot\mathbf 1\{Y_3=0\}$ modulated by the autonomous~$Y_3$; as
$P(Y_3=0)=1-\tfrac38=\tfrac58$, its arrival rate~$3$ is below the mean service rate
$2+2\cdot\tfrac58=\tfrac{13}4$, so it is positive recurrent (again
by~\citealp{loynes1962}) and $Q_5$ is tight. None of these three queues feeds back
into~$b$ or~$Q_4$.

\emph{Queues $1,2$.} On $\{b\ge1\}$ an unmatched class-1 or class-2 arrival
raises~$b$ (rate $4$), a class-1 or class-2 arrival matched to the other class
lowers it (rate $4$), and class~3 clears one of $1,2$ exactly when $Q_4=0$ (rate
$\lambda_3=3$); the net drift of~$b$ is $-3\cdot\mathbf 1\{Q_4=0\}$. Since
$\{Y=0\}\subseteq\{Q_4=0\}$, the extra service is present whenever $Y=0$; couple
$b$ with $\tilde b$ of up-rate~$4$ and down-rate $4+3\cdot\mathbf 1\{Y=0\}$ over the
autonomous chain~$Y$, so that $b(t)\le\tilde b(t)$. The pair $(\tilde b,Y)$ admits
the \emph{exact} Lyapunov function $V=\tilde b+Y$: the linear function $g(Y)=Y$
solves the phase Poisson equation, $\mathcal A_Y\,Y=\lambda_4-\lambda_3\mathbf
1\{Y\ge1\}=-1+3\cdot\mathbf 1\{Y=0\}$, so
\[
	\mathcal A(\tilde b+Y)=-3\cdot\mathbf 1\{Y=0\}
	+\bigl(-1+3\cdot\mathbf 1\{Y=0\}\bigr)=-1\qquad\text{on }\{\tilde b\ge1\}
\]
(while on the boundary $\{\tilde b=0\}$ the drift is $3+3\cdot\mathbf 1\{Y=0\}$).
This exact drift identifies $\tilde b$ as an $M/M/1$ queue of arrival rate~$4$ whose
service rate $4+3\cdot\mathbf 1\{Y=0\}$ is modulated by the autonomous,
positive-recurrent chain~$Y$: its arrival rate is strictly below the
time-stationary mean service rate $4+3\,P(Y=0)=5$, since $P(Y=0)=\tfrac13$, so
$\tilde b$ is positive recurrent by the stability criterion for a queue in a
stationary ergodic environment \citep{loynes1962}. (The additive drift alone does
not close this, as the drift is positive on the infinite boundary $\{\tilde b=0\}$;
what makes $\tilde b$ stable is the negative \emph{mean} drift the identity
exhibits.) Hence $b\le_{\mathrm{st}}\tilde b$ is tight.

\emph{Combining.} Building all five dominators (namely $Y$ for $Q_4$, the
$M/M/1(2,3)$ for $Q_6$, the $M/M/1(3,8)$ for $Q_3$, the nested chain for $Q_5$, and
$\tilde b$ for $b\ge Q_1,Q_2$) on the same arrival clocks as~$Q$ gives the pathwise
bound $Q_i(t)\le D_i(t)$ for each coordinate. Each dominator is positive recurrent,
hence tight, so the marginals $Q_i(t)$, and therefore $Q(t)$, are tight uniformly
in~$t$. The chain has constant total jump rate $\sum_i\lambda_i=18$, so it is
non-explosive, and it is irreducible on the communicating class of the empty state;
a non-explosive irreducible chain on a countable state space whose time-marginals
are tight is positive recurrent, since a transient or null-recurrent chain would
spend an asymptotically vanishing fraction of time in every finite set,
contradicting tightness \citep{srikantbook}. Hence $\Phi^+_0$, and with it every
$\Phi^+_k$, is positive recurrent.
\end{proof}

We now turn to the tail bound. Its heart is a lower bound, uniform in~$k$, on the
mean length of a busy period of class~4 started from a state with a moderate total
backlog.

\begin{lem}[Uniform first-passage bound]\label{lem:fish-firstpassage}
Let $Q(\cdot)$ evolve under $\Phi^+_k$ from a state~$x$ with $Q_4=1$ and $N(x)\le
k/2$, and let $T_+$ be the first time at which $Q_4$ returns to~$0$. There is a
constant $c>0$, independent of~$k$, such that
$\mathbb{E}_x\!\left[T_+\right]\ge c\,k$ for every $k\ge51$ and every such starting
state.
\end{lem}
\begin{proof}
While $Q(\cdot)$ stays in the band $B_k=\{Q_4\ge1,\ N\le k-1\}$ the policy acts as
$\Phi^+_\infty$; there $Q_3=Q_5=0$, $Q_1Q_2=0$, and, writing $b=Q_1+Q_2$, $Q_4$ is
a skip-free birth-death process with up-rate~$2$ and down-rate
$d(b,Q_6)=3\,\mathbf 1\{b=0\}+3\,\mathbf 1\{Q_6=0\}$ modulated by the autonomous,
$k$-independent birth-death chains $b$ and $Q_6$, no arrival changing two of
$Q_4,b,Q_6$.

Set $\Phi=Q_4-c_1\rho^{\,b}-c_2\rho^{\,Q_6}$ with $\rho=0.95$, $c_1=5.4$,
$c_2=19.6$, and write $x_0=c_1(1-\rho)=0.27$, $y_0=c_2(1-\rho)=0.98$,
$A_0=7-4\rho=3.2$, $B_0=3-2\rho=1.1$. Since no arrival moves two of $Q_4,b,Q_6$, the
generator splits over the three coordinates, and on the four regions of the band
\[
  \mathcal A\Phi=
  \begin{cases}
    2-x_0A_0\rho^{\,b-1}-y_0B_0\rho^{\,Q_6-1}, & b\ge1,\ Q_6\ge1,\\
    -1+8x_0-y_0B_0\rho^{\,Q_6-1},          & b=0,\ Q_6\ge1,\\
    -1+2y_0-x_0A_0\rho^{\,b-1},            & b\ge1,\ Q_6=0,\\
    -4+8x_0+2y_0,                        & b=0,\ Q_6=0.
  \end{cases}
\]
Each case increases in $b$ and in $Q_6$ (as $\rho<1$), so its minimum sits at the
corner $b=Q_6=1$ (resp.\ $b=0$ or $Q_6=0$), giving the four values $0.058$, $0.082$,
$0.096$, $0.12$; hence
\[
  \mathcal A\Phi\ge\delta:=0.058>0\qquad\text{on }B_k,
\]
independently of~$k$ and of the magnitude of~$Q_4$. A single transition changes
$\Phi$ by at most~$1$ in absolute value, and the total transition rate is
$\sum_i\lambda_i=18$ at every state. Let $V=e^{-\theta\Phi}$ with $\theta=0.006$.
From $e^{-u}\le1-u+\tfrac12u^2e^{|u|}$ applied to $u=\theta\,\Delta_i\Phi$,
\[
  \mathcal A V=V\sum_i\lambda_i\bigl(e^{-\theta\,\Delta_i\Phi}-1\bigr)
  \le V\bigl(-\theta\delta+9\theta^2e^{\theta}\bigr)<0,
\]
because $9\theta e^{\theta}=0.0543<0.058=\delta$; so $V$ is a bounded nonnegative
supermartingale on the band.

Let $\tau_0$ be the first time $Q_4=0$ and $\tau_{\mathrm{hi}}$ the first time
$N=k$; on $[0,\tau_0\wedge\tau_{\mathrm{hi}})$ the chain stays in~$B_k$. On
$\{Q_4=0\}$ we have $\Phi\in[-(c_1+c_2),0]=[-25,0]$, hence $V\ge1$; at any state
with $Q_4=26$ (a level free of~$k$, chosen above $c_1+c_2=25$) we have $\Phi\ge1$
for every phase, so $V\le e^{-\theta}$. From $Q_4=1$, twenty-five successive
class-4 arrivals raise $Q_4$ to~$26$ without changing $b$ or $Q_6$, hence raising
$N$ by~$25$; since $N(x)\le k/2$ and $25<k/2$ for $k\ge51$, these steps remain in
the band, and they occur with probability at least $(\lambda_4/18)^{25}=(1/9)^{25}$.
Optional stopping of~$V$ at $\tau_0\wedge\tau_{\mathrm{hi}}$ from $Q_4=26$ gives
$P(\tau_0<\tau_{\mathrm{hi}}\mid Q_4=26)\le e^{-\theta}$, so
\[
  q:=P\bigl(\tau_{\mathrm{hi}}<\tau_0\bigr)\ge(1/9)^{25}\bigl(1-e^{-\theta}\bigr)>0,
\]
a bound independent of~$k$ and of the phase.

Finally $N$ is skip-free: every event changes $N$ by exactly~$\pm1$, an arrival
either matching (one waiting item leaves) or queueing (one item joins). Hence
reaching $N=k$ from $N(x)\le k/2$ requires at least $k/2$ upward steps of~$N$, so at
least $k/2$ transitions. On $\{\tau_{\mathrm{hi}}<\tau_0\}$, of probability at
least~$q$, the number of transitions before $T_+$ is therefore at least~$k/2$;
since the total rate is~$18$ at every state, the holding times are i.i.d.\
$\mathrm{Exp}(18)$ and, by Wald's identity,
$\mathbb{E}_x\!\left[T_+\right]\ge q\,(k/2)/18=(q/36)\,k$. Taking $c=q/36$ gives the
claim for every $k\ge51$.
\end{proof}

\begin{prop}[Tail of class~4]\label{prop:fish-tail}
$P(Q_{k,4}=0)=O(1/k)$; in particular $P(Q_{k,4}=0)\to0$ as $k\to\infty$.
\end{prop}
\begin{proof}
Fix~$k$; by \Cref{prop:fish-stability} there is a stationary regime. The visits
of~$Q_4$ to~$0$ split the trajectory into idle sojourns (in $\{Q_4=0\}$) and busy
excursions (in $\{Q_4\ge1\}$), so, writing $p_k=P(Q_{k,4}=0)$ and letting $T_0,T_+$
be the sojourn lengths under the Palm measure of these visits,
\[
	p_k=\frac{\mathbb{E}_{\mathrm{Palm}}\!\left[T_0\right]}
	{\mathbb{E}_{\mathrm{Palm}}\!\left[T_0\right]+\mathbb{E}_{\mathrm{Palm}}\!\left[T_+\right]}
	\le\frac{\mathbb{E}_{\mathrm{Palm}}\!\left[T_0\right]}
	{\mathbb{E}_{\mathrm{Palm}}\!\left[T_+\right]}.
\]
Both crossings are clean: a down-crossing $Q_4\colon1\to0$ leaves $Q_3=Q_5=0$, and
the matching up-crossing is a class-4 arrival that queues, hence also finds
$Q_3=Q_5=0$.

\emph{Idle bound.} On $\{Q_4=0\}$ class~3 cannot match the absent class~4, so its
only threshold-sensitive choice is moot and the generator restricted to $\{Q_4=0\}$
is the \emph{same in both modes}; the boundary dynamics are thus free of~$k$. From
any clean state ($Q_3=Q_5=0$) the next of the class-3, class-4, class-5 arrivals is
a class-4 arrival with probability $\lambda_4/(\lambda_3+\lambda_4+\lambda_5)
=2/8=1/4$, and on that event the class-4 arrival occurs while the state is still
clean and is therefore an up-crossing; between clean states, $Q_3$ and $Q_5$ return
to~$0$ in $k$-free bounded mean time, being dominated by the $k$-independent chains
of \Cref{prop:fish-stability}. Hence $T_0$ is bounded above, uniformly in the entry
state and in~$k$, by the first-passage time to an up-crossing in this fixed
boundary chain, so $\mathbb{E}_{\mathrm{Palm}}\!\left[T_0\right]\le C_0<\infty$ with
$C_0$ free of~$k$. As the up-crossing rate equals $\lambda_4\,P(Q_4=0,\,Q_3=Q_5=0)$,
the renewal-reward identity also reads
\begin{equation}\label{eq:fish-idle}
	\mathbb{E}_{\mathrm{Palm}}\!\left[T_0\right]
	=\frac{1}{2\,P(Q_3=Q_5=0\mid Q_4=0)}\le C_0,
	\qquad\text{so}\qquad
	P(Q_3=Q_5=0\mid Q_4=0)\ge\frac{1}{2C_0}.
\end{equation}

\emph{Escape bound.} At an up-crossing $Q_4=1$ with $Q_3=Q_5=0$, so $N=1+b+Q_6$.
Under the stationary law $b\le_{\mathrm{st}}\tilde b$ and $Q_6\le_{\mathrm{st}}
M/M/1(2,3)$ have exponential tails uniform in~$k$
(\Cref{prop:fish-stability,lem:fish-class6}), so there are $k$-free constants
$M<\infty$ and $z>1$ with
$P(b+Q_6\ge m)\le M z^{-m}$ for every~$m$. Let $A=\{Q_4=Q_3=Q_5=0\}$ and let
$q_k=P_{\mathrm{Palm}}(N>k/2)$ be the Palm probability that an up-crossing carries a
large backlog. Because up-crossings occur at the constant rate $\lambda_4$
throughout~$A$, the Palm entry law is the stationary law conditioned on~$A$; since
$N=1+b+Q_6$ at an up-crossing, and using
$P(A)=p_k\,P(Q_3=Q_5=0\mid Q_4=0)\ge p_k/(2C_0)$ from~\eqref{eq:fish-idle},
\[
	q_k=\frac{P\bigl(A\cap\{b+Q_6> k/2-1\}\bigr)}{P(A)}
	\le\frac{P(b+Q_6\ge k/2-1)}{P(A)}
	\le\frac{2C_0 M z}{p_k}\,z^{-k/2}.
\]
By \Cref{lem:fish-firstpassage} an up-crossing with $N\le k/2$ has mean busy period
at least $c\,k$, so
\[
	\mathbb{E}_{\mathrm{Palm}}\!\left[T_+\right]\ge c\,k\,(1-q_k)
	\ge c\,k-\frac{2cC_0 M z}{p_k}\,k\,z^{-k/2}.
\]

\emph{Combining.} Substituting into $p_k\le\mathbb{E}_{\mathrm{Palm}}\!\left[T_0\right]/
\mathbb{E}_{\mathrm{Palm}}\!\left[T_+\right]\le C_0/
\mathbb{E}_{\mathrm{Palm}}\!\left[T_+\right]$ and rearranging,
\[
	c\,k\,p_k-2cC_0Mz\,k\,z^{-k/2}\le C_0,
	\qquad\text{i.e.}\qquad
	p_k\le\frac{C_0\bigl(1+2cMz\,k\,z^{-k/2}\bigr)}{c\,k}.
\]
As $k\,z^{-k/2}\to0$, the numerator tends to~$C_0$, and $p_k=O(1/k)\to0$.
\end{proof}

Combining \Cref{prop:fish-reduction,prop:fish-stability,prop:fish-tail} proves
\Cref{prop:greedy-fish}: every $\Phi^+_k$ is stable and $P(Q_{k,4}=0)\to0$, so
$\alpha(\Phi^+_k)\to\tfrac12$; the $4\leftrightarrow6$ symmetry gives
$\alpha(\Phi^-_k)\to-\tfrac12$; and convexity of $\La_\Gre$ then yields
$\La_\Gre=(-\tfrac12,\tfrac12)=\Lap$.

\medskip\noindent\emph{On the choice of threshold.}
The result does not depend on the exact trigger: any device that keeps $Q_4$
positive with growing probability while preserving greediness would do. We use a
threshold on the total queue length $N$ rather than on $Q_4$ for one reason, made
explicit in \Cref{prop:fish-stability}: it confines the disagreement with the base
policy $\Phi^+_0$ to the finite set $\{N<k\}$, which turns stability into a finite
perturbation of a single stable policy. A threshold on $Q_4$ produces a numerically
indistinguishable trade-off (see below), but its low-mode region $\{Q_4<k\}$ is
infinite and its class-4 queue is not autonomous, so the same short proof of
stability does not apply.

\paragraph*{Numerical results.}
We corroborate the theorem by exact computation of the stationary distribution,
tractable because greediness confines the support to the independent sets of the
Fish. For the family $(\Phi^+_k)$ the exact solve confirms, over the range $0\le
k\le20$, that $\alpha(\Phi^+_k)$ increases towards~$\tfrac12$ (from $0.41$ at $k=0$
to $0.48$ at $k=20$) while $P(Q_{k,4}=0)$ decreases (from $0.73$ to $0.20$) and
$P(Q_{k,6}=0)$ decreases towards~$\tfrac13$ (from $0.68$ to $0.44$), in line with
\Cref{lem:fish-class6}; the drift of $Q_4+Q_6$ matches \Cref{lem:fish-Sdrift} at
every reachable state, and the bound $\mathcal A\Phi\ge0.058$ of
\Cref{lem:fish-firstpassage} holds throughout the low-mode band. No linear Lyapunov
function certifies the stability of \Cref{prop:fish-stability}: the drift in the
$Q_1,Q_2$ direction vanishes in the high mode, which is why the argument relies on
stochastic domination and the exact nonlinear corrector rather than a single linear
test.

\Cref{perf-fish} shows the delay--regret trade-off. The sum-triggered chain becomes
too large to solve exactly at large~$k$, since its support grows in every
coordinate at once; the figure therefore uses the queue-4-triggered variant
$\widehat\Phi^+_k$, which switches mode when $Q_4\ge k$ rather than when $N\ge k$.
This variant keeps a bounded support in every coordinate except~$Q_4$, so it can be
solved exactly up to $k=128$, and wherever both can be computed it traces a
trade-off numerically indistinguishable from that of $\Phi^+_k$. Since the vertex
at $\alpha=\tfrac12$ is bijective, we include the delay of the reward-optimal policy
$\mles$ for comparison. The regret vanishes as $k\to\infty$, at the cost of a
growing class-4 queue and hence a growing delay, exactly as \Cref{prop:greedy-fish}
predicts.

\begin{figure}[!htb]
	\centering
	 % figures/fish
\begin{tikzpicture}
\definecolor{crimson2143940}{RGB}{214,39,40}
\definecolor{darkgray176}{RGB}{176,176,176}
\definecolor{darkorange25512714}{RGB}{255,127,14}
\definecolor{forestgreen4416044}{RGB}{44,160,44}
\definecolor{lightgray204}{RGB}{204,204,204}
\definecolor{steelblue31119180}{RGB}{31,119,180}

\begin{axis}[width=12cm,
legend cell align={left},
legend style={fill opacity=0.8, draw opacity=1, text opacity=1, draw=lightgray204},
log basis x={10},
log basis y={10},
tick align=outside,
tick pos=left,
x grid style={darkgray176},
xlabel={Delay},
xmin=0.260303984681561, xmax=150.684936867648,
xmode=log,
xtick style={color=black},
xtick={0.01,0.1,1,10,100,1000,10000},
xticklabels={
  \(\displaystyle {10^{-2}}\),
  \(\displaystyle {10^{-1}}\),
  \(\displaystyle {10^{0}}\),
  \(\displaystyle {10^{1}}\),
  \(\displaystyle {10^{2}}\),
  \(\displaystyle {10^{3}}\),
  \(\displaystyle {10^{4}}\)
},
y grid style={darkgray176},
ylabel={Regret},
ymin=1e-08, ymax=2,
ymode=log,
ytick style={color=black},
ytick={1e-10,1e-08,1e-06,0.0001,0.01,1,100,10000},
yticklabels={
  \(\displaystyle {10^{-10}}\),
  \(\displaystyle {10^{-8}}\),
  \(\displaystyle {10^{-6}}\),
  \(\displaystyle {10^{-4}}\),
  \(\displaystyle {10^{-2}}\),
  \(\displaystyle {10^{0}}\),
  \(\displaystyle {10^{2}}\),
  \(\displaystyle {10^{4}}\)
}
]
\addplot [kfilter]
table [row sep=\\] {%
0.353501639 0.267107938199993\\
0.355199268711111 0.261470116799993\\
0.375447997555556 0.213928117199992\\
0.473917173633333 0.126526314599993\\
0.736560342077778 0.0644321465999931\\
1.34942184636667 0.0259297307999924\\
2.82181587975556 0.00582730739999218\\
6.20693396427778 0.000380586599992161\\
13.2946383078222 1.88999999223136e-06\\
27.5165843837556 6.88338275267597e-15\\
55.9608561086444 6.55031584528842e-15\\
112.848973272356 7.66053886991358e-15\\
};
\addlegendentry{$\widehat\Phi^+_k$}
\addplot [filter]
table [row sep=\\] {%
0.431333101055556 1e-08\\
0.431333101055556 2\\
};
\addlegendentry{$\mles$}
\end{axis}

\end{tikzpicture}

	\caption{Approaching the vertex $\alpha=1/2$ in the Fish graph from \Cref{fig:fish} with the queue-4-triggered variant $\widehat\Phi^+_k$ ($r=(2, 2, 2, 1, -1, 0, 1)$); the reward-optimal policy $\mles$ is shown for comparison.\label{perf-fish}}
\end{figure}

\end{document}